# Design, Modeling And Fabrication Of Nanostructure Devices For Optoelectronic Applications

**Thesis submitted for the degree of**

**Doctor of Philosophy (Sc.)**

**In**

**Electronic Science**

**by**

**Subhrajit Sikdar**

**Department of Electronic Science**

**University of Calcutta**

**2021**

***Dedicated to my parents and younger sister***

# Abstract

The current research work encompasses design, modeling and fabrication of vertically aligned nanowire metal-oxide-semiconductor (MOS) based voltage tunable quantum dot (VTQD) devices for optoelectronic applications. A novel device scheme is proposed for developing such VTQDs by combining the effects of nanowire-geometry dependent structural confinement in transverse (radial) directions and the voltage-assisted surface quantization in longitudinal direction. The formation of VTQDs near oxide/semiconductor interface at room temperature is predicted by developing a self-consistent quantum-electrostatic simultaneous solver. Further, the electrostatic performance of such devices is studied in detail and it has been observed that the choice of 'high-$m^*$' plays the crucial role, instead of 'high-$k$', to maintain electrostatic integrity. It allows to maintain simultaneously the electrostatic control as well as the reduction of carrier tunneling probability. The photogeneration phenomenon and carrier transport in nanowire MOS based VTQD devices is analytically modeled by adopting non-equilibrium Green's (NEGF) formalism based on second quantization field operators for incident photons and generated photocarriers. Engineering the combined effects of structural (in the transverse directions) and electrical quantization (in the longitudinal direction) has enabled a route for developing wavelength selective direct color sensors with high spectral resolution. The developed NEGF-based analytical model is further extended to obtain photocurrent which can be utilized for harvesting solar energy. The resonance of incident photon modes with the energy gap between three dimensionally (3D) quantized conduction band electron states and two dimensionally (2D) quantized valence band hole sub-bands has provided a route for achieving very high power conversion efficiency beyond the Shockley-Queisser limit. A design window is also proposed for obtaining the desired values of solar cell performance parameters with the optimized device parameters such as, the nanowire diameter and oxide thickness. Finally, a patterned array of such nanowire MOS based VTQD devices is fabricated by employing electron beam lithography (EBL) where the EBL-process optimization is extensively studied for achieving the desired device dimensions. The step-like behavior in capacitance-voltage (C-V) characteristics of such devices, measured in-situ within the FESEM chamber, confirms the formation of VTQDs at room temperature. Such patterned nanowire MOS based VTQD devices can be utilized for the advanced photosensing and solar energy harvesting applications.

# Acknowledgement

First and foremost, I would like to acknowledge my Ph.D. supervisor Dr. Sanatan Chattopadhyay for his continuous support and involvement towards completing the proposed research work with significant scientific contributions. I am also very much motivated by his rational attitude in establishing and dealing the research problem with proper justifications developed only after a long and healthy discussion. The immense patience and passion he has in developing research laboratories with world class facilities always encourage young scholars to work in multiple domains. Such fabrication facilities create opportunity for the scholars to collaborate with different departments of various academic institutes.

I would like to take the opportunity to express my gratefulness to Dr. Basudev Nag Chowdhury, currently working as a post-doctoral fellow in our Laboratory, for having thoughtful discussions on the philosophical aspects of a research problem.

I am thankful to Dr. Subhadip Chakraborty, Chirantan Das and Rajib Saha, my research colleagues, for their immense support in maintaining appropriate composure towards dealing with the research problem. Apart from academic discussions, the brotherhood we share enrich me as a person.

My sincere thanks to the University Grants Commission and University of Calcutta for funding the fellowship. I would also like to thank the Department of Electronic Science, Center for Research in Nanoscience and Nanotechnology and Center for Excellence in Systems Biology and Biomedical Engineering for proving the infrastructural and research facilities for my research work.

I would like to express my gratitude to Dr. Anupam Karmakar (Dept. of Electronic Science), Prof. Ajay Ghosh (Dept. of Applied optics and Photonics), Prof. Sunanda Dhar (Dept. of Electronic Science), Prof. Abhijit Mallik (Dept. of Electronic Science) and Dr. Anirban Bhattacharya (Dept. of Radiophysics and Electrons) for fruitful discussions on several occasions.

I am thankful to my other Lab members, namely, Dr. Anindita Das, Dr. Mainak Palit, Dr. Avishek Das, Dr. Sulagna Chatterjee, Dr. Somdatta Paul, Dr. Jenifer Sultana, Subrata Mandal, Anuraag Mukherjee, Ananya Bhattacharya and Sandip Bhattacharya for every academic discussion we had throughout the years that helped me to develop as a researcher.

I am grateful to some of my friends, namely, Akant Sagar Sharma, Suvasis Das, Monali Sil, Ayanika De, Emona Datta, Saikat Manna and Chandan Nandi for their encouragement to pursue a research career.

Last but not the least, I am deeply thankful to my parents Mrs. Tapati Sikdar and Mr. Surajit Sikdar, my younger sister Somasree Sikdar and also my school friend Mahul Bhattacharyya for creating a comfort zone for me through optimistic discussions and standing by me in all aspects of life both mentally and financially.

# Contents

# Chapter 6

## Fabrication of the patterned Ge-nanowire array based voltage tunable quantum dot (VTQD) devices on p-Si substrate [96-131]

# List of Figures

# List of Tables

# List of symbols

# (Unless mentioned otherwise)

## Universal constants

| | |
|---|---|
| $\varepsilon_0$ | : Free space permittivity. |
| $c$ | : Speed of light in vacuum. |
| $e$ | : Electronic charge. |
| $\hbar$ | : Planck's constant. |

## Indices

| | |
|---|---|
| *i, j,…* | : Indices for active device. |
| *r, s,…* | : Indices for reservoir. |
| *b,b'…* | : Indices for substrate. |
| *NW* | : Values corresponding to nanowire. |
| *e-pht* | : Electron-photon. |
| *ISO* | : Isolated. |

## Device parameters

| | |
|---|---|
| $H_{3D}$ | : Device Hamiltonian. |
| $H_{ISO}^{C}$ | : Hamiltonian for conduction band in isolated condition. |
| $H_{ISO}^{v}$ | : Hamiltonians for valence band in isolated condition. |
| $H_{ISO}^{p-Si}$ | : Hamiltonian for substrate in isolated condition. |
| $E_n$ | : Energy eigenvalue. |
| $E_{n_T}$ | : Eigenvalue components in transverse directions. |
| $E_{n_L}$ | : Eigenvalue components in longitudinal direction. |
| $E_{sub}$ | : $E_{n_T} + E_{n_L}$ |
| $\lvert n_T \rangle$ | : Transverse sub-space. |
| $\lvert n_L \rangle$ | : Longitudinal sub-spaces. |
| $\varepsilon_r^v$ | : Valence sub-bands. |
| $\varepsilon_i^c$ | : Conduction band states. |
| $\tilde{c}$ | : Speed of light in material. |
| $\lambda$ | : Wavelength of light. |
| $\eta^{\lambda}$ | : Refractive index of material for corresponding wavelengths. |
| $\phi(x_k)$ | : Potential as a function of Cartesian coordinate space $x_k \equiv (x_1, x_2, x_3)$ |
| $\rho$ | : Carrier density. |
| $m_0$ | : Mass of free electron. |
| $m_{ij}^*$ | : Components of effective mass tensor in electronic mass unit. |
| z | : Direction towards the nanowire axis. |
| $U(z)$ | : Conduction band profile along the nanowire axis. |
| $n_{1D}(z)$ | : One dimensional carrier distribution along nanowire axis. |
| $n_{3D}$ | : Three dimensional carrier density. |
| $n_{1D}^{ph}$ | : Photogenerated carriers in conduction band. |
| $E_f$ | : Semiconductor Fermi-level. |
| $V_{app}$ | : Applied voltage. |
| $C$ | : Dark capacitance. |
| $C_{ph}$ | : Photo capacitance. |
| $C^c$ | : Capacitance due to conduction band electrons. |
| $C^v$ | : Capacitance due to valence band holes. |
| $l_{ext}$ or $l_Q$ | : Spatial extension of the quantum well along the nanowire axis. |
| $l_z$ | : Spatial extension of valence band. |
| $D_{NW}$ | : Nanowire diameter. |

$t_{OX}$ : Oxide thickness.

$k_{OX}$ : Oxide dielectric constant.

$T$ : Carrier tunneling probability.

$C_i^c$ : Electron annihilation operators in i[th] state of conduction band.

$V_s^v$ : Electron annihilation operators in s[th] state of the valence band.

$b_\alpha^{pht}$ : Annihilation operator for incident photon in $\alpha$ mode.

$\omega_\alpha$ : Angular frequency.

$n_{ik}^c(t,t')$ : Two-time correlation function for filled state of electrons in conduction band.

$n_{rs}^v(t,t')$ : Two-time correlation function for filled state of electrons in valence band.

$n_{bb'}^{p-Si}$ : Substrate correlation function.

$N_{\alpha\beta}^{ab-pht}$ : Two-time correlation function for absorbed photons.

$p_{ik}^c(t,t')$ : Two-time correlation function for empty state of electrons in conduction band.

$p_{rs}^v(t,t')$ : Two-time correlation function for empty state of electrons in valence band.

$N_{\alpha\beta}^{em-pht}$ : Two-time correlation function for emitted photons.

$\sum$ : Self energy function.

$\Sigma_D / \Sigma^{ij}_{p-Si}$ : Self energy due to coupling between nanowire and substrate.

$\sum_{hole}$ : Hole self energy function.

$\sum_{sc}^{In-pht}$ : Photon in-scattering function.

$\Sigma_{elec}^{in}$ : Electron in-scattering function.

$\Sigma_{hole}^{in}$ : Hole in-scattering function.

$\Gamma_{pht}^{out}$ : Photon out-scattering function.

$\beta g \beta$ : Strength of coupling.

$N_{pht}$ : Number of absorbed photons.

$n(E)$ : Number of electrons per unit energy in each discrete energy level.

$n_{elec}^{pht}$ : Photogenerated electrons per second.

$I_\alpha$ : Intensity of incident light.

$V_{ab}$ : Absorbing nanowire volume.

$\Lambda$ : Physical cross-section of nanowire.

$\tau / \tau_{ir}^{\alpha}$ : Interaction potential.

$|i\rangle$ : Initial state of electron before scattering.

$|f\rangle$ : Final state of electron after scattering.

$\vec{k}_r^{hole}$ : Hole momentum.

$\vec{k}_i^{elec}$ : Electron momentum.

$E_i$ : Energy eigenvalue for state $|i\rangle$

$H_f$ : Device Hamiltonian after scattering.

$V_{sc}$ : Perturbing potential.

$A$ : Vector potential.

$\hat{\vec{p}}$ : Momentum operator.

$A_{abs}$ : Absorption cross-section.

$\eta_{eff}$ : Absorption efficiency.

$D^C$ : Conduction band density of states.

$D^v$ : Valence band density of states.

$I_{Pht}^{hole}$ : Phtotocurrent.

$\langle t \rangle$ : Carrier lifetime.

$v$ : Carrier velocity.

$R$ : Leaky resistance.

$R_C$ : Contact resistance.

$C_R$ : Real capacitance.

$C_m$ : Measured capacitance.

$R_S^{'}$ : Series resistance.

$C_E$ : Series capacitance.

$C_{OX}$ : Oxide capacitance.

$C_D$ : Device capacitance.

$Y_{it}$ : Device admittance.

$C_c$ : Corrected capacitance.

$C_{ma}$ : Measured accumulation capacitance.

$G_{ma}$ : Measured accumulation conductance.

## Mathematical parameters

$\vartheta(t)$ : Step function.

$\Delta$ : Grid size.

$\delta_{ij}$ : Kronecker delta.

$\delta(t)$ : Delta function of 't'.

$[\ ]$ : Matrix.

$[I]$ : Unity matrix.

$G^D / G^c_{ij}$ : Electron Green's function.

$G^{hole} / G^v_{rs}$ : Hole Green's function.

$G_b^{p-Si}$ : Substrate Green's function.

$G^{V\ Rtd}$ : Retarded Green's function.

$\mathrm{Re}[\ ]$ : Real part of matrix elements.

$\mathrm{Im}[\ ]$ : Real part of matrix elements.

## Abbreviations

FESEM : Field emission scanning electron microscopy.

TEM : Transmission electron microscopy.

AFM : Atomic force microscopy.

MBE : Molecular beam epitaxy.

MOCVD : Metal organic chemical vapor deposition.

| | |
|---|---|
| PECVD | : Plasma etched chemical vapor deposition. |
| RF | : Radio frequency. |
| VLS | : Vapor-liquid-solid. |
| ITRS | : International Technology Roadmap for semiconductors. |
| NWFET | : Nanowire field effect transistors. |
| GAA | : Gate all-around. |
| SS | : Sub-threshold slope. |
| STEG | : Solar thermoelectric generators. |
| EUV | : Extreme ultra-violet. |
| EBL | : Electron beam lithography. |
| ALD | : Atomic layer deposition. |
| CVD | : Chemical vapor deposition. |
| PVD | : Plasma vapor deposition. |
| MOS | : Metal-oxide-semiconductor. |
| VTQD | : Voltage tunable quantum dot. |
| NEGF | : Non-equilibrium Green's function. |
| SQ | : Shockley-Queisser. |
| C-V | : Capacitance-voltage. |
| LED | : Light emitting diodes. |
| EHP | : Electron-hole pair. |
| LPE | : Liquid phase epitaxy. |
| CW | : Continuous-wave. |
| VPE | : Vapor phase epitaxy. |
| VCSEL | : Vertical cavity surface emitting laser. |
| LDOS | : Local density of states. |
| MEG | : Multiple exciton generation. |
| EQE | : External quantum efficiency. |
| MIS | : Metal-insulator-semiconductor. |
| EOT | : Equivalent oxide thickness. |
| EBR | : Excitonic Bohr radius. |
| EMA | : Effective mass approximation. |
| CMOS | : Complementary metal-oxide-semiconductor. |
| CCD | : Charge-couple-device. |
| APS | : Active pixel sensors. |
| CFA | : Color filter array. |
| PSA | : Pixel sensor array. |
| ADC | : Analog to digital converter. |
| RGB | : Red-Green-Blue. |
| UV | : Ultraviolet. |
| NIR | : Near infrared spectrum. |
| FWHM | : Full-width-at-half-maximum. |
| PMT | : Photo-multiplier tube. |
| VIBGYOR | : Violet-indigo-blue-green-yellow-orange-red. |
| FF | : Fill factor. |
| RC | : Resistor-capacitor. |
| LH | : Light hole. |
| HH | : Heavy hole. |
| PCE | : Power conversion efficiency. |
| QIP | : Quantum information processing. |
| WF | : Write field. |
| TEM | : Transmission electron microscopy. |
| XRD | : X-ray diffraction. |
| SAED | : Selected area electron diffraction. |
| EDS | : Energy dispersive x-ray spectroscopy. |
| SCS | : Semiconductor characterization system. |
| NWMOS | : Nanowire MOS. |

# List of Publications

## A. Journal publications (Related to thesis)

1. **Subhrajit Sikdar**, Basudev Nag Chowdhury, Rajib Saha, Sanatan Chattopadhyay, *Voltage-tunable quantum-dot array by patterned Ge-nanowire-based metal-oxide-semiconductor devices*, **Physical Review Applied** 05/2021, 15, 054060.

2. **Subhrajit Sikdar**, Basudev Nag Chowdhury, Sanatan Chattopadhyay, *Design and modeling of high efficiency GaAs-nanowire MOS solar cells beyond Shockley-Queisser limit: An NEGF approach*, **Physical Review Applied** 02/2021, 15, 024055.

3. **Subhrajit Sikdar**, Basudev Nag Chowdhury, Sanatan Chattopadhyay, *Understanding the electrostatics of top-electrode vertical quantized Si nanowire metal–insulator–semiconductor (MIS) structures for future nanoelectronic applications*, **Journal of Computational Electronics** 03/2019, 18, 465.

4. **Subhrajit Sikdar**, Basudev Nag Chowdhury, Ajay Ghosh, Sanatan Chattopadhyay, *Analytical modeling to design the vertically aligned Si-nanowire metal-oxide-semiconductor photosensors for direct color sensing with high spectral resolution*, **Physica E: Low-dimensional Systems and Nanostructures** 11/2016; 87, 44.

## B. Conference publications (Related to thesis)

1. **Subhrajit Sikdar**, Basudev Nag Chowdhury, Sanatan Chattopadhyay, *Investigating the performance of Si-nanowire based metal-oxide-semiconductor (NW-MOS) devices for solar energy harvesting applications*, XXth International Workshop on Physics of Semiconductor Devices: **IWPSD 2019**.

2. **Subhrajit Sikdar**, Basudev Nag Chowdhury, Sanatan Chattopadhyay, *Modeling of Si-nanowire based vertical metal-oxide-semiconductor device for solar cell applications by employing NEGF formalism*, International seminar cum research colloquium on MEMS based Sensors and Smart Nanostructured Devices: **MSSND 2019**.

# C. Book chapters (Related to thesis)

1. **<u>Subhrajit Sikdar</u>**, Basudev Nag Chowdhury, Sanatan Chattopadhyay, *Designing InP-nanowire based vertical metal-oxidesemiconductor capacitors for wavelength selective visible light sensing*, **The Physics of Semiconductor Devices**, 12/2017: chapter 146: pages 1-6; Springer Nature Switzerland AG 2018., ISBN: 978-3-319-97603-7., DOI:10.1007/978-3-319-97604-4_146.

2. **<u>Subhrajit Sikdar</u>**, Basudev Nag Chowdhury, Sanatan Chattopadhyay, *Analytical modeling of vertically oriented standalone Si-nanowire metal-oxide-semiconductor capacitors for wavelength selective near-infrared sensing applications*, **Advances in Optical Science and Engineering**, 09/2017: pages 173-179; , ISBN: 978-981-10-3907-2, DOI:10.1007/978-981-10-3908-9_20.

# D. Journal publications (Not included in thesis)

1. Rajib Saha, **<u>Subhrajit Sikdar</u>**, Basudev Nag Chowdhury, Anupam Karmakar, Sanatan Chattopadhyay, *Catalyst-modified vapor-liquid-solid (VLS) growth of single crystalline β-Gallium Oxide ($Ga_2O_3$) thin film on Si-Substrate*, **Superlattices and Microstructures** 10/2019; 136:106316.

2. Sulagna Chatterjee, **<u>Subhrajit Sikdar</u>**, Basudev Nag Chowdhury, Sanatan Chattopadhyay, *Investigation of the performance of strain-engineered silicon nanowire field effect transistors (ε-Si-NWFET) on IOS substrates*, **Journal of Applied Physics** 02/2019; 125(8):082506.

3. Sarmista Sengupta, **<u>Subhrajit Sikdar</u>**, Soumya Pandit, *Substrate bias effect of epitaxial delta doped channel MOS transistor for low power applications*, **International Journal of Electronics** 05/2016; 104(1).

4. Anannya Bhattacharya, Jenifar Sultana, **<u>Subhrajit Sikdar</u>**, Rajib Saha, Sanatan Chattopadhyay, *Investigating the impact of thermal annealing on the photovoltaic performance of chemical bath deposited $SnO_2$/p-Si heterojunction solar cells*, **Microsystem Technologies** 10/2019.

5. Mainak Palit, Basudev Nag Chowdhury, **Subhrajit Sikdar**, Krishnendu Sarkar, Pallab Banerji, Sanatan Chattopadhyay, *Band splitting induced by momentum-quantization in semiconductor nanostructures: Observation of emission lines in Indium Phosphide (InP) nanotubes*, **Physics Letters A** 2/2021**;** 388:127056.

6. Anindita Das, Basudev Nag Chowdhury, Rajib Saha, **Subhrajit Sikdar**, Jenifar Sultana, Goutam Kumar Dalapati, Sanatan Chattopadhyay, *Formation of high-pressure phase of Titanium Dioxide ($TiO_2$-II) thin films by vapor-liquid-solid growth process on GaAs substrate*, **Physica status solidi** 11/2018.

7. Anindita Das, Basudev Nag Chowdhury, Rajib Saha, **Subhrajit Sikdar**, Satyaban Bhunia, Sanatan Chattopadhyay, *Ultrathin vapor–liquid–solid grown Titanium Dioxide-II film on bulk GaAs substrates for advanced metal–oxide–semiconductor device applications*, **IEEE Transactions on Electron Devices**, 04/2018; 65(4):1466-1472.

8. Chirantan Singha, Sayantani Sen, Alakananda Das, Anirban Saha, **Subhrajit Sikdar**, Pallabi Pramanik, Anirban Bhattacharyya, *Spontaneous growth of III-nitride 1D and 0D nanostructures on to vertical nanorod arrays*, **Materials Research Express** 09/2019; 6(10):1050b2.

9. Jenifar Sultana, Somdatta Paul, Rajib Saha, **Subhrajit Sikdar**, Anupam Karmakar, Sanatan Chattopadhyay, *Optical and electronic properties of chemical bath deposited p-CuO and n-ZnO nanowires on silicon substrates: p-CuO/n-ZnO nanowires solar cells with high open-circuit voltage and short circuit current*, **Thin Solid Films** 4/2020; 699:137861.

10. Chirantan Das, Basudev Nag Chowdhury, Subhadip Chakraborty, **Subhrajit Sikdar**, Rajib Saha, Anuraag Mukherjee, Anupam Karmakar, Sanatan Chattopadhyay, *A diagrammatic approach of impedimetric phase angle-modulus sensing for identification and quantification of various polar and nonpolar/ionic adulterants in milk*, **LWT** 01/2021; 136: 110347.

## E. Conference publications (Not included in thesis)

1. **Subhrajit Sikdar**, Basudev Nag Chowdhury, Sanatan Chattopadhyay, *Energy band-structure estimation of semiconductor nanotubes with consideration of momentum space quantization*,

IEEE International Symposium on Devices, Circuits and Systems: **ISDCS 2019**, 03/2018, DOI:10.1109/ISDCS.2018.8379630

2. Anannya Bhattacharya, Jenifar Sultana, **Subhrajit Sikdar**, Rajib Saha, Sanatan Chattopadhyay, *Investigating the chemical bath deposited n-$SnO_2$/p-Si heterojunction devices for optoelectronic applications*, 5th International Conference on Opto-Electronics & Applied Optics: **OPTRONIX-2019**, 03/2019, DOI:10.1109/OPTRONIX.2019.8862381

3. Subrata Mandal, Sraboni Dey, **Subhrajit Sikdar**, Basudev Nag Chowdhury, Rajib Saha, Anupam Karmakar, Sanatan Chattopadhyay, *Parameter Optimization in Electron Beam Lithography for fabricating Patterned Nanostructures*, XXth International Workshop on Physics of Semiconductor Devices: **IWPSD 2019**.

4. Anannya Bhattacharya, **Subhrajit Sikdar**, Susomon Dutta, Sreeparna Paul and Sanatan Chattopadhyay, *Growth of $ZnSnO_3$ nano-crystalloids on Si-substrate by employing chemical bath deposition (CBD) technique for self-powered UV-light sensing applications*, International Symposium on Devices, Circuits and Systems: **ISDCS 2020**.

5. Subrata Mandal, **Subhrajit Sikdar**, Rajib Saha, Anupam Karmakar and Sanatan Chattopadhyay, *Investigating the impact of growth time on the electrical performance of vapour-liquid-solid (VLS) grown Ge/n-Si hetero-junction*, International Symposium on Devices, Circuits and Systems: **ISDCS 2020**.

6. Sandip Bhattacharya, Rajib Saha, **Subhrajit Sikdar**, Subrata Mandal, Chirantan Das and Sanatan Chattopadhyay, *Investigation of density and alignment of ZnO-nanowires grown by double-step chemical bath deposition (CBD/CBD) technique on metallic, insulating and semiconducting substrates*, International Symposium on Devices, Circuits and Systems: **ISDCS 2020**.

# Chapter: 1

# Introduction: Semiconductor optoelectronic devices

## 1.1. CMOS technology and its impact on the optoelectronic industry

The past century has witnessed several breakthroughs and innovations in the domain of semiconductor materials and devices which have played a pivotal role for the sustained progress and reshaping of modern civilization. The transistors, since their first successful implementation in 1947 **[1]**, have been the workhorse of semiconductor industries to develop the devices with high speed and low operating power that actually revolutionized the entire world of electronics. Such progress of the electronic systems achieved tremendous momentum after the inception of metal-oxide-semiconductor field effect transistors (MOSFETs) and the development of first planar integrated circuit in 1959 **[2-4]**. The invention of complementary-metal-oxide-semiconductor (CMOS) technology in 1963 with reduced power consumption dramatically increased the importance of using MOSFETs for developing advanced circuits and systems **[5]**. The twin well CMOS technology developed by Hitachi Central Laboratories in 1978 established the superiority of such CMOS devices to fabricate high speed - low power microprocessors **[6]**. Consequently, this success of CMOS technology was gradually expanded and implemented in all branches of electronic circuits and systems. Such devices have created further research interests due to their underlying fascinating physics and immeasurable practical applications.

Accordingly, the global consumer market of electronics is dominated by CMOS based electronic and optoelectronic devices. For instance, it constituted a mammoth sharing of U$ 729.11 billion in 2019 which has reached to a value of U$ 1 trillion by the following year. The optoelectronics market alone has been projected to grow from U$ 41.4 billion in 2020 to U$ 52.7 billion by 2025. The steady growth of such optoelectronics market is primarily driven by the ever increasing demand for various optical sensing and imaging devices for telecommunication, healthcare, residential, military and aerospace purposes **[7-15]**. For instance, the CMOS image sensors have been successfully utilized by smartphone industries to develop high quality mobile cameras with more and more functionalities. All the modern day high resolution, high speed and high end digital cameras are based on such CMOS imagers. Further, the photovoltaic cells are monolithically integrated in CMOS microchips to develop compact wireless autonomous microsystems **[16-18]**. The CMOS technology has also been utilized to fabricate on-chip light emitting devices with significantly high yield **[19, 20]**. Thus, the complementary MOS technology has provided a common platform to integrate electronic and optical properties together for the development of advanced optoelectronic integrated circuits (OEICs) and systems.

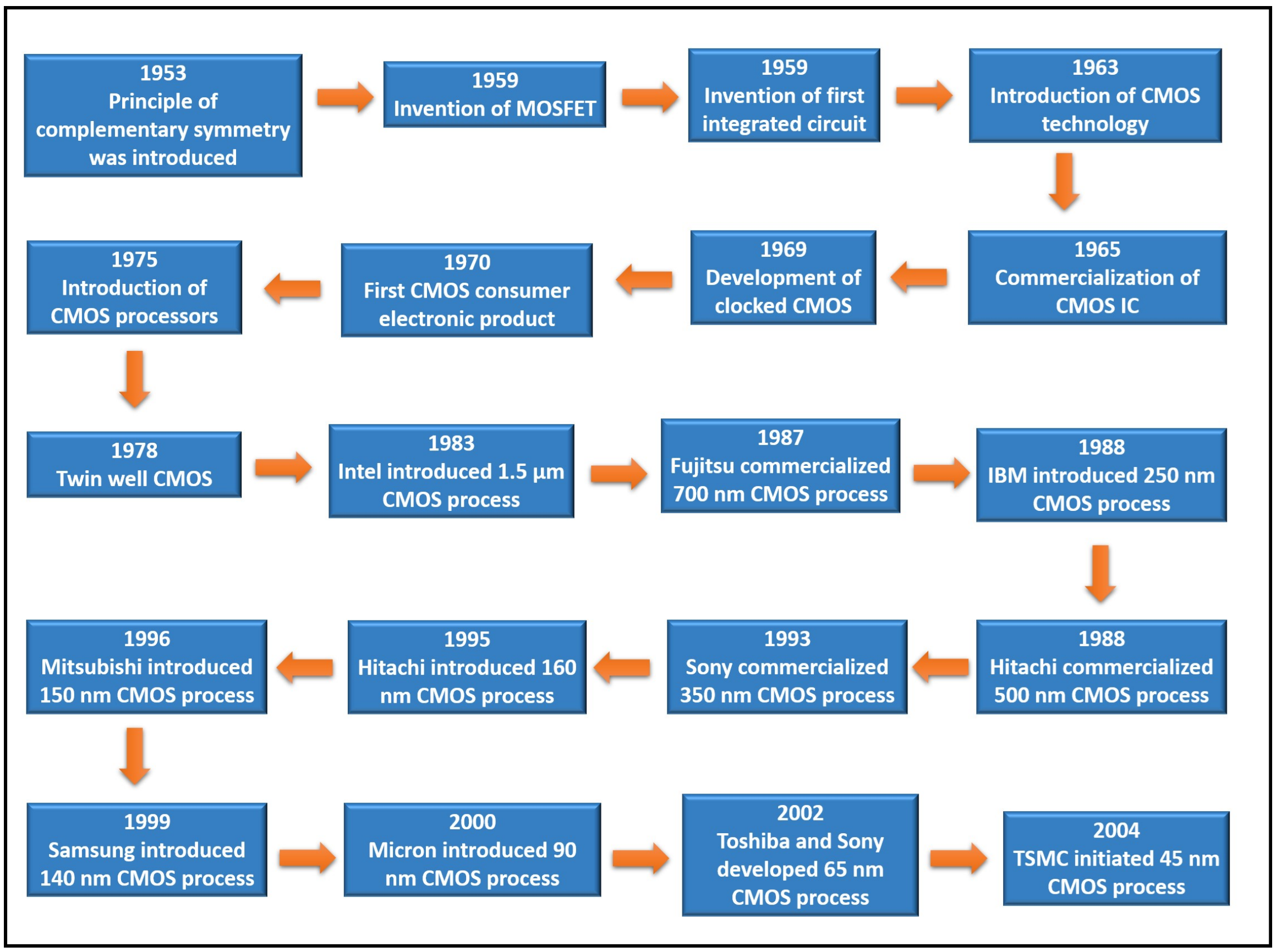

Fig. 1.1. Chronological progress of the complementary-metal-oxide-semiconductor (CMOS) technology with gradual down scaling of device dimensions.

## 1.2. Emergence of 'Nano': Route for performance improvement of electronic and optoelectronic devices

From the industrial point of view, it is profitable to predict the device dimensions for subsequent years, since the manufacturing process requires relevant technology to implement them in reality. In this context, Gordon E. Moore pointed out in 1965 that the number of components in an integrated circuit doubles every two years **[21]**. This very prediction has been the golden rule to define the trajectory of semiconductor technology by miniaturization of device dimensions till date. For instance, the present day Intel core-i7 processor is made of ~3 billion transistors by employing 14 nm CMOS process technology where the minimum feature size is reported to be 8 nm **[22, 23]**. The trend of such miniaturization of primary electronic components over the years is shown in Fig. 1.2.

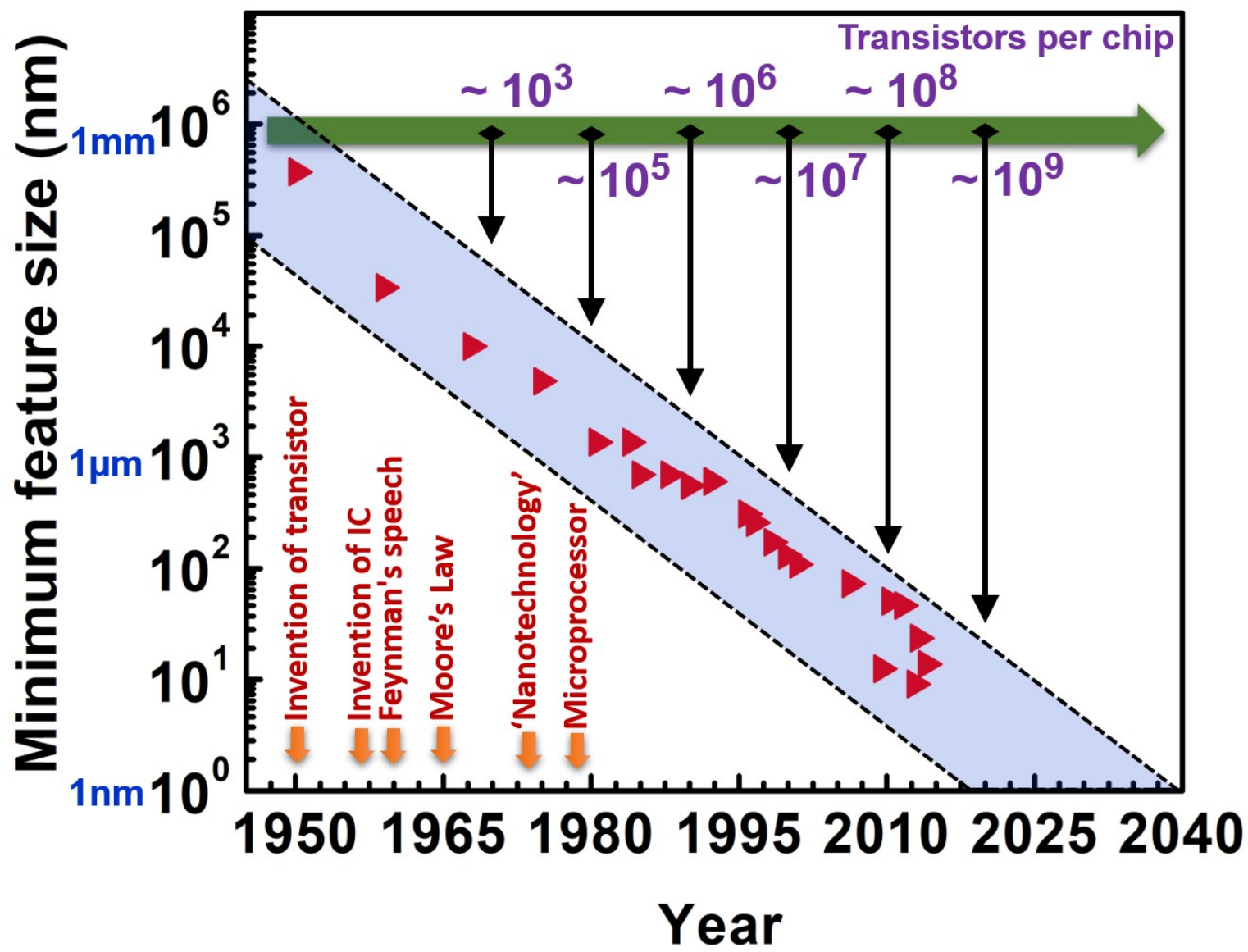

Fig. 1.2. Gradual miniaturization of minimum feature size of the complementary-metal-oxide-semiconductor (CMOS) technology since 1950.

The journey of sustained downscaling approach has led to the emergence of 'nanoscale' device dimensions. Although, Albert Einstein estimated the size of a sugar compound to be ~1 nm in 1905, and the first prototype of electron microscope at its primary level was discovered in 1935, the concept of 'nano' was formally introduced by Richard Feynman in 1959 during his famous lecture entitled "*There's Plenty of Room at the Bottom*" **[24]**. His hypothesis asking "*Why can't we write the entire 24 volumes of the Encyclopedia Britannica on the head of a pin*" has envisioned the construction of novel instruments for observing materials at atomic/molecular level. However, the term 'Nanotechnology' was used for the first time by the Japanese scientist Norio Taniguchi in 1974. The milestones in timeline of the emergence and evolution of nanoscience and nanotechnology are represented in Fig. 1.3.

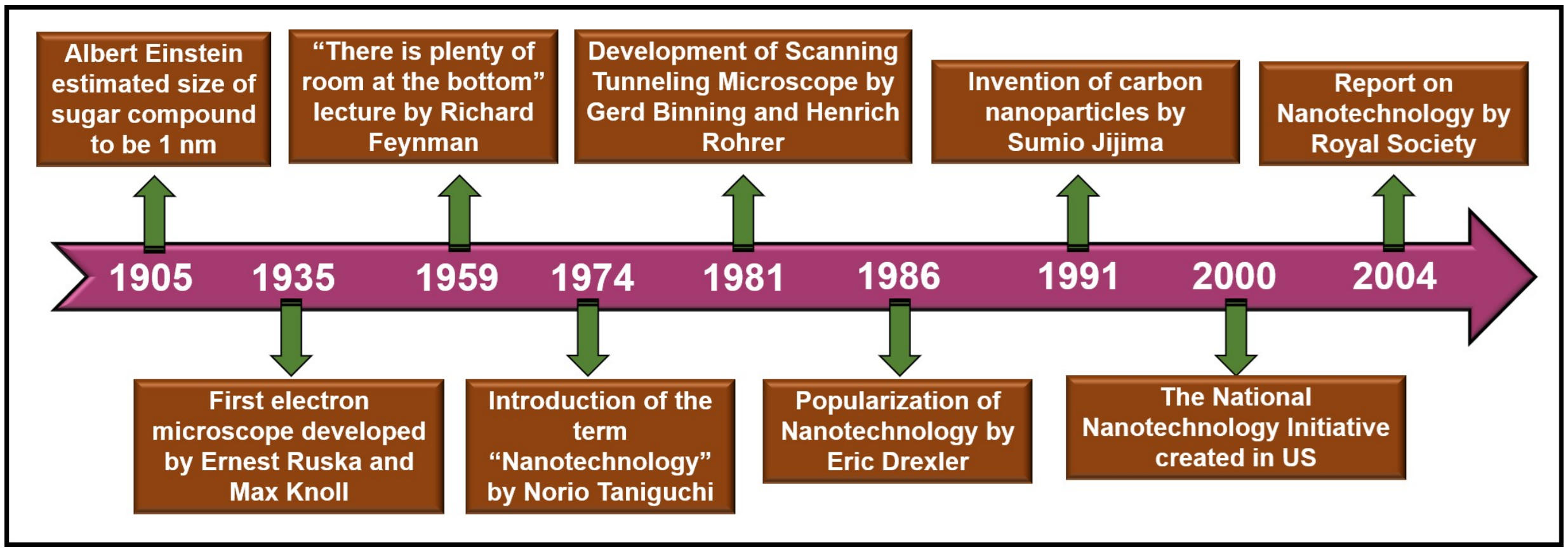

Fig. 1.3. Timeline showing the milestones in the history of nanoscience and nanotechnology development.

In past two decades, 'Nanoscience and Nanotechnology' has drawn tremendous attention of the global research community where 'Nanoscience' leads to achieve scientific breakthroughs and 'Nanotechnology' creates new opportunities to realize it in practice. Particularly, nanotechnology makes it possible to utilize novel phenomena originating at nano-domain by manipulating, assembling and controlling matter at nanometer scale. Primarily, the nano-domain has attracted significant research interest due to the novel properties of nanoscale materials in comparison to their bulk counterparts. The surface-to-volume ratio increases rapidly with shrinking the size of respective materials resulting to such unique material properties. Thus, understanding the underlying physics of nano-dimensional materials and their controlled manipulation can be exploited for the development of novel devices, circuits and systems with superior performance. In this context, 'Nanoelectronics' has emerged as the stream for connecting science, engineering and technology for the electronic devices with physical dimensions of 'nanoscale' order, typically <100 nm. However, the 'nanoscale' is defined by the order of device dimensions beyond which the physical, electronic and optoelectronic properties of the respective materials starts to modify in comparison to their bulk counterparts.

Indeed, the nature is fundamentally composed of structures and processes of nano-dimensions and, in that sense, several 'nanoscale' phenomena have inherently been utilized for centuries by the mankind. However, the invention of several imaging techniques including the transmission electron microscopy (TEM), field emission scanning electron microscopy (FESEM), scanning tunneling microscopy (STM), atomic force microscopy (AFM), and several growth or deposition techniques such as molecular beam epitaxy (MBE), metal organic chemical vapor deposition (MOCVD), plasma etched chemical vapor deposition (CVD) and radio frequency (RF) magnetron sputtering, have enabled researchers to understand and control such physical properties of nanomaterials.

The domain of optoelectronics has experienced a significant boost with the emergence of nano-materials/devices for advanced optoelectronic applications. Fig. 1.4 schematically represents the route for performance improvement of optoelectronic devices by utilizing the benefit of 'Nanotechnology'.

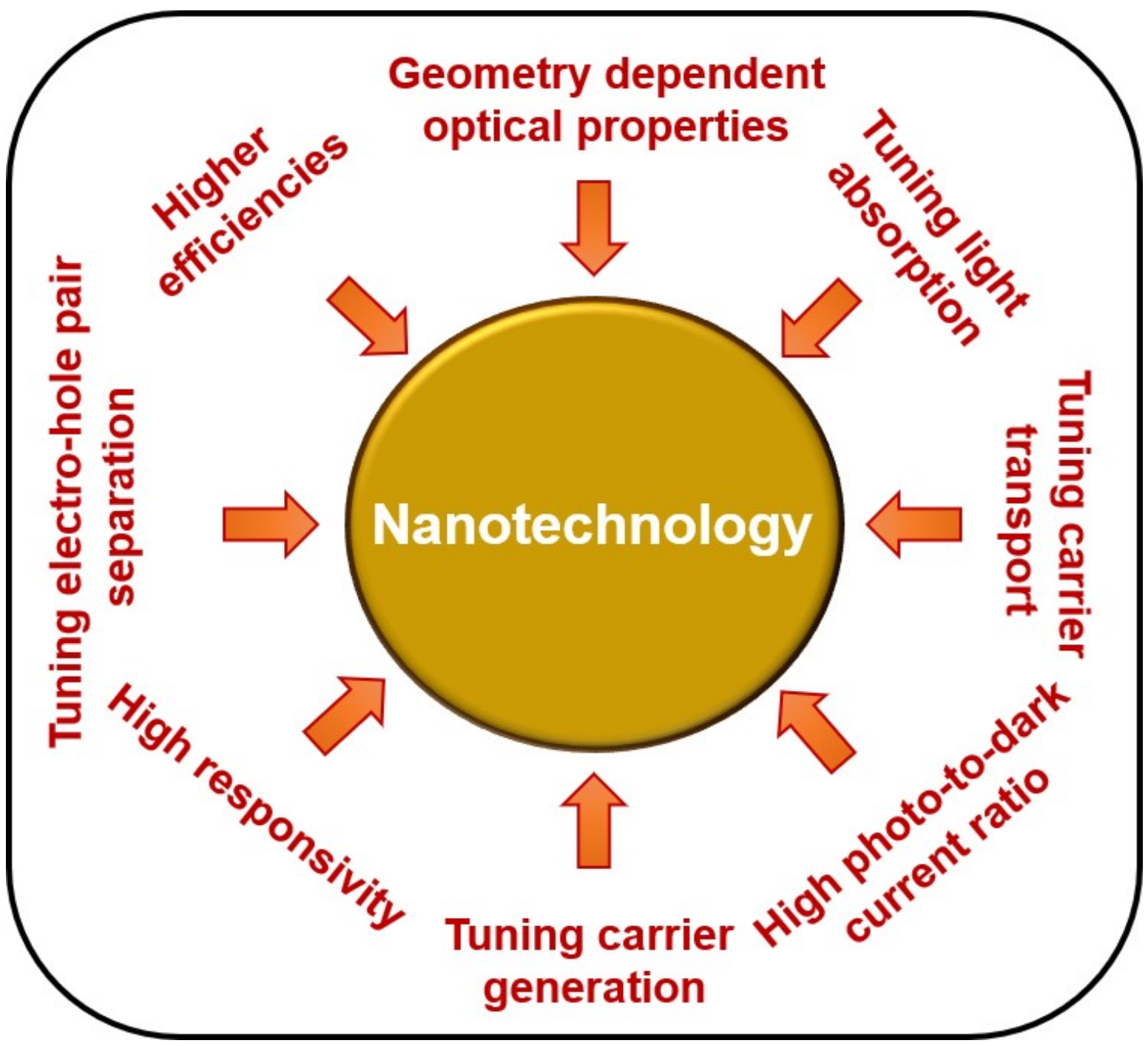

Fig. 1.4. Schematic representation of the route for performance improvement of optoelectronic devices using 'Nanotechnology'.

The geometry dependent tunable optical properties of such nanomaterials make it possible to engineer the light absorption, carrier generation, separation and transport mechanisms **[25-31]**. Such engineering approach provides a systematic route to conceptualize, design and fabricate novel nano-optoelectronic devices. In recent times, light management inside the nanostructures has been implemented to develop novel light trapping devices to achieve much higher efficiencies in comparison to the conventional bulk devices **[32-37]**. Such results indicate that engineering the device architecture and selection of relevant materials to trap light inside nanostructures can lead to achieve major scientific breakthroughs in near future. Thus, the novel nanoscale devices are expected to be the potential candidates for development of advanced optoelectronic systems with superior performance and higher functionalities.

## 1.3. Quantum dots and its optoelectronic applications: "The next big thing is really small"

Semiconductor quantum dot offers an additional route to engineer the optoelectronic properties of nanomaterials due to the presence of strong three dimensional (3D) quantum confinement. Such 3D-confinement creates discrete energy states which increases the effective energy band gap of respective materials. Novel optoelectronic properties of such quantum dots including widely tunable light absorption, multi-color detection, bright and narrowband light emission can be exploited to achieve device performance beyond the conventional limits. The wide spectral tunability in the range of near infrared (NIR) to visible makes quantum dot based devices the potential candidates for multi-color photosensing and imaging with high spectral responsivity. Such devices are also integrated in standard CMOS based cameras to improve the sensor performance **[38-41]**. Most importantly, the devices have relatively higher photo-to-dark current ratio and faster response time which makes them an inevitable choice for several applications such as, multispectral imaging, spectroscopy, aerospace instruments, health monitoring and thermal imaging **[42-51]**. Photovoltaic devices are also expected to experience a 6% to 12% increase of power conversion efficiency by exploiting the properties of quantum dots **[52]**. The presence of discrete energy states in such quantum dots provides an opportunity to generate multiple excitons at the cost of only a single photon which can lead to achieve power conversion efficiencies beyond the Shockley-Queisser limit **[52-54]**. The discrete energy states are also advantageous for light emitting devices (LEDs) with wavelengths of narrow linewidth which is indispensable for the development of QLED-displays **[52, 55-57]**. Further, the quantum dot based lasing devices have frequent applications in optical communications, surgical instruments, manufacturing, digital projection systems and on-chip interconnects **[52, 58-60]**. Therefore, semiconductor quantum dots can be projected for the development of next-generation optoelectronic devices, circuits and systems with enormous potential of scientific discoveries, even beyond the conventional limit. The optoelectronic applications of semiconductor quantum dots are schematically represented in Fig. 1.5.

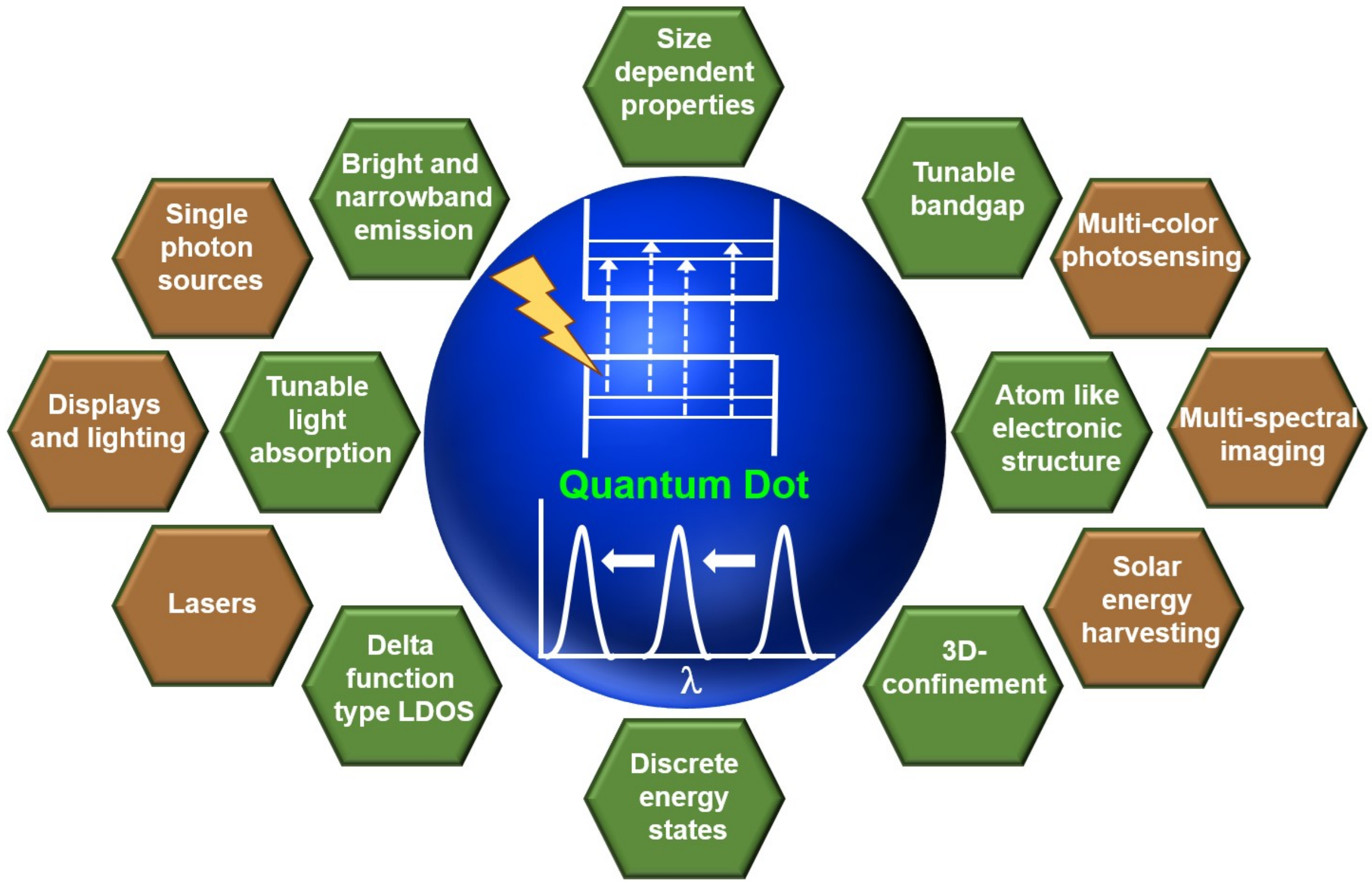

Fig. 1.5. Schematic representation for showing wide range of optoelectronic applications of the quantum dots.

Achieving precise control on the dimensions of such quantum dots requires special attention since it determines the degree of confinement. Several technologies are adopted to achieve appropriate control over the size of quantum dots. For instance, colloidal quantum dots can easily be synthesized with different sizes by following the chemical route **[61-63]**, however, patterned array of such colloidal quantum dots is difficult to arrange on a platform due to its randomness. On the other hand, molecular beam epitaxy (MBE), which is considered to be the most advanced growth technology, is efficient for controlled fabrication of quantum dots with desired size variation **[64-66]**. However, the process complexity is very high for MBE to achieve such size variations that reduces the cost-effectiveness of this growth technique. It is imperative to mention that scalability and reproducibility are the two key challenges which should be addressed appropriately in such fabrication methods.

## 1.4. Fabrication methods of nano- /quantum- scale devices: Challenges and opportunities

The discussions of sections 1.2 and 1.3 on nano- and quantum- scale devices and their applications in the optoelectronics domain indicate that such devices have the potential to contribute significantly in achieving scientific or technological breakthroughs in near future. However, essential requirement for all such applications is to develop the relevant nanofabrication techniques. Significant progress in developing such techniques has enabled the researchers to investigate structural, electronic and optoelectronic properties of different materials with atomic level precision. Two approaches, the 'top-down' and 'bottom-up', are generally adopted to fabricate such nano- /quantum- scale devices. In the 'top-down' approach, a relatively larger geometry is reduced to a smaller dimension by an external means including selective etching and different nanoimprinting techniques **[67-69]**. On the other hand, the 'bottom-up' approach comprises of the growth of nano- /quantum- structures starting from atoms/molecules to form complex self-assemblies **[70-72]**.

Selective lithography techniques are used in the 'top-down' approach where some part of the material is covered and the other part is left exposed by suitably designing the lithographic mask. Optical lithography is the most commonly used technique in microelectronic industry to transfer pattern of the desired structures. Generally, the feature size of patterned devices ~1 μm can be achieved by using the conventional photolithographic technique **[67, 73, 74]**. Such photolithographic technique is predominantly used for the fabrication of both inorganic and organic material based optoelectronic devices due to its high speed, high resolution and parallel patterning capability **[75-77]**.

Electron beam lithography (EBL) has drawn significant interest as the advanced nanofabrication tool to fabricate patterned nano- /quantum- structures with much higher resolution. The primary idea of using EBL is to overcome the diffraction limit of light wave which indeed helps to develop nanostructures and thereby making it superior over the photolithographic technique. In such a system, an accelerated electron beam (diameter <= 10 nm) is used to develop patterned structures directly on the substrate coated with e-beam sensitive materials **[78-80]**. Such direct write e-beam lithography (DWEB) technique utilizes computer controlled scanning electron microscopes (SEMs) for the writing process. Schematic diagrams of such EBL systems are shown in Fig. 1.6. Here, an e-beam sensitive material, *i.e.*, e-resist is used as the mask to achieve nanostructures with desired shape and size. The transformation of e-resist under e-beam exposure can be of two types depending on its nature, *i.e.*, whether it is positive or negative. For the positive e-resist, solubility of the exposed region increases and thus etched away by immersing into the developer solution, and for the negative e-resist, the exposed region becomes cross-linked under e-beam exposure and thus the masked region is etched away.

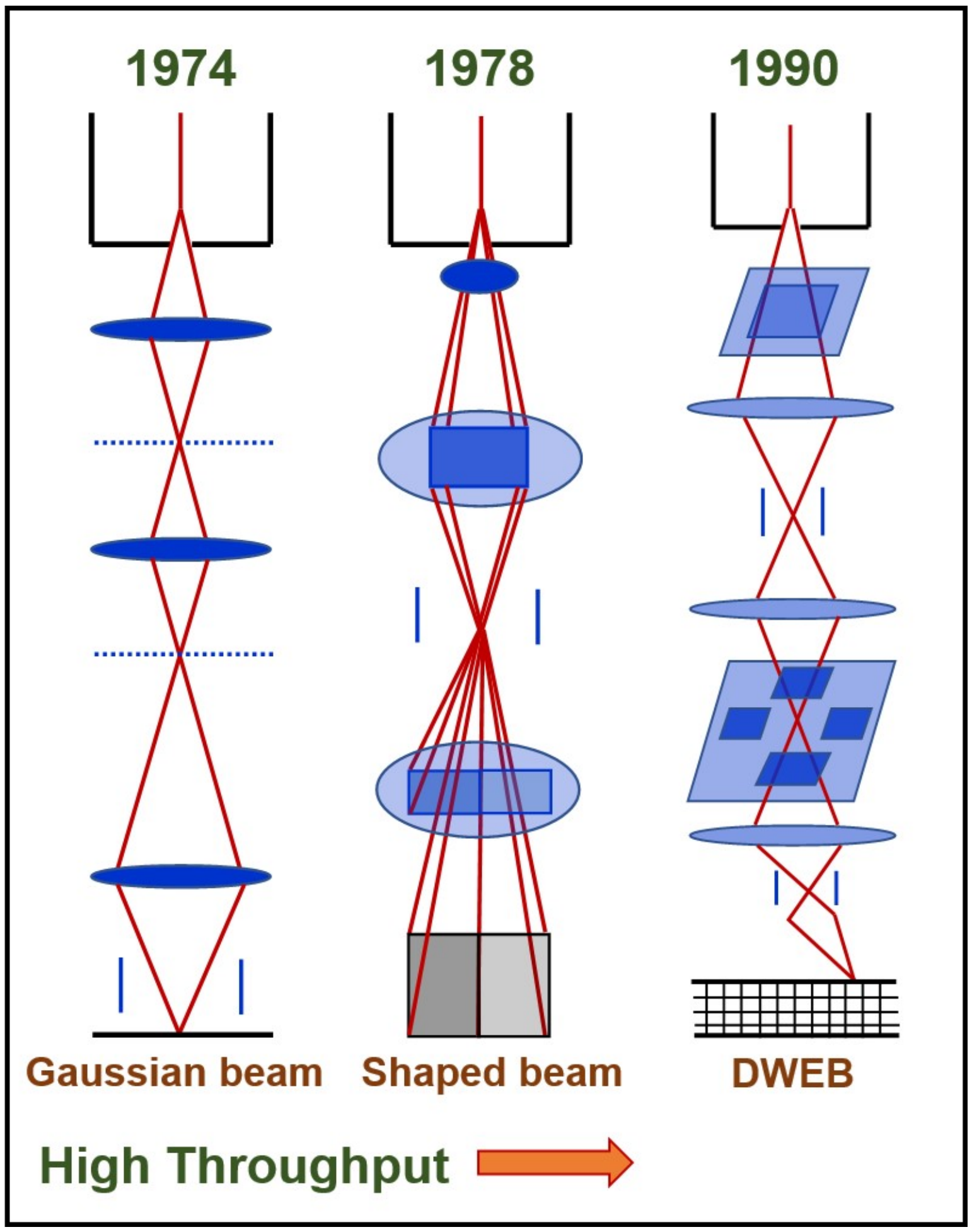

Fig. 1.6. Schematic representation of electron beam engineering technique in a typical electron beam lithography (EBL) system.

Several nano- /quantum- scale optoelectronic devices are fabricated by employing EBL to observe novel phenomenon despite having the comparatively slower writing process **[81-88]**. Particularly, the fabrication of patterned nano/quantum devices with the capability of light trapping and waveguiding requires precision better than the conventional optical lithographic technique **[88, 89-92]**. Such requirements make electron beam lithography an indispensable choice to meet the stringent demands of the state-of-the-art optoelectronic device fabrication and patterning.

The primary aim of 'bottom-up' approaches is to fabricate nano- /quantum- scale devices by self-assembly of atoms or molecules. Atomic layer deposition technique (ALD) has emerged as one of the potential technologies for its highly precise control on the thickness of deposited layer **[93-95]**. The vapor deposition techniques namely, chemical vapor deposition (CVD) and physical vapor deposition (PVD), are used extensively for the fabrication of several nano/quantum structures **[96-98]**. The chemical bath deposition (CBD) method has also been exploited for the growth of high density nanostructures **[99-101]**. Since the first-time demonstration of vapor-liquid-solid (VLS) method, such process has gained significant attention for the growth of vertically aligned nanowires **[102-105]**. Among several 'bottom-up' approaches, the most sophisticated method is the molecular beam epitaxy (MBE) due to its precise degree of control on the growth direction of the material. The MBE technique has also been reported to be very effective in fabricating self-assembled quantum dots with higher efficiency and functionality **[64-66, 106-108]**. Such, 'bottom-up' assembled nano- /quantum- scale devices have attracted significant research interest for the development of nanoscale electronic and optoelectronic devices due to its favorable vertical growth direction **[70, 72]**.

The selection of process technique for patterned nanofabrication is very crucial. In this context, both the 'top-down' and 'bottom-up' approaches have been successfully employed for the formation of various high quality nanomaterials and corresponding fabrication of numerous quantum-scale devices. However, both of the processes have their inherent merits and demerits. The electron-beam lithography has been very useful to generate patterned nanostructures of desired shape and size (~100 nm); however, it is indeed very expensive and has low throughput due to its slow serial writing process. Moreover, achieving patterned nanowires of sub-80 nm diameter is still a challenge. In the 'bottom-up' approaches, ALD is very effective due to its control over the layer thickness; however, it is also very expensive. Sol-gel and CBD, on the other hand, are very cost-effective chemical synthesis processes where various nanomaterials can be successfully grown; however, scalability of such processes is a serious concern. Vapor deposition processes (CVD and PVD) are very effective for the fabrication of complex structures with different materials, however; it is also very expensive and the use of toxic/corrosive gases is a concern. Such 'bottom-up' approaches have been successfully utilized to fabricate nano/quantum devices, however achieving patterned structures has faced serious challenges due to the inherent randomness of nanostructure growth processes. In future, the nanofabrication techniques are expected to progress further and will provide a systematic route for the realization of desired patterned quantum-scale devices. The merits and demerits of various 'top-down' and 'bottom-up' approaches are summarized in Table. 1.1 and a comparison of relevant material parameters is shown in Table. 1.2.

Table 1.1. Summary of the merits and demerits of 'top-down' and 'bottom-up' nanofabrication techniques.

| **Top–down method** | **Merits** | **Demerits** | **Reference** |
|---|---|---|---|
| Optical lithography | Established microfabrication tool for chip production with high throughputs. | Trade off between resist, process sensitivity and resolution. | **[109]** |
| E-beam lithography | Effective in fabrication of patterned nano- /quantum- scale devices ~ 100 nm. | Expensive, low throughput and slow serial writing process. | **[109]** |
| Soft & nanoimprint lithography | Pattern transfer tool for < 10 nm nano/quantum structure fabrication. | Difficult for large-scale production of densely packed nano/quantum structures. | **[109]** |
| Block co-polymer lithography | High-throughput, low-cost method, suitable for parallel assembly of large scale densely packed nanostructures. | Difficult to fabricate self-assembled nanopatterns, high defect densities are observed. | **[109]** |
| Scanning probe lithography | High resolution chemical, molecular and mechanical nanopatterning capability. | Expensive, limited for high throughput applications and manufacturing. | **[109]** |
| **Bottom–up method** | **Merits** | **Demerits** | **Reference** |
| Atomic layer deposition | Digital thickness control with atomic level precision, good reproducibility and adhesion. | Usually a slow process, expensive due to the involvement of vacuum components. | **[109]** |
| Sol-gel | Low-cost chemical synthesis process. | Scalability, patterning capability. | **[109]** |
| Chemical Bath Deposition | Low cost, fabrication of high density nanostructures is possible | Scalability, patterning capability due to inherent randomness of nanostructure growth. | **[99-101]** |
| Vapor-liquid-solid | Fabrication of vertically aligned nanowires, low cost process. | Patterning capability due to randomness of nanowire growth mechanism. | **[102-105]** |
| Molecular self-assembly | Allows the self-assembly of molecular nanostructures ~ 20 nm. | Process complexity, difficult to design and fabricate nano-systems. | **[109]** |
| Physical and chemical vapor | Versatile nanofabrication tools for fabricating complex multi-layered | Expensive due to the requirement of expensive | **[109]** |
| phase deposition | nanosystems. Controlled simultaneous deposition of several materials, scalable process. | vacuum components, requirement of toxic and corrosive gases. | |

Table. 1.2. Comparison of the material parameters for various nanofabrication techniques.

| Methods | Resolution /size | Fabrication complexity | Nanostructural defectivity | Materials range | Reference |
|---|---|---|---|---|---|
| Optical lithography | ~ 200 nm | High | Low | Inorganic and organic | **[109]** |
| E-beam lithography | ~ 10 nm | High | Extremely low | Mostly inorganic | **[109]** |
| Nanoimprint lithography | ~ 5-10 nm | Higher for roll-to-roll processes | Low | Inorganic and organic | **[109]** |
| Scanning probe lithography | < 5 nm | High | Extremely low | Mostly organic | **[109]** |
| Block co-polymer lithography | ~ 10 nm | Low | Low | Inorganic and organic | **[109]** |
| Atomic layer deposition | < 5 nm | High | Extremely low | Mostly inorganic | **[109]** |
| Sol–gel | ~ 50 nm | Low | High | Mostly inorganic | **[109]** |
| CBD | ~ 100 nm | Extremely low | High | Mostly inorganic | **[99-101]** |
| VLS | ~ 100 nm | High for obtaining patterns | Low | Mostly inorganic | **[102-105]** |
| Molecular, layer-by-layer and directed self assembly | ~ 5 nm | High for obtaining patterns | Low | Inorganic and organic | **[109]** |
| Physical and chemical vapor phase deposition | ~ 5 nm | High for obtaining patterns | Low–moderately high | Inorganic, organic, ceramic, insulator, semiconductor | **[109]** |

It is evident from the discussion that the development of advanced optoelectronic integrated circuits and systems relies on the fabrication of ‘patterned’ ‘nano- /quantum- scale’ devices. In this context, the ‘top-down’ method is successfully utilized to fabricate ‘patterned’ devices; however, achieving ‘nano- /quantum- scale’ structures experienced several challenges in terms of its scalability and reproducibility. On the other hand, the 'bottom-up' method meets the demand for 'nano- /quantum- scale' structures; however, achieving 'patterned' structure is still a challenge due to its inherent randomness. Such contradictory results of 'top-down' and 'bottom-up' methods require new approaches to circumvent this issue. In this context, the 'hybrid' approach has emerged for the fabrication of 'patterned' 'nano- /quantum- scale' devices combining the benefits of 'top-down' and 'bottom-up' methods **[110]**.

## 1.5. Modeling of transport properties of nano-/quantum-scale devices: the NEGF formalism

The development of a comprehensive understanding on the novel physical phenomena in nano- / quantum- scale devices is very crucial and therefore the appropriate analytical models are required to predict the device performance prior to their practical implementation. During past two decades, numerous attempts have been made to model the transport properties of such nano- /quantum- scale devices. In this context, the drift-diffusion model has been extensively used due to its relatively less complexity and robustness from the perspectives of both analytical modeling and numerical study. Such model has provided results of satisfactory accuracy as compared to the experimentally obtained values for physical dimensions ~150 nm along the direction of carrier transport **[111-116]**. However, the model lacks satisfactory description of carrier transport properties of nanoscale devices where quantum effects arise, especially in the ballistic regime of non-equilibrium phenomena. Moreover, when charge carriers in such devices interact with photon (or phonon), the phase-breaking phenomena start to dominate in their transport properties. Thus, to model such events a self-consistent quantum-electrostatic description of the entire system in the non-equilibrium process is required. In this context, the non-equilibrium Green's function (NEGF) formalism has emerged to be the most prominent tool to encounter such issues **[117, 118]**. In such approach, the Green's function for electrons, holes and photons (and/or phonons) play the crucial role to provide the description of physical quantities characterizing the relevant transport properties. The 'device' Green's function represents contributions of the system itself and response of the system under external perturbations such as, coupling of the system with 'reservoirs' (*i.e.,* contacts) and electron-photon (and/or electron-phonon) interactions for optoelectronic devices. The contribution of such perturbations is included in the 'device' Green's function by calculating the appropriate self-energies **[117, 118]**. Absorption/emission of photons (and/or phonons) and generation-recombination-transportation processes of the photogenerated carriers are theoretically modeled from such self-energies. Fig. 1.7 shows the schematic representation of NEGF formalism with corresponding self-energy functions.

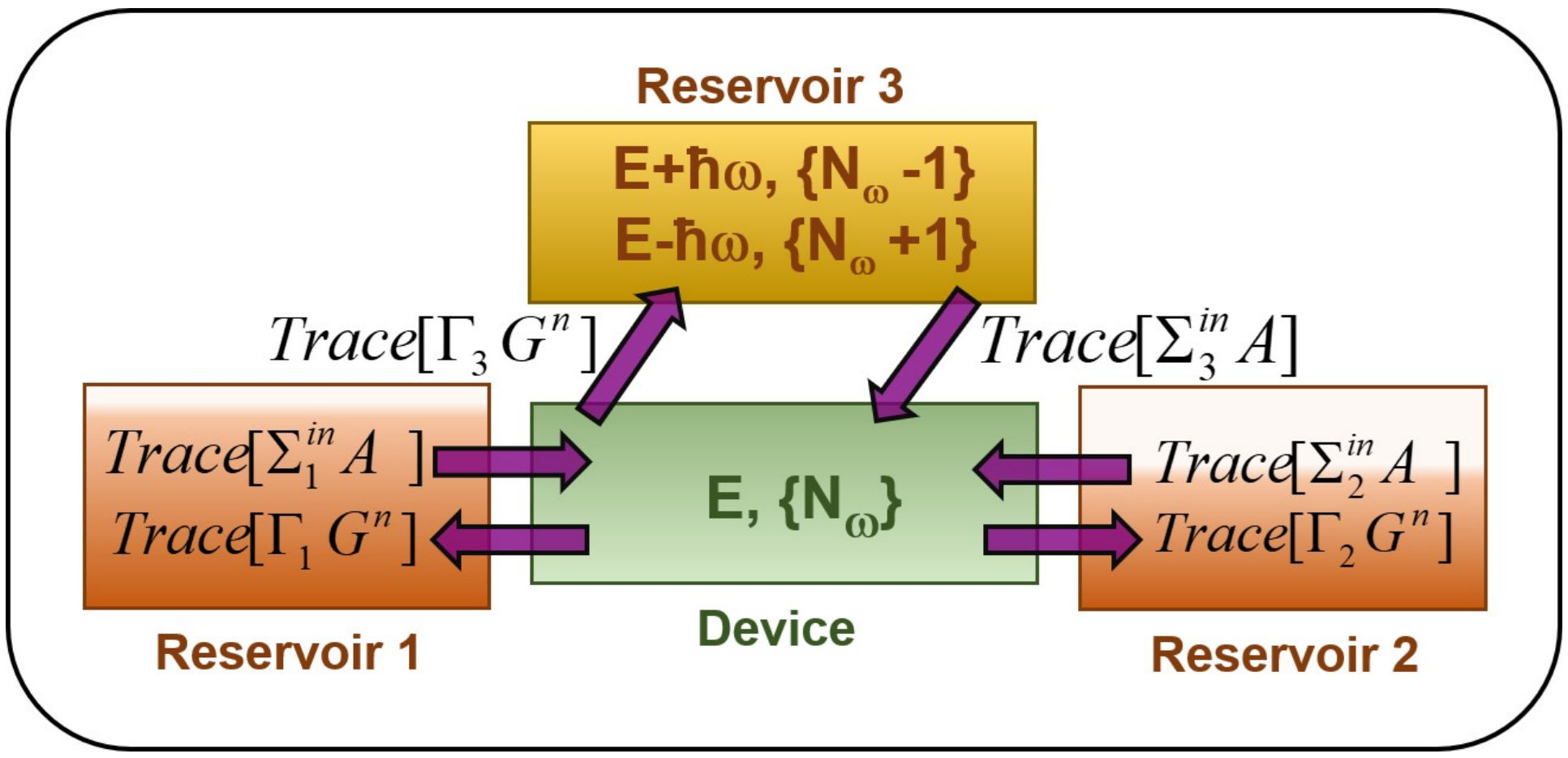

Fig. 1.7. Schematic representation of the scheme of NEGF formalism used in the current dissertation.

In order to numerically study the NEGF formalism for quantum scale device modeling, emphasis needs to be given on increasing the computational efficiency. In this context, a recursive algorithm (following Dyson's equation) was implemented to avoid solving the entire retarded Green's function matrix **[119]**. The significant contribution has been made by introducing the 'mode space' approach in the place of 'real space' approach to improve the computational efficiency by utilizing the quantum discreteness of nanoscale devices **[120]**. In the 'mode space' approach, position coordinate is transformed into energy space where only the modes below Fermi level are considered. It changes the large number of position grids into a few energy grids and makes it significantly effective in terms of computational efficiency. Therefore, the intricate issues of nano- /quantum- scale devices can be successfully addressed in the NEGF formalism based on a self-consistent simultaneous quantum-electrostatic solver following the mode-space approach.

## 1.6. Choice of materials and device architecture to realize quantum effects in optoelectronic applications

Being the backbone of microelectronics industry, Silicon (Si) has shared 97% of the electronics market due to its abundance in nature, mechanical strength, non-toxicity and ability to form high quality native oxide layer ($SiO_2$). However, the realization of quantum effects in Si is very difficult regarding the issues related to fabrication procedures. It is worthy to mention that the observation of quantum effects at room temperature requires the size of corresponding material to go below the value of excitonic Bohr Radius (EBR). In this regard, the electron (and/or hole) effective mass of the respective materials also plays an important role to determine the dimension to observe quantum effects. Si has relatively higher effective mass values of $0.19m_0$, $0.91m_0$ and very small EBR of 5 nm. Such material parameters of Si indicate that quantum effects can be observed only if its size goes below 5 nm. Achieving dimensions beyond 5 nm actually increases the fabrication complexities which in turn also challenges reproducibility of the fabrication procedure. Thus, the materials with relatively lower effective mass values and higher EBR are required to observe quantum effects at relatively higher physical dimensions. Ge, GaAs, InP, PbSe are some of the materials which meet such criteria to realize quantum confinement effects with reducing fabrication complexities. For instance, Ge has relatively lower effective mass of $0.08m_0$ in the transverse direction and relatively higher EBR of 25 nm which makes it suitable for realization of quantum confinement effects at relatively higher physical dimensions. However, both Ge and Si have limited optoelectronic applications due to its indirect energy band gap property. In this context, GaAs and InP fulfill the criteria of direct energy band gap and suitable material parameters ($m_e$=$0.063m_0$ and EBR=12 nm for GaAs; $m_e$=$0.082m_0$ and EBR=15 nm for InP) to observe quantum effects at room temperature. It is also to be mentioned that PbSe has all the required properties of direct energy band gap, relatively lower effective mass ($0.04m_0$) and significantly higher EBR (42 nm) however, utilization of PbSe in optoelectronic applications is limited by its toxic nature. Therefore, the choice of materials with suitable electron effective mass and EBR values is very crucial to develop efficient optoelectronic systems with reduced fabrication complexities.

Such fabrication complexities also depend on the architecture of nano- /quantum- scale devices. Designing the device architecture mainly consists of two different approaches such as, horizontal and vertical configurations. The 'pick-and-place' technique is generally followed to align nano/quantum structures horizontally on the substrate. However, it is an extremely complex process to adopt in terms of its scalability and reproducibility. On the other hand, nano/quantum structures are allowed to grow in the vertical direction by employing 'bottom-up' fabrication methods. The vertical device integration scheme is the most favorable choice for fabrication since it is their natural direction of growth. Thus, the vertical configuration with suitable materials maintaining the criteria of electron effective mass and EBR values is the most feasible solution to realize quantum effects in practice.

## 1.7. Motivation and organization of the thesis

In section 1.1, the revolutionary innovation of CMOS architecture as basic building block for the advanced electronic and optoelectronic devices is briefly discussed. In the following sections, effectiveness of the nano- and quantum- scale devices, achieved by aggressive downscaling approach, to develop novel optoelectronic systems is illustrated. In section 1.4, merits and demerits of the 'top-down' and 'bottom-up' approaches for the fabrication of nano- /quantum-scale devices are briefly discussed. Section 1.5 comprises of the discussion on the emergence of NEGF formalism which provides the appropriate platform to analytically model the electronic and optoelectronic performances of such novel devices. The choice of materials and device architecture to observe quantum effects in practice at room temperature is briefly discussed in section. 1.6.

It is evident from the discussions of above sections that the semiconductor quantum dots have the potential to be the primary electronic component of next generation optoelectronic integrated circuits and systems. The development of such integrated circuits depends on the adaptability of complementary MOS technology since it is the most mature technology available at the moment. However, successful integration of semiconductor quantum dots in the MOS technology has several challenges to face. The underlying physics of such MOS based quantum dot devices is different from the bulk counterpart and this has to be unfolded to develop comprehensive understanding on the electro-optical behavior of such devices. Also, realization of the novel physical phenomenon of quantum domain requires new experimental techniques to go beyond the 'proof-of-concept' based devices to large scale implementation.

'Modeling' and 'fabrication' are the two key components of research which guide scientists to improve the device performance and develop advanced integrated systems. Prediction of the performance of devices prior to fabrication is crucial in terms of designing the fabrication procedure. In this regard, analytical modeling has played the pivotal role in determining the qualitative and quantitative aspects of the devices. As discussed earlier, the NEGF formalism has emerged as the prominent modeling tool to encounter the issues of quantum domain. Few research groups are working in the area of modeling of quantum scale devices by following such formalism **[121-127]**. However, modeling of electro-optical behavior of quantum scale devices by employing NEGF formalism needs special attention for the successful implementation of quantum dot based optoelectronic integrated circuits and systems. Modeling of such devices also helps in selecting materials with appropriate dimensions, prior to fabrication, for the realization of quantum effects in practice. Designing the quantum dot based devices with suitable materials in terms of electron effective mass and EBR is essential to develop a scalable and reproducible fabrication procedure.

In this context, the current dissertation deals with the design, modeling and fabrication of nanowire MOS based quantum dot devices where the quantum dot is formed in axial (longitudinal) direction of the nanowire beneath the oxide/semiconductor junction. It is worthy to mention that the transport of electrons in such quantum dot region along the transverse dimensions (radial) is quantized since its dimensions are chosen to be smaller than the relevant EBR. The longitudinal transport of such electrons along the nanowire axis is quantized due to surface quantization created by the applied voltage at the top electrode. Thus, the transport along two dimensions is quantized due to structural confinement whereas the axial motion can be controlled by varying external voltage, thereby leading to the formation of a voltage tunable quantum dot (VTQD). The primary aim of the research work is to first understand the formation of such quantum dots at room temperature from the basic understanding of quantum transport phenomenon followed by its experimental verification. Applicability of such devices in the optoelectronics domain as wavelength selective photodetectors and high-efficiency solar cells is also analyzed in detail by formulating the relevant theoretical model. Finally, such nanowire MOS based VTQD devices are fabricated by employing EBL and characterized by performing electrical measurements. The current dissertation is organized as the following:

Chapter 1 is dedicated to provide a brief introduction about the invention of CMOS technology and highlights the emergence of quantum scale devices as the fundamental unit of modern optoelectronic circuits. Emphasis has also been given on the fabrication methods and analytical modeling of nano- /quantum- scale devices. An extensive literature review is provided on the evolution of optoelectronic devices from 'bulk' to 'quantum' through 'nano' in Chapter 2. Electrostatics of the nanowire MOS based VTQD devices is theoretically analyzed in detail in Chapter 3 by formulating an analytical model based on self-consistent Schrodinger-Poisson solver. The electrostatic integrity in such devices in terms of maintaining appropriate electrostatic control and reducing the tunneling assisted leakage current has been discussed in detail. In Chapter 4, the process of photogeneration in Si and InP nanowire MOS based VTQD device is studied in detail by formulating a non-equilibrium Green's function (NEGF) based theoretical model where the transport equations are formed by considering the second quantization field operators for electrons, holes and photons. From the understanding of such analytical model, a device scheme is proposed for the wavelength selective multicolor detection with high spectral resolution. Chapter 5 deals with a conceptual framework by extending the already developed NEGF based model to calculate hole current in GaAs nanowire MOS based VTQD devices upon illumination. Such device can also be utilized for high efficiency solar cell applications beyond the Shockley-Queisser (SQ) limit. In Chapter 6, the patterned array of Ge-nanowire MOS based VTQD devices are fabricated by employing EBL. Formation of such voltage tunable quantum dots in the nanowire is characterized by measuring its in-situ capacitance-voltage (C-V) characteristics within the field emission scanning electron microscopy (FESEM) chamber. Finally, concluding remarks of the current research work has been provided in Chapter 7 and the future scopes/applications of the nanowire MOS based VTQD devices are briefly discussed.

# Chapter: 2

# Literature review: Chronological development of optoelectronic devices

## 2.1. Journey of optoelectronic devices: A historical perspective

Optoelectronic devices have been the integral part of any modern electronic gadget in the form of sensors, displays and energy sources. The basic operating principle of such devices relies on the interaction between electrons and photons occurring within the material where the primary aim is to generate, modulate and detect electron-hole pairs in an effective manner. The advantage of such devices is their optoelectronic properties which can be tailored to meet the application dependent specific requirement by varying composition of associated materials and designing the device architecture. The optoelectronic devices are commonly classified into two different types depending on its light absorption and light emission capabilities. Such devices primarily include photodetectors, solar cells, light emitting diodes and lasers. In the present chapter, the chronological development of photodetectors and solar cells is discussed in detail since the current dissertation deals with development of novel photosensing and photovoltaic devices as mentioned in section 1.7 of Chapter 1.

Scientific understanding of light-matter interaction plays an instrumental role for the development of optoelectronic devices, circuits and systems. The journey of light-matter interaction has started long time back in 1865 when James Clerk Maxwell formulated the electromagnetic wave equations in his famous paper entitled *"A Dynamical Theory of the Electromagnetic Field"* **[128]**. Following the overarching development of electromagnetic theory by Maxwell, the next significant contribution was made in 1887 by Heinrich Rudolf Hertz who discovered the generation of electric sparks under ultraviolet illumination **[129]**. Later, in 1900 Max Planck suggested (later known as, Planck's postulate) that the energy carried by electromagnetic waves is released as 'packets' of energy known as 'quanta' which laid the foundation of quantum theory **[130]**. Following the previous report of electric spark generation upon illumination by Heinrich Hertz, Albert Einstein proposed the theoretical model of photoelectric effect in 1905 which is also in consistent with Planck's hypothesis **[131]**. In 1916, Robert Millikan experimentally verified Einstein's model and also provides the first direct measurement of Planck's constant **[132]**. By that time, it was established that light is made of quantum particles however, the term 'photons' was coined later in 1926 by Gilbert N. Lewis. The statistical behavior of such photons was discovered by Satyendra Nath Bose in 1924 however, his paper entitled "*Planck's Law and the Hypothesis of Light Quanta*" was rejected by the journal. Consequently, Bose asked Einstein for his consideration and, Einstein readily agreed to Bose's work and recommended it to be published in *Zeitschrift für Physik* **[133]**.

### 2.1.1. Chronological development of photodetectors

The very first demonstration of a photodetecting device was the photomultiplier tube (PMT) which was invented by L.A. Kubetsky in the year 1930 known as, 'Kubetsky's tube' **[134]**. However, some controversies exist on the invention of PMT since it was also claimed that PMT was discovered by V.K. Zworykin at RCA laboratories in 1936 **[135]**. Such PMTs have been extensively used in various domains such as, space research, medicine, chemistry, agriculture etc. to detect UV, visible and NIR wavelengths due to its high gain, fast response time and low noise properties **[136]**. Further, following the invention of bipolar transistors in 1947, the phototransistor was invented by John N. Shive at Bell laboratories in 1948. Such invention of transistors also accelerated the development of infrared (IR) detectors and consequently, Texus Instruments developed the first advanced infrared system in 1963.

During 1960s, mainly the p-n and n-p-n junction devices were developed to convert incident light into electrical signals. The '*computational sensor*' and '*scanistor*' were reported to generate an output pulse proportional to incident light intensity following the photoconductivity effect **[137, 138]**. However, such devices were lacked the intentional integration of optical signals. In parallel, efforts have also been made to read the electrical signals generated by photosensing devices commonly known as, pixel sensor arrays. Consequently, a phototransistor based 2500 pixel sensor array was reported by Shuster *et. al.* in 1965 with 100 leads for the readout operation **[139]**. Two years later, a significant contribution has been made by Weckler *et. al.* by demonstrating the operation of p-n junction photodiode in 'photo-flux integrating mode' which has provided several advantages such as, linear dependence of output signal with incident light intensity, electronically controllable sensitivity and ease of integration into arrays for sensing applications **[140]**. In 1968, the same research group developed a 100×100 array of silicon photodetectors, named as '*reticons*', to commercialize such sensor devices **[141]**. In the same year, RCA laboratories also reported a completely integrated thin-film transistor (TFT) based solid state image sensor comprising of 32400 element photosensitive array with sequential addressing of pixels **[142]**. A television camera was also constructed by using such 180×180 sensor array. Following the success of photodetecting devices, several configurations of silicon image sensor arrays based on surface and buried photodiodes have been demonstrated for performance improvement of such devices **[143-145]**.

From the architectural point of view, the pixel circuitry of a photodetecting system is divided into, i) passive pixels and ii) active pixels. The photodiode type passive pixel was first introduced by Weckler *et. al.* in 1967 and it has been widely used for various photosensing applications due to its large fill factor and high quantum efficiency **[140]**. However, major problem arises in scalability of such devices. Consequently, active pixel sensors were proposed where the photosensor device is coupled with an active amplifier. In 1968, the photodiode based active pixel sensor was fabricated by Noble *et. al.* **[143]** and following the successful fabrication, research on the development of such devices has gained a significant momentum **[143-145]**.

The year 1970 has witnessed emergence of the charge coupled devices (CCDs) which instantaneously have attracted significant interest due to its high fill-factor and high output uniformity **[146]**. Since the inception of CCD sensors, significant efforts have been made during 1970s - 1980s to improve the performance of such devices in terms of higher quantum efficiency, lower dark current, higher charge transfer efficiency and faster readout operation **[147-154]**. In such devices, every pixel is utilized to capture light and the charge of every pixel is transferred to the output end as an analog voltage which helps to maintain the output uniformity. In contrast, the CMOS imager has different charge to voltage conversions for each pixel and thus often required amplifiers, noise correction methods and digitization circuits. Such additional functionalities increased the design complexity of CMOS imagers. It is to be mentioned that both the CCD and CMOS imagers were invented almost at the same time however, initially the CCD imagers have played the dominant role since it produced superior images by utilizing the available fabrication technology. The CMOS imagers, on the other hand, required relatively smaller features than the semiconductor industry (silicon wafer foundries) could deliver at that time. In the late 1970s to early 1980s, CMOS image sensors were sporadically investigated mainly to increase the functionality of CCD technology by integration of such sensors in a common platform **[155, 156]**. Further, the development of CMOS image sensors was continued by Hitachi and Matsushita for few more years however, they abandoned the approach due to its residual temporal noise and poor performance in low light conditions **[157-159]**. The first all-CMOS active pixel sensor array was fabricated in 1990s using the commercially available 2-µm CMOS process technology **[160, 161]**.

The renewed interest to fabricate such CMOS image sensors was developed due to the market demand for miniaturized, low-power and cost-effective camera-on-a-chip imaging systems. Resurgence in the development of CMOS sensors was generated due to mainly two driving factors such as, low-cost highly functional single chip imaging system and the highly miniaturized instruments for next-generation space exploration spacecrafts (initiated by NASA).

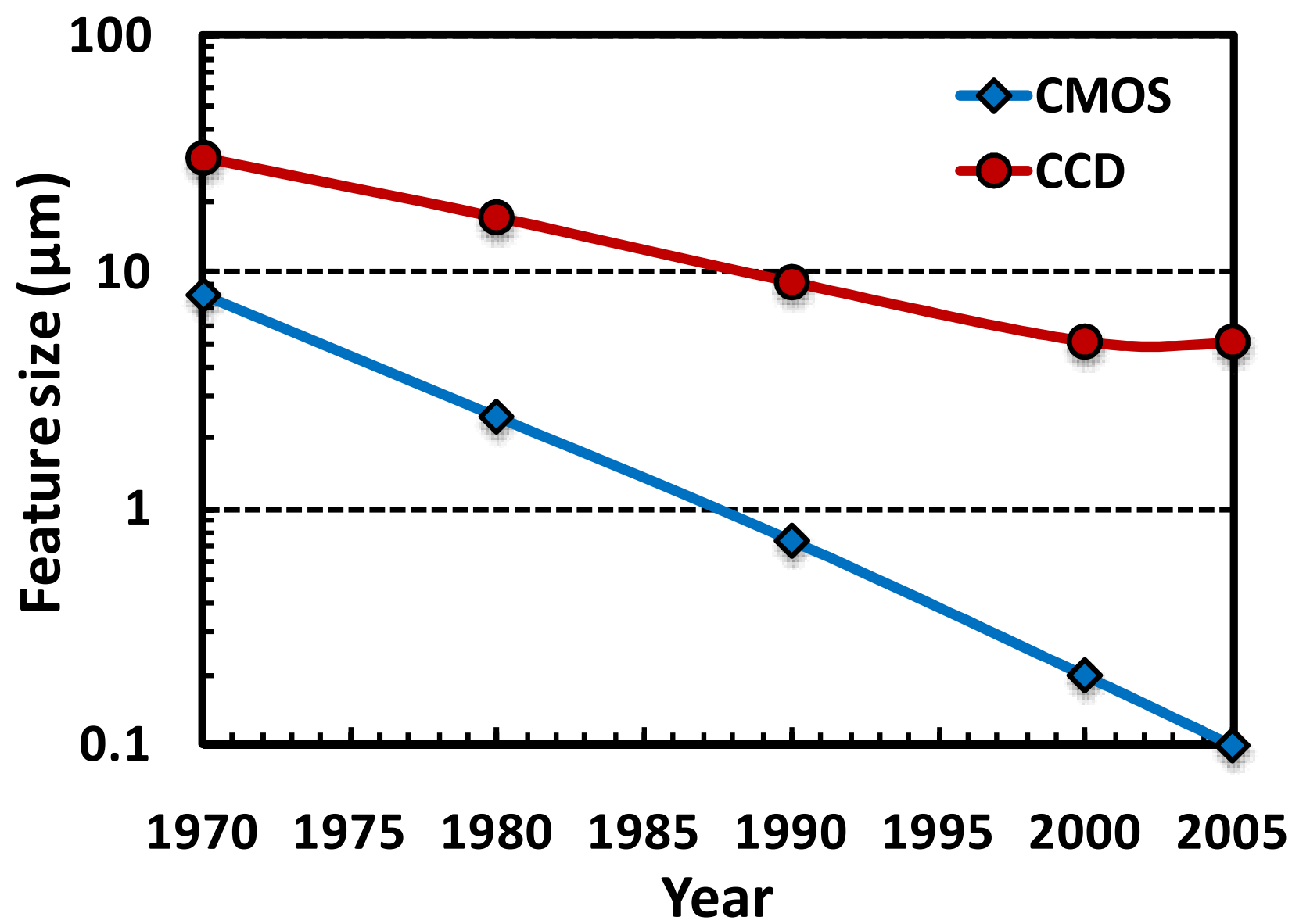

Fig. 2.1. Plot of minimum feature size of CCD and CMOS technology from the year 1970 to 2005 **[162]**.

Fig. 2.1 shows the plot of minimum feature size of CCD and CMOS sensors and it is observed from such plot that the miniaturization of feature size in CMOS technology outpaced the CCD sensors. The impact of scaling trend in CMOS technology on the improvement of device performance was predicted by Fossum *et. al* in 1997 following the industry standard technology roadmap **[162]**.

Such sustained downscaling approach for gradual performance improvement has led to the emergence of 'Nanoscience and Nanotechnology'. From 2000 onwards, the semiconductor nanowire-based devices have drawn significant attention for photodetecting/sensing applications. Numerous reports are available on the detection of a wide range of wavelengths from ultraviolet to infrared region **[163-183]**. The photodetectors using ZnO nanowires have been extensively studied to detect the ultraviolet region due to its high photo-to-dark current ratio, fast response time and high internal gain **[163-166]**. Wavelengths in the visible region are mostly detected by GaAs and InP nanowire based devices and InAs, InAsSb, GaAsSb nanowires are explored to detect the infrared region **[170, 178-183]**. Tuning the optoelectronic properties of such nanowires by varying their composition enables the detection of a wide range of wavelengths **[184-187]**. Such wavelength selective multi-color photodetectors have multifaceted applications in the domain of optical sensing and imaging **[187]**.

Several vertical and horizontal configurations including homo- /hetero- junctions, metal-semiconductor Schottky junctions and core-shell structures have been studied for developing the photodetectors. In this line, the axial homo-junction devices have been investigated for sensing ultraviolet light using GaN nanowires **[188]** and the axial hetero-junction devices have also been fabricated for the detection of infrared light **[189]**. The radial core-shell junction devices have gained significant interest due to its efficient charge carrier separation in comparison to the planar devices **[190, 191]**. Reports are also available on the nanowire based avalanche photodiodes with enhanced photo-responsivity due to its inherent carrier multiplication phenomenon on application of a large reverse bias **[192]**. The Schottky junction devices consisting of n-ZnO nanowires with noble metals (Au, Ag, Pd) have been investigated for UV-light detection **[193]**. Also, the Si-nanowire based phototransistors have been demonstrated for such photosensing applications **[194]**. Thus, over the years such semiconductor nanowires have been extensively used to improve the performance of photodetecting devices in comparison to their bulk counterparts. Table. 2.1 shows the performance metrics of some of the commonly used nanowire based photodetectors.

Table. 2.1. Summary of the performance parameters for a few commonly used nanowire based photodetectors.

| Material | Morphology | Growth/Deposition technique | Dark-current or Conductance | Photocurrent or Conductance or Responsivity | Ref. |
|---|---|---|---|---|---|
| ZnO | Nanowires | CVD | >10 G Ω | 1-5 orders lower | [195] |
| | Nanowires | Vapor phase transport process | 3.5 M Ω-cm | 4-6 orders lower | [196] |
| | Nanowires | CVD | 1-10 nA | 100 µA | [197] |
| | Nanowires | MOCVD | ~110 µA | ~170 µA | [198] |
| | Nanowire film | Sol-gel method | 4.32 x $10^{-9}$ A | 5.11 x $10^{-7}$ A | [199] |
| | Nanowires | Thermal evaporation method | 1.35 x $10^{-5}$ A | $2x10^{-7}$ A | [165] |
| $SnO_2$ | Nanowires | CVD | 0.12 µS | 83.9 µS | [200] |
| | Nanowires | CVD | 7.7 $x10^{-5}$ A | 1.3 x $10^{-4}$ A | [201] |
| | Nanowires | Laser-assisted CVD | 0.66 nS | 760 nS | [202] |
| CdO | Nanoneedles | CVD | 13.3 nS | 114.5 nS | [203] |
| $Ga_2O_3$ | Nanowires | CVD | 15 pA | 10 nA | [204] |
| $Cu_2O$ | Nanowires | Thermal oxidation | 0.7 nS | 4.3 nS | [205] |
| $CeO_2$ | Nanowires | Wet chemical route | 22.8 nA | 0.25 nA | [206] |
| $ZnSnO_3$ | Nanowires | Thermal evaporation | 0.3 nA | 162 nA | [207] |
| InP | Single horizontal nanowire | MOCVD | ~0.3 nA @ -5V | 6.8 A/W | [208] |
| GaAs/AlGaAs core-shell | Single horizontal nanowire | MOVPE | ~pA | $10^{-4}$ A/W | [209] |
| InAs | Single horizontal nanowire | MBE | -1 nA @ 2V | 5.3 × $10^{3}$ A/W | [176] |
| InSb | Single horizontal nanowire | Electro-chemical | -7 µA @ 10V | 8.4 × $10^{4}$ A/W | [210] |
| InAsP | Single horizontal nanowire | VLS | -0.2 µA @ 0.5V | 5417 A/W | [186] |
| InGaAs | Single horizontal nanowire | CVD | 144 nA @ 0.5V | 6.5 × $10^{3}$ A/W | [211] |
| GaAsSb | Single horizontal nanowire | CVD | -200 nA @ 1V | 1.7 × $10^{3}$ A/W | [183] |

Semiconductor quantum dots have emerged as a consequence of sustained downscaling approach enabling the transition of device dimensions from bulk to nanoscale. Last two decades have witnessed significant progress on the fabrication of semiconductor quantum dots for photodetecting applications due to its tunable energy band gap, high quantum efficiency, high operating temperature, reduced dark current and high detectivity **[51, 212, 213]**. Since the first report on inter sub-band transitions in quantum devices **[214]**, significant efforts have been made for the fabrication of quantum dot based photodetectors to achieve superior performances. The overall research on such devices is largely focused to improve the two fundamental performance parameters: i) photo-to-dark current ratio and, ii) responsivity. Efforts have been made for the performance improvement of such photodetectors, focusing on these two parameters by ensuring the formation of highly dense, defect free and uniform quantum dots **[45-51]**. For the implementation of such quantum dots several material systems including GaAs/AlGaAs, InAs/GaAs, GaSb/GaAs, InAs/AlGaAs, InAs/InAlAs etc. have been extensively studied **[215-217]**.

Alongside the development of photodetectors with quantum dots the heterogeneous integration of such structures on Si-substrate has also gained considerable attention since it is the key technology for developing large scale, cost-effective, on-chip photodetector arrays. Only the monolithic integration process has offered opportunities for the development of fully functional CMOS compatible optoelectronic/photonic IC-technology. In this context, several semiconductor quantum dot based structures, using Ge, GaAs and InAs are fabricated on Si-substrate for photosensing applications **[215, 218-220]**. Such reports are expected to provide an efficient route for the monolithic integration of several semiconductor quantum dots on Si-substrate.

The appropriate growth technology plays an important role to ensure the formation of dislocation-free quantum dots during such hetero-integration processes. Huang *et. al.* recently demonstrated the epitaxial growth of InGaAs quantum dots directly on (001) Si-substrate to ensure dislocation-free devices and achieved a very low dark current density of 5.3 pA **[219]**. Inoue *et. al.* reported the fabrication of p-i-n photodiodes using InAs/InGaAs quantum dots on Si-substrate by engineering the strained layer superlattices to achieve significantly lower defect density and dark current **[221]**. The InGaAs/GaAs quantum dot-in-a-well structure has also been investigated for monolithic integration on Si-substrate where a relatively lower dark current density of $2.03\times10^{-3}$ $mAcm^{-2}$ and a higher responsivity of 10.9 $mAW^{-1}$ have been reported **[215]**.

Also, the sub-monolayer quantum cascade structure has been attempted to monolithically grow on Si-substrate where the dark current density of $2.11\times10^{-8}$ $Acm^{-2}$ at -0.1 V bias has been obtained **[222]**.

Such lower dark current and higher responsivity values are also obtained by following a separate route by embedding the semiconductor quantum dots in various materials. Successful incorporation of Si quantum dots in $SiO_2$ (on Si-substrate) has enabled to achieve the internal quantum efficiency of 200% and relatively higher values of responsivity in the range of 0.4 to 2.8 A/W **[223, 224]**. Similarly, the devices by incorporating Ge quantum dots in $SiO_2$ (on Si-substrates) has also been investigated for photodetecting applications and a responsivity in the range of 0.13 A/W to 1.8 A/W with improved internal quantum efficiency have been obtained **[225, 226]**.

In recent times, the progress of photodetectors using quantum dots emphasizes the development of direct, full-color, multiband sensors/imagers so that the sophisticated color filters are not required. Such multi-color photosensors are highly useful for detecting specific wavelengths with superior spectral sensitivity **[213, 227-230]**. In this context, the quantum dot based indium-gallium-zinc-oxide (IGZO) phototransistors are actively investigated for the applications of full-color photodetection **[213]**. However, the monolithic integration of quantum dots in IGZO transistors suffers from the architectural issues, control on the interface and most importantly, realization of a CMOS compatible design. Therefore, the development of next generation photodetectors of high efficiency and functionalities with CMOS compatible quantum dots requires to circumvent such challenges.

### 2.1.2. Chronological development of solar cells

The journey of sustained progress of the solar cells has been now more than 180 years with the discovery of photovoltaic effect in 1839 by Alexandre-Edmond Becquerel. He observed the current in an electrolytic cell to enhance upon illumination with sunlight **[231]**. In 1873, Willoughby Smith discovered the photoconductive effect of selenium **[232]** and subsequently, the first solar cell was developed in 1883 by using Se/Au device structure with ~1% efficiency **[233]**. Following the very first demonstration of solar cell, several patents were filed on the development of such devices during 1887 to 1900 **[234, 235]**. Progress has been continued up to 1940s for the growth of single crystalline materials (Si, Ge) and to explore the novel materials (CdSe) for photovoltaic applications. Major advancement in solar cell research was made by Russell Shoemaker Ohl at Bell Laboratories in the year 1940 when he inadvertently created a p-n junction and observed a flow of current upon illumination. Thirteen years later in 1953, Bell laboratories demonstrated the first practical solar cell using Si with almost 6% efficiency **[236]**. Later, in April 25 1954, Bell Labs officially announced the invention of solar cell by using such converted energy to power a small toy and *The New York Times* published the news as "*may mark the beginning of a new era, leading eventually to the realization of one of mankind's most cherished dreams-the harnessing of the almost limitless energy of the sun for the uses of civilization*".

Next few years have witnessed tremendous progress in the performance improvement of solar cell efficiency from 6% to ~20%. In 1958, following the improvements in power conversion efficiency (PCE), solar energy was used for the first time in space. The first highly efficient GaAs heterostructure based solar cell was reported with ~17% PCE in 1970 by Zhores Alferov **[237]**. In the late 1970s, several remarkable achievements were made such as, the solar cell research institute was established in Colorado (1977), production of photovoltaic cells exceeded 500 kW in the world (1977) and first solar power calculators were made (1978).

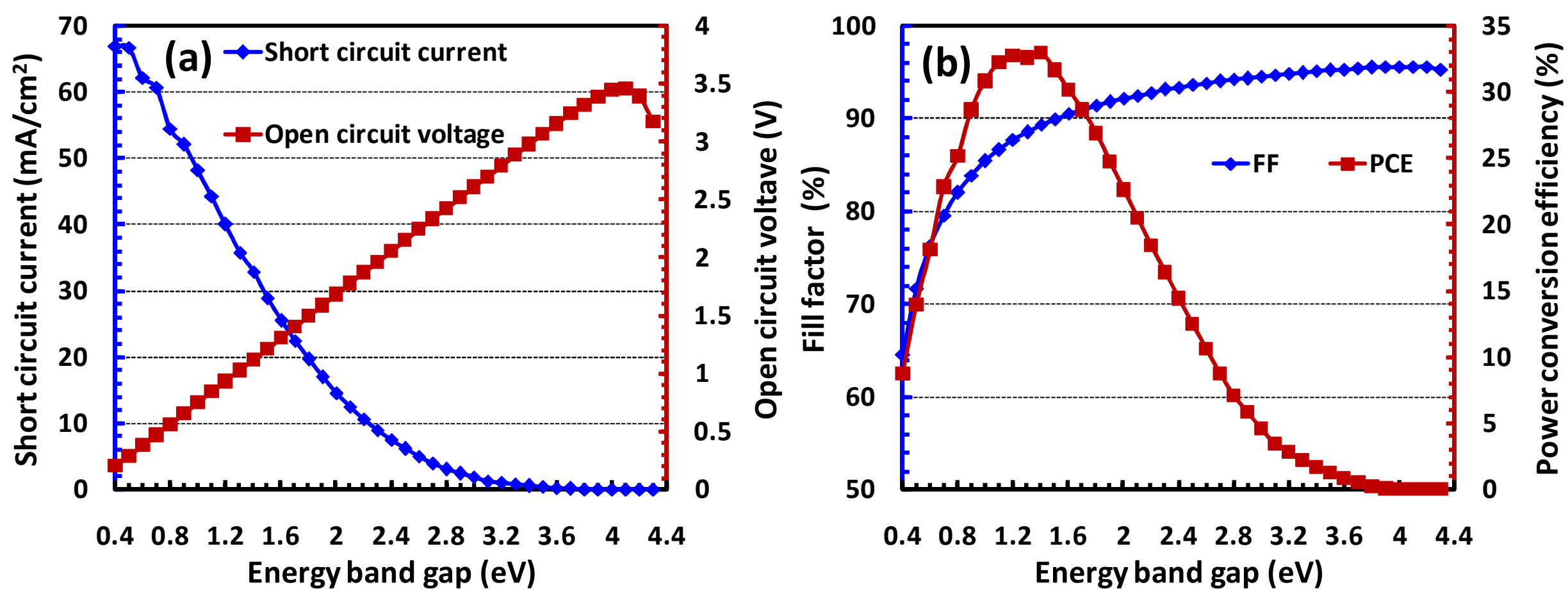

Fig. 2.2. Plot of single junction solar cell performance parameters with band gap of the corresponding material: (a) $J_{sc}$ and $V_{oc}$, and (b) FF and PCE **[238]**.

In 1989, Blakers *et. al.* demonstrated the Si-based passivated emitter and rear (PERC) devices for solar cell applications and reported a PCE of 22.8% with open circuit voltage ($V_{oc}$ = 696 mV), short circuit current ($J_{sc}$ = 40.3 mA/cm$^2$) and fill factor (FF = 81.4%) improved by a large margin **[239]**. It was already predicted that the traditional single-junction solar cell with optimal energy band gap for the solar spectra can achieve maximum efficiency of 33.16% according to the Shockley-Queisser limit. The variation of such conversion efficiency and other solar cell performance parameters is plotted in Fig. 2.2 (a-b) with energy band gap of the corresponding material. The observations of Fig. 2.2 indicate that the four inter-linked parameters are to be optimized for finding a cost-effective solution for large scale production.

Multi-junction solar cells are widely investigated to achieve power conversion efficiency beyond the Shockley-Queisser limit. The vertically stacked devices are designed in such a way that the blue photons are absorbed by top junction and red photons by the bottom junction. It is to be mentioned that such multi-junction solar cells have been studied since 1960s **[240]** however, until late 1980s, high efficiencies could not be achieved due to the unavailability of matured fabrication process.

The high efficiencies, in the range of ~30% to ~50%, have been achieved by developing the double- to six-junction tandem solar cells **[241]**. The higher efficiencies depend on several issues, such as, the selection of material, relevant material engineering to achieve the required band gap, transparency of the corresponding materials, conductivity of the transparent metal contacts and most importantly, the integration process of different materials. Such engineering approach has attracted significant research interest of the material and device scientists to optimize the solar cell performance parameters depending on the specific requirements **[241]**.

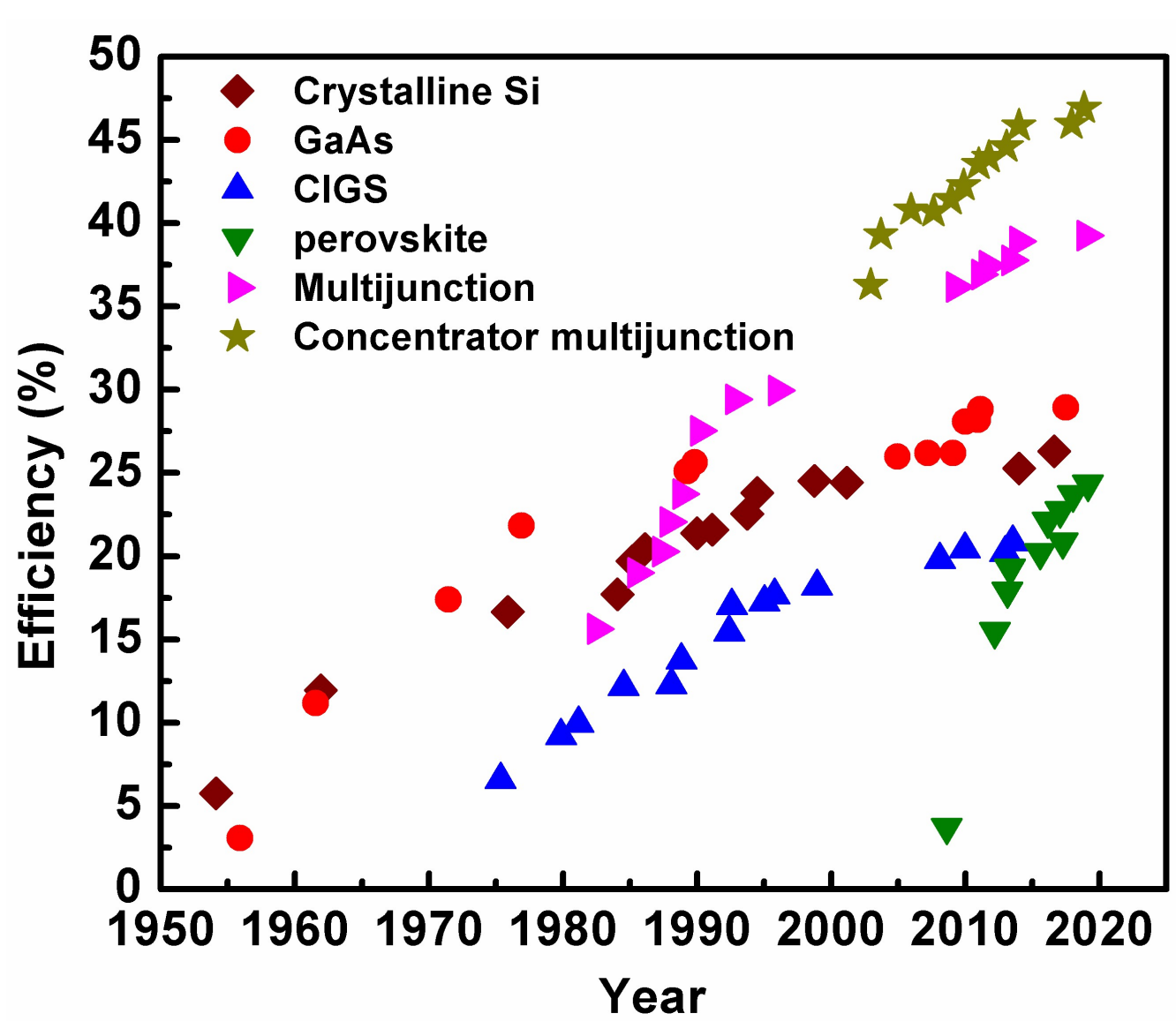

Fig. 2.3. Chronological improvement of power conversion efficiency of single and multi-junction

solar cells **[241]**.

The chronological improvement of PCEs over the years for commonly used semiconductors is shown in Fig. 2.3.

Semiconductor nanowires have emerged in the last two decades as the new class of materials for performance improvement of photovoltaic devices. The group III-V semiconductor nanowires are primarily investigated for solar cell applications to achieve high power conversion efficiency. The conversion efficiency of such nanowire solar cells has experienced a significant improvement from 5% in early 2000s to 15% in the late 2010s. Solar cells using GaAs and InP nanowire array have been fabricated to achieve relatively higher efficiencies of ~15% **[242, 243]**. Nanowire geometry of other materials has also been investigated for solar cell applications such as, Si, CdS, CdTe, $CuS_2$, perovskites and III-nitrides **[244-251]**. Designing the nanowires for efficient light absorption, charge carrier separation and collection are the important aspects to be investigated for achieving higher PCE. The geometry dependent absorption characteristics of nanowires have been very crucial for optimizing the solar cell performance parameters **[252-254]**. Most importantly, the absorption cross-section of such nanowires has been reported to be relatively higher than its physical cross-section **[255]**. Optical modes in such nanowires strongly depend on the nanowire diameter and thus wave-optics plays a crucial role to understand its absorption characteristics. The resonance-controlled absorption enhancement is considered to be the prominent design rule to optimize the solar cell performance parameters **[37]**. Such absorption enhancement is achieved by engineering the diameter of nanowires to create resonance between the mode of incident light with nanowire waveguide mode. The impact of other geometrical parameters of the nanowires including its length and pitch, on the light absorption properties, has been studied to find an optimum solution for specific applications **[256]**.

Alongside the absorption properties, charge carrier separation and collection in such nanowire based solar cells demand special attention to achieve improved performances. Such separation-collection process significantly depends on several parameters such as, the crystal structure, surface passivation, junction geometry, cleaning processes and formation of top/bottom contacts. Presence of mixed crystal phases (zincblende segments in wurtzite, twin planes in zincblende, stacking fault in wurtzite) in nanowires are reported to create scattering centers which adversely affects the electrical transport characteristics and degrade the PCE **[257-260]**. Consequently, the nanowires with high phase-purity are grown by employing several ‘bottom-up’ techniques to circumvent such issue **[261-263]**. Junction geometries in the nanowire solar cells also play an important role to tailor the carrier separation and transportation properties. Radial and axial p-n junctions are commonly fabricated to meet the criteria for efficient conversion of sunlight into electrical energy. Kayes *et. al.* have shown by developing a theoretical model that the radial p-n junctions have the benefit of charge carrier generation and separation since generation happened along the longitudinal direction and separation along the transverse direction whereas, such processes occurred in the same direction for axial p-n junctions **[264]**. Apparently, the radial distance is much smaller than (or comparable to) the minority carrier diffusion length and hence it reduces the recombination possibilities to lead to increase the solar cell power conversion efficiency **[255]**. However, the fabrication of such radial junctions comprises of several process complexities in comparison to the axial structures. The open circuit voltage has also been reported to be strongly affected in radial junction devices due to the presence of defects/traps in the quasi-neutral region **[264]**. The short-circuit current of radial junction devices are comparable to the axial ones however the fill factor and power conversion efficiency values are significantly lower in comparison to the axial junction devices **[256, 265]**. Thus, the vertical configuration of nanowire devices seems to be the most promising for the harvesting of large amount of solar energy. The performance of some of the commonly used axial and radial devices is listed in Table 2.2.

Table 2.2. Performance comparison of some typical axial and radial nanowire based solar cells.

| Nanowire | Patterning | Substrate | Growth | Junction | Jsc ($mA/cm^2$) | Voc (V) | FF (%) | PCE (%) | Ref. |
|---|---|---|---|---|---|---|---|---|---|
| GaAs | EBL | GaAs | MOVPE | Radial p-n | 17.6 | 0.39 | 37 | 2.5 | **[266]** |
| GaAs | EBL | Si | MBE | Radial p-i-n | 18.2 | 0.39 | 46.5 | 3.3 | **[267]** |
| GaAs | EBL | GaAs | MOVPE | Radial n-i-p | 18.9 | 0.57 | 69.0 | 7.4 | **[268]** |
| GaAs | EBL | GaAs | MOVPE | Axial n-i-p | 21.1 | 0.57 | 63.7 | 7.6 | **[269]** |
| InP | EBL | InP | MOVPE | Radial | 11.1 | 0.67 | 58.5 | 4.2 | **[265]** |
| | | | | p-n | | | | | |
| InP | NIL | InP | MOVPE | Axial p-n | 21.0 | 0.73 | 73.0 | 11.1 | **[270]** |
| InP | NIL | InP | MOVPE | Axial p-i-n | 24.6 | 0.78 | 72.4 | 13.8 | **[243]** |
| InGaAs | --- | Graphene | MOVPE | Radial p-n | 17.2 | 0.26 | 55.3 | 2.5 | **[271]** |
| GaAs on Si | UV-lithography | Si | MOVPE | Axial n-i-p | 20.6 | 0.96 | 57.8 | 11.4 | **[272]** |

It is observed from Table 2.2. that despite of having intense research on the performance improvement of nanowire solar cells, the power conversion efficiencies are still significantly lower than the Shockley-Queisser limit. The issues related to thermodynamic efficiency losses in such devices need to be systematically addressed to achieve efficiencies beyond the conventional limit. In this regard, efforts are suggested on the: i) reduction of deficit between the energy bandgap and electron-hole quasi-Fermi level splitting, and ii) minimization of thermal and absorption losses of the sub-bandgap energies of incident light **[37]**. The energy difference between electron and hole quasi-Fermi levels decides the value of open circuit voltage which needs to be maximized for achieving higher efficiencies. Several efficient light trapping strategies are adopted by using nanostructures for achieving higher PCEs **[37]**. Consequently, controlling the propagation of light inside the solar cells has emerged as the new engineering approach to achieve desired efficiency. Light management inside the solar cell can be achieved by engineering the density of states of the corresponding nanowires. Such engineering approach is predicted to achieve efficiencies in the range of 50 % to 70% **[37]**. Krogstrup *et. al.* in 2013 demonstrated a GaAs nanowire based p-i-n solar cell with efficiency beyond the Shockley-Queisser limit **[255]**. In such a structure the high efficiency of 40% is achieved by enhancing the absorption cross-section of several folds than the physical cross-section of the nanowire by utilizing the resonances similar to Mie resonances.

Semiconductor quantum dots have emerged as the potential candidates for next-generation solar cells since the size dependent tunable properties make them suitable to efficiently absorb light of a wider solar spectrum. The multiple exciton generation (MEG) mechanism of such quantum dots is expected to surplus the conventional single junction efficiency limit. In this context, quantum dots are attempted to integrate in various device architectures such as, Schottky quantum dot solar cells, depleted hetero-junction solar cells, thin absorber solar cells, hybrid polymer solar cells and organic-inorganic hetero-junction solar cells for performance investigation of its photovoltaic property **[273-279]**. In early 1990s, Vogel *et. al.* and Weller *et. al.* demonstrated the incorporation of CdS and PbS quantum dots in $TiO_2$ to study photovoltaic properties of the devices **[280]**. Consequently, several semiconductor quantum dots are utilized to fabricate devices for solar cell applications with high quantum yield reaching to ~80% **[281-283]**. Generally, two different strategies are developed, such as, direct growth of quantum dots in the semiconductor (known as, in-situ sensitization) and deposition of pre-sensitized quantum dots inside the semiconductor (known as, ex-situ sensitization) for the performance investigation of quantum dot solar cells.

In such devices the choice of quantum dot material plays an important role of injecting the electrons towards electron transport layer and the holes to the hole transport layer. Optimization of electron and hole transport layers in terms of band alignment, photo-stability and conductivity is essential since these layers are responsible for efficient transmission of photogenerated carriers to the metal contacts. $TiO_2$, ZnO, $SnO_2$, $Zn_2SnO_4$ are commonly used for electron transport layers and iodide electrolyte, Polysulfide electrolyte, Ferrocene electrolyte, Cobalt based electrolyte, Spiro-MeOTAD are explored for the purpose of transportation of holes **[284-301]**. The interfaces of different layers also play a critical role in determining the performance of such quantum dot solar cells. Recombination processes take place at the transparent conducting oxide/electron transport material interface, electron transport material/quantum dot interface and quantum dot/hole transport layer interface which need a systematic engineering approach to reduce the probability of loss photogenerated carriers. Several methodologies are adopted for successful manipulation of photo-carrier transport processes by controlling the recombination probabilities **[302-306]**.

Most importantly, strong 3D-confinement in quantum dots makes the dynamics of photocarrier generation interesting by giving an opportunity to generate more than one electron hole pair at the cost of a single photon. The multiple exciton generation phenomenon in quantum dots has been widely studied for achieving internal quantum efficiency greater than 100% **[53, 54]**. The requirement of appropriate energy (several times the energy band gap) to produce such phenomenon is the subject for further investigation. Reports are available on estimating the threshold energy to produce multiple excitons on the basis of exciton lifetime, Coulomb interaction and impact ionization **[307, 308]**. In future, focused efforts are required to develop CMOS compatible quantum dot based solar cells for large scale implementation with efficiency beyond the conventional limit. Performance comparison of some commonly used quantum dot solar cells is shown in Table 2.3 in terms of quantum dot absorbers, electron transport materials and hole conductors.

Table 2.3. Comparative summary of performances for some of the quantum dot based solar cells.

| **Based on the performance of absorption of the QDs** | | | | | | |
|---|---|---|---|---|---|---|
| **QD absorber** | **Cell configuration** | **$J_{sc}$ ($mA/cm^2$)** | **$V_{oc}$ (V)** | **FF (%)** | **PCE (%)** | **Ref.** |
| CdS | $FTO/TiO_2/CdS/ZnS/S^{2-}$-$Sn^{2-}$/FGO-$Cu_2S$/FTO | 7.2 | 0.496 | 46 | 1.63 | **[309]** |
| CdS-Mn | $FTO/TiO_2$/CdS-Mn/ZnS/$S^{2-}$-$Sn^{2-}$/FGO-$Cu_2S$/FTO | 8.9 | 0.583 | 49 | 2.52 | **[309]** |
| CdSe | $TiO_2$/CdSe-MPA/ZnS/ $S^{2-}$-$Sn^{2-}$/$Cu_2S$/Brass | 16.96 | 0.561 | 56.6 | 5.42 | **[299]** |
| CdS/CdSe | $TiO_2$/CdS/CdSe/ZnS/ $S^{2-}$-$Sn^{2-}$/Pt/FTO | 18.23 | 0.489 | 54 | 4.81 | **[310]** |
| CdTe/CdSe | $TiO_2$/CdTe/CdSe/ZnS/ $S^{2-}$-$Sn^{2-}$/$Cu_2S$/Brass | 19.59 | 0.606 | 56.9 | 6.76 | **[311]** |
| CdSeTe | $TiO_2$/CdSeTe/$TiCl_4$/ZnS/ $S^{2-}$-$Sn^{2-}$/$Cu_{2-x}S$/Brass | 20.69 | 0.700 | 62.2 | 9.01 | **[312]** |
| ZnTe/CdSe | $TiO_2$/ZnTe/CdSe/ZnS/ $S^{2-}$-$Sn^{2-}$/$Cu_{2-x}S$/Brass | 19.35 | 0.646 | 55.1 | 6.89 | **[313]** |
| Zn-Cu-In-Se | $TiO_2$/Zn-Cu-In-Se/ZnS/ $S^{2-}$-$Sn^{2-}$/MC/Ti | 25.18 | 0.742 | 62.4 | 11.66 | **[284]** |
| **Based on the electron transport layer (ETL) performance** | | | | | | |
| **ETL** | **Quantum dot** | **Jsc ($mA/cm^2$)** | **Voc (V)** | **FF (%)** | **PCE (%)** | **Ref.** |
| $TiO_2$ | Zn-Cu-In-Se | 25.25 | 0.739 | 62.2 | 11.61 | **[284]** |
| $TiO_2$ | CdS | 12.64 | 0.41 | 42 | 2.16 | **[285]** |
| $TiO_2$ | PbS | 12.94 | 0.43 | 56 | 3.11 | **[286]** |
| $TiO_2$ | CdS/CdSe | 13.85 | 0.54 | 54 | 4.05 | **[287]** |
| ZnO | CdS/CdSe | 15.42 | 0.62 | 49 | 4.68 | **[288]** |
| ZnO | CdS/CdSe | 10.74 | 0.61 | 50 | 3.28 | **[289]** |
| ZnO | Mn-CdSe | 12.6 | 0.74 | 44 | 4.64 | **[290]** |
| ZnO | PbS | 17.9 | 0.60 | 40 | 4.3 | **[291]** |
| $SnO_2$ | CdS | 11.56 | 0.6 | 43 | 3.00 | **[292]** |
| $Zn_2SnO_4$ | CdS/CdSe | 11.32 | 0.49 | 37 | 2.08 | **[293]** |
| Zinc titanate | CdS/CdSe | 5.96 | 0.59 | 56 | 1.95 | **[294]** |
| Stransium titanate | CdS | 1.53 | 0.76 | 67 | 0.78 | **[295]** |

**Based on the hole transport layer (HTL) performance**

| **HTL** | **Cell configuration** | **Jsc ($mA/cm^2$)** | **Voc (V)** | **FF (%)** | **PCE (%)** | **Ref.** |
|---|---|---|---|---|---|---|
| Iodine/Iodide electrolyte | $TiO_2/CdS_6Se/S^{2-}$-$Sn^{2-}$/Pt/FTO | 1.38 | 0.565 | 34 | 0.32 | **[296]** |
| Cobalt electrolyte | $TiO_2/CdSe_5Te$/cobalt electrolyte /Pt/FTO | 4.94 | 0.67 | 54 | 4.18 | **[297]** |
| Polysulfide electrolyte | $TiO_2$/Zn-Cu-In-Se/ZnS/$S^{2-}$-$Sn^{2-}$/MC/Ti | 25.18 | 0.742 | 62.4 | 11.66 | **[284]** |
| Ferrocene electrolyte | $TiO_2$/CdS/ZnS/Ferrocene/Pt/FTO | 2.45 | 0.68 | 60 | 1.0 | **[298]** |
| Spiro-MeOTAD | $TiO_2/CdSe_5Te$/spiro-MeOTAD /Pt/FTO | 2.15 | 0.70 | 55 | 0.84 | **[299]** |
| Poly 3-hexylthiophene | $TiO_2$/CdS/P3HT/Au/Polymer | 4.31 | 0.67 | 55 | 1.42 | **[300]** |
| Dextran-polysulfide gel | $TiO_2$/CdS/CdSe/ZnS/dextran-$S^{2-}$-$Sn^{2-}$/Pyrolized-Pt | 15.86 | 0.466 | 44 | 3.23 | **[301]** |

### 2.1.3. Development of light emitters (LEDs and lasers): Brief history

The journey of light emitting devices started in 1960s when Holonyak *et. al.* first successfully demonstrated the GaAsP-alloy based p-n (homo-) junction red LEDs through spontaneous emission **[314]**. On the other hand, the stimulated emission was predicted by Albert Einstein in 1917 in the famous article "*Zur Quantentheorie der Strahlung*" (*i.e.*, On the quantum theory of radiation) **[315]**. In 1961, Bernard and Duraffourg, for the first time, derived the condition for obtaining lasing action by such stimulated emission of radiation in the semiconducting materials **[316]**. The following year has witnessed three reports on the fabrication of GaAs lasers where its direct band gap played the key role **[317-319]**. Such lasing devices have significantly higher threshold current densities and therefore it can only be appropriately operated in the pulsed form. To overcome such limitation, intense efforts have been made to fabricate the continuous-wave (CW) semiconductor lasers **[320]**. During this time, there has been a remarkable progress in the growth of heterostructures using liquid phase epitaxy (LPE) technique which enabled the growth of high quality multi-junction (for *e.g.,* $AlGa_{1-x}As_x$/GaAs/$AlGa_{1-x}As_x$) heterostructures. Such devices have two key advantages over the homojunction structures such as: a) the lower band gap of GaAs confines carriers and thus enhances the possibility of recombination across the band gap, and b) the relatively higher refractive index of GaAs than $AlGa_{1-x}As_x$ layers confines the light more efficiently. Such advantages led to generate the room temperature CW-mode semiconductor lasers for the first time with almost two orders of reduced threshold current density **[321, 322]**. In 1980s, the single mode lasers have been introduced as the distributed feedback lasers for long haul optical communications **[323]**. Vertical cavity surface emitting laser (VCSEL) has also been demonstrated in 1980s for low-cost single mode lasing applications, primarily for the 1.55 µm telecommunication wavelength **[324]**.

In early 2000's, the domain of nanowire optoelectronics/photonics has experienced a significant progress with the demonstration of room temperature ultraviolet lasing action from ZnO nanowires **[325]**. Following such progress, nanowires were used to develop sub-wavelength optical waveguides, tunable coherent source of visible light and electrically driven lasers **[326-330]**. Moreover, the light extraction efficiency of such nanowire based devices has been reported to be ~ 5 times higher than the thin film counterparts **[331, 332]**. Further, the core-shell structures have gained enormous interest due to its relatively higher active area and the possibility of obtaining various emission wavelengths of LEDs by tuning the active region material composition which also leads to achieve relatively lower onset voltage, superior rectification and higher internal quantum efficiency **[333-337]**. However, the axial LED structures are favorable for creating contacts since active region is far away from the contacts and such geometry is beneficial due to its controlled carrier injections into the active region **[338-340]**.

The one-dimensional columnar structure of nanowires is advantageous for lasing applications due to the difference of refractive indices between the nanowire and its surrounding dielectric medium which promotes photonic confinement in such nanowires **[330]**. The atomically smooth sidewalls ensure the photons to propagate along the nanowire axis without suffering any losses. The end facets of such nanowires are also reported to act as the Fabry-Perot type cavity **[341, 342]**. The semiconductor nanowires can perform simultaneously as a waveguide, optical cavity and gain media which is indispensable for light amplification **[343]**. Thus, the smaller device dimension, higher optical gain and stronger carrier confinement of nanowires suggest these to be the potential candidates for fabricating nanowire based lasing devices with higher efficiency, lower threshold, less power consumption, high packing density and high modulation speed. Despite of such commendable progress it is worthy to mention that the associated diffraction limit restricts scaling down of the nanowire dimension below half of its operating wavelength and thus, it indicates the need of advanced optical confinement schemes for the miniaturization of nanowire dimension into the deep sub-wavelength range **[344]**.

The introduction of vapor phase epitaxy (VPE) and molecular beam epitaxy (MBE) techniques enabled the growth of very thin (~100 Å) layers of material which in turn provided the reliable route to realize quantum confinement effects. The quantum well heterostructures improved the lasing performance by increasing carrier confinement, decreasing the line width and showing composition dependent wavelength tunability **[345]**. In the mean time, it was predicted that the 3-dimensional confinement of charge carriers in the active region could improve the device performance in comparison to the quantum well based devices **[346]**. Such quantum dot devices are very important, particularly in reducing the threshold current density and increasing thermal stability **[347]**. In such devices, carrier (*i.e.*, electron) cascades through several fabricated quantum wells, which give rise to many photons per injected electron. The quantum cascade lasers have also been fabricated for near-infrared emission by choosing suitable materials **[348]**.

In laser applications, the 3-dimensional confinement of carriers in such quantum dots congregates the injected carriers within a relatively narrower energy range in comparison to the bulk devices, and it enables to attain a much steeper dependence of optical gain on the injection current **[349]**. The corresponding delta-function type local density of states (LDOS) and the relatively higher inter sub-band gap of quantum dots significantly suppress the temperature dependence of quantum dot lasers. Moreover, the localization of charge carriers in quantum dots significantly reduces surface recombination which in turn gives an advantage to the fabrication of quantum dot laser with high output power **[350]**. After the first recognition of such fundamental advantages of quantum dot lasers, it is experimentally realized for the first time in 1994 **[351]**. In 1998, the 1.3 µm quantum dot laser is fabricated by using a single layer of $In_{0.5}Ga_{0.5}As$ **[352]** and consequently in 1999, dot-in-a-well structure is demonstrated to achieve the near-1.3 µm emission **[353]**. Such reports indicate a relatively lower threshold current density for quantum dot lasers which is appropriate for optical interconnects with lower power consumption. Subsequent years have witnessed numerous reports on the development of several quantum dot based device structures with improved threshold current density, output power, gain, reliability, dynamic characteristics and lifetime **[354-359]**.

## 2.2. Modeling of nano-optoelectronic devices: From drift-diffusion model to NEGF formalism

Many years of intense research in the domain of semiconductor device physics has led to the development of mathematical models for describing the operation of optoelectronic devices such as, photodetectors, solar cells, LEDs and lasers. The models comprise a set of fundamental equations namely, Poisson’s equation, continuity equations and transport equations. Poisson’s equation relates the variation of electrostatic potential to charge density where the continuity and transport equations describe the variation of charge density as a result of generation-recombination processes. The generation-recombination processes under illumination are modeled by developing appropriate guiding equations to predict the performance of various optoelectronic devices. Such equations help to understand qualitative and quantitative behavior of such semiconductor optoelectronic devices. The fundamental and other important guiding equations are summarized in Table 2.4.

Table 2.4. Fundamental and other important guiding equations of semiconductor devices **[360]**.

| | | |
|---|---|---|
| 1. | Maxwell's equations | $\nabla \times E = -\frac{\partial B}{\partial t}$, $\nabla \cdot B = 0$, $\nabla \times H = J + \frac{\partial D}{\partial t}$, $\nabla \cdot D = \rho$ |
| 2. | Poisson's equation | $\frac{d^2\phi}{dx^2} = \frac{q}{\varepsilon_s}(p - n + N_d - N_a)$ |
| 3. | Continuity equations | $\frac{\partial n}{\partial t} = \frac{1}{q}\nabla . J_n + G_n - R_n$, $\frac{\partial p}{\partial t} = \frac{1}{q}\nabla . J_p + G_p - R_p$ |
| 4. | Fermi-Dirac and Bose-Einstein statistics | $f_{FD}(E) = \frac{1}{1 + \exp\left(\frac{E - E_F}{kT}\right)}$, $f_{BE}(E) = \frac{1}{\exp\left(\frac{E}{kT}\right) - 1}$ |
| 5. | Drift-diffusion model | $J_n = qn\mu_n E_n + qD_n \frac{dn}{dx}$, $J_p = qp\mu_p E_p - qD_p \frac{dp}{dx}$ |
| 6. | Shockley-Read-Hall recombination | $R_{SRH} = \frac{pn - n_{ie}^2}{\tau_p\left[n + n_{ie}\exp\left(\frac{E_{tr}}{kT_L}\right)\right] + \tau_n\left[p + n_{ie}\exp\left(\frac{-E_{tr}}{kT_L}\right)\right]}$ |
| 7. | Auger Recombination | $R_{Auger} = C_n\left(pn^2 - nn_{ie}^2\right) + C_p\left(np^2 - pn_{ie}^2\right)$ |
| 8. | Beer-Lambert law | $A = \varepsilon cl$ |
| 9. | Electron-hole generation rate | $G_e(x) = \Phi_0 \alpha \exp(-\alpha x)$ |
| 10. | Shockley equation | $I = I_s\left[\exp\left(\frac{qV}{\eta kT}\right) - 1\right] - I_L$ (for p-n junction solar cell) |

The traditional drift-diffusion approach has been well-established for the analytical modeling of optoelectronic devices to predict the performance with adequate results **[111-116, 360]**. Reports are available on appropriate estimation of the device performances in terms of optical generation-recombination and transport of electron-hole pairs within the drift-diffusion formulation **[361]**. However, the drift-diffusion model is not efficient to model hot-carrier effects and non-local effects (*e.g.*, the velocity overshoot) for the nano- /quantum- scale devices. In this context, the hydrodynamic transport model, energy transport model and a six-moments model are proposed to circumvent the limitations **[362, 363]**. However, the six-moments model has a large number of adjustable parameters which makes the approach relatively complex and cumbersome. On the other hand, compact mobility models are capable of finding the appropriate mobility values, however, requires precise methods for model validation and calibration which leads to the manifestation of Monte Carlo technique **[364-367]**. The Monte Carlo method is based on the phase space simulation constituted of the corresponding position and momentum coordinates and such method deals with the free motion of carriers unless disturbed by scattering phenomenon. The origin of such scattering, duration of free paths between two successive scattering events and post-scattering states are selected on the basis of given probability distribution functions. The generation of electron-hole pairs upon illumination can be incorporated in the Monte Carlo method by using other mathematical models **[368]**. In such models, it is assumed that photons are equally absorbed at the surfaces and exponentially absorbed along the depth of the device. The entire trajectory of optically generated carriers until reaching the contacts is considered in the Monte Carlo method to obtain the device photocurrent. However, such approach is fundamentally based on the thermodynamic statistical distribution which cannot be defined for nano-/quantum-structures since there exist only few numbers of atoms.

It is apparent from the discussions of Chapter 1 (section 1.5) that the NEGF formalism has emerged as the most versatile and powerful tool to study the performance of nano- /quantum-scale optoelectronic devices. Such NEGF formalism has been introduced by Prof. S. Dutta in the year 2000 for the modeling of nanoscale devices **[117]**. This pioneering work primarily focused on the dialectics between 19$^{th}$ century physics (i.e., electrostatics) and 20$^{th}$ century physics (i.e., quantum physics) to determine the transport properties of nano- /quantum- scale devices. Consequently, researchers are motivated to implement it in their work and thus, several nanoscale devices are investigated to predict electronic and optoelectronic performance of the particular device **[121-127]**. It is worthy to mention that the NEGF formalism was first proposed by P. C. Martin and J. Schwinger in 1959, G. Baym and L. P. Kadanoff in 1962, and separately by L. V. Keldysh in 1965 **[369-371]**.

The model was first developed as a generic approach by R. Lake *et. al.* in 1997 **[372]**. In such model, a device Hamiltonian is first developed to describe the impact of reservoir and the phase breaking processes (electron-photon and/or electron-phonon interaction in optoelectronic devices) on the device characteristics. Such impact of reservoir and photons/phonons can be incorporated by considering relevant self-energies. Scattering events (*e.g.*, electron-photon) in such devices are generally addressed by considering a chain of ballistic nanowires where the connecting junctions of such chain are considered as the scattering centers **[373-375]**. The concept of Büttiker probes has been proposed to study the device performance in presence of such scattering events. It is conceptualized with similar analogy of reservoir contacts where Büttiker probes can be assumed as 'special' contacts that extract carriers from the device and reinject those with a different distribution of energy-momentum states. In such nano- /quantum-scale devices, the equations of motion (involving all the scattering events) are solved by NEGF formalism considering second quantization field operators for the corresponding carriers to obtain charge density and device current.

The NEGF formalism has been successfully utilized for the modeling of optical transition processes (including charge carrier generation and recombination) and carrier transport properties of nano- /quantum- scale optoelectronic devices **[117, 118, 121-127]**. Photosensing and photovoltaic performances of such devices can appropriately be predicted by employing the NEGF formalism **[122-124, 126, 127]**. Also, the phonon-mediated optical transition process in semiconductor nanostructures has been described by following the NEGF formalism **[125]**. In recent times, the NEGF formalism is utilized to study the multiple exciton generation (MEG) process which generally occurred in quantum dot based optoelectronic devices **[376, 377]**.

It is worthy to be mentioned that in quantum scale devices the photogeneration current is dominated by the presence of resonance phenomenon which increases the absorption cross-section by several fold in comparison to the physical cross-section of quantum devices. Such phenomenon cannot appropriately be modeled by the conventional Beer-Lambert's law since it considers linear dependence of absorbance with the optical path of incident light. Also, it has already been discussed in Chapter 1 (section 1.7) that the charge carriers are confined in transverse directions in a nanowire based device which makes the energy spectra in such directions discrete and thereby forming different modes of sub-bands. The energy sub-bands/discrete states play a crucial role in determining the performance of nano- /quantum- scale optoelectronic devices since multiple transitions are occurred upon illumination due to the presence of such discrete quantum states. It is essential to incorporate the multiple transition effect in the developed theoretical model to obtain photocurrent/photocapacitance of the considered device.

## 2.3. Summary

In this chapter, the journey of optoelectronic devices from bulk to quantum structures has been discussed with an emphasis on the chronological development of such devices. The advantages and applicability of semiconductor nanowires to develop various device structures for performance improvement have been cited. With the sustained miniaturization of device structures, semiconductor quantum dots have eventually emerged as the basic unit for the devices of next generation due to their unique properties. Significant contributions in the domain of novel devices using quantum dots for various optoelectronic applications have also been discussed in detail.

It is apparent from the overall discussions that the semiconductor quantum dot based devices are the most feasible choice for developing next-generation high performance optoelectronic devices. Also, the progress of modern fabrication technology enables to go beyond the 'proof-of-concept' based devices to large scale production. However, despite of remarkable progress in the application domain of such devices, there are still significant challenges to fabricate CMOS compatible quantum dot based devices with desired dimensions and improved performances. In the subsequent chapters a novel nanowire-MOS quantum dot device has been designed and modeled to investigate its electronic and optoelectronic properties by using NEGF formalism and finally such device has been fabricated and electrically characterized.

# Chapter: 3*

# Electrostatics of the vertical nanowire metal-insulator-semiconductor (MIS) structures

## 3.1. Introduction: From horizontal to vertical device architecture

The metal-insulator-semiconductor (MIS) structures are the key of modern semiconductor technology due to their multifaceted applications in the domain of electronics and optoelectronics **[378-387]**. The emergence and implementation of nanostructures in the fabrication of advanced devices have created further opportunity to explore the novel physical phenomena. In this context, the nanowire MIS structures appear to be an extremely promising choice for the wide spread technological applications including non-volatile memories **[388]**, gas sensors **[389]**, photodetectors/multispectral imagers **[390]** and solar cells **[391]**. Significant efforts are being made on the fabrication of such nanowire based devices by adopting 'bottom-up' and 'top-down' approaches **[67-69, 70-72, 109, 99-101, 102-105]**, either to grow the vertical structures or to place those nanowires horizontally on the desired location of a substrate.

However, it is observed from several reports that the fabrication techniques using nanowires along the horizontal direction adopt 'pick-and-place' approaches which are extremely complex, contamination prone and limited by reproducibility **[392, 393]**. In comparison, the vertical device integration scheme using nanowires encompasses relatively less process complexities with superior control since it is their natural growth direction **[394-397]**. Moreover, such vertical configuration with relatively higher areal density of the nanowires is reported to be compatible with both the 'top-down' and 'bottom-up' approaches **[393, 398, 399]**. An array of such vertical Si-nanowire metal-oxide-semiconductor (MOS) capacitors has recently been fabricated with precise control over their location, diameter, height and areal density of the nanowires **[400]**.

Therefore, the vertically aligned nanowires can practically be the feasible choice to realize future generation 'working devices' for nanoelectronics in comparison to the horizontal structures. Hence, the conventional approach of wrap-gate structure for MIS devices is possibly required to

***Most of the content of Chapter: 3 is published in the article "Understanding the electrostatics of top-electrode vertical quantized Si nanowire metal–insulator–semiconductor (MIS) structures for future nanoelectronic applications", by Subhrajit Sikdar, Basudev Nag Chowdhury and Sanatan Chattopadhyay, 'Journal of Computational Electronics' 03/2019, DOI:10.1007/s10825-019-01321-7***

change or re-design to a vertical nanowire with top-electrode structures. Such a vertical device architecture has already been experimentally demonstrated **[390, 400]** for optoelectronic applications. The device structures using vertical nanowires comprise of the deposition of an insulating layer on top of the nanowire, followed by a subsequent formation of metal contact on top of such insulating layer. Generally, Si-technology uses silicon dioxide ($SiO_2$) as the natural insulating layer for its process compatibility, thermodynamic feasibility and interfacial reliability. However, as the thickness of such $SiO_2$ layer is thinned to atomic scale for maintaining electrostatic integrity in the nanoscale MOS devices, its insulating nature is seriously challenged by the enormous amount of tunneling-assisted leakage current **[401]**. In other words, electrostatic integrity of the system with very thin oxide layer ($SiO_2$) is lost. In the state-of-the-art complementary-metal-oxide-semiconductor (CMOS) technology such limitations are overcome by replacing gate $SiO_2$ with an insulating material of high dielectric constant (high-*k*). The use of such high-*k* materials effectively increases the oxide layer thickness by introducing the concept of 'equivalent oxide thickness' (EOT) to achieve similar electrostatic performance **[402]** where the thickness of such high-*k* materials is obtained as, $t_{high-k} = EOT\left(k_{High-k} / k_{SiO_2}\right)$.

## 3.2. Concerns regarding utilizing the conventional 'high-*k*' technology in nanowire vertical MIS structures to maintain electrostatic integrity

The fundamental concept of using high-*k* gate dielectric materials (depending on an identical EOT) in MIS devices has been developed on the assumption of linear relationship between the insulator thickness and its dielectric constant by considering the model of a parallel-plate capacitor for such structures. In such model, generally for the bulk structures, the oxide and semiconductor capacitors are assumed to be in series combination and the insulator capacitance dominates since it is relatively smaller than the semiconductor capacitance. However, due to the quantization effect in nano-/quantum-wire MIS capacitors, the semiconductor capacitance is actually smaller than the insulator capacitance and thus the impact of such semiconductor capacitance is too small to neglect **[403]**. Hence, the linear relationship between the insulator dielectric constant and insulator thickness (as per the conventional EOT formula) is required to be modified to incorporate the effects of quantum confinement in the quantum-scale devices. Also, at this point it is essential to further investigate the design issues relevant to maintain the electrostatic integrity in such low dimensional devices using vertical quantum wires.

In this chapter, electrostatic control of the vertically oriented top-electrode Si-nanowire MIS devices is studied in detail by considering nanowire diameter within the limit of excitonic Bohr radius (EBR) of Si (*i.e.*, 5 nm) **[404]**. Several commonly used dielectric materials, with different dielectric constant (*k*) values, such as silicon dioxide ($SiO_2$) (*k*=3.9), silicon nitride ($Si_3N_4$) (*k*=7.0), aluminum oxide ($Al_2O_3$) (*k*=9.0), yttrium oxide ($Y_2O_3$) (*k*=15.0), hafnium dioxide ($HfO_2$) (*k*=25.0) and lanthanum oxide ($La_2O_3$) (*k*=30.0) have been considered as gate insulators in the present work. The investigation is conducted by formulating an analytical model where Schrodinger and Poisson equations are solved self-consistently to obtain the band alignment and carrier concentration profile of the device. The capacitance-voltage (C-V) characteristics and carrier tunneling probability through the insulating layer are also extensively investigated in such top-electrode MIS structures for the considered gate dielectrics. Most importantly, the impact of several controlling parameters such as, applied voltage, barrier height at the insulator/semiconductor junction, carrier effective mass of the insulator and nanowire diameter on the carrier tunneling probability is analyzed in detail to develop a comprehensive understanding to maintain electrostatic integrity in such devices.

## 3.3. Analytical modeling of the vertical nanowire MIS devices

The schematic representation of vertical nanowire-MOS device, considered for the current work, under positive bias is shown in Fig. 3.1(a). The charge carriers in such a device are confined in the two transverse directions (radial) since diameter for the corresponding nanowire is chosen to be less than the relevant EBR limit. However, these carriers are free to move along the nanowire axis towards the substrate since no constraint for confinement is imposed on those in the longitudinal direction. However, on application of a positive bias on the top-metal electrode, bands bend in downward direction and form a quantum well at the oxide/semiconductor interface, as shown in Fig. 3.1(b). As a result, electrons accumulate within the well and holes are pushed back to the substrate. Therefore, under such a condition, the electrons are confined both in the radial as well as in the axial (*i.e.*, in all the three) directions which is equivalent to the electrons confined within a quantum dot. Thus, the combined effect of radial confinement due to geometric dimension less than EBR and the voltage induced surface quantization leads to the formation of a voltage tunable quantum dot (VTQD) near the oxide-semiconductor interface on such top-electrode nanoscale MOS structures, as schematically shown in Fig. 3.1(b).

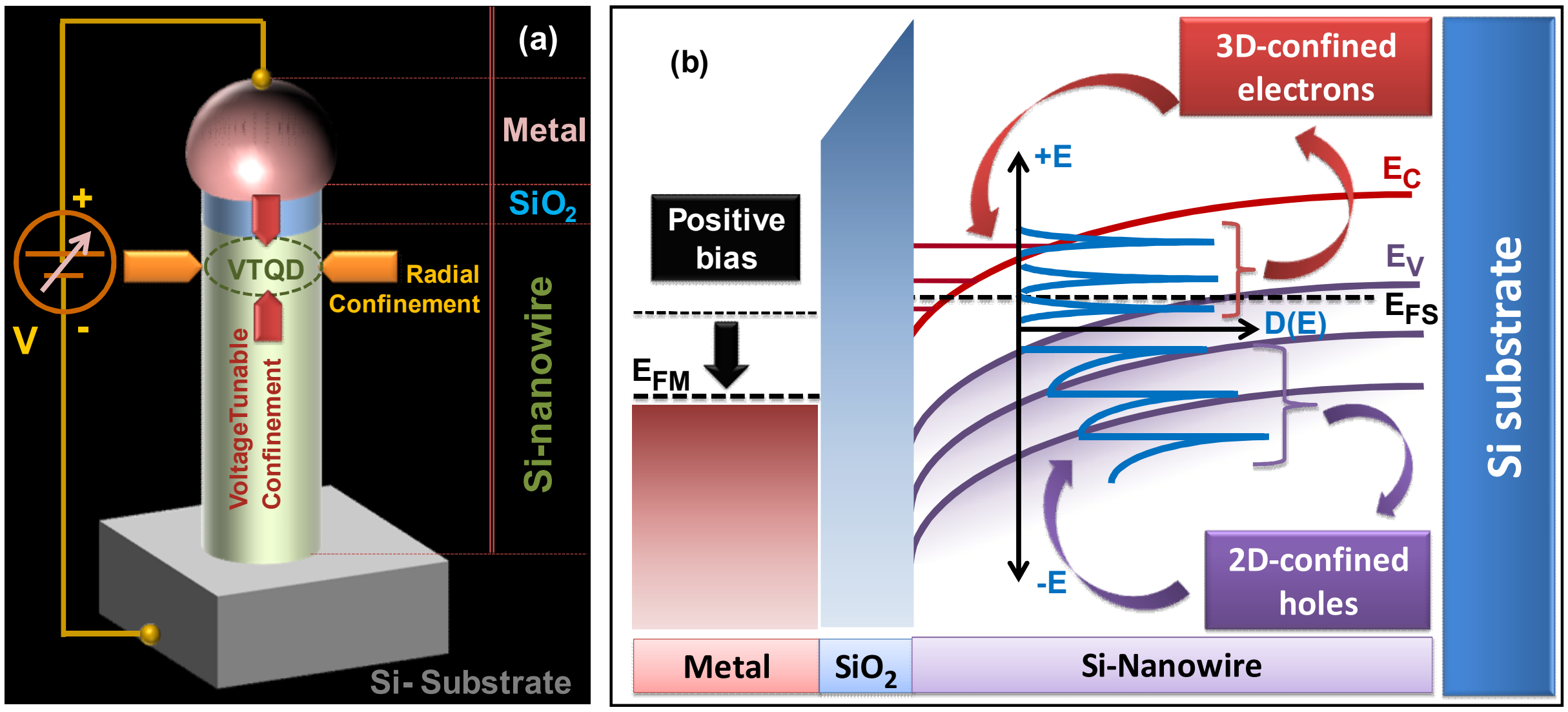

Fig. 3.1. (a) Schematic representation of the formation of voltage tunable quantum dot at the top of nanowire due to the combined effect of structural and electrical quantization; (b) Energy-band diagram of the device showing the regions of 3D-confined (structural and electrical) electrons and 2D-confined (structural only) holes in conduction and valence band, respectively. The corresponding local density of states (LDOS) for such electrons and holes are also shown.

The equations of motion for charge carriers in nanowire MOS devices using Schrodinger and Poisson's equations can be expressed as,

$$H_{3D}|n\rangle = E_n|n\rangle \tag{3.1}$$

and

$$\partial_i \varepsilon_{ij} \partial_j \phi(x_k) = -(-e)\rho(x_k) \tag{3.2}$$

where, $H_{3D}$ is the device Hamiltonian and $E_n$ is the energy eigenvalue at $|n\rangle^{th}$ eigenstate. $\varepsilon$ represents the permittivity tensor for the entire system, $\phi(x_k)$ is the potential as a function of the Cartesian coordinate space $x_k \equiv (x_1, x_2, x_3)$, $e$ is the value of electronic charge and $\rho$ represents the carrier density.

In Eqn. (3.1), the 3-D Hamiltonian can be expressed as,

$$H_{3D} = \frac{-\hbar^2}{2m_0}\sum_{i,j}\left(\partial_i m_{ij}^{*-1}\partial_j\right)+U(z) \tag{3.3}$$

where, $\hbar$ is the reduced Planck's constant, $m_0$ is the mass of free electron, $m_{ij}^*$ are the components of effective mass tensor in electronic mass unit and $U(z)$ represents the conduction band profile along the nanowire axis (*i.e.,* $z$ -axis) from metal/insulator junction towards the substrate.

At this point, it is imperative to mention that the device Hamiltonian is constructed in accordance to the effective mass approximation (EMA) theory. It should also be noted that the components of $[\varepsilon]$ and $[m^*]$ tensors are different at the insulator/semiconductor and metal/insulator junctions. Since the potential in such device varies only along its longitudinal axis, the electronic state can be split up into the transverse ($|n_T\rangle$) and longitudinal sub-spaces ($|n_L\rangle$) as,

$$|n\rangle = |n_T\rangle \otimes |n_L\rangle \tag{3.4}$$

The solution of equation of motion in the transverse sub-space leads to the eigenvalue components $E_{n_T}$ and that in the longitudinal sub-space results in the eigenvalue components $E_{n_L}$, which are added together to give rise the set of energy eigenvalues for the present device. The solution for longitudinal sub-space needs to be consistent with the Poisson's equation given in Eqn. (3.2). Now, the 1-D carrier distribution along the nanowire axis is given by,

$$n_{1D}(z) = \sum_{n_L,n_T} f\left(\left(E_{n_L}+E_{n_T}\right)-E_f\right)\langle n_L|z\rangle\langle z|n_L\rangle\oint\langle n_T|x,y\rangle dxdy\langle x,y|n_T\rangle \tag{3.5}$$

where, $x$ and $y$ axes represent two transverse directions, $E_f$ is the semiconductor Fermi-level and $f\left(\left(E_{nL}+E_{nT}\right)-E_f\right)$ is the corresponding Fermi-Dirac (FD) distribution function. Further, Eqn. (3.5) can be represented as,

$$n_{1D}(z) = \sum_{n_L,n_T}\left|\langle z|n_L\rangle\right|^2 f\left(\left(E_{n_L}+E_{n_T}\right)-E_f\right) \tag{3.6}$$

since, $\oint\langle n_T|x,y\rangle dxdy\langle x,y|n_T\rangle = \oint\left|\langle x,y|n_T\rangle\right|^2 dxdy = 1$ due to the closure property. The charge carrier density (*i.e.*, $\rho = n_{1D}/Area$) is then calculated by using Eqn. (3.6) which is incorporated into Eqn. (3.2) to obtain the electrostatic potential along the nanowire axis. The obtained potential value is then compared with the potential value of the previous loop and the iteration process is continued until a satisfactory solution is achieved self-consistently where such difference between two consecutive loop potential is negligible (< 25 meV) at room temperature.
It is also worthy to mention that in nanowire based devices where the diameter is less or similar to EBR of the corresponding material, the charge transport phenomenon cannot be explained by considering electrostatic interactions only. The Poisson's equation describes interaction between the number of charge carries where more the number of charges, more is the potential energy which governs the charge carriers to stay at the particular state. However, the quantum properties restrict the charge carriers to stay at the same state by following the condition that more is the potential energy, less is the probability of the charge carriers to stay at the same state. This contradictory requirement of the charge carriers in the quantum domain 'to stay' or 'not to stay' at the particular quantum state is actually the origin of their motion.
In the current study, the entire system including the Si-nanowire and $SiO_2$-layer up to the metal/insulator junction has been taken into consideration to account for the penetration of carrier waves into the metal through the insulating layer, which is necessary to obtain the carrier tunneling probability. The energy eigenvalues relevant to both the longitudinal and structural confinements, the 1-D carrier density and conduction band potential profile are obtained only after the self consistency is achieved. Eqn. (3.6) suggests that the longitudinal variation of such parameters strongly depends on the transverse dimensions of nanowires, which is usually not observed in bulk-structures/devices. The total number of carriers is obtained by integrating the 1-D carrier density profile within the quantum well formed near the interface and finally the capacitance of such nano-MIS structure is obtained by dividing the total charge by the applied voltage $\left(V_{app}\right)$ as,

$$C = \left( e \int_0^{l_{ext}} n_{1D} \, dl_Z \right) \Big/ V_{app} \tag{3.7}$$

where, $l_{ext}$ is the spatial extension of the quantum well along the nanowire axis.

## 3.4. Carrier concentration profile and capacitance-voltage (C-V) characteristics of the vertical nanowire MIS devices

In the current work, Si nanowire is assumed to be grown in [100] direction on an intrinsic Si substrate and an insulating layer is considered to be deposited on top of the nanowire, followed by the formation of metal electrode, as schematically shown in Fig. 3.1(a). The values of carrier effective mass components for Si nanowires with sub-5 nm diameters are taken from ref. **[405]** and the device is assumed to operate at room temperature (300 K).

### 3.4.1. Conduction band potential and carrier concentration with distance from the metal/oxide interface

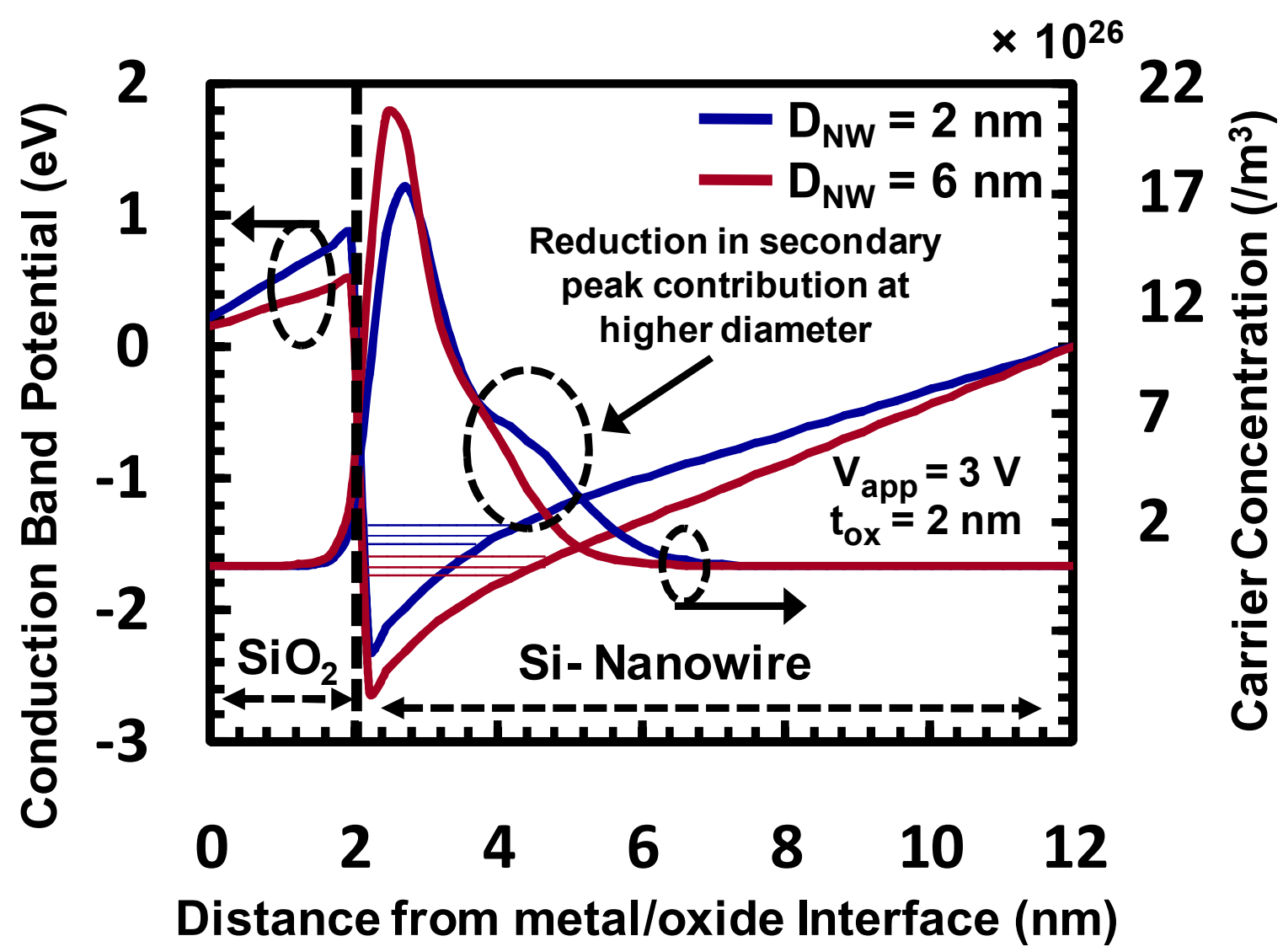

Fig. 3.2. Plots of the conduction band potential and carrier concentration profile with distance from metal/oxide interface for the nanowire diameters ($D_{NW}$) of 2 nm and 6 nm with oxide thickness ($t_{OX}$) of 2 nm at an applied bias ($V_{app}$) of 3 V.

The conduction band potential and carrier concentration along the nanowire axis from metal/insulator interface toward the substrate are shown in Fig. 3.2, for an applied bias of 3 V at the top electrode metal terminal. For the present study, the insulator thickness for $SiO_2$ is taken to be 2 nm to maintain electrostatic integrity of the Si/$SiO_2$ system **[406]**. It has already been discussed that in such a device with nanowire diameter smaller than EBR, carriers are confined in the transverse direction due to geometrical confinement. It is also worthy to mention that similar confinement effects can be obtained in other devices with different dimensions depending upon the excitonic Bohr radii of the corresponding semiconducting materials. For instance, for InP nanowire-MIS device, such confinement effects will be obtained at relatively higher diameter devices in comparison to Si since the EBR of InP is (~15 nm) is 3 times higher than that of Si.

The data for nanowires with diameter of 2 nm and 6 nm are compared in the figure (Fig. 3.2) to investigate the effects of radial confinement. It is observed from such plot that the carrier distribution in both the devices shows multiple convoluted peaks; however, the smaller diameter device exhibits prominent secondary peak when compared to the larger diameter device, which is attributed to the higher degree of confinement in the lower dimensions. Such peaks are originated from the discrete energy states created due to the simultaneous quantum effects of radial geometry as well as the voltage-assisted surface quantization within the quantum well.

### 3.4.2. C-V characteristics of the device and deviation from conventional results regarding the implementation of 'high-*k*' technology

Fig. 3.3 represents the C-V characteristics of Si nanowire MIS capacitors for different nanowire diameters, ranging from 2 nm to 6 nm. The variation of turn-on voltage of such nanowire MOS capacitors with diameter is also shown in the inset of Fig. 3.3. The C-V curve for the device with 2 nm diameter shows step-like behavior which gradually reduces with increasing the nanowire diameter and, finally gets smoother for diameters larger than the EBR of Si (*i.e.*, > 5 nm).

The origin of such step-like behavior is attributed to the sequential filling of consecutive energy eigenstates below the Fermi level which contributes to the accumulation of charge carriers for the MIS capacitor. Similarly, the origin of turn-on voltage is the minimum voltage required to fill up the first available eigenstate which increases with the reduction of nanowire diameter as observed from the plot of inset in Fig. 3.3. Since the device with smaller nanowire diameter experiences a relatively stronger 3-D quantization and thus a higher value of the first sub-band energy, it requires comparatively higher applied bias to occupy this state, leading to the higher turn-on voltage.

Fig. 3.4(a)-(c) represent the impact of high-*k* dielectrics on the electrostatic control of such nanowire MIS devices maintaining an identical EOT of 2 nm. The commonly used high-*k* materials, such as silicon nitride ($Si_3N_4$) **[407]**, aluminum oxide ($Al_2O_3$) **[408]**, yttrium oxide ($Y_2O_3$) **[409]**, hafnium dioxide ($HfO_2$) **[410, 411]** and lanthanum oxide ($La_2O_3$) **[412-413]** are considered as the insulators for studying their comparative performances. The relevant material parameters of such insulators are summarized in Table. 3.1.

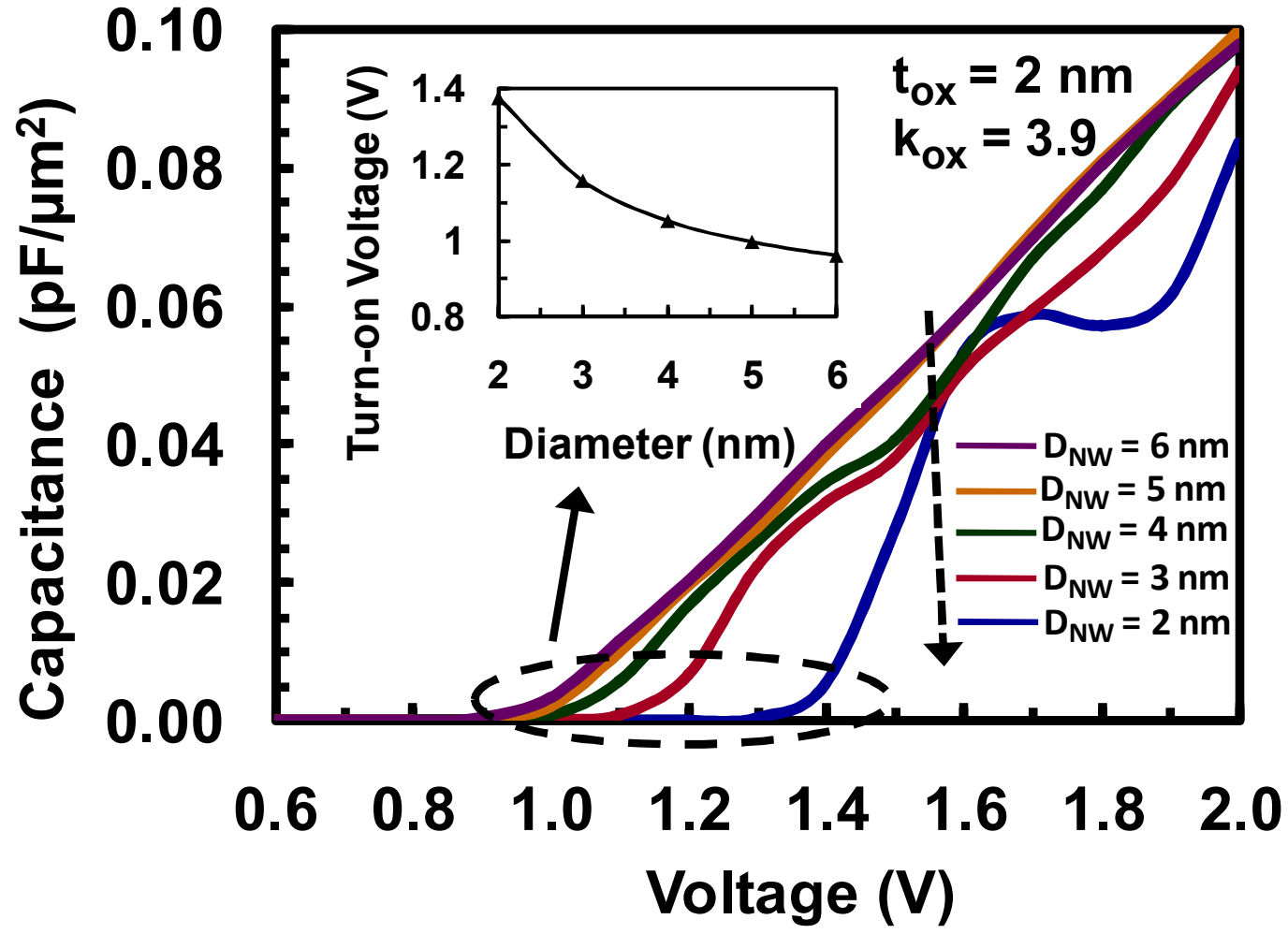

Fig. 3.3. Plot of C-V characteristics of the nanowire MIS structures for the diameters ($D_{NW}$) in the range of 2 nm to 6 nm with a 2-nm thick $SiO_2$ ($k_{OX} = 3.9$) layer as the insulator. Inset shows the variation of turn-on voltage with nanowire diameter.

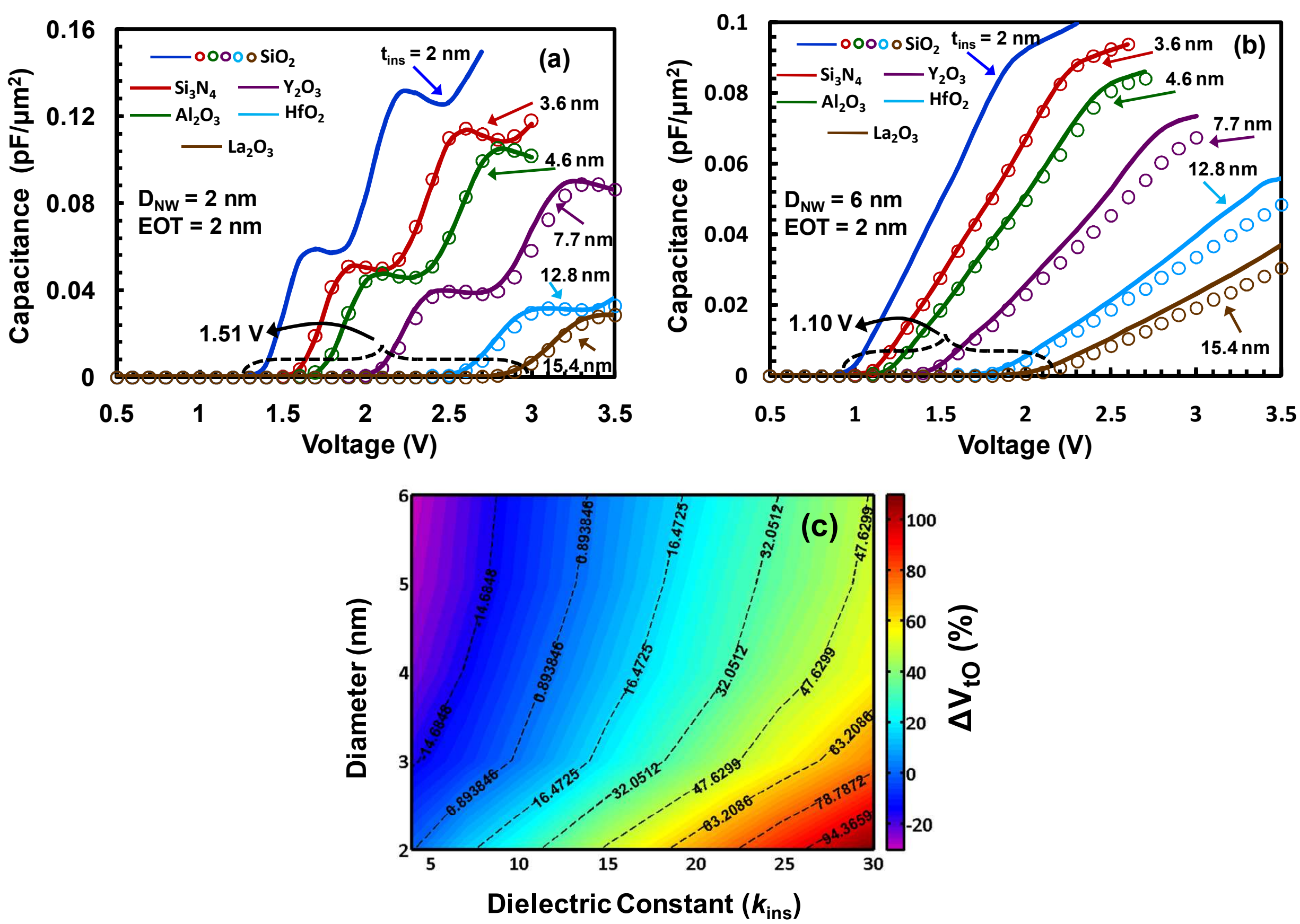

Fig. 3.4. Plots of C-V characteristics for various high-*k* dielectrics with an EOT of 2 nm for: (a) diameter, $D_{NW}$ = 2 nm, (b) diameter, $D_{NW}$ = 6 nm; and (c) Contour plot of the percentage shift of turn-on voltage with nanowire diameter and dielectric constant of the insulator.

Table 3.1. Summary of the dielectric constant, barrier height and effective mass of the different insulating materials considered in the current work **[402, 414-416]**. Tunneling probability for the NW-MIS device using 2-nm diameter nanowires for an applied bias of 2 V is also included.

| **Materials** | **Dielectric Constant** | **Barrier Height with Si (eV)** | **Effective Mass ($\times m_0$)** | **Tunneling Probability in NW-MIS (%)** |
|---|---|---|---|---|
| **$SiO_2$** | 3.9 | 3.20 | 0.40 | $\sim 10^{-6}$ |
| **$Si_3N_4$** | 7.0 | 2.10 | 0.50 | $\sim 10^{-5}$ |
| **$Al_2O_3$** | 9.0 | 2.80 | 0.35 | $\sim 10^{-5}$ |
| **$Y_2O_3$** | 15.0 | 2.30 | 0.25 | $\sim 10^{-3}$ |
| **$HfO_2$** | 25.0 | 2.05 | 0.11 | $\sim 10^{-3}$ |
| **$La_2O_3$** | 30.0 | 2.30 | 0.26 | $\sim 10^{-4}$ |

Fig. 3.4(a) and (b) show the comparative electrostatic performances for the 2-nm and 6-nm nanowire diameters for similar high-*k* combinations. It is observed from Fig. 3.4(a) that the C-V characteristic curves (for 2 nm) are shifted towards higher voltages for higher dielectric constants to attain the same capacitance values. Such observation is not in consonance with the conventional understanding of achieving same capacitance for a given (identical) EOT. For instance, the turn-on voltage ( $V_{to}$ ) for $La_2O_3$ (*k*=30) is increased by ~1.5 V when compared with $SiO_2$ of same EOT, which indicates a significant deterioration of electrostatic control in such nanowire MIS devices.

It is apparent that the physical thickness of insulators with identical EOT increases with *k*-values, and when $SiO_2$ of such thicknesses (symbols in Fig 3.4(a)) are considered, then the capacitance values exhibit no shift from that for the high-*k* materials (lines). Similar effects are also observed in the 6-nm diameter device (Fig. 3.4(b)); however, the shift in the turn-on voltage, with increasing *k*-values, reduces and the impact of using $SiO_2$ of similar thickness is not trivial. Thus, the observed results suggest that the dielectric constant of the considered materials have negligible impact on the nanowire capacitance. Hence, the impact of high-*k* materials in the vertically oriented top-electrode quantized nanowire MIS devices is insignificant since the non-linearity of EOT gets prominent.

Fig. 3.4(c) shows the plots of percentage-shift of turn-on voltage ( $\Delta V_{to} = \left(V_{high-k} - V_{SiO_2}\right)/V_{SiO_2}$ ) for different diameter-dielectric constant combinations (in reference to the 2-nm diameter device). Here, the turn-on voltage is extracted at the capacitance value of 0.01 F/m$^2$. It is observed from Fig. 3.4(c) that the turn-on voltage for 2-nm diameter device is increased by almost 15% for $Si_3N_4$, 22% for $Al_2O_3$, 50% for $Y_2O_3$, 89% for $H_fO_2$ and 110% for $La_2O_3$ as the insulating material, in comparison to $SiO_2$. Such a variation is attributed to the optimum voltage required to occupy the first available energy eigenstate of the quantum well formed near the insulator/semiconductor interface depending on the nanowire diameter and insulator dielectric constant. Thus, the 'turn-on voltage shift' for higher dielectric constant increases with decreasing diameter which indicates the limitation of conventional EOT model for maintaining electrostatic control at the ultra-scaled dimensions unlike the bulk MIS devices.

It is interesting to note that the insulator capacitance dominates in bulk architecture; however, at nano-/quantum-scale the nanowire itself contributes significantly to the net capacitance of the device. Therefore, the conventional parallel plate capacitor model for EOT that shows linearity with dielectric constant is not valid in such quantized nanowire MIS structures. In bulk MIS devices, charge at the insulator/semiconductor interface is distributed in accordance with the electrostatic interaction induced by the voltage applied at the metal electrode, whereas for quantum wire based MIS devices the distribution and hence the net amount of charges is governed by the number of induced quantum states. The number of such quantum states created below the Fermi level is shown in Fig. 3.5 with various nanowire diameters and applied bias combinations. It is observed from the plot that the number of generated quantum states varies in a step-like fashion with increasing applied voltage. Thus, the generation of quantum states is a non-linear function of applied voltage, which leads to such non-linear dependence of EOT with dielectric constant.

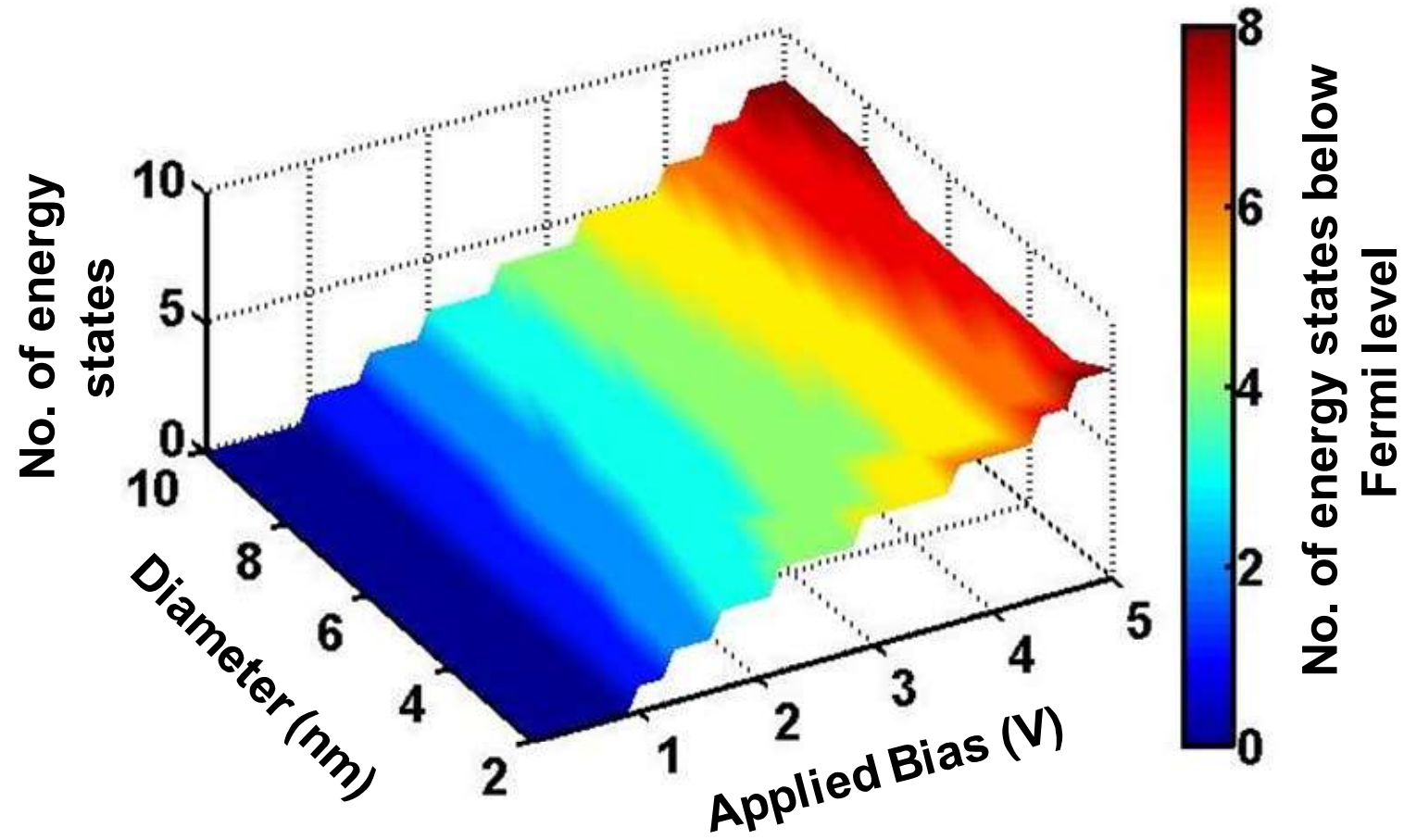

Fig. 3.5. Contour plot of the number of quantum states, created due to the combined effect of structural and longitudinal confinement below the Fermi level, with different nanowire diameter and applied bias combinations. The diameter is varied in the range of 2 nm - 10 nm and the voltage is considered in the range of 0 V - 5 V.

## 3.5. Finding of alternative route for maintaining electrostatic integrity in the vertical nanowire MIS devices

The high-*k* dielectrics as the gate insulator in MOS devices has been introduced primarily to reduce the carrier tunneling probability while maintaining the appropriate electrostatic control. However, the results of Fig. 3.4(a)-(c) suggest that it is not possible to keep the turn-on voltage constant for a given EOT in such quantized vertical nanowire MIS devices. Therefore, for maintaining the superior electrostatic control in such devices the insulator thickness must have to be reduced. However, such reduction will also lead to the enhancement of tunneling-assisted leakage current. Thus, contradictory requirements are needed to fulfill the demand of the present situation in terms of maintaining the electrostatic control while reducing the tunneling current. In this context, the impact of other controlling parameters is investigated to find an alternative route to maintain electrostatic integrity in such quantum wire MIS devices.

The barrier height of oxide/semiconductor junction plays a significant role in controlling the carrier tunneling probability and nanowire diameter also affects the tunneling event by shifting the relevant charge centroid away from the oxide/semiconductor interface leading to an increased effective insulator thickness **[417]**. Apart from these two parameters it is interesting to note that the carrier effective mass at the insulator also plays a significant role in controlling the tunneling process **[418]**. Hence, the impact of such parameters (*i.e.*, the nanowire diameter, carrier effective mass at the insulator and the insulator/semiconductor barrier height) in reducing the tunneling-assisted leakage current is extensively investigated.

Carrier tunneling probability ( $T$ ) in the present nanowire MIS structure is calculated to be,

$$T = \frac{n_{3D}\big|_{z=0}}{n_{3D}\big|_{z=t_{OX}}} \tag{3.8}$$

where, z is the direction along the nanowire axis (as mentioned above), $t_{OX}$ is the insulator thickness and $n_{3D}$ is the three dimensional carrier density. It is worthy to mention that the probability of each (higher and lower order) eigenstate to participate in the tunneling phenomenon at room temperature is not identical due to the difference in their occupancy according to the Fermi-Dirac distribution. Therefore, it is necessary to calculate the tunneling probability from $n_{3D}$, computed by considering the Fermi-Dirac distribution function, instead of obtaining it solely from the wave functions, which is incapable to incorporate the temperature dependent effects and thus, cannot differentiate the participation probabilities of the higher and lower eigenstates. Further, $n_{3D}$ can be expressed as,

$$n_{3D} = \sum_{n_L, n_T} f\left((E_{n_L} + E_{n_T}) - E_f\right) \left|\langle z | n_L \rangle\right|^2 \left|\langle x, y | n_T \rangle\right|^2 \tag{3.9}$$

Fig. 3.6(a) shows the variation of tunneling probability with applied voltage for various nanowire diameters with a 2-nm thick $SiO_2$ as the gate insulator. It is apparent from the plots of Fig. 3.6(a) that the tunneling probability increases significantly with increasing applied voltage, however, it increases marginally with the nanowire diameter. This is attributed to the net increase of number of discrete occupied states below the Fermi level with increasing applied voltage; however, for larger diameters, more closely-spaced transverse energy states get occupied at room temperature and thus contributing less significantly in the tunneling phenomenon. Also, small undulations are observed in the profile of such carrier tunneling probability for the relatively smaller diameter devices due to strong transverse confinement. The penetration of electron waves into the insulator is depicted in the inset of Fig. 3.6(a) by plotting the electron probability densities in the lowest four occupied eigenstates for a 6-nm diameter device at 2 V applied bias.

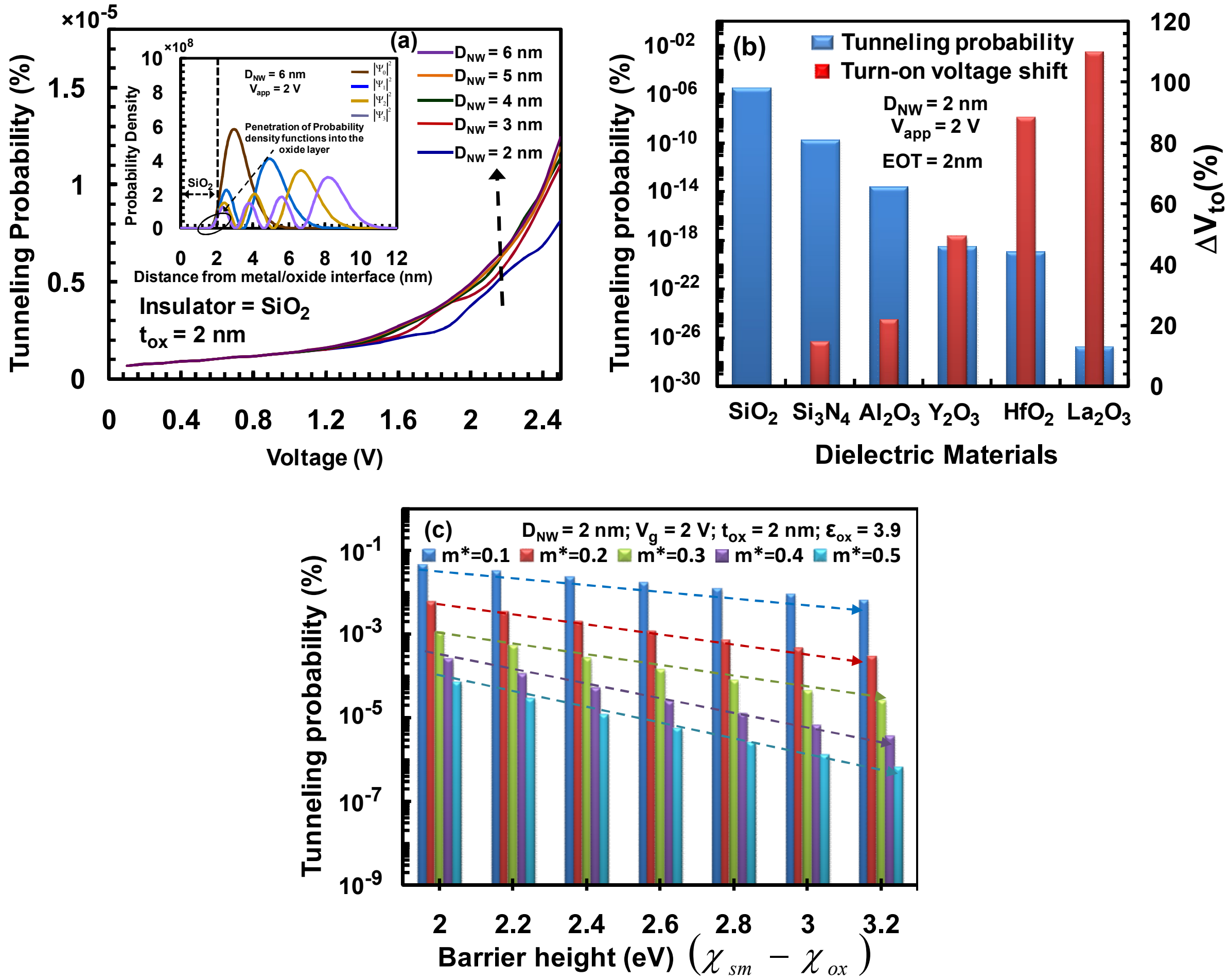

Fig. 3.6. (a). Plot of carrier tunneling probability through the insulator with the applied voltage for different nanowire diameters ($D_{NW}$) while keeping insulator thickness ($t_{OX}$) fixed at 2 nm. The penetration of electron probability density wave functions into the oxide layer is also shown in the inset; (b) Plot of tunneling probability and turn-on voltage shift for different insulators with 2 nm nanowire diameter at an applied bias of 2 V ($V_{app}$) where the barrier heights and insulator effective masses of different insulators are considered from Table.3.1; (c) Plot of carrier tunneling probability for the combinations of various barrier heights and effective masses for the insulator thickness of 2 nm at 2 V applied bias.

Fig. 3.6(b) shows the comparative plots of tunneling probability and turn-on voltage-shift for the MIS devices with different dielectric materials for an identical EOT of 2 nm. The tunneling probability is observed to decrease for the devices with higher dielectric constant; however, the shifts of their turn-on voltage (with respect to $SiO_2$) increase gradually. Thus, higher thicknesses

of dielectric layers degrade electrostatic control in such quantum-wire based MIS devices and therefore, it indicates that the reduction of insulator thickness is the only means to sustain its electrostatic performance.

On the other hand, such thinner insulating layer may lead to higher leakage current and thus, to maintain electrostatic integrity the impact of other relevant controlling parameters such as barrier height and insulator (electron) effective mass should be investigated. In this context, Fig. 3.6(c) represents the variation of carrier tunneling probability with different barrier height and insulator effective mass combinations. It is observed from such plots that the tunneling probability decreases significantly with increasing carrier effective mass in the insulator. It also decreases with increasing barrier height, however, the influence of such barrier height is very small in comparison to that of the insulator effective mass. As apparent from such figure, the tunneling probability decreases by an order of $\sim 10^4$ for increasing the effective mass value from 0.1 to 0.5. However, the tunneling probability decreases by maximum $10^2$ order for increasing the barrier height in the range of 2 eV - 3.2 eV for a given value of effective mass. The order of tunneling probability for different high-*k* materials with different carrier effective mass and barrier height with Si is summarized in Table 3.1. It is interesting to note that the tunneling probability for $SiO_2$ with 3.2 eV barrier height (with Si) and relatively higher insulator effective mass of $0.4m_0$ attains the lowest value among all of the insulating materials considered in the present work. Therefore, it is observed from such study that the carrier tunneling probability can be reduced by increasing both the barrier height and insulator effective mass, where the latter one is observed to be the significant controlling parameter. Thus, it can be concluded from such observations that the selection of high-$m^*$ insulating materials instead of high-*k* dielectrics is more effective to maintain appropriate electrostatic control and also to reduce the tunneling assisted leakage current of the quantized vertical nanowire MIS devices.

## 3.6. Summary

In this chapter, electrostatics of the top-electrode vertically aligned quantized Si-nanowire MIS devices is investigated in detail. An analytical model is developed by considering the self-consistent simultaneous solution of Schrodinger and Poisson equations to study its electrostatic performance. The conduction band-alignment, distribution of 3D-confined charge carriers, C-V characteristics and carrier tunneling probability (through the insulating layer) are calculated to study the effect of quantization (structural and electrical) in such devices. Several insulating materials are also considered to develop a comparative understanding on the impact of such materials on the electrostatic control of the device. The C-V characteristic curves are observed to shift towards higher voltages for high-*k* insulators when $SiO_2$ is replaced; however, such shift is not usually observed in bulk MIS structures. Such shift in turn-on voltages are found to increase with decreasing nanowire diameter and hence suggests that to maintain the electrostatic control by using high-*k* technology is not effective in the quantized nanowire MIS devices. Moreover, at smaller diameter (*i.e.*, diameter less than EBR), the high-*k* materials and $SiO_2$ of identical physical thickness of insulating layers provide similar results which indicate negligible impact of the insulator dielectric constant on nano-/quantum-wire capacitance. Such non-linearity between dielectric constant (of the insulator) and EOT in the quantized nanowire MIS devices is observed due to the non-linear dependence of induced quantum states with the applied voltage. Therefore, in such devices, it is necessary to reduce the insulator thickness to maintain electrostatic control; however, such reduction will enhance the carrier tunneling probability through it. In this context, it has been observed that the applied voltage, barrier height of the insulator/semiconductor junction, carrier effective mass (at the insulator) and the nanowire diameter have considerable impact on the carrier tunneling probability. However, the insulator effective mass has been found to have relatively stronger impact among all such parameters to reduce the carrier tunneling probability. Therefore, in the top-electrode quantized nanowire MIS devices, the selection of high-$m^*$ materials should be of prime interest instead of high-*k* materials for maintaining both the electrostatic control and reduction of carrier tunneling probability.

The current study prescribes a device scheme based on semiconducting nanowire MIS structure which is capable of realizing the voltage-tunable 3D-quantum confinement effects at room temperature. Such scheme can be utilized for fabricating the future quantum dot devices with multitude of applications including the emerging area of quantum information processing. Therefore, the present chapter deals with the design issues of nanowire based VTQD devices for maintaining electrostatic integrity. In the following chapter, the capability of such devices to detect the visible wavelength region will be investigated by formulating an analytical model using non-equilibrium Green’s function (NEGF) formalism.

# Chapter: 4*

# Vertically aligned Si-nanowire MOS photosensors for direct color sensing: Analytical modeling and device design

## 4.1. Introduction: Need for direct color sensing

Sensors have been synonymous as the marker for progress of modern civilization for their palpable uses in almost all application sectors ranging from industry, healthcare, consumer electronics, automotive to the aerospace and defense industries. The contribution of global market values of such sensors was ~U$166.69 billion in 2019 and it is projected to be ~U$188.4 billion by 2022 and ~U$345.77 billion by 2028. The future of sensors' market is thus promising with remarkable potential of expansions. In this context, the complementary-metal-oxide-semiconductor (CMOS) devices have been the workhorse of sensor industries to meet the ever increasing demands for high-definition image-capturing with required speed and providing high-quality images. The compatibility of such CMOS sensors with the existing semiconductor process technology and meeting the essential criteria of *'low-cost low-power high-speed'* have made these the preferable choice over the charge-couple-device (CCD) based active pixel sensors (APS) [**419**]. The schematic of a typical CMOS based sensor is shown in Fig. 4.1 which comprises of a color filter array (CFA), an array of pixel sensors (PSA), an analog to digital converter (ADC) and other relevant digital circuits. The purpose of color filter (*e.g.*, Bayer filter in Red-Green-Green-Blue (RGGB) configuration) is to allow only certain wavelengths

***Most of the content of Chapter: 4 is published in,***
***(a) "Analytical modeling to design the vertically aligned Si-nanowire metal-oxide-semiconductor photosensors for direct color sensing with high spectral resolution", by Subhrajit Sikdar, Basudev Nag Chowdhury, Ajay Ghosh and Sanatan Chattopadhyay, 'Physica E Low-dimensional Systems and Nanostructures' 11/2016; 87:44-50., DOI:10.1016/j.physe.2016.10.039.***
***(b) "Analytical Modeling of Vertically Oriented Standalone Si-Nanowire Metal-Oxide-Semiconductor Capacitors for Wavelength Selective Near-Infrared Sensing Applications", by Subhrajit Sikdar, Basudev Nag Chowdhury and Sanatan Chattopadhyay, 'Advances in Optical Science and Engineering', 09/2017: pages 173-179; , ISBN: 978-981-10-3907-2, DOI:10.1007/978-981-10-3908-9_20.***
***(c) "Designing InP-nanowire based vertical metal-oxide-semiconductor capacitors for wavelength selective visible light sensing", by Subhrajit Sikdar, Basudev Nag Chowdhury and Sanatan Chattopadhyay, 'The Physics of Semiconductor Devices,' 12/2017: chapter 146: pages 1-6; Springer Nature Switzerland AG 2018., ISBN: 978-3-319-97603-7, DOI:10.1007/978-3-319-97604-4_146.***

(primarily RGB) to the sensor array since such PSA can only detect the intensity of light, but not its color. The PSA uses millions of photodevices to convert the light into an electrical signal depending on its intensity. Such signal is then converted into a digital voltage signal by the ADC and finally transported to the image processing circuitry at the output end. The digital controlling circuit, on the other hand, is responsible for making the crucial decisions by sending timing signals to run the process in synchronization to each other. In such conventional CMOS sensors, the photosensing devices split the incident light into a linear combination of RGB-signals and subsequently convert those into the corresponding photocurrents, followed by a color correction method by superposition to find the actual color of the incident wavelength **[420]** (*i.e.,* split-and-superposition principle).

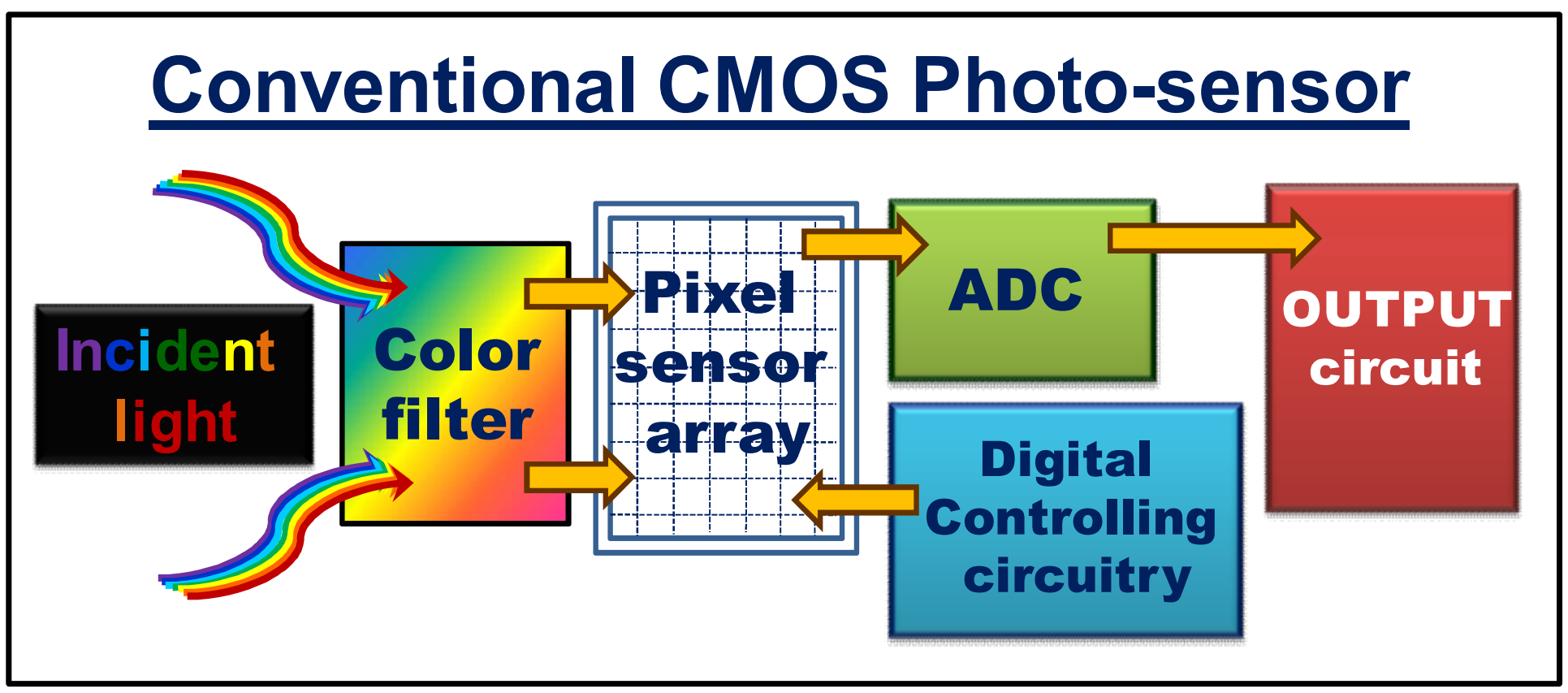

Fig. 4.1. Schematic representation of a typical CMOS photo-sensor.

However, such conversion of photo-impulse into the electrical signal is not a linear function of intensity of the incident light for these three basis (RGB) wavelengths. Therefore, error arises in accuracy of determining the actual color when the same image is captured under illuminating light with different intensities since all the color pixels in that particular image suffer red/blue shift depending on the variation of illumination intensities. Such differences in the illumination conditions violate the color constancy of the image **[421, 422]** and required further image processing technology to retrieve the actual color of the image. Therefore, there is a need of designing novel devices which would be capable to detect the individual incident wavelengths distinctly within the visible spectrum avoiding the conventional split-and-superposition technique.

In the present chapter, a vertical nanowire MOS based photosensing device is designed for direct

(multi-) color sensing with high spectral resolution. The relevant physics and working principle of such devices is explained in detail by developing an analytical model where the process of photogeneration is studied by solving a set of coupled quantum field equations associated with the second quantization of electron and photon field operators, using the non-equilibrium Green's function (NEGF) formalism **[117, 118]**.

## 4.2. Key concern regarding sustained evolution of photosensing devices

In recent times, there has been a remarkable progress in extensive research on the semiconductor nanostructure based devices for optical sensing and imaging applications with a coverage of ultraviolet (UV) to near infrared (NIR) wavelengths **[423-432]**. Such nanoscale photosensors find their applications in almost all sorts of digital imaging tools including the microscopy and spectrometry **[433-436]**. The detection of NIR region is also essential for the field of astronomy **[437]**, agriculture **[438]**, remote sensing **[439]**, material science **[440]** and medical imaging **[441]**. Several reports are available on developing theoretical models for building up a comprehensive understanding on the operation of such nanoscale photodetectors **[442-446]**. However, the high spectral resolution with smaller full-width-at-half-maximum (FWHM) still remains the key concern for sustained progress of such photosensing devices to achieve high quality multispectral sensing/imaging **[447, 448]**. In this context, it is important to mention that the commercially available imaging instruments offer a spectral resolution (in terms of number of spectral bands) of 2 nm to 5 nm in the visible range, however, the FWHM of such spectral bands are reported to be ~10 nm **[449-451]**. Thus, there exists an unavoidable overlap between each spectral line which in turn leads to the loss of spectral information in such photosensing devices. Moreover, the conventional instruments offer detection of the broader spectrum by employing two devices, such as, the photo-multiplier tube (PMT) for UV-region and Indium-Gallium-Arsenide (InGaAs) for the IR-region. Thus, a noisy behavior is observed while the detector is changed from IR-to-UV which affects the detection mechanism significantly if the material responses are expected to occur in that particular wavelength range. Also, the high spectral resolution in such multispectral imagers is achieved by integrating organic dye filters at the wafer level **[452]** which increases complexity of such devices. Therefore, the architectural complexities of such commercially available multispectral imagers and the scalability/durability issues **[421, 453]** of conventional hyperspectral filters demand novel device concepts with substantial improvement over the current technology.

In this context, several approaches have been proposed to fabricate fully integrated wavelength selective photosensors for enhancing the quality of color sensing/imaging **[420, 454-458]**. However, as mentioned earlier, such proposed devices used the split-and-superposition principle to detect the color of a particular image and since such detection process highly depends on the light intensity, one cannot eliminate the possibility of distortion in the image color. In this chapter, a novel Si-nanowire MOS based color detection scheme is proposed which provides a specific route to address such issues.

## 4.3. Development of the theoretical model

In the theoretical model, generation of an electron-hole pair (EHP) between conduction and valence bands of the Si-nanowire due to an incident single photon is considered as an equivalent scattering picture of a single electron by a single photon, where the incoming electron with the effective mass of a hole is deviated from its trajectory with the effective mass of an electron after interacting with the incident photon, as schematically shown in Fig. 4.2.

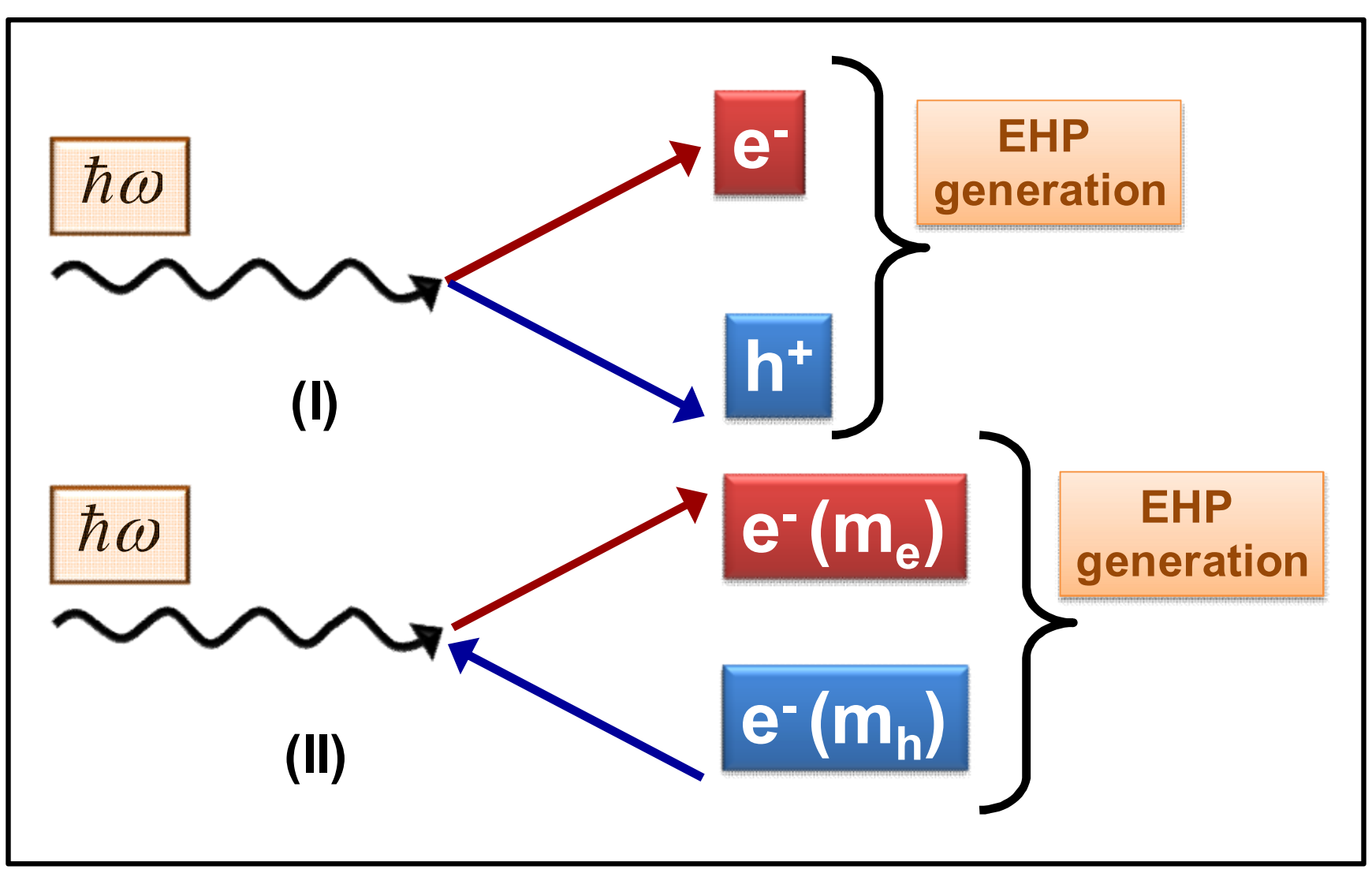

Fig. 4.2. Schematic representation of the electron-hole pair (EHP) generation: (**I**) single photon incident ($\hbar\omega$) and; (**II**) model used in the theoretical development by considering an incident electron with effective mass of holes ($e^{-}(m_h)$) and reflected electron with effective mass of electrons ( $e^-(m_e)$ ).

In the analytical model such phenomenon is considered as annihilation of incoming electron (with effective mass of hole) and incident photon, and subsequently creation of the outgoing electron (with effective mass of electron). It was mentioned in Chapter 3 (section 3.3) that the conduction band electrons are 3D-confined due to the combined effect of structural and electrical confinement whereas, valence band holes are 2D-confined due to the effect of structural confinement only. When such a system is illuminated with light, EHPs are created where electrons are trapped in the quantum well and holes are pushed away towards the substrate. Thus, in the present device scheme the probability of recombination is negligible. Hence, the valence band acts here only as the *'source'* or *'reservoir'* of charge carriers whereas the conduction band is supposed to be the *'device'*. Such scattering phenomenon, *i.e.*, transition of a Fermion (electron) from valence band to the conduction band after interacting with a Boson (incident photon) is described by the following equations **[118]**:

$$i\hbar\frac{d}{dt}C_i^c = H_{ISO}^c C_i^c + \sum_{r,\alpha}(\tau_{ir}^{\alpha} V_r^v b_{\alpha}^{pht} + \tau_{ir}^{\alpha*} V_r^v b_{\alpha}^{pht+}) \quad (4.1.a)$$

$$i\hbar\frac{d}{dt}V_s^v = H_{ISO}^v V_s^v + \sum_{j,\alpha}(\tau_{sj}^{\alpha} C_j^c b_{\alpha}^{pht} + \tau_{sj}^{\alpha*} C_j^c b_{\alpha}^{pht+}) \quad (4.1.b)$$

$$i\hbar\frac{d}{dt}b_{\alpha}^{pht} = \hbar\omega_{\alpha}^{pht} b_{\alpha}^{pht} + \sum_{j,r}(\tau_{jr}^{\alpha*} C_j^c V_r^{v+} + \tau_{jr}^{\alpha*} C_j^{c+} V_r^v) \quad (4.1.c)$$

where $C_i^c$ and $V_s^v$ represent electron annihilation operators in i^th^ state of the conduction band and s^th^ state of the valence band, respectively. $b_{\alpha}^{pht}$ is the annihilation operator for the incident photon in $\alpha$ mode with the angular frequency $\omega_{\alpha}$. Such equations (Eqn. 4.1.a to Eqn. 4.1.c) are framed in consideration that electrons are injected from ($r^{th}$ or) $s^{th}$ mode of valence band to the $i^{th}$ (or $j^{th}$) mode of conduction band during interaction with the incident photons, where τ's represent the corresponding interaction potential. $H_{ISO}^C$ and $H_{ISO}^v$ are the Hamiltonians for conduction and valence band, respectively, in isolated conditions. All the b's and C's follow the Bose-Einstein (BE) commutation and Fermi-Dirac (FD) anti-commutation relations, respectively as shown below,

$$[b_{\alpha}, b_{\alpha}^+] = \delta_{\alpha\beta} \quad (4.2.a)$$

$$\left\{C_i^c, C_k^{c+}\right\} = \delta_{ik} \tag{4.2.b}$$

$$\left\{V_r^v, V_s^{v+}\right\} = \delta_{rs} \tag{4.2.c}$$

All the other commutation and anti-commutations are zero. The corresponding two-time correlation functions are calculated by using the Keldysh formalism as **[371]**,

$$n_{ik}^c(t,t') = \left\langle C_k^{c+}(t')C_i^c(t)\right\rangle \tag{4.3.a}$$

$$n_{rs}^v(t,t') = \left\langle V_s^{v+}(t')V_r^v(t)\right\rangle \tag{4.3.b}$$

$$N_{\alpha\beta}^{ab-pht}(t,t') = \left\langle b_\beta^{pht+}(t')b_\alpha^{pht}(t)\right\rangle \tag{4.3.c}$$

where, $n_{ik}^c(t,t')$ and $n_{rs}^v(t,t')$ are two-time correlation functions for the filled state of electrons in conduction band and valence band regions, respectively. $N_{\alpha\beta}^{ab-pht}(t,t')$ represents the two-time correlation function for the absorbed photons. Similarly, the corresponding two-time correlation functions for empty state of electrons in conduction or valence band and emitted photons are given by, $p_{ik}^c(t,t'), p_{rs}^v(t,t')$ and $N_{\alpha\beta}^{em-pht}(t,t')$.

In the process of solving Eqn. (4.1.a), at first the value of $C_r^v b_\alpha^{pht}$ needs to be evaluated,

$$i\hbar\frac{d}{dt}\left[V_r^v b_\alpha^{pht}\right] = \left[i\hbar\frac{d}{dt}V_r^v\right]b_\alpha^{pht} + V_r^v\left[i\hbar\frac{d}{dt}b_\alpha^{pht}\right]$$

$$= \left[H_{Iso}^v + \hbar\omega_\alpha^{pht}\right]V_r^v b_\alpha^{pht} + \sum_{j,\beta}\tau_{rj}^{\beta*}C_j^c b_\beta^{pht+}b_\alpha^{pht} + \sum_{j,s}\tau_{js}^{\alpha*}C_j^c V_r^v V_s^{v+} \tag{4.4}$$

(Using Eqn. (4.1.b), Eqn. (4.1.c) and applying the commutation relations as stated in Eqn. ((4.2.a) to (4.2.c)).

Now, simplifying second and the third term of Eqn. (4.4) by applying the commutation/anti-commutation relations, we get,

$$X_{r\alpha}^{pht} = \sum_{j,\beta}\tau_{rj}^{\beta*}C_j^c b_\beta^{pht+}b_\alpha^{pht} + \sum_{j,s}\tau_{js}^{\alpha*}C_j^c V_r^v V_s^{v+}$$

$$= \sum_j\left[\sum_{s,\beta}\tau_{js}^{\beta*}\left(\left[b_\beta^{pht+}b_\alpha^{pht}\right]\delta_{rs} + \delta_{\alpha\beta}\left[V_r^v V_s^{v+}\right]\right)C_j^c\right]$$

$$= \sum_j\left[\sum_{s,\beta}\tau_{js}^{\beta*}\begin{pmatrix}\left[b_\beta^{pht+}b_\alpha^{pht}\right]\left[V_r^v V_s^{v+} + V_s^{v+}V_r^v\right] + \\ \left[b_\alpha^{pht}b_\beta^{pht+} - b_\beta^{pht+}b_\alpha^{pht}\right]\left[V_r^v V_s^{v+}\right]\end{pmatrix}C_j^c\right]$$

$$=\sum_{j}\left[\sum_{s,\beta}\tau_{js}^{\beta *}\left(\left[b_{\beta}^{pht+}b_{\alpha}^{pht}\right]\left[V_{s}^{v+}V_{r}^{v}\right]+\left[b_{\alpha}^{pht}b_{\beta}^{pht+}\right]\left[V_{r}^{v}V_{s}^{v+}\right]\right)C_{j}^{c}\right] \tag{4.5}$$

Further, Eqn. (4.4) can be represented in a simpler form as,

$$i\hbar\frac{d}{dt}\left[V_{r}^{v}b_{\alpha}^{pht}\right]-\left[H_{Iso}^{v}+\hbar\omega_{\alpha}^{pht}\right]V_{r}^{v}b_{\alpha}^{pht}=X_{r\alpha}^{pht} \tag{4.6}$$

The solution of Eqn. (4.6) is obtained by introducing the Green's function $(G_{r\alpha}^{e-pht}(t,t'))$ for electron-photon interaction as,

$$\left[V_{r}^{v}b_{\alpha}^{pht}\right]=\left[V_{r}^{v}b_{\alpha}^{pht}\right]^{Iso}+\int dt' G_{r\alpha}^{e-pht}(t,t')X_{r\alpha}^{pht}(t') \tag{4.7}$$

To the lowest order approximation, taking average over all of the valence band electrons and photons, using Eqn. (4.3.a) and Eqn. (4.3.c), we get,

$$\left[V_{r}^{v}b_{\alpha}^{pht}\right]=\left[V_{r}^{v}b_{\alpha}^{pht}\right]^{Iso}+\int dt'\vartheta(t-t')\sum_{j}\left[\sum_{s,\beta}\tau_{js}^{\beta *}\left(\left[N_{\alpha\beta}^{ab-pht}(t,t')\right]\left[n_{sr}^{v}(t,t')\right]+\left[N_{\alpha\beta}^{em-pht}(t,t')\right]\left[p_{sr}^{v}(t,t')\right]\right)C_{j}^{c}\right] \tag{4.8}$$

Similarly, the value of $C_{r}^{v}b_{\alpha}^{pht+}$ is obtained to be,

$$\left[V_{r}^{v}b_{\alpha}^{pht+}\right]=\left[V_{r}^{v}b_{\alpha}^{pht+}\right]^{Iso}+\int dt'\vartheta(t-t')\sum_{j}\left[\sum_{s,\beta}\tau_{js}^{\beta}\left(\left[N_{\alpha\beta}^{em-pht}(t,t')\right]\left[n_{sr}^{v}(t,t')\right]+\left[N_{\alpha\beta}^{ab-pht}(t,t')\right]\left[p_{sr}^{v}(t,t')\right]\right)C_{j}^{c}\right] \tag{4.9}$$

Further, substituting the values of $C_{r}^{v}b_{\alpha}^{pht}$ and $C_{r}^{v}b_{\alpha}^{pht+}$ from Eqn. (4.8) and Eqn. (4.9) in the Eqn. (4.1.a) results to the equation of *'device'* as,

$$i\hbar\frac{d}{dt}C_{i}^{c}=H_{Iso}^{c}C_{i}^{c}+\sum_{r,\alpha}\left(\tau_{ir}^{\alpha}\left[V_{r}^{v}b_{\alpha}^{pht}\right]^{Iso}+\tau_{ir}^{\alpha *}\left[V_{r}^{v}b_{\alpha}^{pht+}\right]^{Iso}\right)+\sum_{j}\int dt'\left(\Sigma_{sc}^{pht}(t,t')\right)_{rs}C_{j}^{c}$$

$$=H_{Iso}^{c}C_{i}^{c}+\sum_{j}\int dt'\left(\Sigma_{sc}^{pht}(t,t')\right)_{rs}C_{j}^{c}+\left(S_{sc}^{pht}\right)_{r} \tag{4.10}$$

where,

$$\left(S_{sc}^{pht}\right)_{r}=\sum_{r,\alpha}\left(\tau_{ir}^{\alpha}\left[V_{r}^{v}b_{\alpha}^{pht}\right]^{Iso}+\tau_{ir}^{\alpha *}\left[V_{r}^{v}b_{\alpha}^{pht+}\right]^{Iso}\right) \tag{4.11}$$

$$\left(\Sigma_{sc}^{pht}(t,t')\right)_{rs}=\vartheta(t-t')\times\left(\Gamma_{sc}^{pht}(t,t')\right)_{rs} \tag{4.12}$$

$$\left(\Gamma_{sc}^{pht}(t,t')\right)_{rs}=\sum_{r,\alpha,j,\beta}\left[\begin{array}{l}\left\{\tau_{ir}^{\alpha}\left(\left[N_{\alpha\beta}^{ab-pht}(t,t')\right]\left[n_{sr}^{v}(t,t')\right]+\left[N_{\alpha\beta}^{em-pht}(t,t')\right]\left[p_{sr}^{v}(t,t')\right]\right)\tau_{js}^{\beta *}\right\}\\+\left\{\tau_{ir}^{\alpha *}\left(\left[N_{\alpha\beta}^{em-pht}(t,t')\right]\left[n_{sr}^{v}(t,t')\right]+\left[N_{\alpha\beta}^{ab-pht}(t,t')\right]\left[p_{sr}^{v}(t,t')\right]\right)\tau_{js}^{\beta}\right\}\end{array}\right] \tag{4.13}$$

Finally, introducing the *'device'* Green's function $(G^{D})$, solution of Eqn. (4.10) is obtained to be,

$$C_{i}^{c}(t)=\sum_{i'}dt' G_{i,i'}^{D}(t,t')\left(S_{sc}^{pht}(t')\right)_{i'} \tag{4.14}$$

The corresponding correlation function for the filled state of electrons in conduction band (as shown in Eqn. (4.3.a)) in the presence of electron-photon interaction is obtained to be (using Eqn. (4.14)),

$$n_{ik}^{c}(t,t') = \sum_{i',k'} \int\int dt_1 dt_2 \left[G_{i,i'}^{D}(t,t_1)\Sigma_{sc}^{In-pht}(t_1,t_2)G_{k',k}^{D*}(t_2,t')\right] \tag{4.15}$$

where, the in-scattering function ($\Sigma_{sc}^{In-pht}$) is expressed as,

$$\Sigma_{sc}^{In-pht}(t_1,t_2) = \sum_{r,\alpha,j,\beta} \begin{pmatrix} \tau_{ir}^{\alpha}\left[N_{\alpha\beta}^{ab-pht}(t_1,t_2)n_{sr}^{v}(t_1,t_2)\right]^{Iso}\tau_{js}^{\beta*} + \\ \tau_{ir}^{\alpha*}\left[N_{\alpha\beta}^{em-pht}(t_1,t_2)n_{sr}^{v}(t_1,t_2)\right]^{Iso}\tau_{js}^{\beta} \end{pmatrix} \tag{4.16}$$

The equations are further transformed into energy domain by Fourier Transformation which gives rise to (in matrix form),

$$n_{ik}^{c}(E) = \left[G^{D}(E)\right]\left[\Sigma_{sc}^{In-pht}(E)\right]\left[G^{D+}(E)\right] \tag{4.17}$$

where, the photon in-scattering function is obtained to be,

$$\Sigma_{sc}^{In-pht}(E) = \sum_{\alpha}(N_{pht}+1)\left[\tau^{\alpha}\right]\left[n(E+\hbar\omega_{\alpha})\right]\left[\tau^{\alpha+}\right] + (N_{pht})\left[\tau^{\alpha+}\right]\left[n(E-\hbar\omega_{\alpha})\right]\left[\tau^{\alpha}\right] \tag{4.18}$$

Here, $N_{pht}$ represents number of the absorbed photons and $n(E)$ is the number of electrons per unit energy in each discrete energy level in the conduction band. The photon occupation ($N_{pht}$) depends on the intensity of incident light as given by **[443]**,

$$N_{pht} = \left[(I_{\alpha}/\hbar\omega_{\alpha})*V_{ab}\right]/\tilde{c} \tag{4.19}$$

where, $I_{\alpha}$ and $V_{ab}$ are the intensity of incident light and the absorbing nanowire volume, respectively. The speed of light within the considered material is given by $\tilde{c} = c/\eta^{\lambda}$, where $c$ is the speed of light in vacuum and $\eta^{\lambda}$ represents refractive index of the material for corresponding wavelengths ($\lambda$) of the incident light. It is worthy to mention that the *'device'* is an 'isolated system' unless illuminated and thus the relevant *'device'* Green's Function can be represented as,

$$G^{D}(E) = Lt_{\Sigma(E)\to 0}\left[EI - E_{sub} - \Sigma(E)\right]^{-1} \tag{4.20}$$

where $E_{sub}$s are the discrete energy states created due to both the structural and electrical quantization. $\Sigma(E)$ represents the corresponding self energy function which is considered to be negligibly small since electrons are confined from all of the three directions in the quantum well.

Thus, for computing the relevant carrier density (n) in the conduction band region, the *'device'* Green's function is considered to be the delta function as,

$$\delta(E-E_{sub})=(1/2\pi\hbar)\int_{-\infty}^{\infty} e^{i/\hbar(E-E_{sub})t}\,dt \tag{4.21}$$

$$\therefore n=\sum_{allsub-bands}\int_{Ec}^{\infty} G^{D}\,\Sigma_{sc}^{In-pht}\,G^{D+}dE=\sum_{allsub-bands}\left[\int_{Ec}^{\infty}\delta(E-E_{sub})\Sigma_{sc}^{In-pht}(E)\left\{\int_{-\infty}^{\infty}(1/2\pi)\,e^{-i/\hbar(E-E_{sub})t}\,dt\right\}dE\right]/\hbar$$

$$=\sum_{allsub-bands}\int_{-\infty}^{\infty}\left(\int_{Ec}^{\infty}\left[\delta(E-E_{sub})\Sigma_{sc}^{In-pht}(E)(1/2\pi\hbar)\,e^{-i/\hbar(E-E_{sub})t}\right]dE\right)dt$$

$$=\sum_{allsub-bands}\left(\int_{-\infty}^{\infty}\left[(1/2\pi\hbar)\Sigma_{sc}^{In-pht}(E_{sub})\right]dt\right)$$

Consequently, the total number of photo-generated carriers in the conduction band region per second ($n_{1D}^{ph}$) is obtained to be,

$$n_{1D}^{ph}=\left[\sum_{allsub-bands}\left(1/2\pi\hbar\right)\Sigma_{sc}^{In-pht}\left(E_{sub}\right)\right] \tag{4.22}$$

It is important to mention that the spatial argument is included in the above equation through the spatial distribution of such discrete energy states ($E_{sub}$) in the conduction band regime.

### 4.3.1. Calculation of interaction potential

It is apparent from Eqns. (4.1.a), (4.1.b) and (4.1.c) that the interaction of valence band holes with the conduction band electrons due to illumination of light with specific wavelengths depend significantly on the interaction potential ($\tau$). As mentioned earlier, the photogeneration process in the present device scheme has been conceptualized by an equivalent scattering picture (Fig. 4.2) which gives rise the interaction potential to be,

$$\tau\equiv\langle f|V_{sc}|i\rangle \tag{4.23}$$

where the initial state ($|i\rangle$) represents state of the 'virtual electron' just before the scattering with effective mass of hole and equal and opposite momentum, *i.e.,* $|i\rangle=\left|-\vec{k}_r^{hole},m_h^*\right\rangle$, and $|f\rangle$ represent the final state of the 'real electron' just after the scattering with $|f\rangle=\left|\vec{k}_i^{elec},m_e^*\right\rangle$.

To develop the expression of such interaction potential, here, it is considered that before the scattering is initiated the system is in a particular state and the equation of motion for such isolated system can be written as,

$$H_{ISO}\left|\psi_i(r,t)\right\rangle = E_i\left|\psi_{ini}(r,t)\right\rangle \tag{4.23.1}$$

where, $\psi_i(r,t)=\phi_i(r)\exp(-iE_it/\hbar)$ and $E_i$ is the energy eigenvalue for the state $\left|i\right\rangle$. Further, after the interaction of such an isolated system with the externally applied perturbations the equation of motion for such interacting-system is modified to,

$$H_f\left|\psi_f(r,t)\right\rangle = i\hbar\frac{\partial}{\partial t}\left|\psi_f(r,t)\right\rangle \tag{4.23.2}$$

where, $\psi_f(r,t)=\sum_i C_i(t)\phi_i(r)\exp(-iE_it/\hbar)$ and the Hamiltonian for such system is sum of the Hamiltonians for the isolated system and the perturbing scattering potential as, $H_f = H_{ISO}+V_{sc}(t)$ where, $V_{sc}$ is the small perturbation potential representing the interaction process. Now, putting the value of $\psi_f$ into the Eqn. (4.23.2) we get,

$$\sum_i C_i(t)V_{sc}\phi_i(r)\exp(-iE_it/\hbar)=\sum_i i\hbar\dot{C}_i(t)\phi_i(r)\exp(-iE_it/\hbar) \tag{4.23.3}$$

Then, multiplying by $\phi_f^*(r)$ on the both sides of the above equation and using the orthogonal relation $\int\phi_f^*(r)\phi_i(r)dr=\left\langle f\middle|i\right\rangle=\delta_{fi}$, such equation results into,

$$i\hbar\frac{\partial}{\partial t}C_f=\sum_i C_i(t)\left\langle\phi_f^*\middle|V_{sc}\middle|\phi_i\right\rangle\exp(-i(E_i-E_f)t/\hbar) \tag{4.23.4}$$

Further, assuming that the isolated system initially is in a particular state, *i.e.,* say, $j=i$, which gives rise to the value of $C_i=1$, for $j=i$ and $C_j=0$, for $j\neq i$ which finally results to,

$$C_f=2\pi i\sum_i\left\langle\phi_f^*\middle|V_{sc}\middle|\phi_i\right\rangle\delta(E_i-E_f) \tag{4.23.5}$$

where, $\left|C_f\right|^2$ represents the probability of finding the system after interaction in the state *final* and $\left\langle\phi_f^*\middle|V_{sc}\middle|\phi_i\right\rangle$-term represents the interaction potential as function of valence band holes, conduction band electrons and the incident photons.

Now, incorporating the vector potential ($A$) due to incident photons in the Schrodinger equation, the interaction potential is obtained to be,

$$\tau = \frac{q}{2m^*}\left\langle \phi_f^* \middle| V_{sc} \middle| \phi_i \right\rangle = \frac{q}{2m} \int_{volume} dV_{ab} \phi_f^* \left(\vec{p}.\vec{A} + \vec{A}.\vec{P}\right)\phi_i \tag{4.23.6}$$

where, $q$ is the electronic charge, $m^*$ represents the combination of the effective masses of electrons and holes and $V_{ab}$ is the absorbing volume.

Here, the expression for the wave vector of 2D-confined valence band holes is obtained from the E-k dispersion relationship for the nanowire coupled with substrate **[117, 118]** as,

$$k_r^{hole} = (1/a)\cos^{-1}\left(1 - \frac{\left(E - \varepsilon_r^v\right)}{2\left(\hbar^2/2m_h^* a^2\right)}\right) \tag{4.24}$$

The relevant expression for wave vector of conduction band electrons is very trivial (until the processes like carrier tunneling or hot electron injection start to dominate) since such electrons are 3D-confined from all the three directions inside the quantum well,

$$k_i^{elec} = \sqrt{\frac{2m_e^* \varepsilon_i^c}{\hbar^2}} \tag{4.25}$$

Now, considering optical permittivity and volume of the absorbing (nanowire) medium to be $\varepsilon$ and $V_{ab}$, respectively, and vector potential of the incident photons is given by,

$$\vec{A} = \vec{A}_0 (b_\alpha^{pht} e^{i\vec{k}_\alpha^{pht}.\vec{r}} + b_\alpha^{pht+} e^{-i\vec{k}_\alpha^{pht}.\vec{r}}) \tag{4.26}$$

where, $\vec{k}_\alpha^{pht}$ is the wave vector for incident photons and $\vec{A}_0 = \sqrt{\frac{\hbar}{2\omega_\alpha^{pht} \varepsilon V_{ab}}}$. The equivalent scattering potential, neglecting the higher order term ($\frac{q^2 A^2}{2m}$) in Schrödinger's equation, is given by,

$$V_{sc} = \left[\hat{\vec{p}}.\left(\frac{q}{2m}\vec{A}\right) + \left(\frac{q}{2m}\vec{A}\right).\hat{\vec{p}}\right] \tag{4.27}$$

where, $q$ is the electronic charge and $\hat{\vec{p}}$ is the momentum operator. Thus, Eqn. (4.23) and Eqn. (4.27) finally give rise to,

$$\tau_{ir}^\alpha = \frac{q\hbar}{2}\left[\left(\frac{\vec{k}_i^{elec}}{m_e^*} + \frac{\vec{k}_r^{hole}}{m_h^*}\right) \cdot \vec{A}_0 b_\alpha^{pht}\right] \tag{4.28}$$

with the restriction of $\vec{k}_i^{elec} + \vec{k}_r^{hole} = \vec{k}_\alpha^{pht}$.

### 4.3.2. Calculation of device capacitance

The developed capacitance comprises of charge carriers in both of the illuminated and dark condition. The number of charge carriers in the illuminated condition ($n_{1D}^{ph}$) is obtained from Eqn. (4.22) whereas, the number of charge carriers in dark condition ($n_{1D}$) is obtained by developing a Schrodinger-Poisson simultaneous solver. The development of such a quantum-electrostatic simultaneous solver along with its importance in nano-/quantum-dimensions is extensively discussed in Chapter 3 (section 3.3). Here, the iterative process of finding an appropriate solution for carrier density is shown as a flowchart in Fig. 4.3. Further, after solving Eqn. (3.1) and Eqn. (3.2) (Chapter 3) the carrier density ($n_{1D}$) is obtained to be,

$$n_{1D} = \sum_{allsub-bands} |\psi_{normalized}|^2 \times f((E_L + E_T) - E_f) \quad (4.29)$$

where, $E_L$ and $E_T$ represent the discrete energy levels due to longitudinal and transverse confinement, respectively. $E_f$ is the intrinsic Fermi-level and $f((E_L + E_T) - E_f)$ represents the corresponding Fermi-Dirac (FD) distribution function.

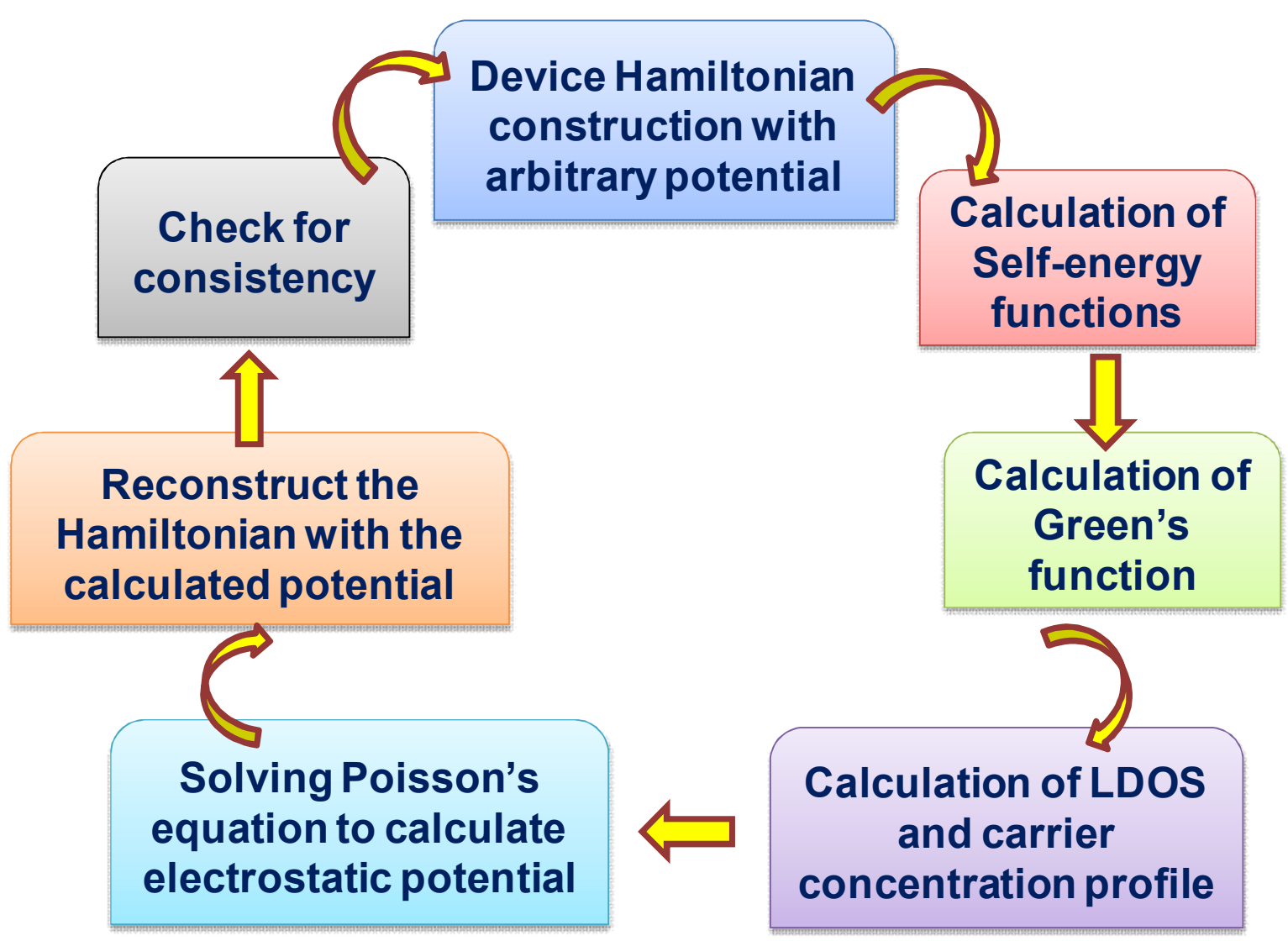

Fig. 4.3. Schematic representation of the iterative process followed in the current work to develop the quantum-electrostatic simultaneous solver.

Finally, the device photo-capacitance ($C_{ph}$) is obtained by integrating the total number of photogenerated electrons ($n_{1D}^{ph}$) and the electrons in dark condition ($n_{1D}$) within the voltage tunable quantum dot (VTQD) extension region and subsequently divided by the top-electrode applied voltage as,

$$C_{ph} = \left[ \int_0^{l_{ext}} (n_{1D} + n_{1D}^{ph}) dl_{ext} \right] \Big/ V_{app} \tag{4.30}$$

where, $l_{ext}$ is the length of VTQD extension and $V_{app}$ is the applied voltage.

## 4.4. Device scheme and carrier concentration at dark condition

Schematic representation of the vertically oriented Si-nanowire MOS device is shown in Fig. 4.4(a) where Si nanowires are assumed to be grown in [100] direction on a Si-substrate and a $SiO_2$ layer of 2 nm thickness is considered to be deposited on top of such nanowires, followed by the deposition of metal electrode on top of $SiO_2$. The values of carrier effective mass for such Si nanowires with sub-5 nm diameter are taken from ref. **[405]** and as mentioned in Chapter 3 the MOS device is assumed to operate at room temperature (300 K). It is worthy to mention that the analytical model is developed by employing coupled mode space approach **[120]** to increase its computational efficiency where the transverse modes for both electrons and the holes are considered according to the respective non-negligible value of Fermi-Dirac distribution function.

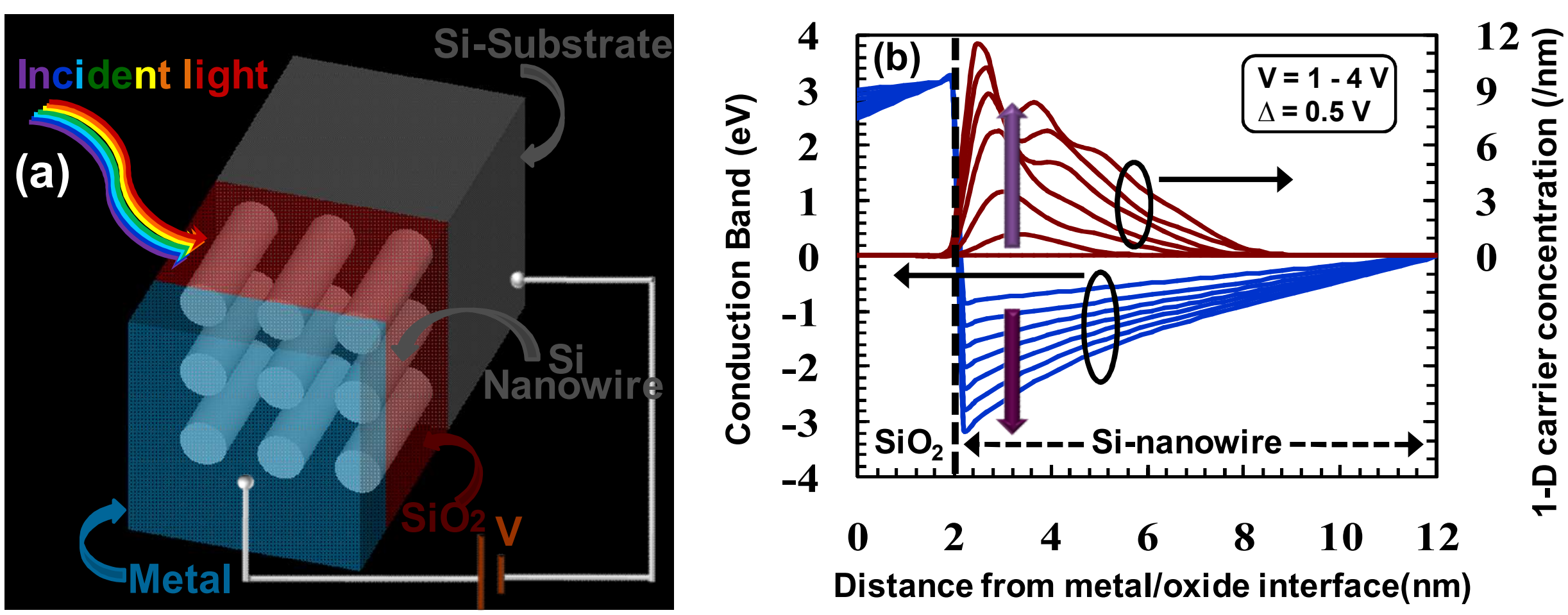

Fig. 4.4. (a) An angled top view of Si-nanowire based vertical MOS devices on Si substrate where the thickness of $SiO_2$-layer is assumed to be 2 nm, (b) Plots of conduction band potential profile and carrier concentration under dark condition along the nanowire axis for 3 nm nanowire diameter at an applied voltage range of 1V to 4V.

The variation of conduction band potential and the corresponding one dimensional (1D) carrier concentration profile under dark condition along the nanowire axis (from the metal/oxide interface) is shown in Fig. 4.4(b) for different applied voltages in the range of 1 V to 4 V. It is observed from such plot that the conduction band is bent in downward direction due to the application of positive voltages and thus, discrete energy states are created in the wedge-shaped potential well. Consequently, the (dark) carrier concentration is observed to increase with increasing positive bias and most importantly troughs and crests are created in the carrier profile which confirms the presence of such discrete quantum states created due to the combined effect of structural and electrical quantization.

## 4.5. Electron-photon interaction in vertical MOS device

Fig. 4.5(a) depicts the variation of energy gap between two definite interacting states of conduction and the valence band with distance from the oxide/semiconductor interface. It is apparent from the figure that there is no spatial variation of energy states in the conduction band region, however, valence sub-band energy varies with the distance along the nanowire axis. Thus, the energy gap between such two interacting states is observed to decrease due to the different nature of charge carriers in conduction (3D-confined) and valence (2D-confined) band regions. Interaction between such conduction energy states with valence sub-bands commences only after the MOS device is illuminated with light of a specific wavelength and as a result, EHPs are created throughout the length of the nanowire. However, the positive bias applied at the metal terminal pushes back all the holes away from the oxide/semiconductor interface whereas electrons are attracted towards it. Such a situation signifies that in the conduction band region near the interface, at the top of nanowire, photogenerated carriers do not suffer recombination processes while all the EHPs in the rest of the device region are neutralized by it. It is worthy to mention that the power of incident light has been considered to be 50 mW for the present work and the absorption cross-section ($A_{abs}$) is defined as, $A_{abs} = \left( \pi D_{NW}^{2} / 4 \right) \times \eta_{eff}$, where $\eta_{eff}$ is the absorption efficiency and $D_{NW}$ is diameter of the nanowire **[454]**. Here, multiple interactions between both conduction and valence band region are possible, however, such interaction or transition probability is different for different modes since the lower conduction energy states have higher transition probability due to relatively smaller energy gap, although,

their effective interaction area (spatial extension) is smaller in comparison to their higher order counterparts.

Fig. 4.5(b) pictorially shows the variation of photogeneration rate in the conduction band region along the nanowire axis for different incident wavelengths of light. It has already been mentioned that multiple transitions within the conduction- and valence- band regions are possible to occur when the system is illuminated.

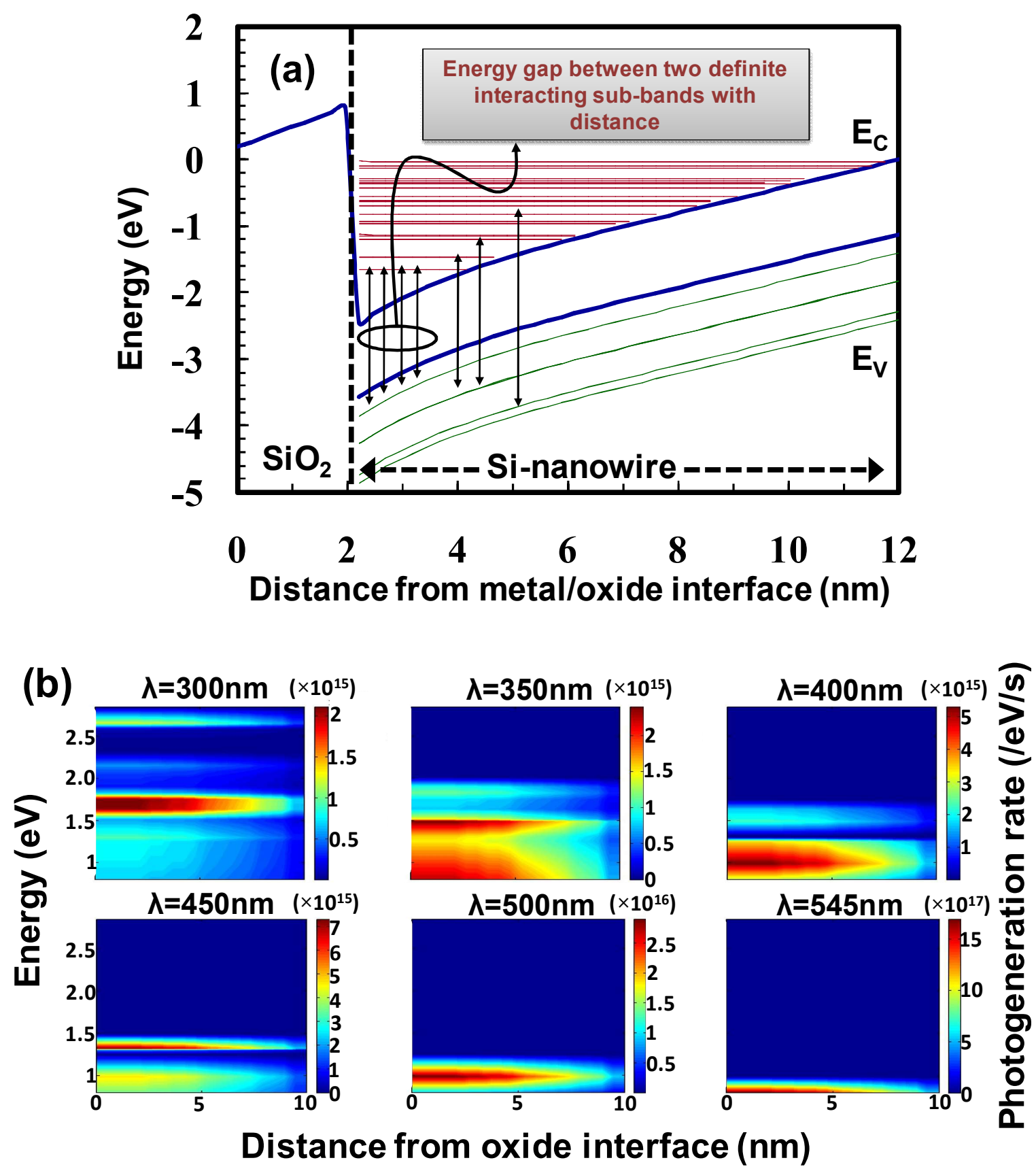

Fig. 4.5. (a) Variation of energy gap between two interacting states in conduction and valence band regions with distance from the metal/oxide interface for the nanowire diameter of 2.5 nm at an applied bias of 3.1 V, (b) Plot of the photogeneration rate along the nanowire axis for different incident wavelengths (300 nm to 545 nm) in the conduction band regime with the same combination of nanowire diameter and applied bias (*i.e.,* 2.5 nm and 3.1 V).

Thus, it is apparent from the figure that for 300 nm incident wavelength, highest photogeneration rate (~$10^{15}$ /eV/s) is observed to occur near the energy level of 1.75 eV which also confirms the participation of higher order energy states in the interaction. However, when the wavelength is increased such interactions of the higher order energy states are observed to decrease and consequently, for 500 nm and 545 nm illumination, interactions due to lower order energy states start to dominate. Such reallocation of interacting states from higher to lower order energy states is evident since for higher wavelengths the energy of the incident photon is less which in turn, also signifies the domination of lowest available energy state in the interaction.

It is observed from the plot that at 545 nm, the interaction is fully concentrated at the lowest available energy state of 0.8 eV and such interaction contributes to a significantly higher photogeneration rate of ~$10^{18}$ (/eV/s). Moreover, the broadening of photogeneration spectrum is also observed to decrease for the 545 nm illumination when compared to the other incident wavelengths.

## 4.6. Vertical Si-nanowire MOS device as multicolor photosensor

The color selection ability of the vertical nanowire MOS device is investigated by choosing different combinations of nanowire diameters and applied voltages for detecting the entire range of visible spectrum. In this regard, Fig. 4.6(a) represents photocapacitance peaks for the red-green-blue (RGB) photo-impulses obtained for the devices with relevant diameter-voltage combinations. Here, such photoresponse for two different illumination intensities is also compared. The photocapacitance profiles corresponding to the RGB bands show sharp peaks at the wavelengths of 465 nm, 545 nm and 670 nm respectively, where the amplitudes of secondary peaks are negligible in comparison to the primary peak. Such negligible magnitudes of other secondary peaks lead to a very high selectivity of a particular wavelength to be detected. It is also observed from the plot that with increasing the power of incident light from 50 mW to 100 mW only the primary peaks are amplified; however, such amplification is also not directly proportional to the increment of source power (SP). The table at the inset elaborates such nonlinearity in the relation between the increment of incident light source power and amplification of the photocapacitance peaks for different wavelengths. At 50 mW light source power the occupancy of photocapacitance peaks for the red (R) and blue (B) spectral bands with respect to the green (G) spectral band is 67.7% and 79.5% respectively, whereas it changes to 74.1% and 76.2% after doubling the source power to 100 mW. Therefore, when a particular image is captured under the illumination of 50 mW and 100 mW source power, color constancy is not maintained due to the differences in color occupations for RGB spectral bands. Hence, such nonlinear behavior indicates inconsistency of the photosensing devices to identify/detect the color of incident light precisely, independent of the incident light source intensity.

The capability of the present device to detect wavelengths in the NIR-region is also investigated as shown in Fig. 4.6(b). For this purpose, the nanowire diameter is considered to be 30 nm. The choice of such relatively higher diameter to detect the NIR wavelengths depends on the fact that with increasing the diameter, the level of structural confinement is reduced which reduces the inter-sub band energy gap and consequently helps to detect longer wavelengths. Thus, the key is to select appropriate diameter-voltage combinations to detect the desired wavelength of the EM spectra. It is observed from the figure that the photocapacitance profile shows a sharp peak at 810 nm wavelength for a combination of 30 nm diameter and 2 V applied bias. When such applied voltage is increased to 2.2 V the photocapacitance peak is observed to be shifted towards the shorter wavelength region (blue shift) and results a sharp peak at 760 nm. Similarly, on further increasing the bias to 2.3 V such a profile is shifted more towards the blue and results a peak at 740 nm. Such a shifting of photocapacitance peaks occurs with increasing the bias (*i.e.,* increasing the electrical confinement) since the energy gap between two interacting states from conduction and valence band regions is increased. Corresponding FWHMs for these three spectral bands are calculated by fitting such spectra with Gaussian distribution function and the values are obtained to be 14.35 nm, 10.56 nm and 9.90 nm, respectively. Thus, the proposed device is also able to detect wavelengths in the NIR-region with high selectivity.

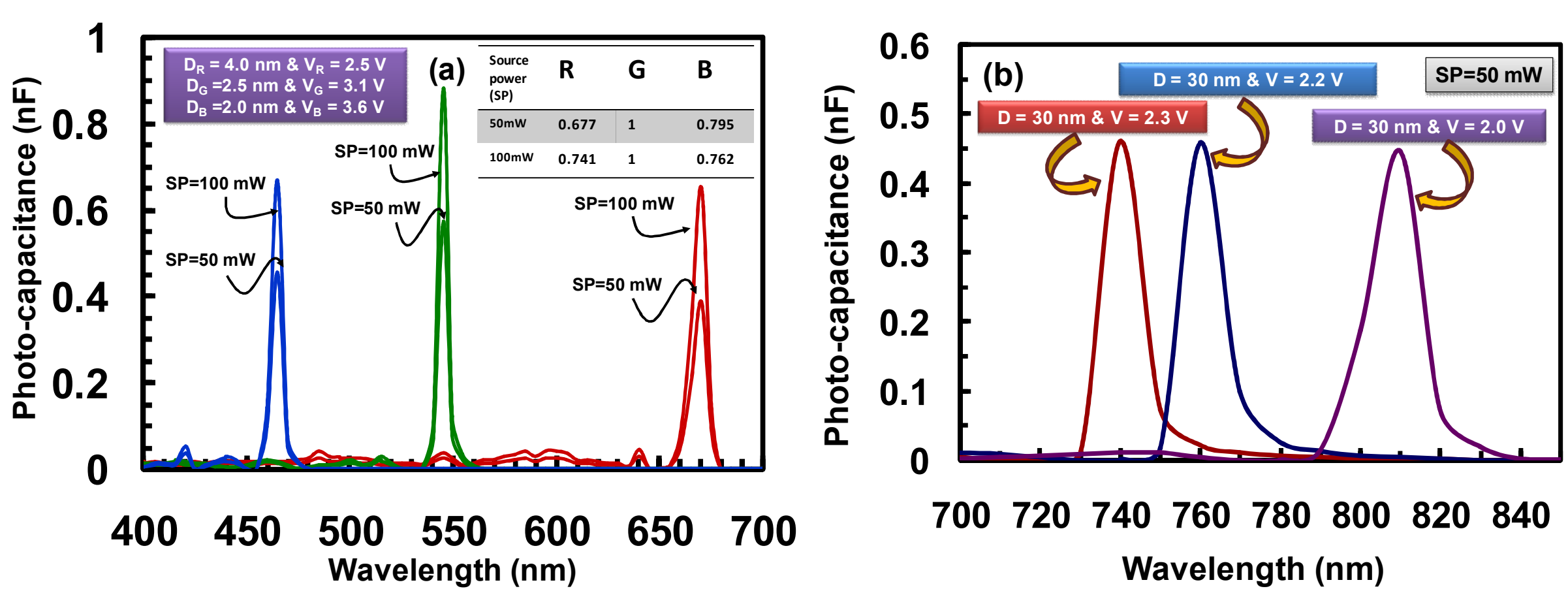

Fig. 4.6. (a) Plot of photocapacitance with illumination wavelength in visible region for the MOS capacitors with various combination of nanowire diameter (D) and applied voltage (V). To capture the conventional RGB signals the required window of diameter-voltage combination is obtained to be, $D_R$=4.0 nm & $V_R$=2.5 V, $D_G$=2.5 nm & $V_G$=3.1 V and $D_B$=2.0 nm & $V_B$=3.6 V

for red, green and blue, respectively. Table at the inset shows the nonlinear behavior in the relation between the increment of illumination intensity and amplification of the photocapacitance peaks. (b) Plot of photocapacitance profile with incident light wavelength for the detection of NIR-wavelengths with nanowire diameter of 30 nm and applied bias in the range of 2 V to 2.4 V.

## 4.7. Design window for multicolor detection with high spectral resolution

It was previously mentioned in section 4.1 that in conventional photosensing devices, colors of a particular image are reproduced by using CFAs which conversely acts as the detrimental source of color aliasing effects due to unavoidable overlap between each spectral band **[453]**. However, it is apparent from Fig. 4.6 that the proposed nanowire MOS device is capable of detecting color of the incident light with very high selectivity since there is no indication of overlapping between the individual spectral lines of red (R), green (G) and blue (B). Also, the multispectral imaging demands several spectral bands in a particular wavelength region to precisely extract the spectral information of an image. Here, the device scheme is designed in such a way that such device can detect wavelengths of the visible spectrum in several bands with high spectral resolution. The design approach comprises of engineering the energy gap between two interacting states in the conduction and valence band regions by structural and electrical quantization to detect wavelengths of the entire visible spectrum. It is also worthy to mention that the possibility of multiple optical transitions is eliminated by choosing an appropriate combination of nanowire diameter and applied voltage which in turn leads to maximize the photogeneration rate with high spectral selectivity as evident from Fig. 4.5(b). Fig. 4.7(a) shows the design window for the detection of visible region wavelengths by the nanowire MOS based photodevice for different combinations of nanowire diameter and applied voltage. The entire visible region (380 nm to 700 nm) can be divided into 64 spectral bands with 5 nm spectral resolutions by selecting appropriate diameter-voltage combinations in the range of 1.5 nm to 5 nm and 2 V to 5 V. This also leads the photosensing of visible spectrum to be discrete where each combination of nanowire diameter and applied voltage represent a particular color of the incident light. In Fig. 4.7(a), the color variation owing to each spectral band is plotted in z-direction where the apparent discreteness appears due to the selection of diameter-voltage combinations which, in other sense, is a representation of digital imaging. Further, to analyze the broadening of each spectral band in the visible region, FWHMs are calculated by using the Lorentzian distribution function and plotted with the incident wavelengths in Fig. 4.7(b). It is observed from the figure that, smaller FWHMs (< 1 nm), especially in the shorter wavelength region (380 nm to 495 nm), signifies highly monochromatic detection of incident wavelength due to increased inter sub-band energy gap for such low diameter (< 2 nm) devices. Overall, the proposed device scheme offers the FWHMs to be less than 5 nm for the entire range of photodetection which is also less than the spectral resolution (*i.e.,* 5 nm) in terms of the number of spectral bands, which is essential for reducing the overlap between successive spectral bands and thus results to a high quality digital imaging device.

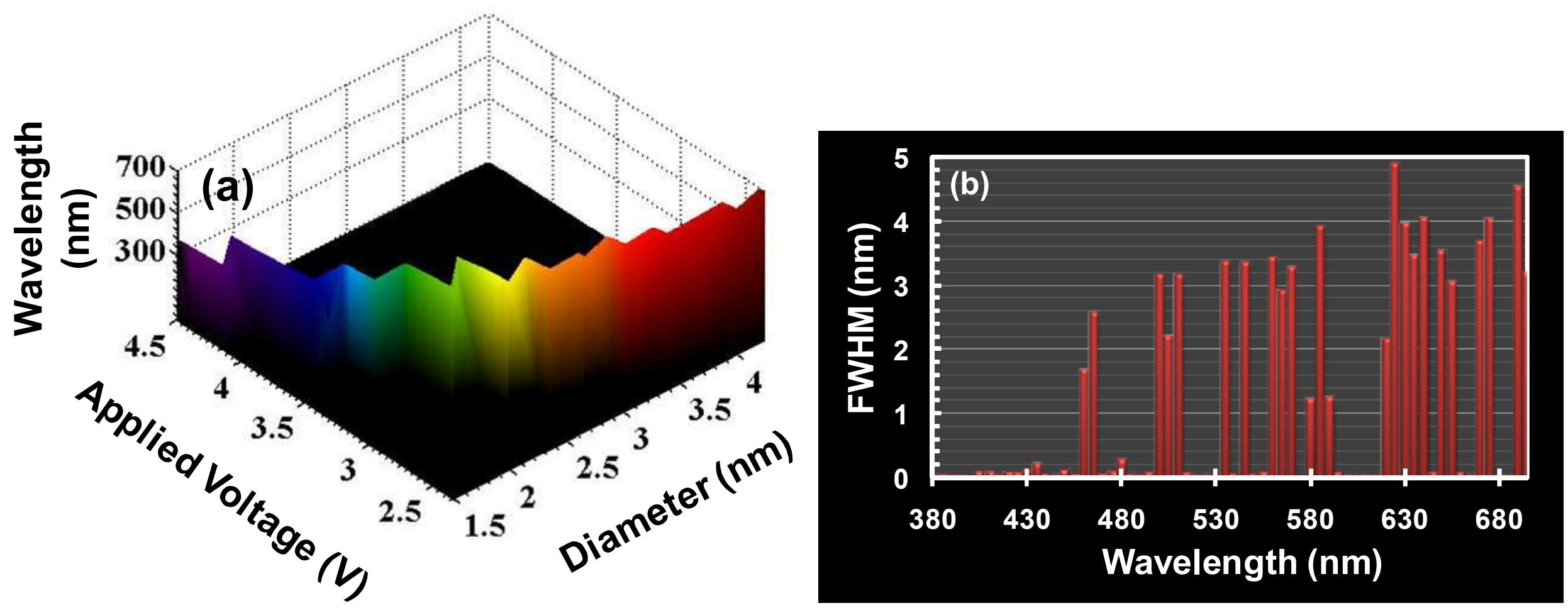

Fig. 4.7. (a) Plot of the visible spectrum (380 nm to 700 nm) detected through 64 spectral bands with 5 nm spectral resolution by specific combinations of nanowire diameter and applied voltage, (b) Plot of FWHM for each spectral band with the incident wavelengths in the visible region. FWHMs for the entire domain of photodetection are designed to be less than the spectral resolution (5 nm) which is essential for digital photosensing.

## 4.8. Selection of InP instead of Si as the nanowire material for wavelength selective visible light sensing

In Fig. 4.7, the Si-nanowire based vertical MOS photocapacitor has been investigated to detect directly the entire visible spectrum (380 nm to 700 nm) with high selectivity avoiding the split and superposition technique. At this point it is important to mention that such devices require sub-5 nm nanowire diameters to achieve color sensing with a sub-0.5 nm control on such diameters which increases complexities associated in fabrication procedures. In contrast, the III-V semiconductors have gained prime interest for nanowire photodetectors due to their high electron mobility and lighter bulk electron effective masses. Such lighter electron effective mass enables the execution of significant level of quantization for relatively higher device dimensions and thereby, reduces the relevant fabrication complexities. Thus, the capability of InP-nanowire based vertical MOS photocapacitor to directly detect the color of incident light can also be investigated by following the same theoretical model and using the values of material properties for InP instead of Si. InP is chosen since it has a relatively higher EBR of 15 nm and its band gap of 1.34 eV is also appropriate for detecting the wavelengths of visible region. Here, the InP nanowires are assumed to be grown on an InP substrate and hafnium oxide ($HfO_2$) is considered as the insulator (instead of $SiO_2$) with an EOT (equivalent oxide thickness) of 0.5 nm. The photocapacitance profile for a 50 nm diameter device with different applied voltages is shown in Fig. 4.8(a) to detect the VIBGYOR (violet-indigo-blue-green-yellow-orange-red) colors.

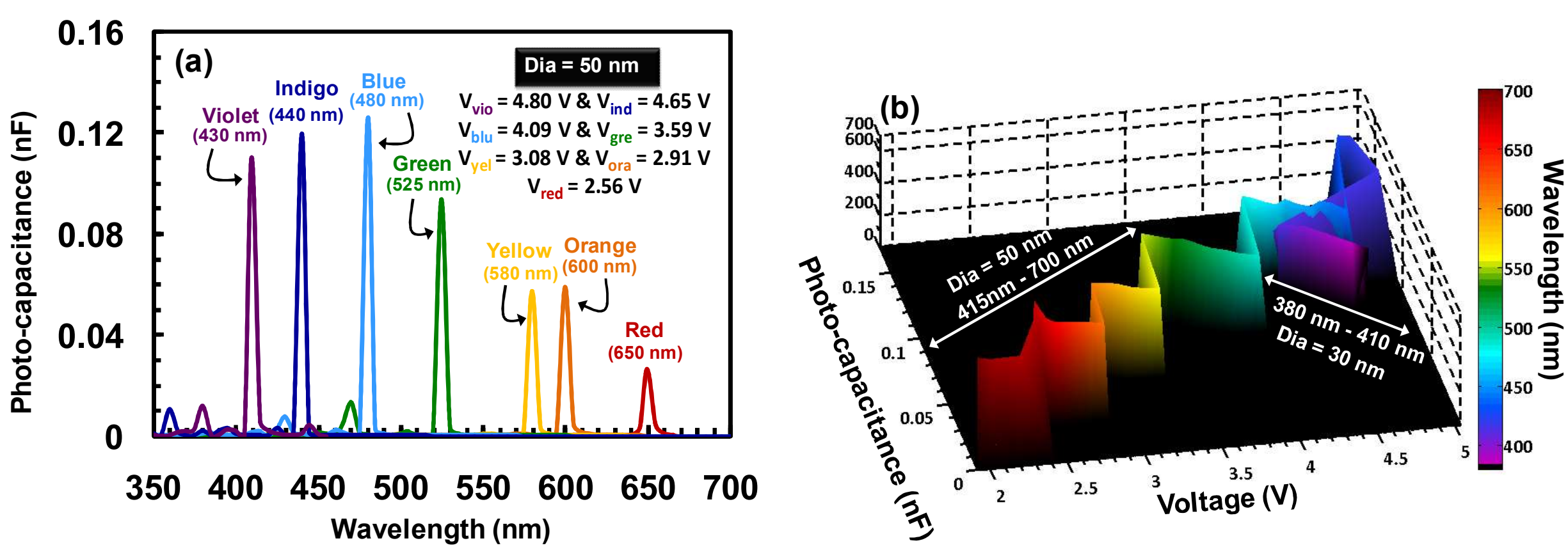

Fig. 4.8. (a) The plot of photocapacitance with illumination wavelength for detecting the VIBGYOR colors with appropriate diameter-voltage combinations, (b) Design window to detect the entire visible spectrum with 5 nm spectral resolutions by InP-nanowire based MOS device.

Here also the sharp photocapacitance peaks for all the diameter-voltage combinations ensure high selectivity of a particular wavelength and such high selectivity is obtained by the combined effect of structural and electrical quantization in terms of engineering the energy gap between two interacting states of the conduction and valence band region of InP nanowire. The design window to detect the entire visible spectrum (380 nm to 700 nm) directly with 5 nm spectral resolution (*i.e.,* 64 spectral bands) is shown in Fig. 4.8(b). It is observed that the InP-nanowire based device is capable of detecting the entire visible spectrum with relatively higher nanowire diameters (50 nm and 30 nm) in comparison to the Si-nanowire based devices.

## 4.9. Summary

In the present chapter, an analytical model has been developed for designing the vertically aligned nanowire MOS devices as multi-color photosensors with high spectral resolution. The photogeneration in such devices has been studied in detail by solving a set of coupled quantum field equations associated with second quantization electron-photon field operators in the NEGF formalism. The photoresponse of such devices show sharp peaks for a particular wavelength which is detected by the appropriate selection of a nanowire diameter-applied voltage combination. Such selection of diameter-voltage combinations depends on engineering the energy gap between conduction and valence energy states. The combination of geometrical dimensions and external voltages is used to tune the level of quantization in such structures. The proposed Si- and InP- nanowire based devices can directly detect all the colors of visible region with high selectivity, however; the similar performance can be achieved in InP-nanowire based devices with relatively higher geometrical dimensions. Thus, the proposed devices can be useful for developing the future nanostructure based high quality digital photosensing or imaging systems.

The development of current NEGF based model in this chapter has been extended in the next chapter to use such vertical nanowire MOS devices for harvesting the solar energy. Thus, the following chapter deals with the working principle of vertical nanowire MOS devices to operate as a solar cell with efficiency beyond the Shockley-Queisser (SQ) limit.

# Chapter: 5*

# Vertically aligned GaAs-nanowire MOS solar cell: Route for achieving high efficiency beyond Shockley-Queisser limit

## 5.1. Introduction: Breaking the efficiency limit of single junction solar cells

The climate action summit 2019, held in New York, addresses the urgency to develop sustainable and renewable energy solutions to fight against the consequences of climate change in human life. Realistic, concrete and visionary actions are expected to be taken since the reports of NASA and NOAA on the temperature increase of last five years are shockingly alarming. In the UN summit, appropriate policies are invited to reduce the green-house gas emissions, caused mainly due to the fossil-fuel driven technologies, by 45% over the period of next ten years. Recently, the Swedish teenager environmental activist raises her voice in demand of immediate actions against the climate change. In this regard, the responsibility of 'Science' is to provide inroads to understand the root causes of such climate change and develop methodologies to mitigate its impact on mankind.

'Sun'- the infinite source of sustainable and renewable energy is evidently the most promising solution available in nature to fight against the world's climate crisis. The development of solar cells for harnessing solar energy has witnessed tremendous progress in terms of its efficiency and economic growth from first generation wafer based technology to second generation thin-film based solar cells and finally, the third and fourth generation emerging technologies, including the organic, polymer and nanostructure based devices **[459-463]**. Enormous efforts are being made by the researchers since last decade to improve the efficiency of nanostructure based solar cells so that it can reach to a competitive level to produce at the large industrial scale **[464-466]**. Semiconductor nanowires have specially gained keen attention of the global research community

***Most of the content of Chapter: 5 is published in the article, "Design and modeling of high efficiency GaAs-nanowire metal-oxide-semiconductor solar cells beyond the Shockley-Queisser limit: An NEGF approach", by Subhrajit Sikdar, Basudev Nag Chowdhury and Sanatan Chattopadhyay, 'Physical Review Applied' 02/2021, 15, 024055, DOI: 10.1103/PhysRevApplied.15.024055.***

to design next-generation solar cells **[242-254]**. The impact of nanowire geometry, crystal structure, controllable doping, junction geometries, surface passivation and cleaning and, most importantly, the contact formation has been rigorously studied to achieve optimal sun light absorption and the generation, separation and collection of charge carriers **[256-263]**. It is also worthy to mention that several approaches have been adopted to surplus the conventional Shockley-Queisser (SQ) efficiency limit, however, breaking such SQ-limit by fabricating a single junction solar cell still appears as a challenge in terms of reducing the fabrication cost or to increase its large scale productions. Recently, a vertical nanowire based p-n junction solar cell has been reported as an apparent solution to achieve higher efficiency beyond the conventional SQ-limit, which is realized by engineering the propagation of particular electromagnetic (EM)-modes inside the nanowire which effectively results in a higher light absorption cross-section in comparison to its geometrical cross-section **[255]**. Such device is reported to achieve a significantly higher current density of 180 $mA/cm^2$ with a 'record' single junction power conversion efficiency of 40%. Also, significantly higher external quantum efficiency (EQE) of 14.5 is reported which has been attributed to the resonances similar to Mie resonances **[255]**. Thus, the vertical p-n junction nanowire device provides a route for achieving several-fold higher power while reducing the footprint. It is also discussed in ref. **[37]** that in terms of achieving next generation high efficiency solar cells, the focus of research should shift from only 'device design and materials' to particularly on the 'light management or light trapping inside the solar cells' which pave the way of developing the solar cells with ultrahigh efficiency in the range of 50% to 70%.

In this chapter, the NEGF based theoretical model is employed to study the working principle of a GaAs-nanowire based vertical MOS solar cell structure. Here, the photogenerated holes are utilized to harvest solar energy with significantly higher power conversion efficiency beyond the SQ-limit. Such high efficiency is achieved due to the resonance occurring between the incident photon modes and the energy difference created by 3D-quantized states (electrons) with the 2D-quantized sub-bands (holes). Moreover, the power conversion efficiency along with the other relevant solar cell performance parameters such as, open circuit voltage ($V_{OC}$), short-circuit current ($I_{SC}$), fill factor (FF), external quantum efficiency (EQE) and responsivity are also studied in detail. Thus, the present chapter indicates an alternate route to design next-generation MOS based solar cells with significantly higher efficiency beyond the SQ-limit.

## 5.2. Scheme of the device

Fig. 5.1(a) shows the schematics of a vertically oriented GaAs-nanowire based MOS device considered for the design of high efficiency solar cell. Such structure has also been elaborated in Chapter 3 and Chapter 4. In such devices, a single nanowire is assumed to be grown or deposited on a (Si-) substrate and covered with a silicon dioxide ($SiO_2$) layer. The tosylate modified poly (3,4-ethylenedioxythiophene) or PEDOT-$T_{os}$ is considered to be deposited on top the oxide layer. Here, GaAs is chosen as the semiconducting material due to its relatively larger excitonic Bohr radius (EBR) of 12 nm **[467]**, and correspondingly, the largest diameter is considered to be 24 nm in order to study the quantization effect in the device. The choice of oxide material ($SiO_2$) and its thickness are chosen depending upon the previously observed results in Chapter 3 where the $SiO_2$ at nanodimension has exhibited relatively higher impact on reducing carrier tunneling probability than the other insulating materials. The PEDOT-$T_{os}$ has been assumed as metallic contact since it has emerged as a potential conducting polymer for optoelectronic applications owing to its high electrical conductivity (~1000 S/cm) and ~90% optical transmittance in the visible region. In the current device, it is advantageous to utilize its relatively lower work function of ~3.8 eV **[468]** compared to that of GaAs (~4.78 eV, considering intrinsic) which also determines the open-circuit voltage of the proposed nanowire MOS solar cell.

Schematic of the photogeneration phenomenon in such vertical nanowire-MOS device is shown in Fig. 5.1(b) at zero bias condition. Here, the work function difference between metal and semiconductor creates a virtual bias which bends the conduction and valence bands in downward direction following the metal Fermi level and thereby, leads to the formation of a quantum well near the oxide/semiconductor interface on top of the nanowire. In such quantum well, charge carriers, *i.e.,* the electrons are two dimensionally confined in transverse directions due to its dimension and the virtual bias, created due to the work function difference of respective materials, confines such charge carriers also in the longitudinal direction. Therefore, the conduction band electrons are 3D-confined whereas, in the valence band region the charge carriers, *i.e.,* holes are 2D-confined since these are pushed away from the interface and free to move along the longitudinal direction. On illumination with energy above the material energy band-gap, the EHPs are created where the optically generated holes are 2D-confined (in valence band) and electrons are 3D-confined (in conduction band). This also indicates the photogeneration phenomenon due to the interaction between incident photons with 3D-confined electrons and 2D-confined holes. It is also worthy to mention that the number of sub-bands and energy states contributes towards such photogeneration process in valence and conduction band primarily depends on the diameter and work function difference of the relevant materials (and also on the applied bias when applied).

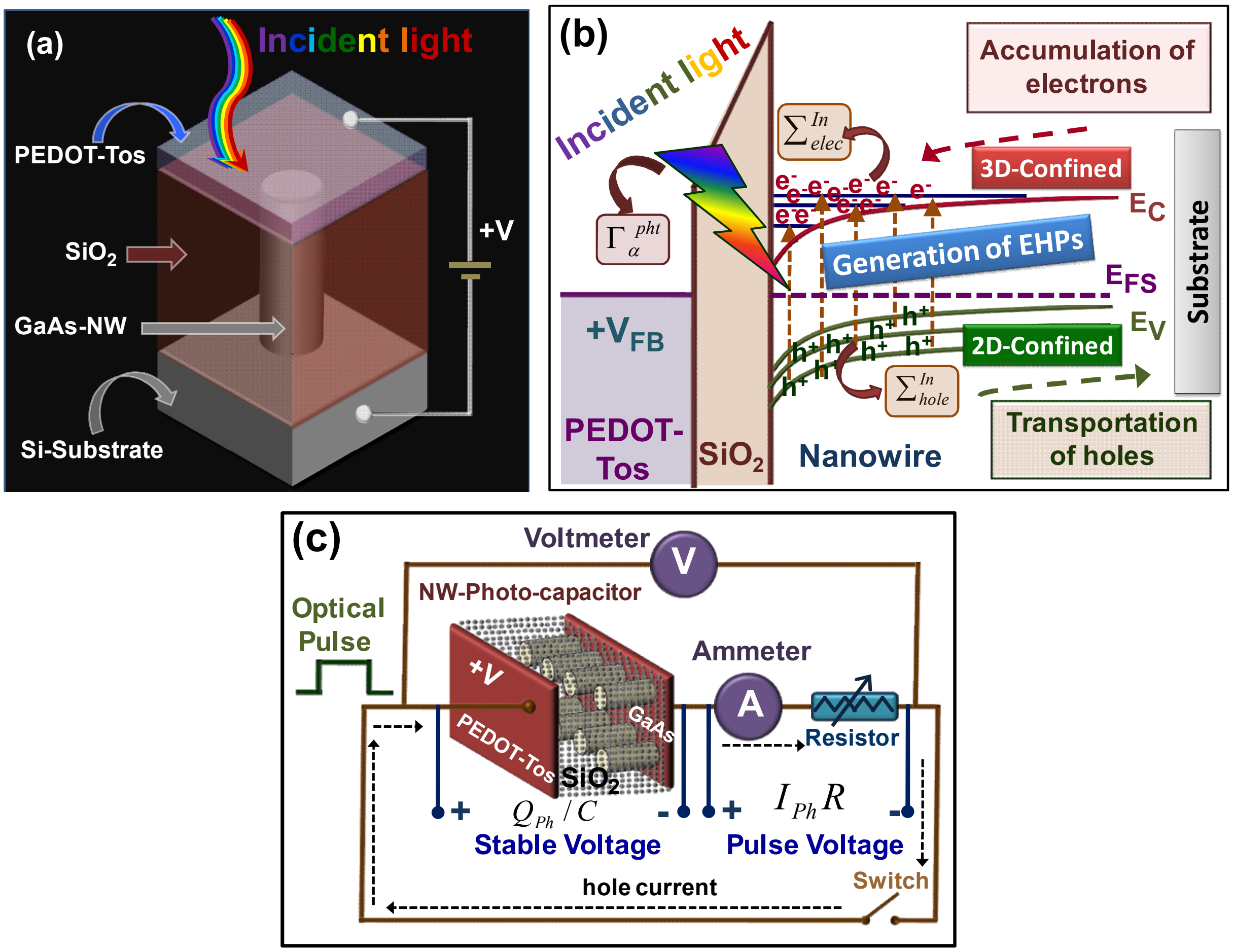

Fig. 5.1. (a) Schematic of the single nanowire vertical MOS device with incidence of light on the metal; (b) Schematic of the photogeneration phenomenon showing the mechanism of interaction between 2D-confined holes with the 3D-confined electrons; (c) Probable circuit diagram to obtain the photogenerated current.

In the present device scheme, the valence band charge carriers, *i.e.,* holes are actually responsible for generating the device current since these are free to move in the longitudinal direction and the conduction band charge carriers, *i.e.,* electrons are accumulated at the oxide/semiconductor interface and develop a capacitance since these are 3D-confined. Therefore, in the current device scheme, photogenerated hole-current can be utilized for solar cell operation whereas the photoelectrons can be utilized as solar-capacitance for storing such energy.

Fig. 5.1(c) shows the possible circuit diagram of the nanowire MOS device, connected in series with a resistor and an ammeter, where the device hole current is monitored by it and, a voltmeter is also connected in parallel with the capacitor and the resistor. On application of an optical pulse, the electron-hole pairs are generated which charge the capacitor, and simultaneously, the holes flow through the nanowire and resistor up to the metal terminal to complete the circuit. Such hole current can be detected by the ammeter and the generated voltage across the capacitor (formed by the accumulated electrons) and resistor can be obtained by the voltmeter. At this point, it is important to mention that the output in terms of current/capacitance obtained from such circuit is basically in pulsed form, *i.e.,* as long as the optical pulse is present, the hole-current and the electron-capacitance exist. Such circuit arrangement may also resemble with the conventional capacitor-discharge (RC) circuit.

## 5.3. Theoretical modeling

The analytical model of photogeneration process in the vertical nanowire MOS devices is developed by assuming an equivalent picture of scattering which has been discussed in detail in section 4.3 of Chapter 4. In such approach, the absorption of a photon to create an electron-hole pair is analogically viewed as the scattering of a *'virtual electron'* (having effective mass of a valence-band-hole with equal and opposite momentum) by the incident photon, leading to the transition of such *'virtual electron'* into *'real electron'* of conduction band (having electron effective mass). Initially, the potential distribution along the nanowire MOS quantum well and the associated quantized states for the electron $(\varepsilon_i^c)$ and hole sub-bands $(\varepsilon_r^v)$ are calculated self-consistently in dark condition (without any optical illumination) by solving the Schrodinger-Poisson equations (discussed in detail in Chapter 3, section 3.3). However, when the device is illuminated with an optical pulse, the resulting changes are obtained from the relevant Hamiltonian of the device, given by **[118]**,

$$H = \sum_i H_{ISO}^c C_i^{c+} C_i^c + \sum_r H_{ISO}^v V_r^{v+} V_r^v + \sum_\alpha \hbar\omega_\alpha^{pht} b_\alpha^{pht+} b_\alpha^{pht} + \sum_{i,r,\alpha} \begin{pmatrix} \tau_{ir}^\alpha C_i^{c+} V_r^v b_\alpha^{pht} + \tau_{ir}^{\alpha*} V_r^{v+} C_i^c b_\alpha^{pht+} \\ + \tau_{ri}^\alpha V_r^{v+} C_i^c b_\alpha^{pht} + \tau_{ri}^{\alpha*} C_i^{c+} V_r^v b_\alpha^{pht+} \end{pmatrix} \quad (5.1)$$

where $C_i^c$ and $V_r^v$ are the second quantized field operators for electron in $i^{\text{th}}$ state of conduction band and hole in the $r^{\text{th}}$ sub-band of valence band, respectively, and $b_\alpha^{pht}$ is the field operator for incident photon in $\alpha$ mode with angular frequency $\omega_\alpha$, while $\tau$ represents the interaction potential governing the photogeneration of electron-hole pairs in the device. It is worthy to mention that the first three terms of Eqn. (5.1) represent the isolated Hamiltonians (*i.e.,* without interaction) of conduction band electrons, valence band holes and incident photons, respectively, while, the last term corresponds to the respective interactions. Following Heisenberg' equation of motion, Eqn. (5.1) leads to the relevant equations of motion for such photogenerated electron, hole and the incident photons as (recapitulating from Chapter 4, section 4.3),

$$i\hbar\frac{d}{dt}C_i^c = H_{ISO}^c C_i^c + \sum_{r,\alpha}(\tau_{ir}^\alpha V_r^v b_\alpha^{pht} + \tau_{ri}^{\alpha *} V_r^v b_\alpha^{pht+}) \tag{5.2.a}$$

$$i\hbar\frac{d}{dt}V_r^v = H_{ISO}^v V_r^v + \sum_{i,\alpha}(\tau_{ri}^\alpha C_i^c b_\alpha^{pht} + \tau_{ir}^{\alpha *} C_i^c b_\alpha^{pht+}) \tag{5.2.b}$$

$$i\hbar\frac{d}{dt}b_\alpha^{pht} = \hbar\omega_\alpha^{pht} b_\alpha^{pht} + \sum_{i,r}(\tau_{ri}^{\alpha *} C_i^{c+} V_r^v + \tau_{ir}^{\alpha *} V_r^{v+} C_i^c) \tag{5.2.c}$$

The Eqns. (5.2(a) – (c)) physically imply that if either of the conduction band electrons, valence band holes or incident photons are considered to compose the *'device'* then the remaining two work as the *'reservoirs'*. Therefore, the equations are mathematically solved by finding the respective Green's functions depending on the corresponding self-energies and in/out-scattering functions (Chapter 4, section 4.3). It is also to be noted that the self-energy physically implies a 'disturbance' to the Hamiltonian of the *'device'* from the *'reservoirs'* during their interactions, whereas, the in- and out-scattering functions respectively indicate the 'disturbances' from the reservoirs just before interaction, and to the reservoirs just after interaction. Further, the probability of recombination of the photogenerated electrons and holes in such nanowire MOS devices reduces significantly since the holes are pushed away from the oxide/semiconductor interface due to the 'virtual bias'. Consequently, the probability of emission of photons also turns out to be negligibly small. Therefore, out-scattering functions for the photogenerated electron-hole pairs and the in-scattering function for photons are neglected in the current model. Thus, the relevant in/out-scattering functions for electrons ($\Sigma_{elec}^{in}$), holes ($\Sigma_{hole}^{in}$) and photons ($\Gamma_{pht}^{out}$) for the nanowire MOS device are obtained to be,

$$\left[\Sigma_{elec}^{in}(t_1,t_2)\right]_{ij} = \sum_{r,s,\alpha,\beta}\left(\tau_{ir}^\alpha \left[N_{\alpha\beta}^{ab-pht}(t_1,t_2) n_{rs}^v(t_1,t_2)\right]^{Iso} \tau_{js}^{\beta *}\right) \tag{5.3}$$

$$\left[\Sigma_{hole}^{in}(t_1,t_2)\right]_{rs} = \sum_{i,j,\alpha,\beta}\left(\tau_{ir}^{\alpha}\left[N_{\alpha\beta}^{ab-pht}(t_1,t_2)n_{ij}^{c}(t_1,t_2)\right]^{Iso}\tau_{js}^{\beta*}\right) \tag{5.4}$$

and,

$$\left[\Gamma_{pht}^{out}(t_1,t_2)\right]^{\alpha\beta} = \sum_{i,j,r,s}\left[\begin{array}{l}\left\{\tau_{ir}^{\alpha}\left(\left[n_{rs}^{v}(t_1,t_2)\right]\left[n_{ij}^{c}(t_1,t_2)\right]+\left[p_{rs}^{v}(t_1,t_2)\right]\left[p_{ij}^{c}(t_1,t_2)\right]\right)\tau_{js}^{\beta*}\right\}\\+\left\{\tau_{ri}^{\alpha*}\left(\left[p_{rs}^{v}(t_1,t_2)\right]\left[n_{ij}^{c}(t_1,t_2)\right]+\left[n_{rs}^{v}(t_1,t_2)\right]\left[p_{ij}^{c}(t_1,t_2)\right]\right)\tau_{sj}^{\beta}\right\}\end{array}\right] \tag{5.5}$$

where, $n_{ij}^{c}(t_1,t_2)$, $p_{ij}^{c}(t_1,t_2)$, $n_{rs}^{v}(t_1,t_2)$ and $p_{rs}^{v}(t_1,t_2)$ are the correlation functions for filled and empty states in the conduction and valence bands, respectively. $N_{\alpha\beta}^{ab-pht}$ represents the number of photons absorbed by the quantum well created in the nanowire MOS device and is given by,

$$N_{\alpha\beta}^{ab-pht} = \left[\left(I_\alpha/\hbar\omega_{\alpha}^{pht}\right)*\Lambda\right]*n(\omega_\alpha)\delta_{\alpha\beta} \tag{5.6}$$

where, $I_\alpha$ is the solar-intensity and $n$ represents the combination of refractive indices of the respective materials, such as PEDOT-$T_{os}$, $SiO_2$ and GaAs, for the respective wavelengths of incident light, and $\Lambda$ denotes the absorbing area (physical cross-section) of the nanowire. It is apparent from Eqns. (5.3), (5.4) and (5.5) that the in/out-scattering functions depend on the interaction potential governing the photogeneration phenomenon. As mentioned earlier in section 4.3.1 of Chapter 4, the photogeneration process in the current model has been conceptualized by an equivalent scattering picture which gives rise the interaction potential to be, $\tau_{ir}^{\alpha} = \frac{q\hbar}{2}\left[\left(\frac{\vec{k}_i^{elec}}{m_e^*}+\frac{\vec{k}_r^{hole}}{m_h^*}\right)\cdot\vec{A}_0 b_\alpha^{pht}\right]$, from Eqn. 4.28. This in turn gives rise to the concentration of photogenerated electrons per second in the device to be,

$$n_{elec}^{pht} = \frac{1}{2\pi\hbar}\sum_i\left[\Sigma_{elec}^{in}\left(\varepsilon_i^{c}\right)\right] \tag{5.7}$$

whereas, the photogenerated hole concentration is obtained to be,

$$n_{rs}^{v}(t,t') = \sum_{r',s'}\int\int dt_1 dt_2\left[G_{r,r'}^{hole}(t,t_1)\left(\Sigma_{hole}^{in}(t_1,t_2)\right)_{r's'}G_{s',s}^{hole*}(t_2,t')\right] \tag{5.8}$$

where, $G^{hole}$ denotes the Green's function of 2D-confined holes given by (on Fourier transform into energy domain),

$$G^{hole}(E) = \left[E - H_{ISO}^{v} - \Sigma_D - \Sigma_{hole}\right]^{-1} \tag{5.9}$$

where, $\Sigma_D$ is the self-energy to represent coupling between the nanowire and substrate, and $\Sigma_{hole}$ is the same for photogeneration of holes. It is interesting to note that the Fourier transform of integrand in Eqn. (5.8) into energy domain gives rise to the local density of states (LDOS) of photogenerated holes in the device. Finally, the device current ( $I_{Pht}^{hole}$ ) due to such photogenerated holes is calculated from the Landauer formula as,

$$I_{Pht}^{hole} = \frac{q}{\hbar}\int Trace\left[\Sigma_{hole}^{In}(E)\left(\frac{i}{2\pi}(G^{hole}(E) - G^{hole+}(E))\right)\right]dE \qquad (5.10)$$

## 5.4. Variation of LDOS of photogenerated carriers in the valence band (*i.e.*, holes) along the nanowire axis

It is apparent that photocurrent in the GaAs nanowire MOS solar cell includes the contributions from both light holes (LH) and heavy holes (HH), *i.e.*, $I_{pht}^{hole} = I_{pht}^{LH} + I_{pht}^{HH}$. Therefore, to study their relative significance in photogeneration process, the LDOS of such photogenerated LHs and HHs available in the device are plotted in Fig. 5.2(a-d) with energy and distance from the oxide/semiconductor interface. In such plots, a comparison of LDOS between the larger (10 nm) and smaller (2 nm) oxide thicknesses is also illustrated. An incident light of 500 nm wavelength has been considered for such comparative analysis since it is energetically higher than the energy bandgap of GaAs (1.42 eV). It is observed from such plots that the value of LDOS near the sub-bands is relatively higher for heavy holes, however, their broadening is comparatively larger for the light holes. This is attributed firstly to the fact that LDOS represents the available area in energy space per energy range, which suggests that more closely spaced the energy eigenstates, higher is the value of LDOS and thus heavy holes exhibit higher LDOS values. On the other hand, the photogeneration induced broadening of sub-bands is larger for smaller effective mass since the corresponding interaction potential is comparatively stronger (see Eq. (4.28)) and thus enhances the photogeneration probability. For GaAs, such values of carrier effective mass for the LH and HH in electronic mass unit are $0.08m_0$ and $0.50m_0$, respectively, resulting to the variation of LDOS as shown in Fig. 5.2. It is interesting to note that the broadening of LDOS in the energy space for both heavy and light holes are larger for the devices with thinner oxide (2 nm) than that of the thicker one (10 nm). This is attributed to the higher penetrability of electric field into the semiconductor nanowire through thinner oxide which leads to larger band-bending and thereby enhances the out-flow of photogenerated holes towards the substrate and consequently creates the sub-band broadening. This in turn improves quantum efficiency of the present NW-MOS based solar cell for reduced oxide thickness.

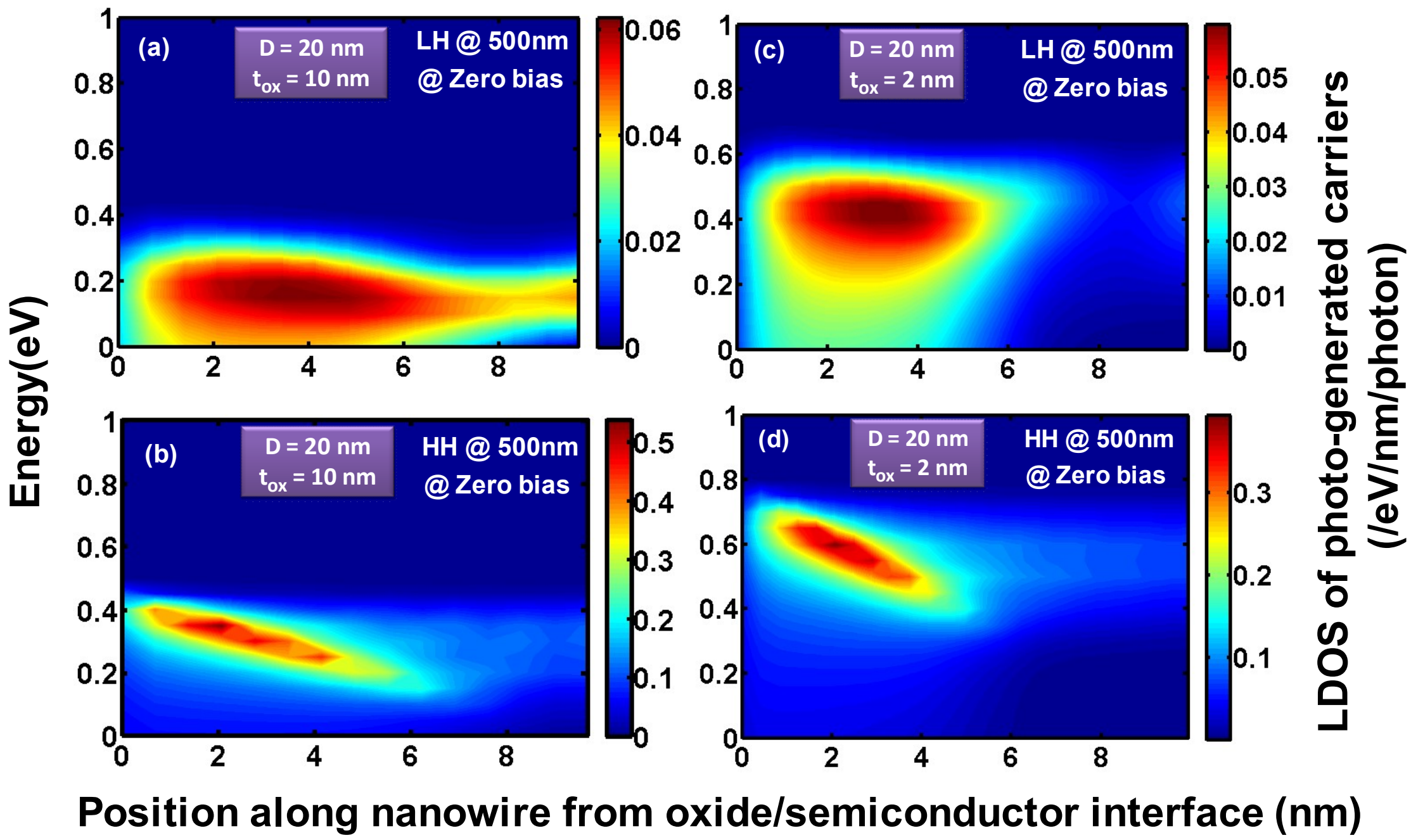

Fig. 5.2. Contour plots of LDOS for the photogenerated holes with energy and position along the nanowire axis from oxide/semiconductor interface for the devices with 10 nm oxide thickness for: (a) light holes and (b) heavy holes; and for the devices with 2 nm oxide thickness for: (c) light holes and (d) heavy holes. The nanowire diameter is considered to be 20 nm.

## 5.5. Variation of external quantum efficiency (EQE)

Fig. 5.3(a) and 5.3(b) depict the variation of EQE with incident wavelength for the device under consideration for different nanowire diameters where the oxide thicknesses are considered to be 10 nm and 2 nm, respectively. The wavelength dependent refractive indices of participating materials have also been incorporated into the present model (Eqn. (5.6)) to take into the account the loss of photons due to reflection and/or absorption by the layers of PEDOT-$T_{os}$, $SiO_2$ and GaAs, which may degrade the relevant EQE. However, even after the consideration of such losses, the present device exhibits significantly high EQE of ~30, compared to the conventional solar cells. It is apparent from such plots that the downscaling of nanowire diameter enhances its EQE and also the thinning of oxide layer further improves it considerably. For instance, EQE for the device with 14 nm diameter and 10 nm oxide thickness at 500 nm illumination wavelength is obtained to be ~20 while such value increases to ~30 for the device considering 2 nm oxide thickness. Such extremely high value of EQE originates due to the resonance between the modes of incident photon with energy-gap between the quantized states/sub-bands of electrons and holes in the present NW-MOS device. Such resonances similar to Mie-resonance have already been observed in the nanowire solar cells which effectively increases the absorption cross-section significantly compared to the geometrical cross-section of such nanowires **[255]**.

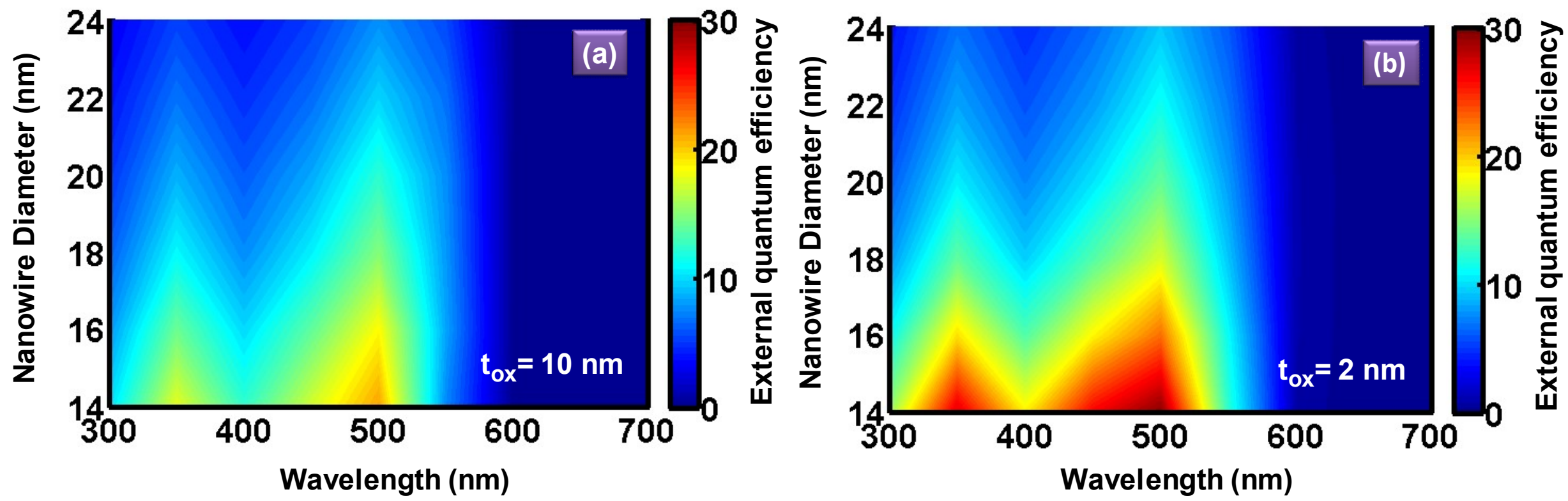

Fig. 5.3. Contour plot of EQE with illumination wavelengths in the range of 300 nm to 700 nm and the nanowire diameter ranging from 24 nm to 14 nm for: (a) the oxide thickness of 10 nm and, (b) 2 nm.

## 5.6. Current-voltage (I-V) characteristics of the nanowire-MOS devices assuming its light illumination from AM 1.5G standard solar spectra

I-V characteristics of the GaAs-nanowire vertical MOS solar cell, for the incident light of AM 1.5G standard solar spectra, are plotted in Fig. 5.4. To provide the design aspect of such devices to achieve superior performance, the I-V characteristic curves for different nanowire diameters (24 nm - 14 nm) are shown for different oxide thicknesses in the range of 10 nm to 2 nm in Fig. 5.4(a)-(e), respectively.

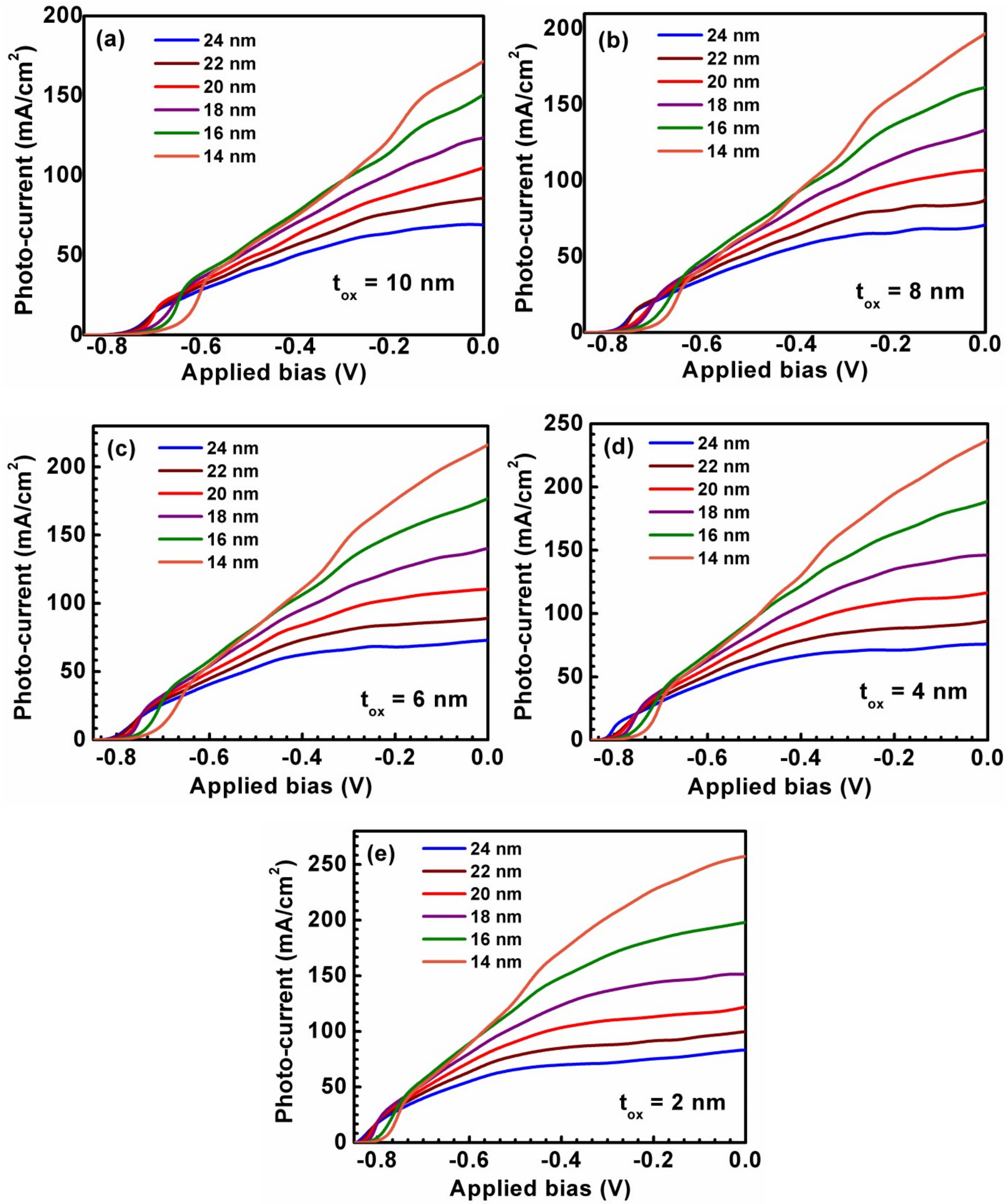

Fig. 5.4: Plots of photocurrent with applied the voltage for different nanowire diameters in the range of 24 nm to 14 nm and varied oxide thicknesses: (a) 10 nm, (b) 8 nm, (c) 6 nm, (d) 4 nm and (e) 2 nm.

It is relevant to mention that in the conventional p-n junction solar cells the energy harvesting is generally performed from the part of I-V curve in 4$^{th}$ quadrant, however, in the current MOS based solar cell this may be obtained from either the 2$^{nd}$ or 4$^{th}$ quadrant, depending on the nature of charge carriers, *i.e.,* either the positively charged holes (as considered in the present device scheme) or negatively charged electrons, respectively. It is observed from the plots of Fig. 5.4 that after certain voltage (*i.e.,* necessary to bend the band enough to create the quantum well for electrons with at least one quantum state), the photocurrent per unit cross-sectional area of the nanowire increases with decreasing nanowire diameter as well as the oxide thickness. For instance, the short-circuit current density, $I_{sc}$, in such GaAs-NW MOS solar cells can be enhanced from ~70 mA/cm$^2$ to ~170 mA/cm$^2$ if the nanowire is diametrically scaled down from 24 nm to 14 nm for a 10 nm oxide thickness. This can be further increased to ~260 mA/cm$^2$ if the oxide thickness is reduced further to 2 nm. Therefore, results suggest that the downscaling of nanowire diameter beyond sub-EBR range to increase quantum efficiency and thinning the oxide to enhance carrier out-flow provide a possible route to achieve higher photocurrent density in the proposed device scheme. However, such increased quantization decreases the 'squareness' of I-V characteristic curves and thereby leads to decline the fill-factor, FF, which is not desirable. Also, the value of open-circuit voltage, $V_{oc}$, is observed to degrade with such miniaturization. Therefore, it is required to optimize the device design parameters to control the quantization effect for achieving the desired output for superior performance.

## 5.7. Variation of solar cell performance parameters with nanowire diameter and oxide thickness

The dependence of short-circuit current and open-circuit voltage with nanowire diameter for the oxide thicknesses of 10 nm and 2 nm is presented in Fig. 5.5(a) and 5.5(b), respectively. The comparative analysis of such plots in the figures suggests $I_{sc}$ to increase rapidly while $V_{oc}$ decreases moderately for scaling down the nanowire diameter. However, both of these parameters get enhanced for thinning the oxide layer since thinner the oxide lesser is the voltage drop across it, leading to larger band bending. The $V_{oc}$ in the proposed device can also be engineered by selecting appropriate materials for the top-metal contact with relevant work-function. The corresponding power conversion efficiency (PCE) and fill-factor (FF) of the present GaAs NW-MOS solar cell as a function of nanowire diameter are plotted in Fig. 5.5(c) and 5.5(d) for the oxide thicknesses of 10 nm and 2 nm, respectively. The most fascinating result, apparent from the plots, is that such a device is capable of providing very high power conversion efficiency of ~ 70%, beyond the SQ-limit due to the resonance phenomenon which has already analyzed. Further inspection into the respective plots reveals that miniaturization of the nanowire diameter improves PCE of the device significantly, whereas degrades its FF. However, both of these performance parameters assume superior value on thinning the oxide layer of such NW-MOS. In the current plots for 10 nm oxide layer, PCE is observed to increase from ~20% to ~30% for downscaling of the diameter to 10 nm from 24 nm, while FF drops from ~40% to ~28%. The corresponding improvement in PCE for the solar cell with 2 nm thick oxide is ~32% to ~70% whereas FF degrades from ~50% to ~35%.

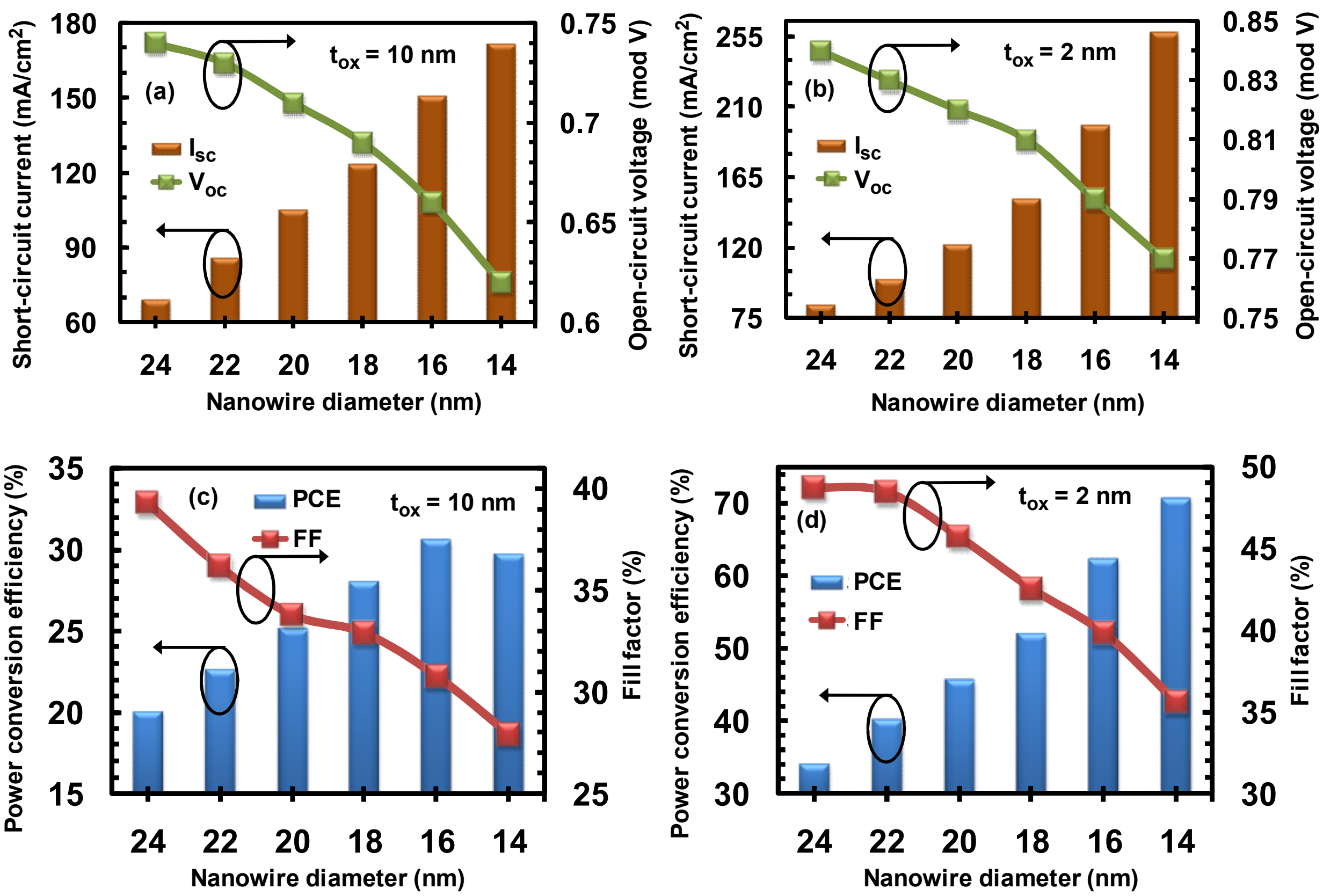

Fig. 5.5. Plot of: (a-b) short-circuit current and open-circuit voltage with nanowire diameter for the oxide thicknesses of 10 nm and 2 nm, respectively. Plot of: (c-d) PCE and FF with nanowire diameters for the oxide thicknesses of 10 nm and 2 nm, respectively.

Such behavior is attributed to the fact that quantization in such vertical nanowire MOS devices depends both on the ‘geometrical confinement’ (radial) as well as on the ‘electrical confinement’ (due to virtual bias) and the bias dependent part varies significantly with oxide thickness.

Table. 5.1. Summary of the solar cell performance parameter ($I_{sc}$, $V_{oc}$, PCE and FF) values for the considered nanowire diameter-oxide thickness combinations.

| *tox (nm)* | *D (nm)* | *Isc (mA/cm$^2$)* | *Voc (V)* | *FF (%)* | *PCE (%)* | *EQE @ 500nm* | *Responsivity @ 500nm(A/W)* |
|---|---|---|---|---|---|---|---|
| **10** | 24 | 69.03 | -0.74 | 39.31 | 20.08 | 7.80 | 3.14 |
| | 22 | 85.84 | -0.73 | 36.19 | 22.68 | 9.83 | 3.96 |
| | 20 | 104.82 | -0.71 | 33.81 | 25.16 | 12.36 | 4.98 |
| | 18 | 123.57 | -0.69 | 32.89 | 28.04 | 15.00 | 6.05 |
| | 16 | 150.64 | -0.66 | 30.78 | 30.61 | 18.65 | 7.52 |
| | 14 | 171.67 | -0.62 | 27.94 | 29.74 | 21.34 | 8.61 |
| **8** | 24 | 70.49 | -0.77 | 43.17 | 23.43 | 7.97 | 3.21 |
| | 22 | 85.98 | -0.76 | 40.86 | 26.70 | 9.86 | 3.98 |
| | 20 | 107.00 | -0.75 | 36.96 | 29.66 | 12.53 | 5.05 |
| | 18 | 133.45 | -0.73 | 33.94 | 33.07 | 15.99 | 6.45 |
| | 16 | 161.48 | -0.70 | 32.85 | 37.13 | 19.65 | 7.92 |
| | 14 | 196.84 | -0.67 | 28.16 | 37.14 | 24.24 | 9.78 |
| **6** | 24 | 72.88 | -0.80 | 45.95 | 26.79 | 8.28 | 3.34 |
| | 22 | 88.59 | -0.79 | 43.67 | 30.56 | 10.23 | 4.12 |
| | 20 | 110.54 | -0.77 | 41.85 | 35.62 | 12.91 | 5.21 |
| | 18 | 140.70 | -0.76 | 36.86 | 39.41 | 16.68 | 6.73 |
| | 16 | 176.84 | -0.73 | 33.59 | 43.37 | 21.29 | 8.59 |
| | 14 | 216.17 | -0.70 | 29.92 | 45.28 | 25.29 | 10.46 |
| **4** | 24 | 75.76 | -0.81 | 47.84 | 29.36 | 8.62 | 3.48 |
| | 22 | 94.12 | -0.80 | 44.36 | 33.40 | 10.80 | 4.35 |
| | 20 | 116.47 | -0.79 | 41.53 | 38.21 | 13.45 | 5.43 |
| | 18 | 146.05 | -0.77 | 38.89 | 43.73 | 17.09 | 6.89 |
| | 16 | 188.65 | -0.75 | 35.33 | 49.99 | 22.30 | 8.99 |
| | 14 | 236.99 | -0.72 | 31.50 | 53.76 | 28.06 | 11.32 |
| **2** | 24 | 83.39 | -0.84 | 48.73 | 34.13 | 9.45 | 3.81 |
| | 22 | 99.91 | -0.83 | 48.46 | 40.19 | 11.39 | 4.59 |
| | 20 | 121.97 | -0.82 | 45.74 | 45.74 | 14.01 | 5.65 |
| | 18 | 151.19 | -0.81 | 42.53 | 52.08 | 17.43 | 7.03 |
| | 16 | 198.06 | -0.79 | 39.82 | 62.31 | 22.83 | 9.21 |
| | 14 | 257.53 | -0.77 | 35.61 | 70.62 | 29.67 | 11.97 |

This in turn opens up a design window for the nanowire diameter-oxide thickness combinations to achieve a desired range of performance parameter values for $I_{sc}$, $V_{oc}$, PCE and FF. The values of all such relevant performance parameters for the nanowire diameters and oxide thicknesses considered in the current work are summarized in Table 5.1.

## 5.8. Impact of phonon scattering on the performance of nanowire MOS solar cells

Generally, the device performance is expected to degrade due to the presence of electron-phonon scattering. However, it has already been discussed that such phonon scattering is not always dissipative in quantum scale devices and rather the phonon energy may be harvested in certain conditions (*e.g.* utilizing effective mass mismatch between quantum 'device' and 'reservoir' or induced strain within the device) **[469]**. Such effective mass mismatch between GaAs nanowire and Si-substrate is already included in the theoretical modeling as presented in section 5.3. Here, the electrical characteristics of such nanowire MOS solar cells is also calculated considering the phonon scattering phenomenon and it is found that such energy may be dissipated or harvested depending on oxide thickness and the applied bias. The comparative plot of such I-V characteristics with and without phonon scattering is shown in Fig. 5.6 (a)-(b).

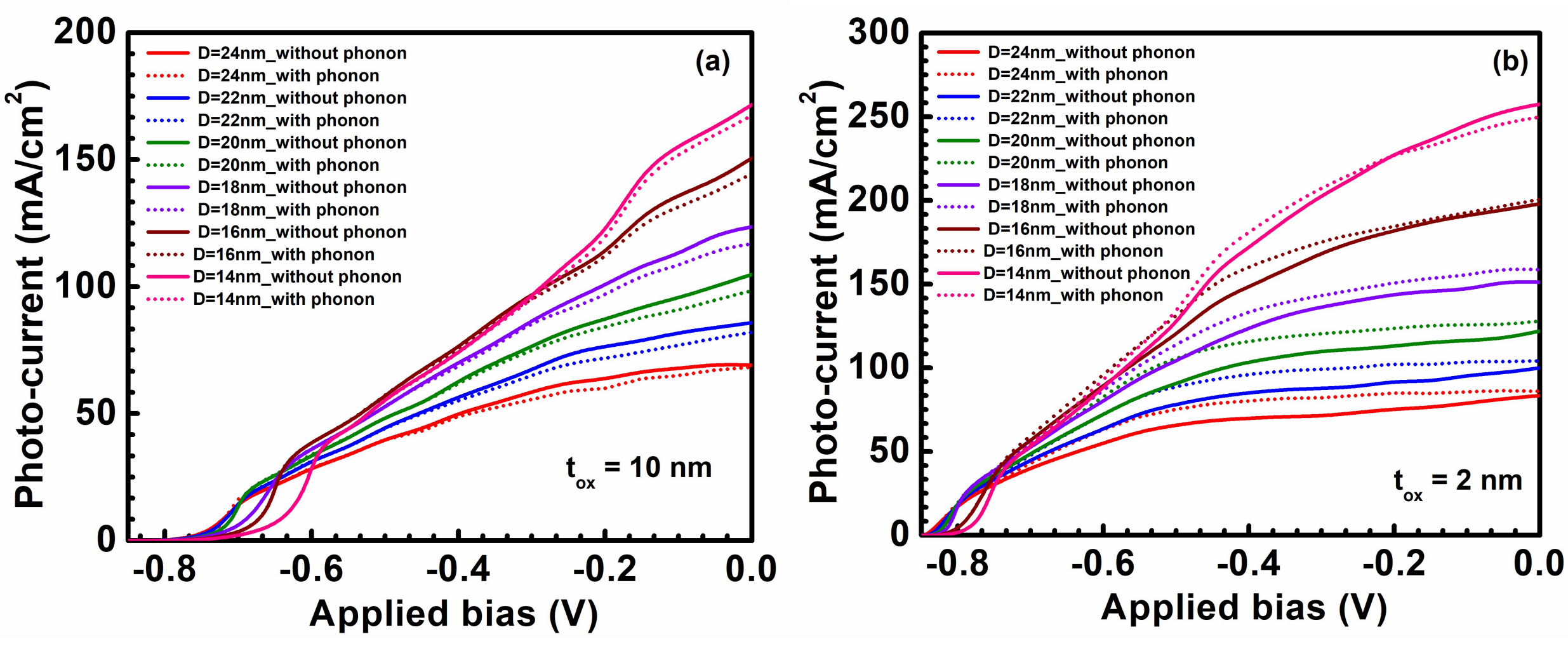

Fig. 5.6. Comparative plot of photocurrent vs. applied bias with and without phonon scattering for different diameters: (a) $t_{ox}$ =10 nm; (b) $t_{ox}$ =2 nm.

It is apparent from the figure that, for 10 nm oxide thickness, devices with all diameters considered exhibit mild degradation in the photocurrent and thus in PCE; however the squareness of such IV-curves improves thereby increasing the FF values. On the other hand for a lower oxide thickness such as 2 nm, phonon energy is harvested thereby increasing FF and PCE correspondingly. For instance, the results show a possible degradation of PCE up to <1% (from 20.08 to 19.89 for 24 nm diameter, 10 nm oxide) whereas a possible improvement in PCE up to >5% (from 70.62 to 74.61 for 14 nm diameter, 2 nm oxide) due to phonon scattering. Such results are summarized in the Table 5.2.

Table 5.2. Summary of performance parameters of the GaAs NW-MOS solar cell with and without phonon scattering.

| **$t_{ox}$ (nm)** | **D (nm)** | **$I_{sc}$ (mA/cm$^2$)** | | **FF (%)** | | **PCE (%)** | | **EQE @ 500 nm** | |
|---|---|---|---|---|---|---|---|---|---|
| | | Without phonon | With phonon | Without phonon | With phonon | Without phonon | With phonon | Without phonon | With phonon |
| 10 | 24 | 69.03 | 68.23 ↓ | 39.31 | 39.40 ↑ | 20.08 | 19.89 ↓ | 7.80 | 7.73 ↓ |
| | 22 | 85.84 | 82.03 ↓ | 36.19 | 37.54 ↑ | 22.68 | 22.48 ↓ | 9.83 | 9.48 ↓ |
| | 20 | 104.82 | 98.43 ↓ | 33.81 | 35.43 ↑ | 25.16 | 24.76 ↓ | 12.36 | 11.70 ↓ |
| | 18 | 123.57 | 116.98 ↓ | 32.89 | 34.25 ↑ | 28.04 | 27.64 ↓ | 15.00 | 14.26 ↓ |
| | 16 | 150.64 | 144.45 ↓ | 30.78 | 31.74 ↑ | 30.61 | 30.26 ↓ | 18.65 | 17.92 ↓ |
| | 14 | 171.67 | 167.35 ↓ | 27.94 | 28.50 ↑ | 29.74 | 29.57 ↓ | 21.34 | 20.79 ↓ |
| 2 | 24 | 83.39 | 85.94 ↑ | 48.73 | 53.91 ↑ | 34.13 | 38.92 ↑ | 9.45 | 9.41 ↓ |
| | 22 | 99.91 | 104.16 ↑ | 48.46 | 53.00 ↑ | 40.19 | 45.82 ↑ | 11.39 | 11.48 ↑ |
| | 20 | 121.97 | 128.01 ↑ | 45.74 | 50.77 ↑ | 45.74 | 53.29 ↑ | 14.01 | 14.28 ↑ |
| | 18 | 151.19 | 158.60 ↑ | 42.53 | 44.33 ↑ | 52.08 | 56.95 ↑ | 17.43 | 17.92 ↑ |
| | 16 | 198.06 | 200.68 ↑ | 39.82 | 42.75 ↑ | 62.31 | 67.77 ↑ | 22.83 | 22.93 ↑ |
| | 14 | 257.53 | 249.84 ↓ | 35.61 | 38.78 ↑ | 70.62 | 74.61 ↑ | 29.67 | 28.53 ↓ |

## 5.9. Summary

In the present chapter, a GaAs-nanowire based vertical MOS device has been designed as potential alternative for the next generation solar cells with ultra-high efficiency (~70%) above the SQ-limit. The generation of photocarriers and relevant transport phenomenon are theoretically modeled by using NEGF approach with the consideration of second quantization field operators for the incident photons and generated EHPs. The focus of this chapter is to study the impact of quantum confinement on the photogeneration process, by a detailed mode-by-mode analysis with the consideration of distribution (or spatial extension) of the transverse and longitudinal energy states, for solar energy harvesting. The relative contributions of LHs and HHs under illumination are also investigated to determine total photocurrent in the device. A very high short-circuit current of ~260 $mA/cm^2$ and photoresponsivity of ~12 A/W are obtained for the device with 14 nm diameter and 2 nm oxide thickness. The highest fill-factor of ~50% for the device with 24 nm diameter and power conversion efficiency of ~70% for the 14 nm diameter are obtained for 2 nm of oxide layer. Also, the impact of phonon scattering on the performance of such nanowire MOS solar cells is investigated. Therefore, the proposed vertical PEDOT-$T_{os}$/$SiO_2$/GaAs- NW MOS-structure offers a novel device scheme for solar energy harvesting with superior efficiency and also provides a route for engineering the other performance parameters by controlling its quantization effect.

In Chapter 4 and Chapter 5, the possible optoelectronic applications of the vertical nanowire MOS device as photodetector and solar cell are discussed in detail by developing a NEGF based theoretical model. In the following chapter, fabrication procedure of such devices will be discussed in detail and emphasis will be given to verify the formation of voltage tunable quantum dot which was predicted in Chapter 3.

# Chapter: 6*

# Fabrication of the patterned Ge-nanowire array based voltage tunable quantum dot (VTQD) devices on p-Si substrate

## 6.1. Introduction: Semiconductor quantum dots (QDs)

Semiconductor quantum dots (QDs) have attracted significant interest among the global research community for sustaining performance improvement of nanoscale devices in the domain of opto-electronics and also in the emerging area of quantum computing. Three dimensional (3D)-confinement in such QDs makes them superior over the 2D-confined quantum wires and 1D-confined thin films in terms of achieving performances beyond the conventional limit. As discussed earlier in Chapter 1 and 2, in the domain of optoelectronics, QD-photodetectors are explored for multi-color photosensing applications with high spectral responsivity. QD-lasers are demonstrated with higher efficiency, high differential gain, high output power and high modulation speed. Size dependent spectral properties with enhanced internal quantum efficiency have been observed in QD-based LEDs. Further, the QD-solar cells have theoretically been predicted to exceed the Shockley-Queisser limit of efficiency in single-junction devices. Multi-exciton generation (MEG) in such QD-solar cells with more than 100% external quantum efficiency due to strong 3D-confinement has also been extensively studied.

QD-based devices have also hold great potential in generating entangled photons for the development of quantum information processing (QIP) technology. The ideal characteristics for entanglement such as, deterministic generation, high fidelity, indistinguishability and high collection efficiency are obtained in such QD-based single photon sources. Charge qubit generation is also demonstrated by using an isolated double-QD (Si) system with detailed device operations (initialization and manipulation of quantum states) and measurements. Thus, the semiconductor QD-based devices are currently regarded as the primary unit of modern advanced

***Most of the content of Chapter: 6 is published in the article, "Voltage-tunable quantum-dot array by patterned Ge-nanowire-based metal-oxide-semiconductor devices", by Subhrajit Sikdar, Basudev Nag Chowdhury, Rajib Saha and Sanatan Chattopadhyay, 'Physical Review Applied', 05/2021, 15, 054060, DOI: 10.1103/PhysRevApplied.15.054060.***

and emerging electronic technologies over a wide range and such multifaceted applications of such QDs have been extensively reviewed in ref. **[470]**.

Several 'bottom-up' and 'top-down' approaches are attempted to fabricate such QDs by employing the MOCVD and MBE techniques; and the QD-arrays are developed by using the sequential process steps of patterning (by EBL) and directional etching (by RIE). However, the nanowire diameters achieved in such reports are of the order of 100 nm which are too large to observe any quantum effects at room temperature **[471-475]**. Therefore, the realization of such controlled quantum effects at room temperature has been a central point of research in recent time for the charge based quantum computing to generate qubits. At this point it is to be mentioned that, achieving uniformity and precision in geometrical dimensions of such QDs still remains a key concern which actually determines their 3D-confinement characteristics. Efforts have been made to realize the voltage tunable quantum dots (VTQDs) with appropriate control over their quantum states **[476-480]**. However, these techniques involve additional experimental steps and complicate the fabrication procedure. Also, the additional steps may introduce further process-induced variability to affect the desired uniformity and precision in geometrical dimensions. Thus, at this point, it is essential to develop novel concepts for creating explicit control over the formation of such QDs.

In this context, it is mentioned, in Chapter 3 (section 3.1), that two fabrication procedures, including the horizontal and vertical device integration schemes, are currently available to develop the nano-/quantum-structure based devices. In the horizontal scheme, nanowires are subjected to a 'pick-and-place' approach where such nano-/quantum-structures are picked up by some mechanical means and placed on the desired location of the substrate which makes the process extremely complex, contamination prone and of limited reproducibility. In contrast, the vertical device integration scheme provides a relatively simpler way of fabricating nano-/quantum-structure based devices with reduced complexity since such nanowires are naturally aligned in the vertical direction. Therefore, the vertical nanowire based devices may be useful for developing the realistic and practical circuits and systems.

Metal-oxide-semiconductor (MOS) capacitor based devices find their multifaceted applications in the domain of nanoscale electronics and optoelectronics due to their usual compatibility with the existing CMOS process technologies. Recently, the array of vertical silicon (Si)-nanowire structures, with silicon dioxide ($SiO_2$) as the dielectrics, has been demonstrated and its capacitance-voltage (C-V) characteristics are also reported **[400]**. However, no quantum confinement effects in such characteristic curves have been observed since the chosen nanowire diameter is 430 nm which is much higher than the excitonic Bohr radius of Si (*i.e.*, ~5 nm).

In this context, fabrication and characterization of the vertically aligned patterned array of germanium (Ge)-nanowire based MOS devices on Si-platform is extensively studied in this chapter. The diameters of fabricated nanowire are kept within the Ge excitonic Bohr radius (*i.e.*, 24.3 nm) to achieve quantum confinement effects at room temperature. Template for the patterned nanowire is generated by employing EBL technique (Raith Elphy Plus system coupled with Zeiss Auriga FESEM). Subsequently, the nanowires (*i.e.*, Ge) and oxide (*i.e.*, $SiO_2$) materials of desired thicknesses are sequentially deposited by using e-beam evaporation technique. The top metal contacts are created by depositing platinum (Pt) on the $SiO_2$ layer by dc sputtering method. It is apparent to mention that the Ge nanowires are used for this study, instead of Si, due to its relatively higher EBR of 24.3 nm than Si (4.9 nm) which suggests an easier size control during the fabrication of Ge nanowires to observe quantum confinement effects. Also, very small difference of electron affinity values (50 meV) between Si and Ge ensures no degradation in maintaining electrostatic integrity of such devices, which has already been elaborated in Table. 3.1 of Chapter 3. The commonly used semiconductors with their EBR and electron affinity values are listed in Table 6.1.

Table 6.1. The summary of electron effective mass, electron affinity and EBR for some of the commonly used semiconductors.

| ***Material*** | ***Electron effective mass*** | ***Electron affinity (eV)*** | ***Barrier height with $SiO_2$ (eV)*** | ***EBR (nm)*** |
|---|---|---|---|---|
| **Si** | **$0.19m_0$ (transverse)** | **4.05** | **3.20** | **4.9** |
| **Ge** | **$0.08m_0$ (transverse)** | **4.00** | **3.15** | **24.3** |
| **GaAs** | **$0.063m_0$** | **4.07** | **3.22** | **12.0** |
| **InP** | **$0.08m_0$** | **4.38** | **3.53** | **15.0** |

The corresponding capacitance-voltage (C-V) characteristics of such fabricated devices are measured *in-situ* within the FESEM system by employing Keithley 4200 semiconductor characterization system (SCS) where the contacts are made by Kleindiek (tungsten) nano-probes. The effect of quantum confinement, observed from the C-V characteristics, is further explained by developing an analytical model based on non-equilibrium Green's function (NEGF) formalism.

## 6.2. Working principle of the Ge-NWMOS based VTQD-devices

The working principle of a vertically aligned VTQD device using NWMOS is schematically shown in Fig. 6.1. It is already discussed in Chapter 3, section 3.3, that such VTQDs form near the $SiO_2$/nanowire interface due to the combined effects of radial quantization due to sub-EBR nanowire radius and by the voltage-assisted surface quantization effect along the nanowire axis. The number of energy states in such VTQDs depends on the amount of surface quantization (*i.e.*, applied voltage at the top electrode) and the nanowire radius (< EBR). The 'degree' of quantization increases for the higher applied voltages or smaller radii and consequently, the number of induced quantum states reduces within such voltage assisted quantum dots. However, only the number of quantum states below the Fermi level, as indicated in Fig. 6.1, contributes to the device capacitance. Charge carriers accumulated in the quantum well also have a possibility to tunnel through the insulating layer and thus may degrade the device electrostatic integrity which has been analyzed in detail, in Chapter 3. The relatively higher barrier height at $SiO_2$/Si interface and higher effective mass of $SiO_2$ have been observed (Table 3.1) to achieve superior electrostatic control in such quantized nanowire MOS devices as compared to the other high-k dielectrics.

Here, the VTQD region (as shown in Fig. 6.1) in the conduction band is considered as the *'device'* and p-Si substrate is considered as the *'reservoir'*. The overall coupling between such *'device'* and *'reservoir'* is very weak since the electrons are trapped in the quantum well due to 3D-confinement. However, holes in the valence band are confined only in the radial directions and free to move away from interfacial region towards the substrate, *i.e.*, along the nanowire axial direction. Thus, the interaction between *'device'* and the *'reservoir'* for such 2D-confined holes is relatively stronger in comparison to the 3D-confined electrons.

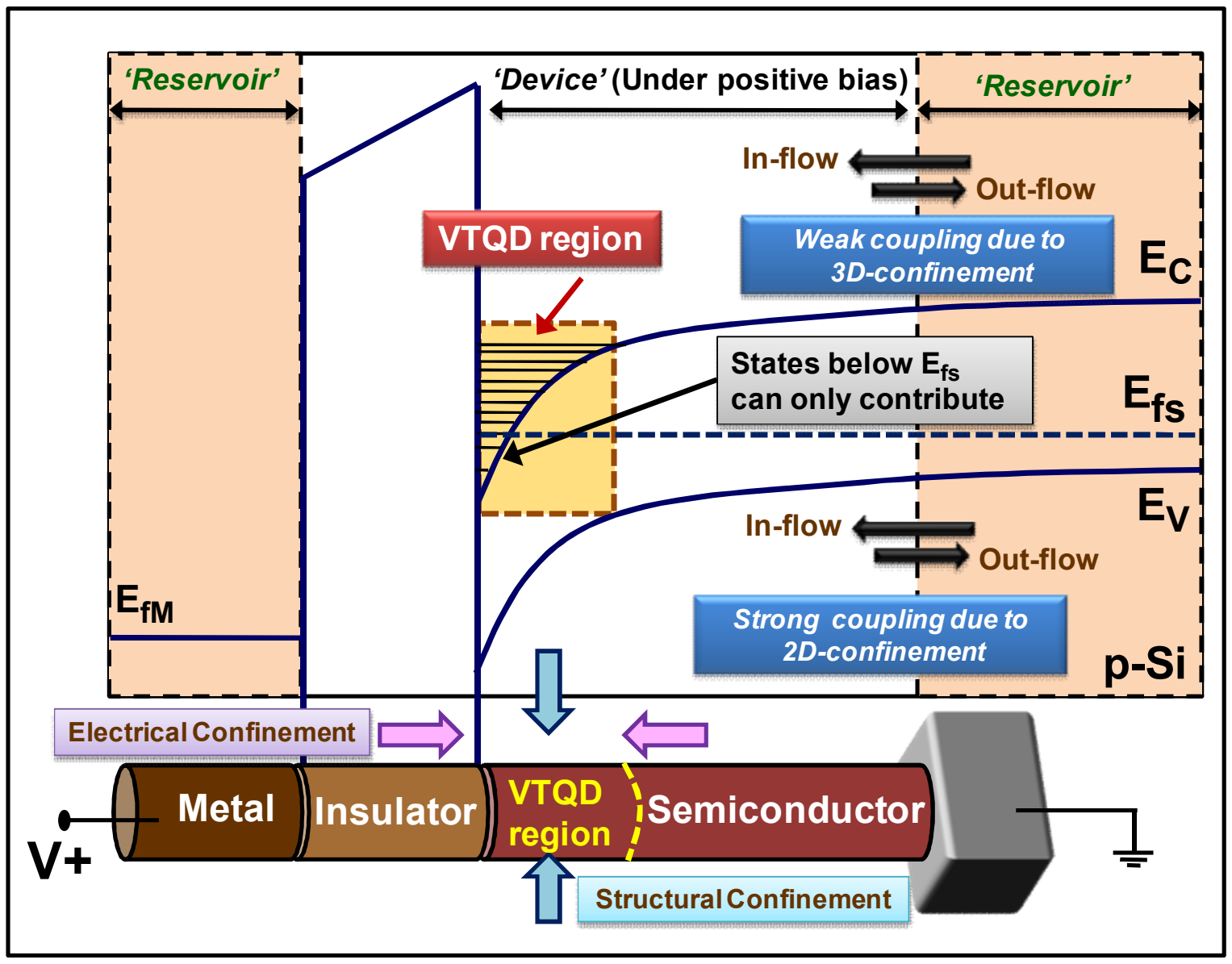

Fig. 6.1. Schematic representation of the working principle of nanowire MOS based VTQD devices.

However, in the present device structure, such holes are also subjected to moderate confinement due to the valence band offset at the junction between Ge-nanowire/p-Si substrate as observed from the energy band diagram of Fig. 6.2. On application of negative voltages at the top metal electrode, holes are accumulated near the oxide/nanowire junction to contribute to its capacitance; and, these accumulated holes gradually decrease and move towards the substrate with the gradual decrease of negative bias. However, there exists a valence band off-set of 0.5 eV at the Ge-nanowire/p-Si interface which restrict such holes to move further and thus, a hole-confined region is created. On application of positive bias at the top electrode, minority carriers (*i.e.*, electrons) are attracted towards the oxide/nanowire junction, surmounting the small barrier of 0.05 eV, and get collected in the conduction band quantum well.

In the current work, the applied voltage is varied to study the entire range of variation from accumulation-to-inversion of the MOS capacitance. Generally, the device doping concentration and its type (p- or n- type) govern such variation. However, fabrication of sub-EBR nanoscale semiconductor nanowires with controlled doping is technologically very challenging, since even a single dopant results in a very high doping concentration (~$10^{16}$/cc). In this context, the array of intrinsic Ge-nanowires is grown on p-Si substrate and forms the hetero-junctions with 0.5 eV valence band offset at the Ge-NW/p-Si interface (Fig. 6.2). Such inherent hetero-junction makes the Ge nanowires to be the degenerated semiconductor and is attributed to convert the intrinsic Ge nanowires to be p-type.

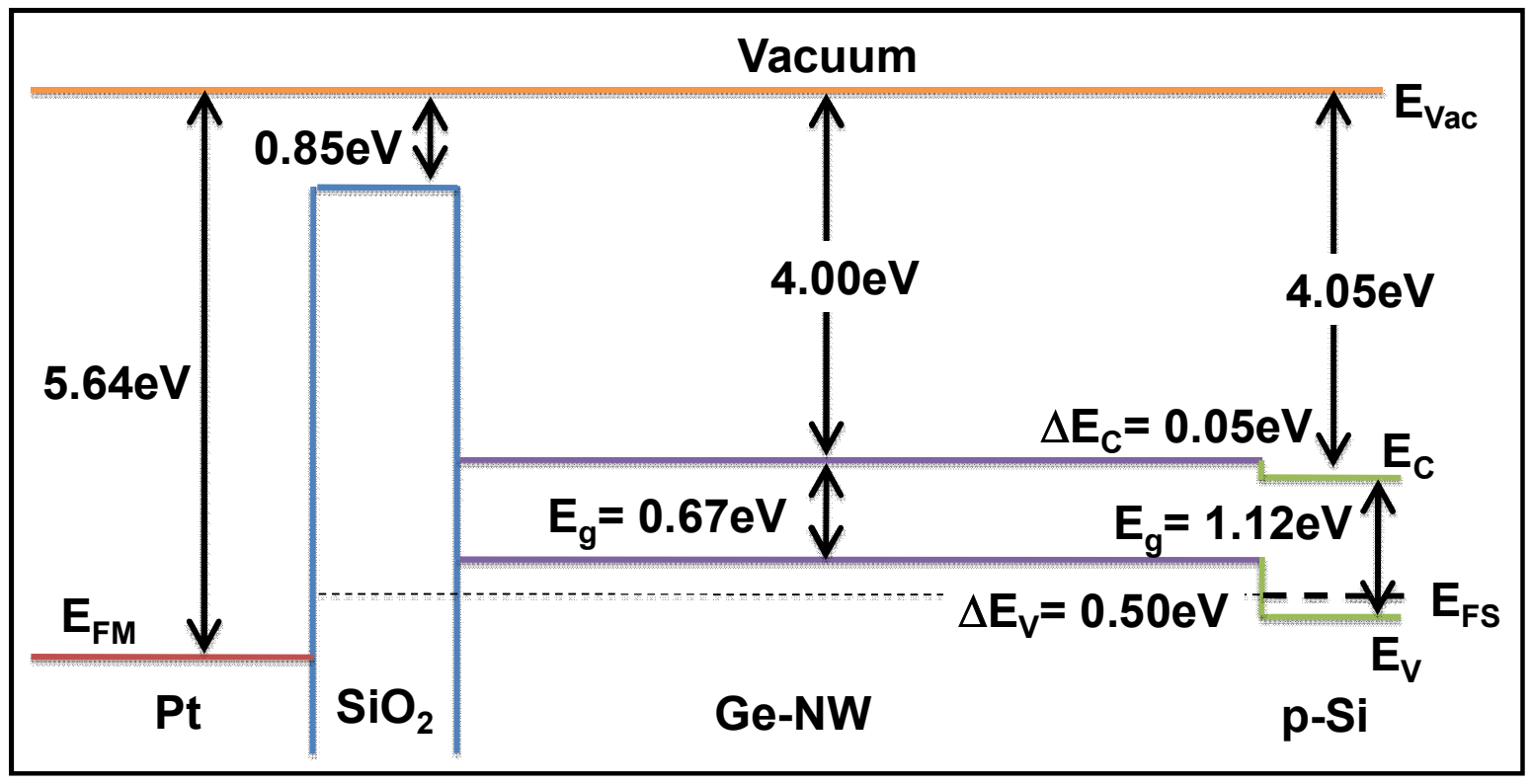

Fig. 6.2. Energy band diagram of the NWMOS device at unbiased condition illustrating the Si/Ge valence and conduction band offsets.

# 6.3. Theoretical modeling

## 6.3.1. Quantum-electrostatic simultaneous solution based on NEGF formalism

It has been already discussed in Chapter 3 (section 3.3) and Chapter 4 (section 4.3) that the motion of conduction band electrons and valence band holes in the vertically oriented nanowire MOS based VTQD devices is governed simultaneously by the quantum and electrostatic conditions of the system. Therefore, a quantum-electrostatic simultaneous solver is required to model charge transport phenomenon in such low dimensional devices where the Schrodinger and Poisson equations are solved self-consistency. For developing the theoretical model, a relevant Hamiltonian of the device is first constructed (Eqn. (3.3) of chapter 3) with an arbitrary potential variation along the nanowire axis. Then the transport equation assuming such Hamiltonian is solved by NEGF formalism to obtain LDOS and carrier concentration profile for the electrons and holes. The obtained carrier density is then used to calculate the electrostatic potential by solving relevant Poisson's equation and compared with the previously assumed value. When the difference between such successive potential values will be insignificant, the relevant value will be used for subsequent calculations. The iterative process used of achieving self-consistency has been shown in Fig. 4.3 of chapter 4.

The equation of motion for the *'device'* in conduction (*i.e.*, 3D-confined electrons) and valence (*i.e.*, 2D-confined holes) band regions interacting with the *'reservoir'* (*i.e.*, p-Si substrate) are given by,

$$i\hbar \frac{d}{dt} C_i^c = H_{ISO}^c C_i + \sum_b \tau_{ib} S_b \tag{6.1}$$

$$i\hbar \frac{d}{dt} S_b = H_{ISO}^{p-Si} S_b + \sum_i \tau_{bi}^* C_i^c \tag{6.2}$$

and,

$$i\hbar \frac{d}{dt} V_r^v = H_{ISO}^v V_r + \sum_b \tau_{rb} S_b \tag{6.3}$$

$$i\hbar \frac{d}{dt} S_b = H_{ISO}^{p-Si} S_b + \sum_r \tau_{br}^* V_r^v \tag{6.4}$$

where, $C_i^c$, $V_r^v$ and $S_b$ are second quantized field operators for electrons (indexed by $i, j$), holes (indexed by $r, s$), and the substrate region (indexed by $b, b'$), respectively. The isolated Hamiltonian for electrons, holes, and substrate are represented by, $H_{ISO}^c$, $H_{ISO}^v$, and $H_{ISO}^{p-Si}$, respectively and $\tau$ is the interaction potential. Generally, the *'device'* Hamiltonian for such electrons and holes is expressed as (from Eqn. (3.3)),

$$H = \frac{-\hbar^2}{2m_0} \sum_{i,j} \left( \partial_i m_{ij}^{*-1} \partial_j \right) + U(z) \tag{6.5}$$

Here, $U(z)$ represents the energy band profile for conduction (putting $U_C(z)$ in place of $U(z)$) and valence (putting $U_V(z)$ in place of $U(z)$) band along the nanowire axis (*i.e.*, $z$-axis) from metal/insulator junction towards the substrate. In the process of obtaining relevant equation for electrons and holes, the *'reservoir'* equations (Eqn. (6.2) and Eqn. (6.4)) are firstly solved by introducing the equilibrium Green's function ($G_b^{p-Si}$) of the substrate, as shown below,

$$S_b(t) = [S_b(t)]^{ISO} + \sum_j \int_{-\infty}^{+\infty} dt_1 G_b^{p-Si}(t, t_1) \tau_{bj}^* C_j^c(t_1) \tag{6.6}$$

and,

$$S_b(t) = [S_b(t)]^{ISO} + \sum_s \int_{-\infty}^{+\infty} dt_1 G_b^{p-Si}(t, t_1) \tau_{bs}^* V_s^v(t_1) \tag{6.7}$$

In Eqn. (6.6) and (6.7), the first term(s) represent isolated *'reservoir'* and the second term(s) in both the equations governs the interaction of such *'reservoir'* with the *'device'*. Further, the solution of *'reservoir'* equations, obtained from Eqn. (6.6) and Eqn. (6.7), are put into Eqn. (6.1) and Eqn. (6.3), respectively, which lead to,

$$i\hbar\frac{d}{dt}C_i^c - H_{ISO}^c C_i^c - \sum_j \int_{-\infty}^{+\infty} dt_1 \Sigma_{p-Si}^{ij}(t,t_1) C_j^c(t_1) = S_i^{'}(t) \tag{6.8}$$

and,

$$i\hbar\frac{d}{dt}V_r^v - H_{ISO}^v V_r^v - \sum_s \int_{-\infty}^{+\infty} dt_1 \Sigma_{p-Si}^{rs}(t,t_1) V_s^v(t_1) = S_r^{'}(t) \tag{6.9}$$

where, $\Sigma_{p-Si}^{ij}(t,t_1) = \sum_b \tau_{ib} G_b^{p-Si}(t,t_1)\tau_{bj}^*$ and $\Sigma_{p-Si}^{rs}(t,t_1) = \sum_b \tau_{rb} G_b^{p-Si}(t,t_1)\tau_{bs}^*$ are the self-energy functions regarding coupling between the *'device'* and *'reservoir'* in conduction and valence band regions, respectively. $S_i^{'}(t) = \sum_b \tau_{ib}\left[S_b(t)\right]^{ISO}$ and $S_r^{'}(t) = \sum_b \tau_{rb}\left[S_b(t)\right]^{ISO}$ of Eqn. (6.8) and Eqn. (6.9) represent the source terms which further lead to the calculation of corresponding in-scattering functions. Solving Eqn. (6.8) and Eqn. (6.9) by introducing the *'device'* Green's functions for electrons ( $G_{ij}^C$ ) and holes ( $G_{rs}^V$ ), we get,

$$C_i^c(t) = \sum_j \int_{-\infty}^{+\infty} dt_1 G_{ij}^C(t,t_1) S_j^{'}(t_1) \tag{6.10}$$

and,

$$V_r^v(t) = \sum_s \int_{-\infty}^{+\infty} dt_1 G_{rs}^v(t,t_1) S_s^{'}(t_1) \tag{6.11}$$

Now, the number of such 3D-confined electrons and 2D-confined holes in the conduction and valence band regions are obtained by calculating the corresponding correlation functions (using Eqn. (6.10) and Eqn. (6.11)) as the following,

$$n_{ij}^C(t,t^{'}) = \left\langle C_j^{c+}(t')C_i^c(t)\right\rangle = \sum_{i'j'} \int_{-\infty}^{+\infty} dt_2 \int_{-\infty}^{+\infty} dt_1 G_{ii'}^C(t,t_1)\Sigma_{i'j'}^{in}(t_1,t_2) G_{jj'}^{C+}(t',t_2) \tag{6.12}$$

and,

$$n_{rs}^V(t,t^{'}) = \left\langle V_s^{v+}(t')V_r^v(t)\right\rangle = \sum_{r's'} \int_{-\infty}^{+\infty} dt_2 \int_{-\infty}^{+\infty} dt_1 G_{rr'}^V(t,t_1)\Sigma_{r's'}^{in}(t_1,t_2) G_{ss'}^{V+}(t',t_2) \tag{6.13}$$

where, the in-scattering functions are evaluated by introducing the substrate correlation function ($n_{bb'}^{p-Si}$) for the isolated condition as,

$$\Sigma_{i'j'}^{in}(t_1,t_2) = \left\langle S_{j'}^{'}(t_2) S_{i'}^{'}(t_1) \right\rangle = \sum_{bb'} \tau_{i'b} n_{bb'}^{p-Si}(t_1,t_2) \tau_{j'b'}^{*} \quad (6.14)$$

and,

$$\Sigma_{r's'}^{in}(t_1,t_2) = \left\langle S_{s'}^{'}(t_2) S_{r'}^{'}(t_1) \right\rangle = \sum_{bb'} \tau_{r'b} n_{bb'}^{p-Si}(t_1,t_2) \tau_{s'b'}^{*} \quad (6.15)$$

The equations for carrier concentration of electrons (Eqn. (6.12)) and holes (Eqn. (6.13)) are further transformed into the energy domain by Fourier Transformation and represented as,

$$n^{C}(E_e) = \left[G^{C}(E_e)\right]\left[\Sigma_e^{in}(E_e)\right]\left[G^{C+}(E_e)\right] \quad (6.16)$$

and,

$$n^{V}(E_h) = \left[G^{V}(E_h)\right]\left[\Sigma_h^{in}(E_h)\right]\left[G^{V+}(E_h)\right] \quad (6.17)$$

where, $E_e$ and $E_h$ , $\Sigma_e^{in}(E)$ and $\Sigma_h^{in}(E)$, $G^{e}(E)$ and $G^{v}(E)$ are the total energy varying from band edges to +/- infinity, the in-scattering functions and the Green's functions for electrons and holes, respectively. The expressions for *'device'* Green's functions for the 3D-confined conduction band electrons (following Eqn. 4.20) and 2D-confined valence band holes are given by,

$$G^{C}(E_e) = Lt_{\Sigma(E)\to 0}\left[E_e I - E^{C} - \Sigma_{p-Si}(E_e)\right]^{-1} \quad (6.18)$$

and

$$G^{V}(E_h) = \left[E_h I - E^{V} - \Sigma_{p-Si}(E_h)\right]^{-1} \quad (6.19)$$

where, $E^{C}$ comprises of all the 3D-quantized energy states created in the voltage tunable quantum well and $E^{V}$ represents the matrix consisting of 2D-confined valence sub-bands.

It is worthy to mention that the 'strength of coupling' between the *'device'* and *'reservoir'* is governed by the term '$\beta g \beta$' of the self-energy matrices ($\Sigma_{p-Si}(E_e)$ and $\Sigma_{p-Si}(E_h)$). The impact of such 'strength of coupling' due to the mismatch in transport effective masses between the *'device'* and *'reservoir'* on the electrical performance of nanostructure devices has already been studied in detail **[481]**. The coupling term '$\beta g \beta$' can be represented as, $\beta g \beta = \dfrac{A \pm \sqrt{A^2 - 4\beta^2}}{2}$, where, $A = \left[E - E_n - E_D(z) + 2\beta\right]$ and $\beta = -\left(\frac{\hbar^2}{2m^*\Delta^2}\right)$, $\Delta$ being the grid size. In the valence band, the term $E_n$ represents the 2D-confined energy sub-bands ($E^V$) while in the conduction band such term represents 3D-confined discrete energy states ($E^C$). The next term, in the valence band region, indicates distribution of the band potential along nanowire axis (*i.e.*, $E_D^V(z)$) and similarly, in the conduction band, such term represents spatial extensions of the discrete energy states ($E_D^C(z)$) within the quantum well. In the present device architecture, bias applied at the top electrode terminal pushed the holes away from interface towards the substrate and thus, these holes are strongly coupled with the *'reservoir'*. However, in the conduction band region, accumulated electrons are trapped in the quantum well from all three directions and therefore the possibility of interaction of such electrons with the substrate reduces significantly. Such weak coupling of electrons leads to the modified form of *'device'* Green's function where $\Sigma_{p-Si}(E_e) \rightarrow 0$ as shown in Eqn. (6.18). The electron self-energy is considerably small for two reasons: i) electrons are the minority carriers in p-Si substrate, and ii) significant mismatch of the quantized energy states between the nanowire and Si-substrate due to the voltage induced band banding. Therefore, the density of states in conduction band, which is equivalent to the imaginary part of Green's function **[117, 118]**, will be,

$$D^C(E_e) = 2\delta(E_e - E^C) \tag{6.20}$$

indicating the 3D-quantization of electrons. However, the corresponding LDOS for the 2D-confined holes is obtained from the Green's function by following its general form as **[117]**,

$$D^V(E_h) = \frac{i}{2\pi\Delta}\left(G^v(E) - G^{v+}(E)\right) \tag{6.21}$$

where, the grid size ($\Delta$) is chosen to be less than the lattice constant of Ge for achieving higher accuracy. Further, total charge density ($\rho$) of the nanowire MOS device is obtained from the number density of electrons ($n^C$) and holes ($n^V$), as calculated from Eqn. (6.16) and Eqn. (6.17), respectively, after integrating them over the distance from metal/oxide interface to the spatial extension of the quantum well in the conduction band and the distance up to the band offset at the nanowire/substrate junction in the valence band, respectively. After obtaining the charge densities, the variation of potential in the conduction and valence band is obtained by using the Poisson's equation, as shown below,

$$\vec{\nabla}.\left(\varepsilon\vec{\nabla}U_C(z)\right) = (-e)\rho_C(z) \tag{6.22}$$

and,

$$\vec{\nabla}.\left(\varepsilon\vec{\nabla}U_V(z)\right) = (+e)\rho_V(z) \tag{6.23}$$

where, $\rho_C(z)$ and $\rho_V(z)$ are the charge densities for electrons and holes and $e$ is the value of electronic charge. The potential energy values ($U_C(z)$ and $U_V(z)$), obtained by solving the Poisson's equation with relevant boundary conditions, as: (a). $U(z)\big|_{Ge-NW/p-Si} = 0$; (b). $U(z)\big|_{Pt/SiO_2} = -V_{app}$ for positive bias (inversion) and $U(z)\big|_{Pt/SiO_2} = +V_{app}$ for negative bias (accumulation); and (c). $\varepsilon_{SiO_2}\left.\frac{dU(z)}{dz}\right|_{z<=t_{SiO_2}} = \varepsilon_{Ge-NW}\left.\frac{dU(z)}{dz}\right|_{z>t_{t_{SiO_2}}}$. The obtained potential values are then put into Eqn. (6.5) to check for consistency and the values for respective charge densities are only taken for further calculation when self-consistency is achieved. Finally, the developed capacitance due to such 3D-confined electrons and 2D-confined holes are obtained by dividing the total accumulated charge with the applied voltage ($V_{app}$),

$$C^C = \left(e\int_0^{l_Q}\rho_C(z)dl_Q\right)\Big/V_{app} \tag{6.24}$$

and,

$$C^V = \left(e\int_0^{l_Z}\rho_v(z)dl_Z\right)\Big/V_{app} \tag{6.25}$$

where, $l_Q$ is the spatial extension of the quantum well whereas $l_z$ covers the distance up to the Ge-NW/p-Si interface.

## 6.3.2. Accumulation region transport

Energy band diagram of the considered vertical nanowire MOS device under negative bias is shown in Fig. 6.3. It is well known for the MOS structures on p-type substrate that the valence and conduction bands bend in the upward direction for negative applied bias and the electrons are pushed away from the oxide/semiconductor interface. The relevant surface potential ($\psi_s$) is negative as per convention. The holes in valence band accumulate at the oxide/semiconductor interface and form the accumulation capacitance of the device. In the present device scheme, the Ge-NW is hole-degenerated since these are grown on the p-Si substrate, as observed from Fig. 6.3.

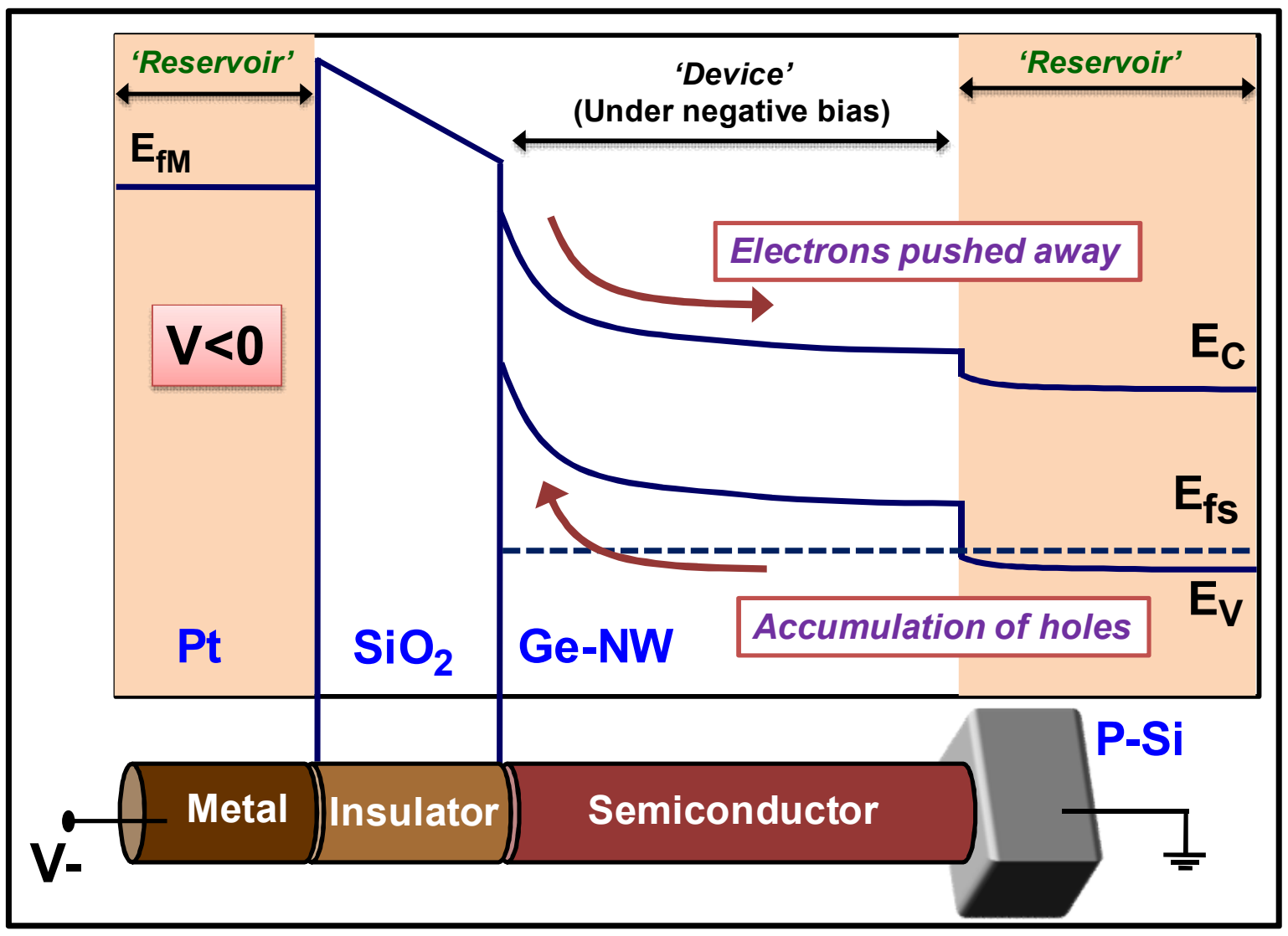

Fig. 6.3. The schematic energy band diagram of the Pt/$SiO_2$/Ge-NW/p-Si MOS device in accumulation condition.

### 6.3.3. Accumulation-to-inversion region

Energy band diagram of the accumulation-to-inversion region is shown in Fig. 6.4 where the valence band offset plays an important role in determining the device capacitance. When the applied voltage is varied from negative to positive, the surface potential gradually increases and consequently, upward bending of the bands reduces, and finally, these become flat ($\psi_s$=0) for an appropriate voltage. Accordingly, holes are gradually swept away from the oxide/semiconductor interface towards the substrate and subjected to a quasi-confinement due to the valence band off-set created at the Ge-NW/p-Si hetero-interface. It is apparent to note that the difference of (hole) effective masses for Ge-NW (0.33$m_0$ for heavy hole and 0.043$m_0$ for light hole) and p-Si substrate (0.49$m_0$ for heavy hole and 0.16$m_0$ for light hole) is also considered in this theoretical model to study its impact on the transport characteristics of the device. Such effective mass mismatch between *'device'* and the *'reservoir'* is included in the theoretical model by modifying the relevant self energy function ($\Sigma_{p-Si}(E_h)$). It has been mentioned earlier that the 'strength of coupling' depends on the parameter '$\beta g \beta$', however, when the 'mismatch between effective masses' is included in the system's (*'device'* and *'reservoir'*) Green's function, such term is modified to '$b(\beta g \beta)$' **[481]**.

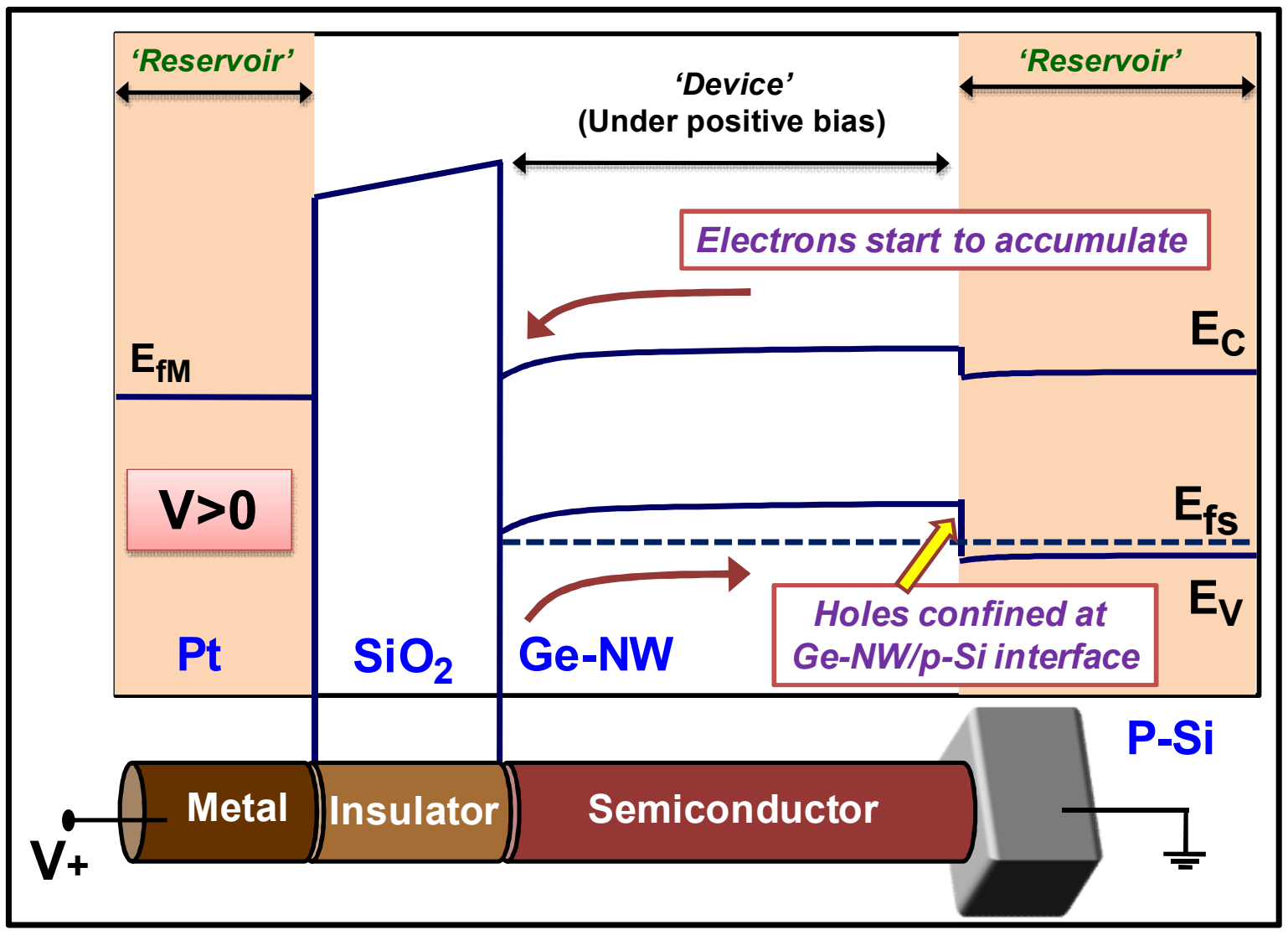

Fig. 6.4. The schematic energy band diagram of the Pt/$SiO_2$/Ge-NW/p-Si MOS device in 'accumulation-to-inversion' condition.

Here, '$b$' is introduced as the factor to determine the 'strength of coupling' as **[481]**,

$$b = \left( \frac{1}{1 - 2\left(\beta_{reservoir} - \overline{\beta}\right)g} \right) \tag{6.26}$$

where, $\overline{\beta}$ determines the interaction at the Ge-NW/p-Si interface and its value is derived to be **[481]**,

$$\overline{\beta} = \frac{\left(\beta_{reservoir} + \beta_{device}\right)}{2} \tag{6.27}$$

It is worthy to mention that if there is no mismatch of effective mass values between the *'device'* and *'reservoir'* then, $\overline{\beta} = \beta_{reservoir}$ which leads to the value of such 'strength of coupling' factor, $b = 1$. In the current model, the differences between effective mass values of holes (both heavy and light) in the Ge-NW and substrate region are included which contributes significantly in perturbing the *'device'* Hamiltonian.

In the 'accumulation-to-inversion' region, the carrier concentration profiles for both the heavy and light holes are plotted in Fig. 6.5(a) and 6.5(b), respectively, with distance from the oxide/semiconductor interface and applied bias. It is observed from such plots that the carrier concentration is significantly higher at the interface due to the accumulation of holes (majority carriers) at negative bias. Moreover, when such bias changes from negative to positive, a 'hump' is observed in the carrier profile at Ge-NW/p-Si interface due to the valence band offset ($\Delta E_V$) and it signifies partial confinement of holes at the Ge-NW/p-Si (substrate) hetero-interface. It is also observed from such plots that the heavy hole ($m_{HH}^* = 0.33m_0$) concentration (Fig. 6.5(a)) is relatively higher in comparison to the light holes (Fig. 6.5(b)). This is attributed to the reduced sub-band energy gaps due to heavy holes, which create more sub-bands to participate in the transport process ($m_{LH}^* = 0.043m_0$).

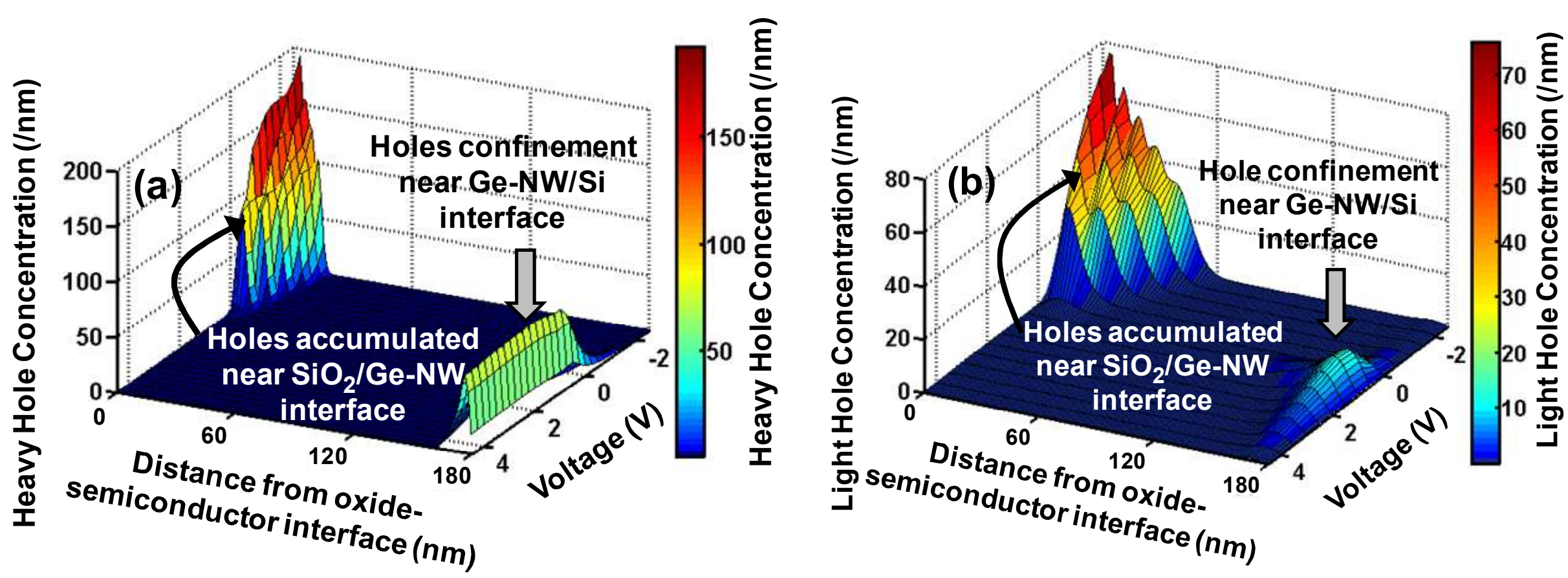

Fig. 6.5. Plots of the carrier concentration profile for: (a) heavy holes, and (b) light holes with distance from the $SiO_2$/Ge-NW interface up to the depth of Ge-NW/p-Si (substrate) junction and the applied voltage.

It is worthy to mention that the frequency responses of such light and heavy holes will be different due to the difference in their transmission lifetimes which in turn also significantly depends on their 'strength of coupling' with the *'reservoir'*. The relation between hole carrier lifetime ($\langle t \rangle$) and the self energy function ($\Sigma_{p-Si}(E_h)$) is developed by performing the inverse Fourier transform of the *'device'* Green's function which gives rise to,

$$\frac{1}{2\pi\hbar}\int_{-\infty}^{+\infty}(E_h[I]-[H]-\Sigma_{p-Si}(E_h))[G^V(E)]\exp(-iE_h t/\hbar)dE = \delta(t)[I] \qquad (6.28)$$

Then, considering the retarded Green's function $G^{V\,Rtd.}(t)$, which is the Fourier transform of $G^{V}(E)$, the above equation can be expressed as,

$$\frac{1}{2\pi\hbar}(i\hbar\frac{d}{dt}[I]-[H]-[\widetilde{\Sigma}_{p-Si}(t)])[G^{V\,Rtd}(t)]=\delta(t)[I] \tag{6.29}$$

The solution of Eqn. (6.27), assuming the self-energy to be explicitly independent of time, is obtained as,

$$[G^{V\,Rtd}(t)]=\vartheta(t)\exp(-i([H]+\mathrm{Re}[\widetilde{\Sigma}])t/\hbar)\exp(-i(\mathrm{Im}[\widetilde{\Sigma}])t/\hbar) \tag{6.30}$$

where, $\vartheta(t)$ represents a step function with non-zero values for $t>0$ only. It is noticed from the above equation that the real part of self-energy function perturbs the Hamiltonian and the imaginary part is responsible for obtaining the carrier lifetime,

$$\langle t\rangle \equiv \frac{\hbar}{2}\mathrm{Im}[-\Sigma_{p-Si}(E_h)]^{-1} \tag{6.31}$$

The imaginary part of self-energy is actually responsible for the broadening of *'device'* energy sub-bands due to the uncertainty (effective mass mismatch) arising from its coupling to the *'reservoir'*. Such determining term ' $\beta g\beta$ ' for the 'strength of coupling' can further be expressed as the following (Eqn. (6.32)) to develop a relation between the self-energy and effective mass of respective regions (*'device'* and *'reservoir'*),

$$\beta g\beta=\frac{A\pm\sqrt{A^2-4\beta^2}}{2}=\frac{(a+2\beta)\pm\sqrt{(a+2\beta)^2-4\beta^2}}{2}, \text{ where } a=\left[E_h-E_n^V-E_D^V(z)\right]$$

$$=\frac{(a+2\beta)\pm i\sqrt{4\beta^2-(a+2\beta)^2}}{2}=\beta\exp(i\theta) \tag{6.32}$$

Now, considering $\theta=k\Delta$ and $\beta=-\left(\frac{\hbar^2}{2m^*\Delta^2}\right)$, we obtain $\cos k\Delta=\frac{(E_h-E_n^V-E_D^V)}{2\beta}+1$, which further implies to the E-k dispersion relation as,

$$E_h=E_n^V+E_D^V(z)+2\beta(\cos k\Delta-1) \tag{6.33}$$

The carrier velocity and effective mass are obtained from such dispersion relation as,

$$v=\frac{1}{\hbar}\frac{\partial E_h}{\partial k}=\Delta\left(\frac{2}{\hbar}\mathrm{Im}[-\Sigma]\right)=\frac{\Delta}{\langle t\rangle} \tag{6.34}$$

and,

$$\widetilde{m}^* = \hbar^2 \left( \frac{\partial^2 E_h}{\partial k^2} \right)^{-1} = \frac{\hbar^2}{2\Delta^2} (\beta[-\cos k\Delta])^{-1} \Rightarrow \langle t \rangle = \left( \frac{\hbar^2}{2v^2 m^*} (\beta[-\cos k\Delta])^{-1} \right)^{\frac{1}{2}} \tag{6.35}$$

Thus, the required modification of carrier effective mass due to the *'device'* - *'reservoir'* coupling is obtained to be,

$$\frac{\widetilde{m}^*}{m^*} = \left| \frac{\mathrm{Re}(\Sigma_{p-Si})}{\Sigma_{p-Si}} \right| \tag{6.36}$$

It is apparent from Eqn. (6.36) that if there is no change in the imaginary part of self-energy, the numerator will be equal to the denominator and this will keep the transport effective mass unaltered. Further, it is evident from Eqn. (6.35) that the light holes (with lower effective mass) have relatively higher carrier lifetime in comparison to the heavy holes and thus, the (low) frequency response of such light holes is significantly higher. On the other side, the (high) frequency response of heavy holes (higher effective mass) is not as significant as the light ones due to their smaller carrier lifetime. The effect of such variations in frequency responses can be observed from the experimental C-V characteristics (discussed in the following sections).

## 6.3.4. Inversion region confinement

Energy band diagram of the device under strong inversion is shown in Fig. 6.6 where the bands are bent in downward direction due to the application of a relatively higher voltage (in the range of +2 V to +4 V). The applied bias creates a quantum well at the oxide/semiconductor junction due to surface quantization in conjunction with the existing radial quantization and consequently, 3D-quantized discrete energy states are formed inside such quantum well. Thus, top of the nanowire now acts as the quantum dot where its formation strongly depends on the applied voltage and dimensions of the nano/quantum wire.

The carrier concentration profile in such inversion region is shown in Fig. 6.7 which extends spatially up to the quantum well created by a very small conduction band off set of 50 mV ($\Delta E_C$ = 50 mV). Such a small band offset does not show any impact on the carrier profile. These charge carriers are responsible for the 'step-like' variation of C-V characteristics, shown in Fig. 3.3 of Chapter 3.

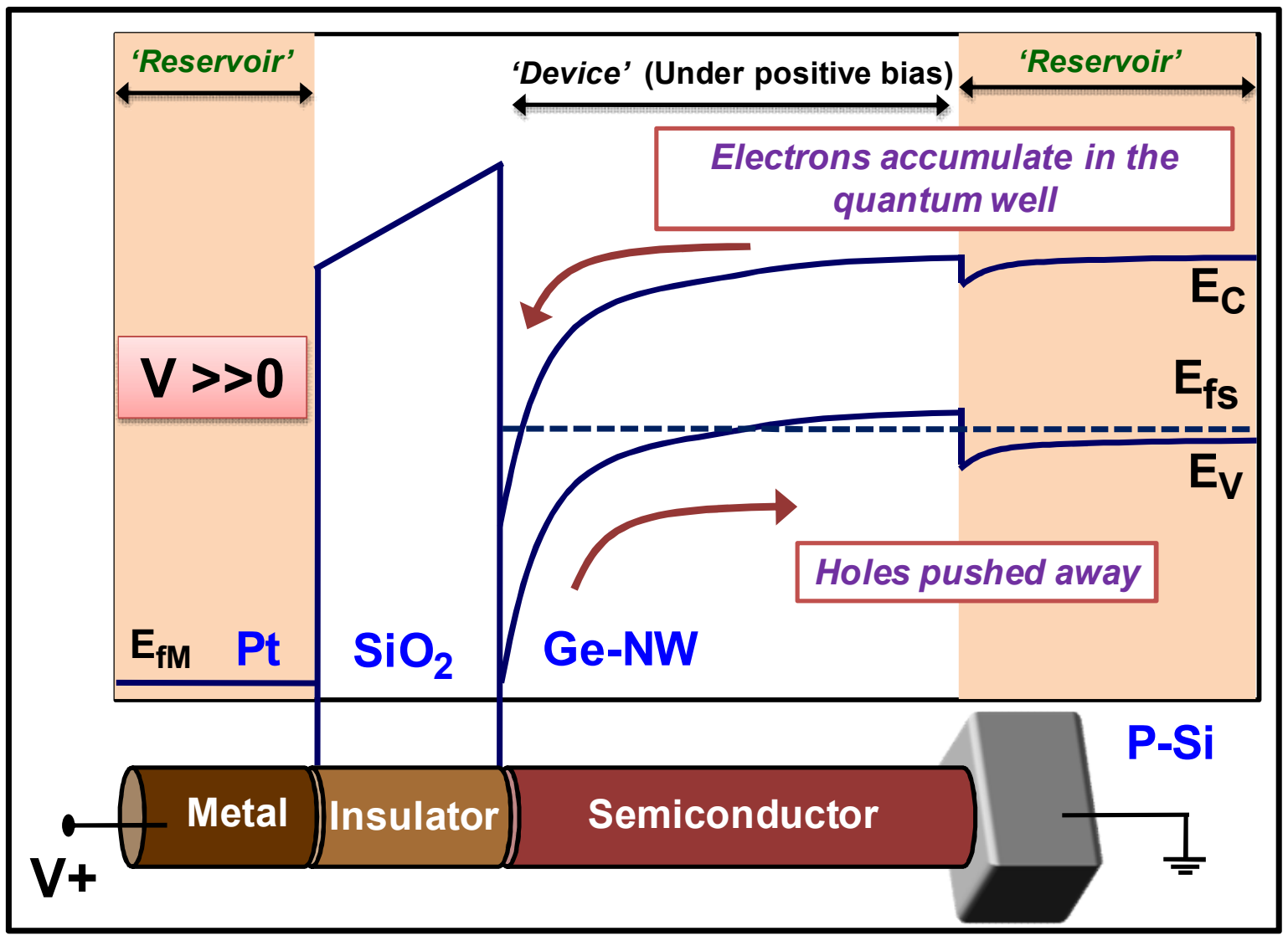

Fig. 6.6. Schematic representation of energy band diagram of the Pt/$SiO_2$/Ge-NW/p-Si MOS device under inversion condition.

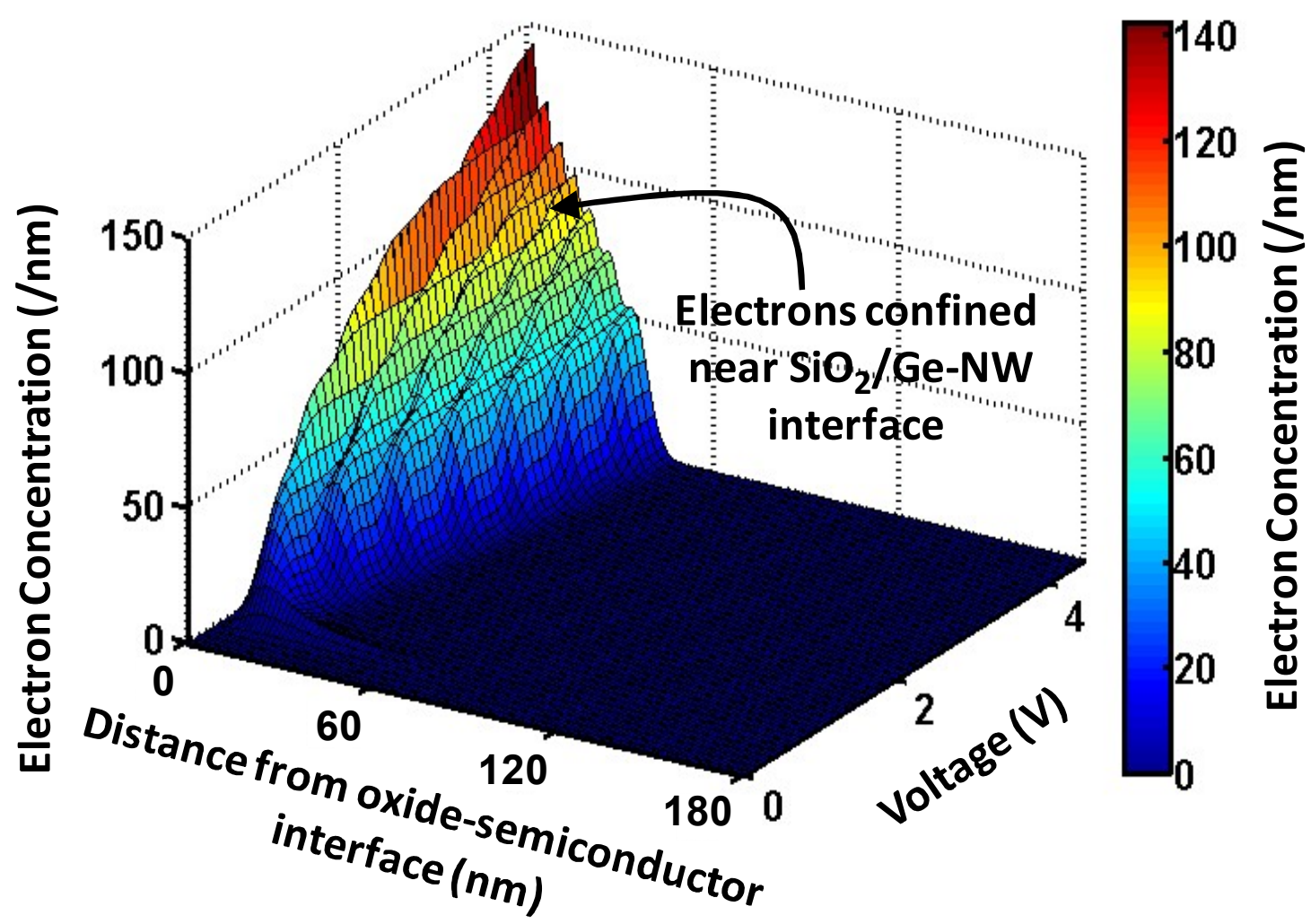

Fig. 6.7. Plot of carrier concentration profile for electrons in the inversion region with distance from the $SiO_2$/Ge-NW interface to Ge-NW/p-Si (substrate) junction and the applied voltage.

## 6.3.5. Plot of LDOS for electrons and holes

The local density of states (LDOS) for both the 2D-confined holes (heavy and light) and 3D-confined electrons for an applied bias of 3 V are plotted in Fig 6.8(a) and 6.8(b), respectively, with distance from the $SiO_2$/Ge-NW interface. It is observed from the plots that the LDOS for electrons show a 'delta-function' like variation due to the formation of discrete energy states in the conduction band. In comparison, energy sub-bands are created in the valence band and the LDOS for holes are broadened due to their free transport along the nanowire axis. The number of such energy sub-bands participating in the transport phenomena is relatively higher for heavy holes in comparison to the light holes due to the mismatch of their effective masses (*i.e.*, $0.33m_0$ for heavy hole and $0.043m_0$ for light hole). The broadening of LDOS also depends significantly on the mismatch of carrier effective masses between the Ge nanowire and p-Si substrate. It has been previously mentioned that the coupling between conduction band electrons of Ge and p-Si substrate is relatively lower since such electrons are 3D-cnfined in the quantum well. Moreover, such coupling also depends on the spatial extension of discrete energy states where the higher order energy states, with larger spatial extension, have relatively higher probability to interact with the substrate. Holes are 2D-confined in the valence band and therefore interact more with the substrate, to lead to a broadened variation in the LDOS profile for both the heavy and light holes.

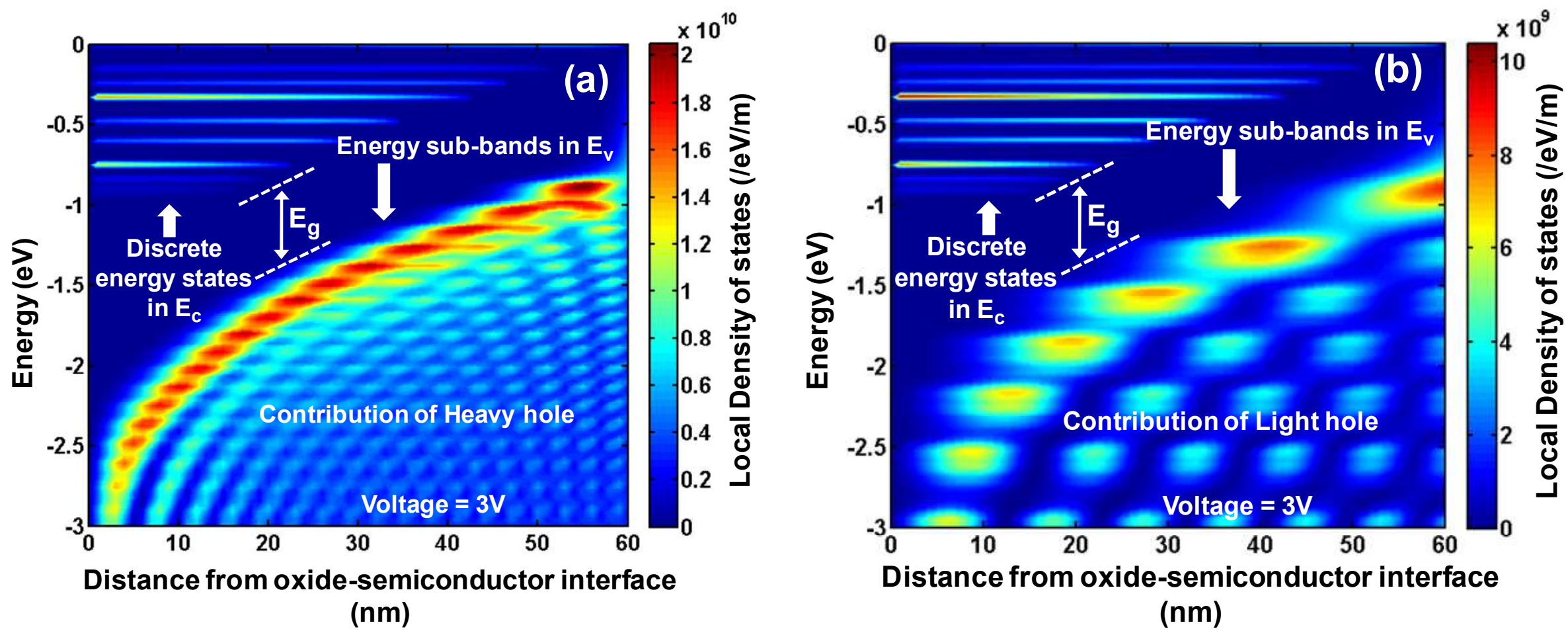

Fig. 6.8. Plots of the local density of states against energy for the 3D-confined electrons and 2D-confined: (a) heavy holes and, (b) light holes with distance from the oxide/semiconductor interface.

## 6.4. Fabrication of the nanowire-MOS device

### 6.4.1. Process flow

The schematic process flow for VTQD-devices, fabricated using the vertically oriented patterned Ge-nanowire array is shown in Fig. 6.9. Prior to the growth/deposition the p-Si wafer is cleaned by following RCA-1 and RCA-2 techniques and then dipped into the hydrofluoric acid solution (with HF:DI-water = 1:50) to remove the native oxide layer (*i.e.*, $SiO_2$). The cleaned samples are then coated with positive electron resist, poly (methyl methacrylate) (PMMA) where the PMMA solution is prepared by dissolving 10 mg of PMMA in 100 ml anisole solvent. The thickness of PMMA layer is maintained to be ~1 μm by spin-coating at 700 rpm and it is measured by ellipsometry. The PMMA thickness has been optimized to achieve desired dimension as preset by the Raith Elphyplus lithography software (details are given in the next section). The PMMA coated samples are then pre-baked at 120 $^0$C for 2 minutes and loaded into the FESEM chamber. Once the desired chamber vacuum is achieved the working distance is set at 10 mm and the accelerating voltage is fixed at 10 kV.

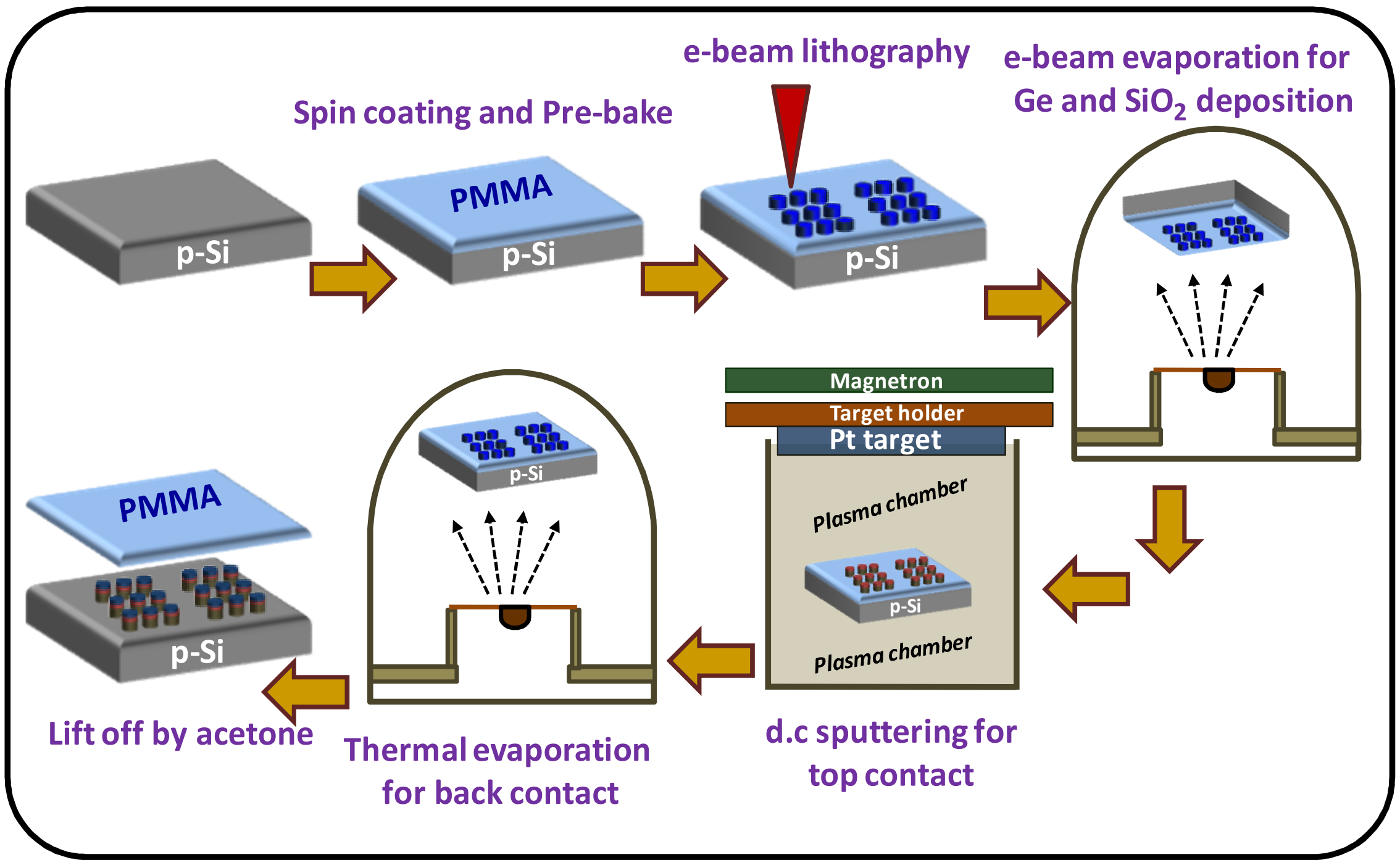

Fig. 6.9. Schematic representation of the fabrication process flow for the QD-devices used in the current work.

The EBL-process parameters such as, magnification, write field and the electron dose, are optimized (discussed in detail in the next section) to reduce the size of patterned sites to the order of EBR of Ge. As has been observed from theoretical prediction this will ensure the realization of quantum effects at room temperature. The lithography process for patterning the PMMA film is performed at standard 30 μm aperture. Such patterned samples are subsequently subjected to the post-bake process at 120 $^0$C for 2 minutes and then these are loaded into a high vacuum e-beam evaporation chamber for the sequential deposition of Ge and $SiO_2$. The deposition thickness of such materials (200 nm for Ge and 20 nm for $SiO_2$) is controlled by the thickness monitor attached into the system. Following the deposition of Ge and $SiO_2$, the samples are loaded into a D.C sputtering chamber to deposit Pt of 50 nm thickness for the top metal contact. The back contact is formed by depositing a thick layer of Al (~ 500 nm; confirmed by Dectak Profilometer) using e-beam deposition technique. Finally, all the samples are emerged into the beaker, containing acetone solution, and placed on a magnetic stirrer unit with 80 rpm rotation speed for controlled lift-off process of the PMMA layer.

### 6.4.2. Optimization of e-beam lithographic process for patterning

A schematic representation of the optimization process carried out in the current work for achieving patterned sites with ~25 nm radii is shown in Fig. 6.10. The key purpose of developing such optimization process is to find the 'optimum dose' of electrons to match the size of fabricated patterned features with the electromagnetic mask pre-designed in Raith ElphyPlus software.

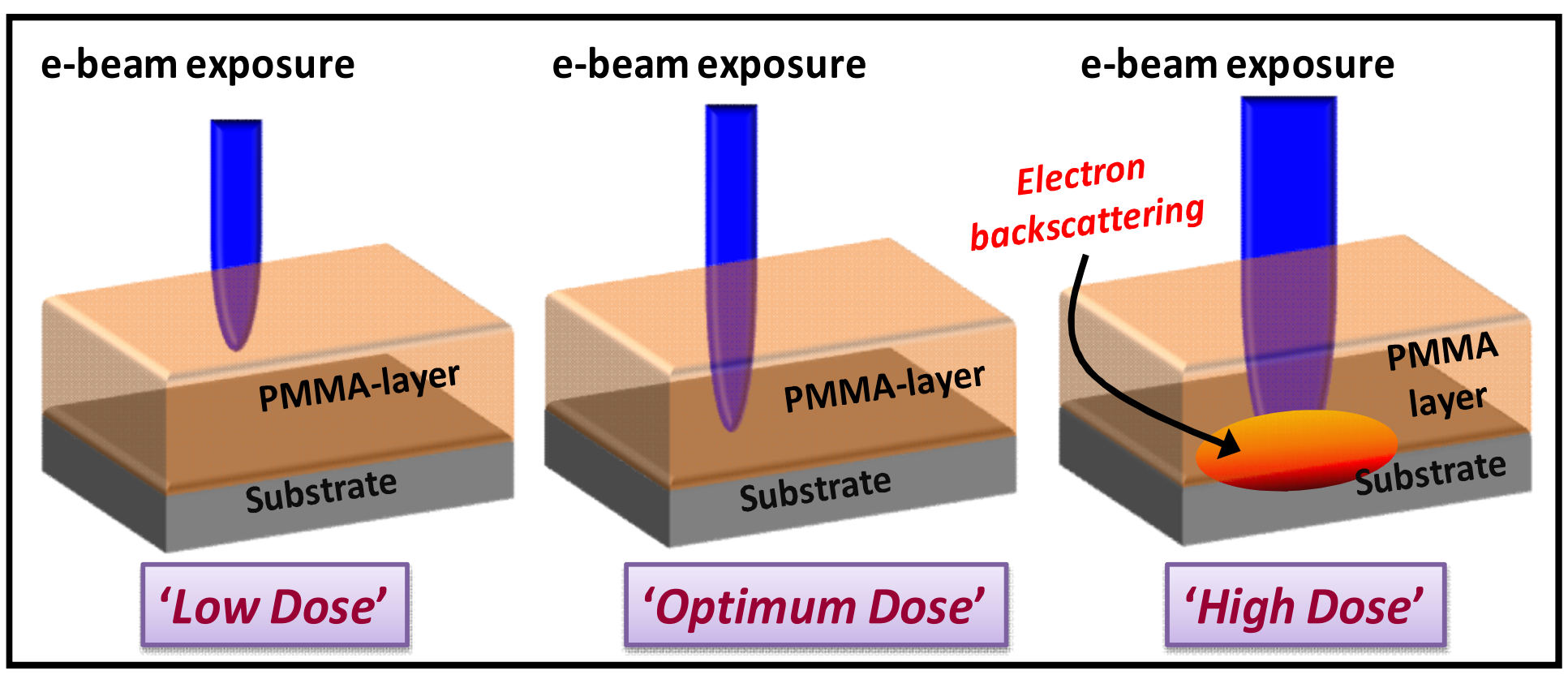

Fig. 6.10. Schematic representation of the optimization process to achieve desired patterned sites.

Three cases may arise following such route: (a) the electron dose is considerably low and thus, electrons cannot reach up to the substrate surface ('Low dose'), (b) the electron dose is significantly high and thus, after reaching the surface such electrons will be backscattered and thereby distorts the pre-designed size ('High dose'), (c) an optimum condition is achieved when electrons reach up to the substrate surface however, these are not backscattered and thus, the undistorted pattern size and shape can be obtained. Along with 'electron dose' the optimization of other parameters, listed in Table 6.2, is also very important to control the quality of lithographic patterns.

Table. 6.2. List of the parameters with their optimized values used for the current work.

| Parameters | **Values (Optimized/Standard)** |
|---|---|
| PMMA concentration | 10 mg in 100 ml anisole (Optimized) |
| PMMA-thickness | 1 μm (Optimized) |
| Pre- /Post- bake temperature and time | 120 $^0$C for 2 minutes (Optimized) |
| Write Field | 50 μm (Optimized) |
| Operating voltage | 10 kV (Standard) |
| Operating distance | 10 mm (Standard) |
| Magnification | 1500X (Optimized) |
| System aperture | 30 μm (Standard) |
| Developer and Stopper concentration with time | MIBK:IPA (1:3) for 45 seconds and IPA for 15 seconds (Optimized) |

First, the PMMA-concentration and spin speed (coating of three successive layers at 700 rpm) are optimized to obtain its thickness of ~1 μm. The time duration of pre-bake, post-bake, pattern development, and their relevant temperature are optimized for the precise lift-off of, PMMA, Ge, $SiO_2$ and Pt layers. The write field (WF) of 50 μm is chosen so that the lithographic system operates at high precision (1 nm) to obtain accurate size of the patterned sites. In the current work, circular patterns with different inter-dot spacing of 200 nm, 500 nm and 1 μm are developed for optimizing the process, as shown in Fig. 6.11. The size of each of the boxes is 5 μm × 5 μm with the spacing between the boxes is 1 μm. Such design is optimized for the 50 μm WF with 1500X magnification at an electron dose of 50 k μC/cm$^2$ and the other parameters are fixed at their respective values, as listed in Table 6.2.

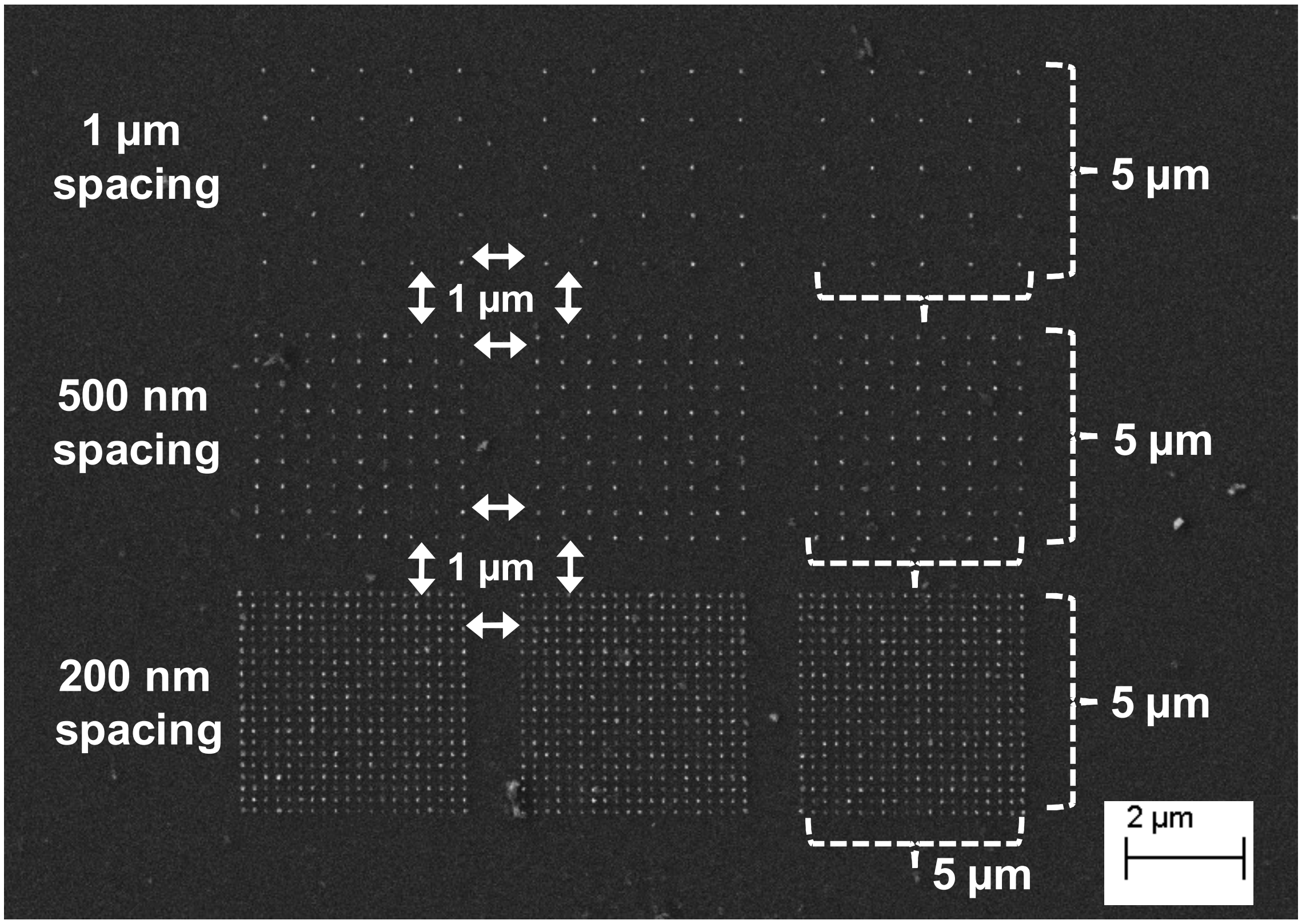

Fig. 6.11. Patterned array of Ge-dots with the inter-dot spacings of 1µm, 500 nm and 200 nm, developed by using the optimized parameters in electron beam lithography.

Now, keeping all such parameters fixed at their respective values, the electron dose remains to be the only parameter which needs to be carefully monitored. For this purpose, the pre-designed pattern with 1 µm spacing is exposed for electron dose values ranging from 100 k µC/cm$^2$ (high dose) to 50 k µC/cm$^2$ (low dose). All such patterned circular Ge-structures are shown in Fig. 6.12(a)-(f). It is to be mentioned that before fabricating the MOS device, the aim was to optimize the size of such circular structures and after achieving the dot radius of ~25 nm the subsequent device fabrication process steps were performed. Such plots indicate the dot radius to decrease from ~90 nm to ~30 nm for decreasing the electron dose from 100 k µC/cm$^2$ to 50 k µC/cm$^2$. The values of electron dose are set over a wide range, depending upon the need for PMMA layer thickness (~1 µm) so that the optimum condition is achieved with site dimension ~ EBR of Ge.

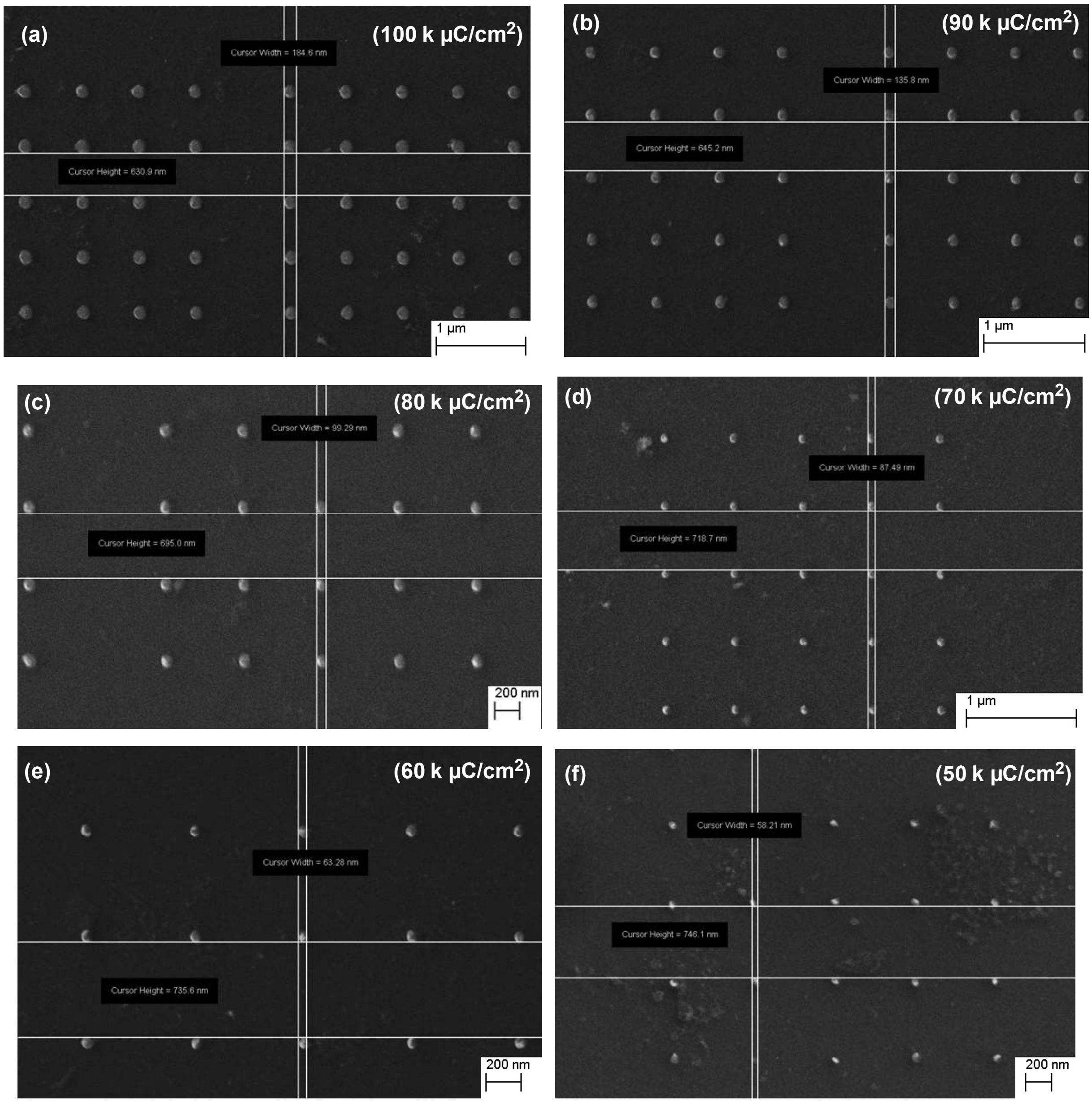

Fig. 6.12. FESEM images of patterned Ge-structures with 1 μm inter-dot spacing for: (a) 100 k μC/cm$^2$, (b) 90 k μC/cm$^2$, (c) 80 k μC/cm$^2$, (d) 70 k μC/cm$^2$, (e) 60 k μC/cm$^2$, and (f) 50 k μC/cm$^2$ electron doses.

Similar structures with 500 nm spacing are shown in Fig. 6.13(a)-(d) for the electron doses ranging from 70 k μC/cm$^2$ to 45 k μC/cm$^2$. The corresponding dot radius is measured to be ~25 nm for both the electron doses of 50 k μC/cm$^2$ and 45 k μC/cm$^2$, however, at the relatively lower dose of 45 k μC/cm$^2$ some of the structures are under-exposed. The under-exposed structures may be wiped off during the lift-off process and therefore, it is also optimized in the current work by reducing rpm of the magnetic stirrer unit.

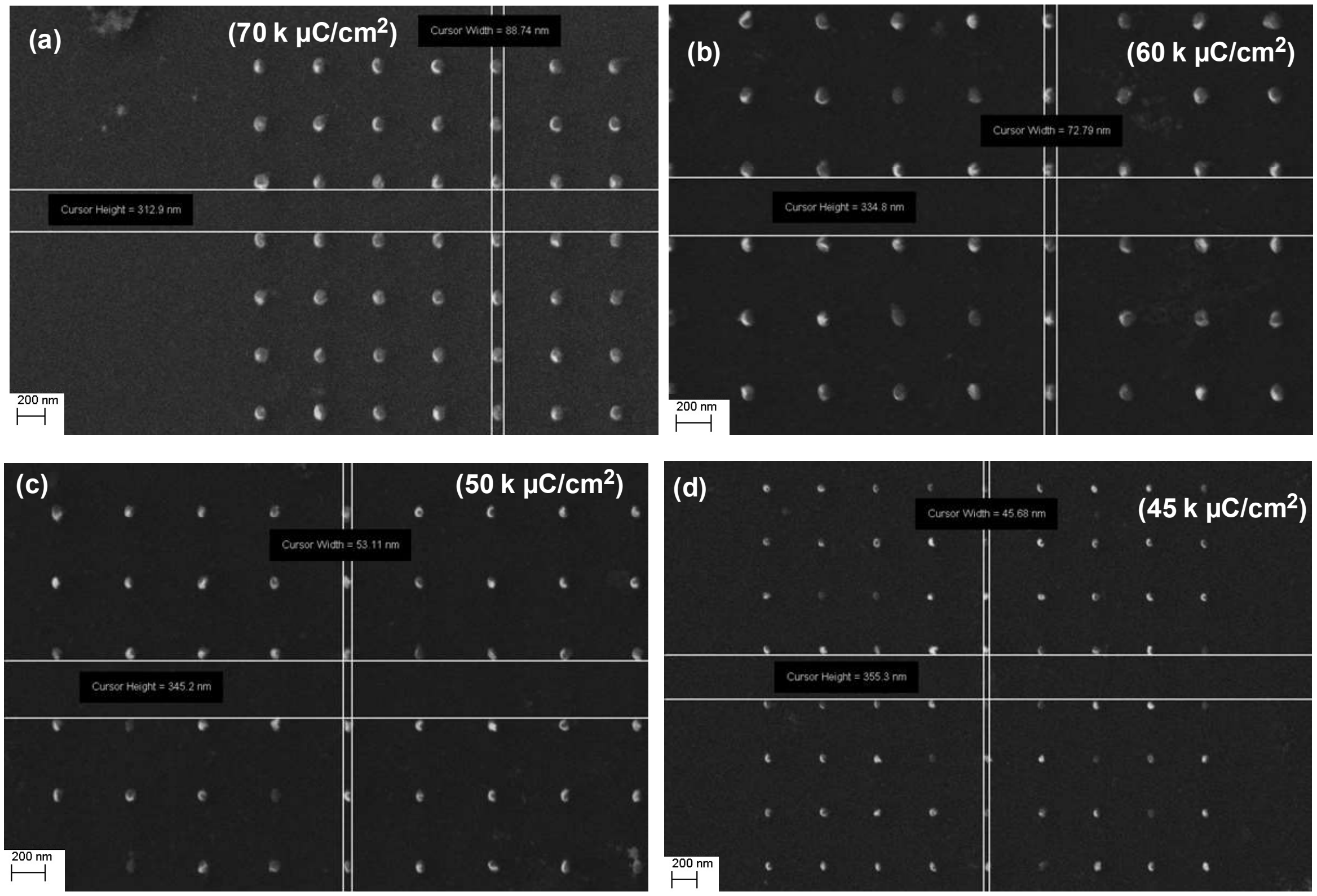

Fig. 6.13. FESEM images of the Ge-dot structures of 500 nm inter-dot spacing with electron doses of: (a) 70 k μC/cm$^2$, (b) 60 k μC/cm$^2$, (c) 50 k μC/cm$^2$, and (d) 45 k μC/cm$^2$.

Finally, FESEM images of the patterned Ge-structures with 200 nm inter-dot spacing is shown in Fig. 6.14(a)-(d) for the electron doses in the range of 60 k μC/cm$^2$ to 40 k μC/cm$^2$. As mentioned earlier, the relatively higher dose of 70 k μC/cm$^2$ also creates an over-exposed pattern since the inter-dot spacing is further scaled down to 200 nm. Most importantly, a site radius of ~20 nm is achieved for the relatively lower electron dose of 40 k μC/cm$^2$. It is also observed from the optimization process that patterned sites with size of ~25 nm can be achieved for the 200 nm inter-dot spacing with relatively lower electron dose. Thus, for the fabrication of MOS devices, the relevant electron doses are chosen to be 50 k μC/cm$^2$, 60 k μC/cm$^2$, 65 k μC/cm$^2$ and 70 k μC/cm$^2$ with an inter-dot spacing of ≤ 200 nm. Such spacing in the fabricated structures is measured to be slightly less than its pre-designed value which may be attributed to the slightly elliptical instead of perfect circular shape of the dots. This arises due to aberrations in the system aperture and hence, the system aperture has been re-calibrated, prior to the fabrication of MOS devices.

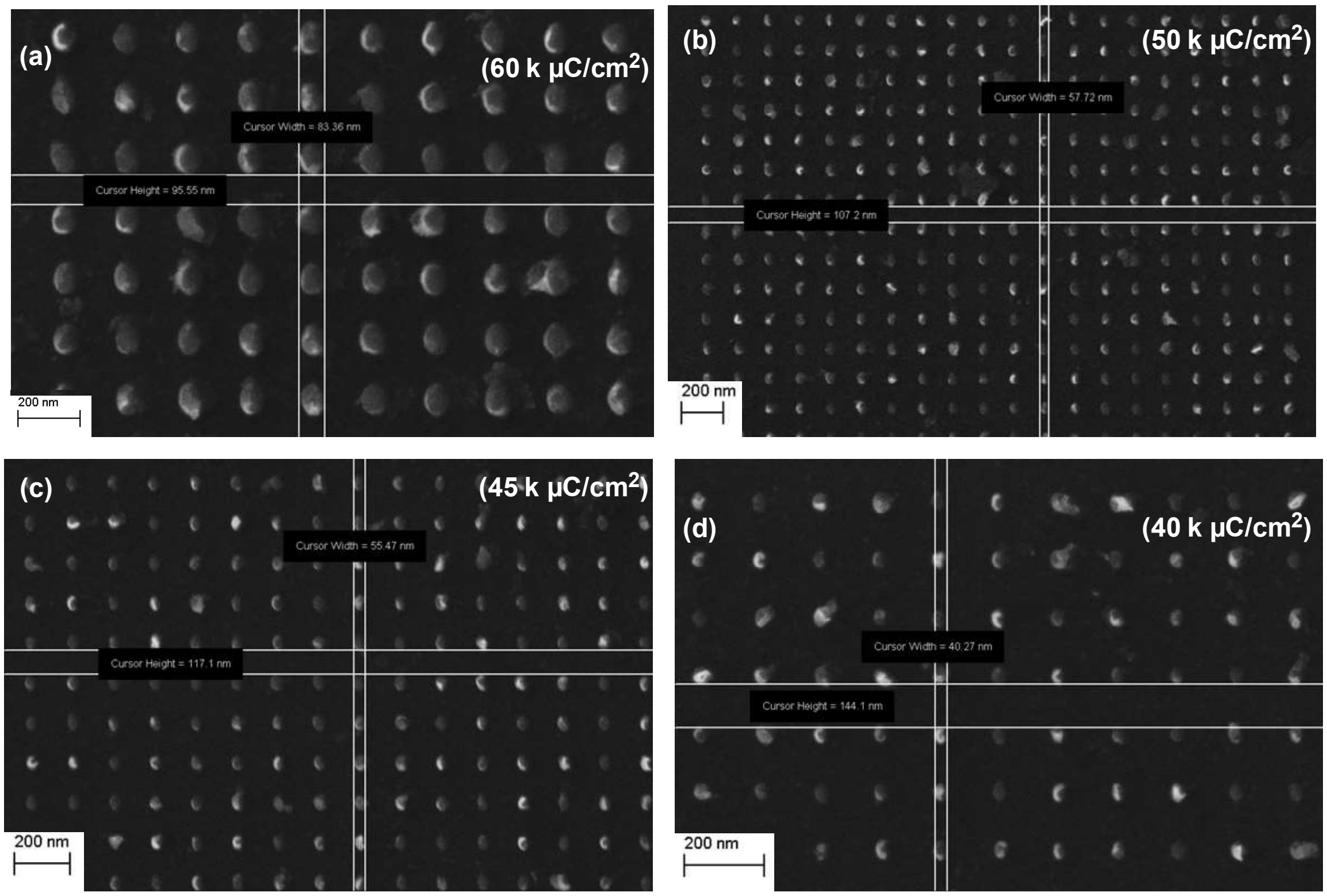

Fig. 6.14. FESEM images of the Ge-nanostructures with 200 nm inter-dot spacing for the electron dose of: (a) 60 k μC/cm$^2$, (b) 50 k μC/cm$^2$, (c) 45 k μC/cm$^2$, and (d) 40 k μC/cm$^2$.

The optimization process for obtaining a dot radius of ~25 nm with different inter-dot spacing and various electron doses is summarized in Fig. 6.15 where the dot radii are measured from FESEM images. It is already mentioned, that the patterned structures with 500 nm and 200 nm inter-dot spacing are over exposed at relatively higher electron dose and such structures with 1 μm and 500 nm spacing are under exposed at the lower electron dose. The structures with all of such three considered spacings are only exposed at 60 k μC/cm$^2$ and 50 k μC/cm$^2$ electron doses. However, the dots with 500 nm and 200 nm spacing are still slightly over exposed for the 60 k μC/cm$^2$ dose as the dot radius is observed to gradually increase with deceasing inter-dot spacing. Therefore, it is essential to find the optimum condition for which the desired device dimensions can be achieved.

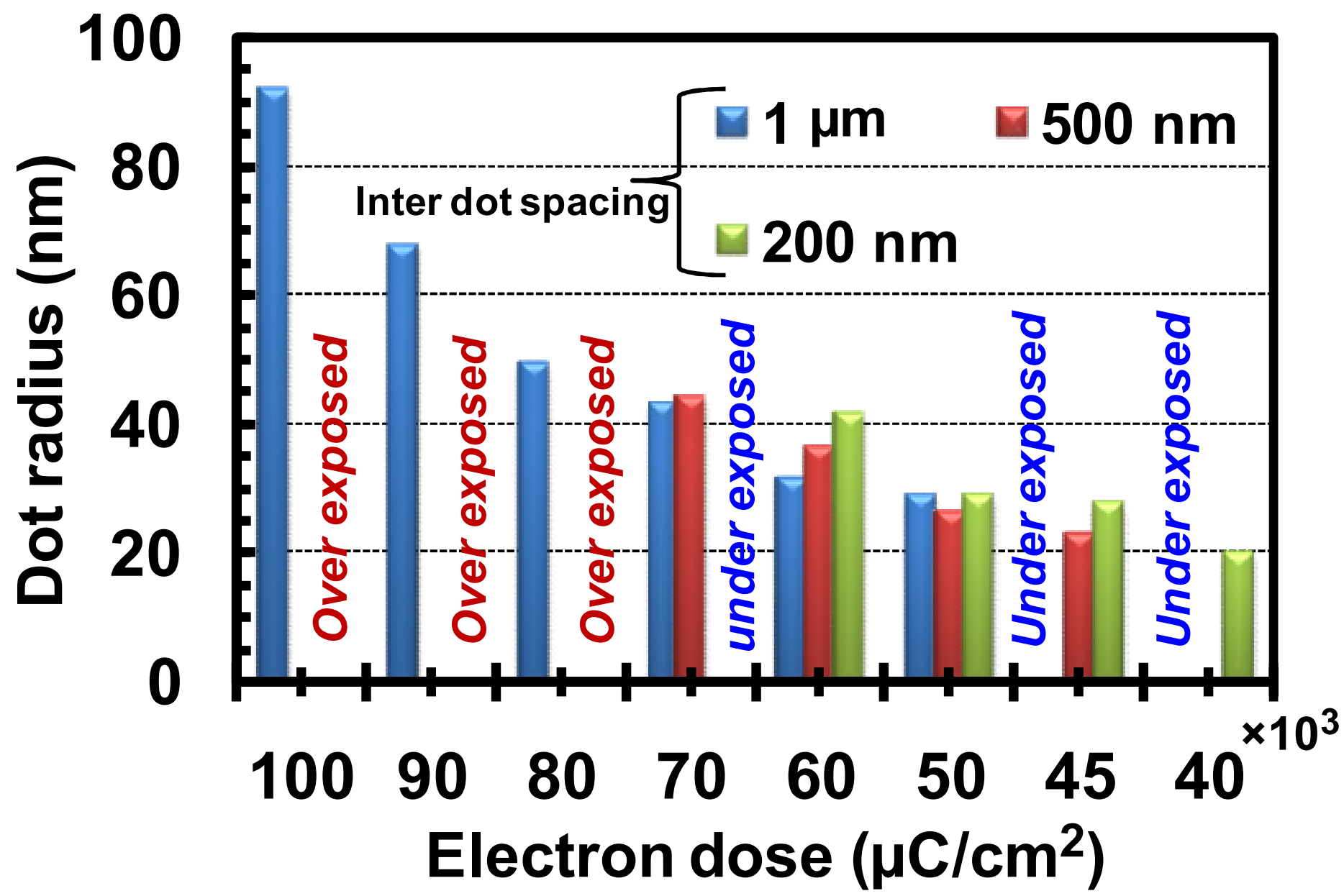

Fig. 6.15. The comparative plot of dot-radius with electron doses for three different inter-dot spacings.

### 6.4.3. Characterization of NWMOS based VTQD-devices

FESEM images of the fabricated patterned array of Pt/$SiO_2$/Ge-nanowire MOS structures on p-Si substrate with different inter-spacing are shown in Fig. 6.16.

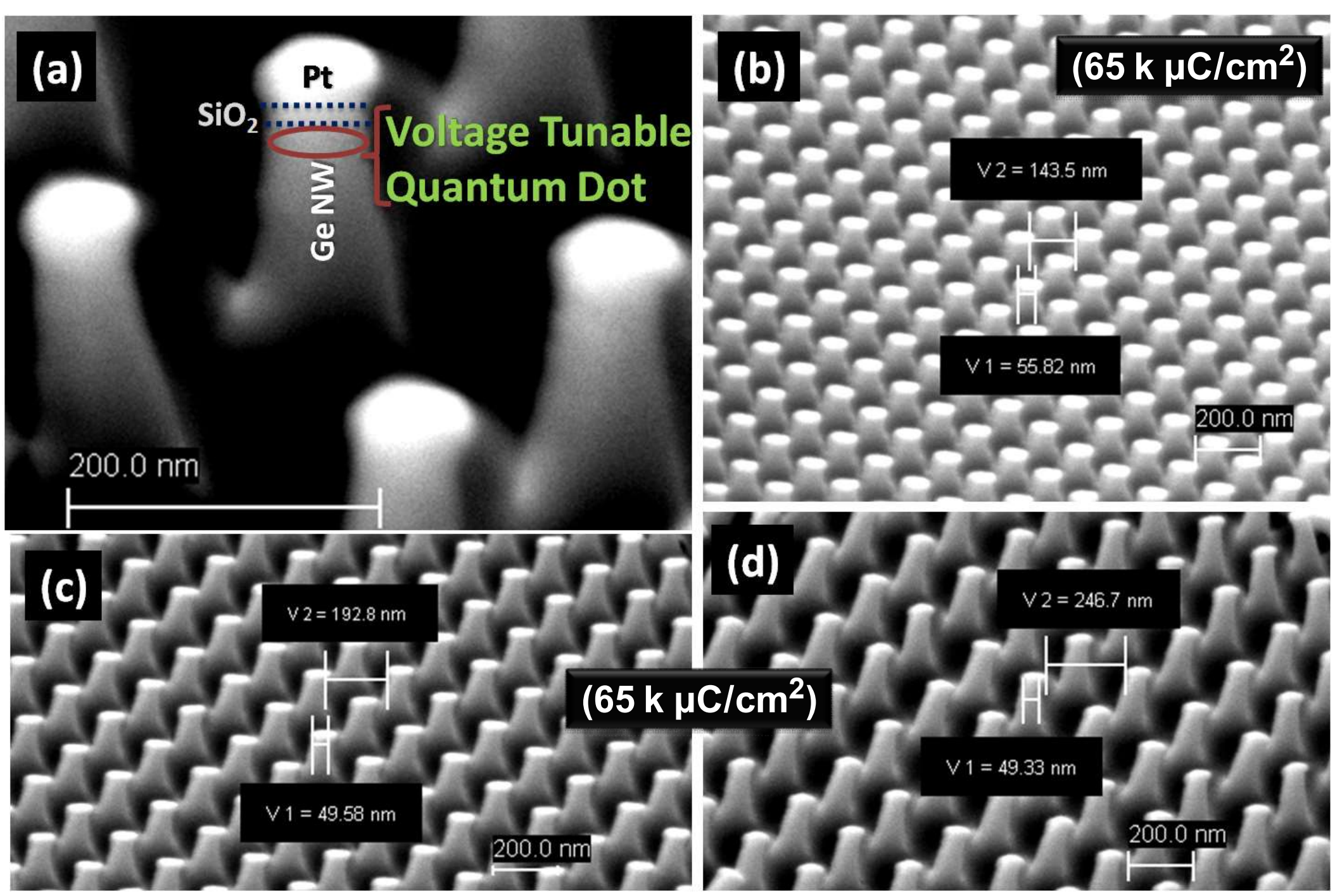

Fig. 6.16. (a) FESEM image of the Pt/$SiO_2$/Ge-NW vertical MOS device, indicating the formation of voltage tunable quantum dot region on top of the Ge-nanowires beneath the oxide/semiconductor interface. The image is captured at a magnification of 120 kX and EHT of 3 kV; FESEM images of the array of NWMOS based VTQD devices, patterned by using an electron dose of 65 k μc/$cm^2$ with the inter-nanowire spacing of: (b) ~150 nm, (c) ~200 nm and (d) ~250 nm.

The different material regions, forming the VTQDs, are indicated in Fig. 6.16(a), and Fig. 6.16(b)-(d) depict the MOS device arrays with inter-dot spacing of ~150 nm, ~200 nm and ~250 nm, respectively. The size distribution of such devices, patterned by using an electron dose of 65 k μC/$cm^2$ with 200 nm inter-nanowire spacing is shown in Fig. 6.17(a) where the average nanowire radius is obtained to be 25.3 nm (which corroborates to the excitonic Bohr radius of Ge *i.e.*, 24.3 nm), with a standard deviation of 5.2 nm. The undesired sizes (e.g. ≠ 25 nm) appear only at the edges of array as shown in Fig. 6.17(b). Such size distortion at the edges of array is possibly attributed to spherical aberration of lithographic e-beam. Therefore, the distorted devices are not subjected to C-V measurement; only the devices at the central regions of arrays are considered for C-V characterization.

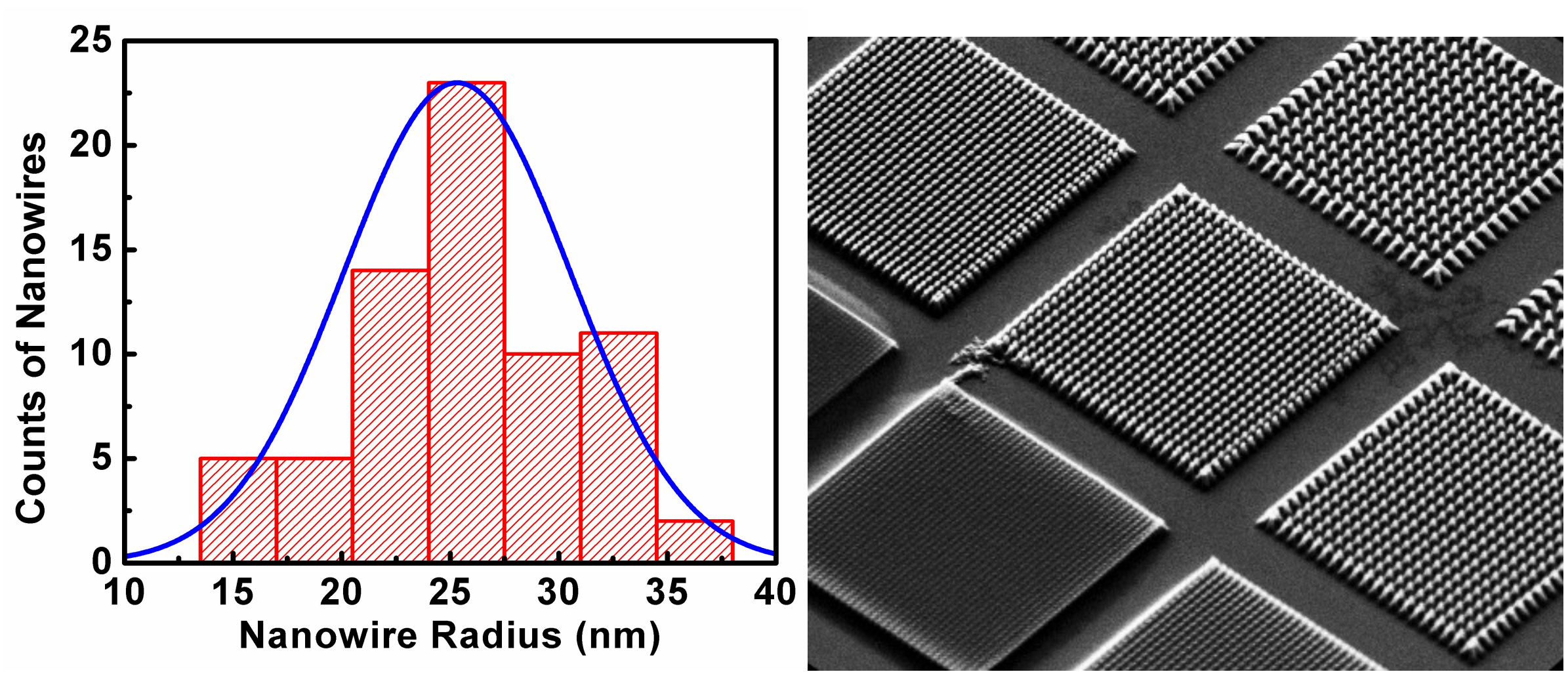

Fig. 6.17. (a) Size distribution of the nanowires of 200 nm spacing, fabricated by using the electron dose of 65 k μC/$cm^2$; (b) Arrays of devices with ~25 nm radii with different inter-spacings showing size distortion near the edges.

Fig. 6.19(a) represents the TEM image of a single NWMOS structure where different regions of it including the Ge-nanowire, $SiO_2$ and Pt are indicated. The relevant selected area electron diffraction (SAED) pattern of the Ge-nanowire is shown in Fig. 6.19(b). The TEM-samples are prepared by ultra-sonication of the fabricated device in ethanol solution and pouring it drop-by-drop on the copper grid.

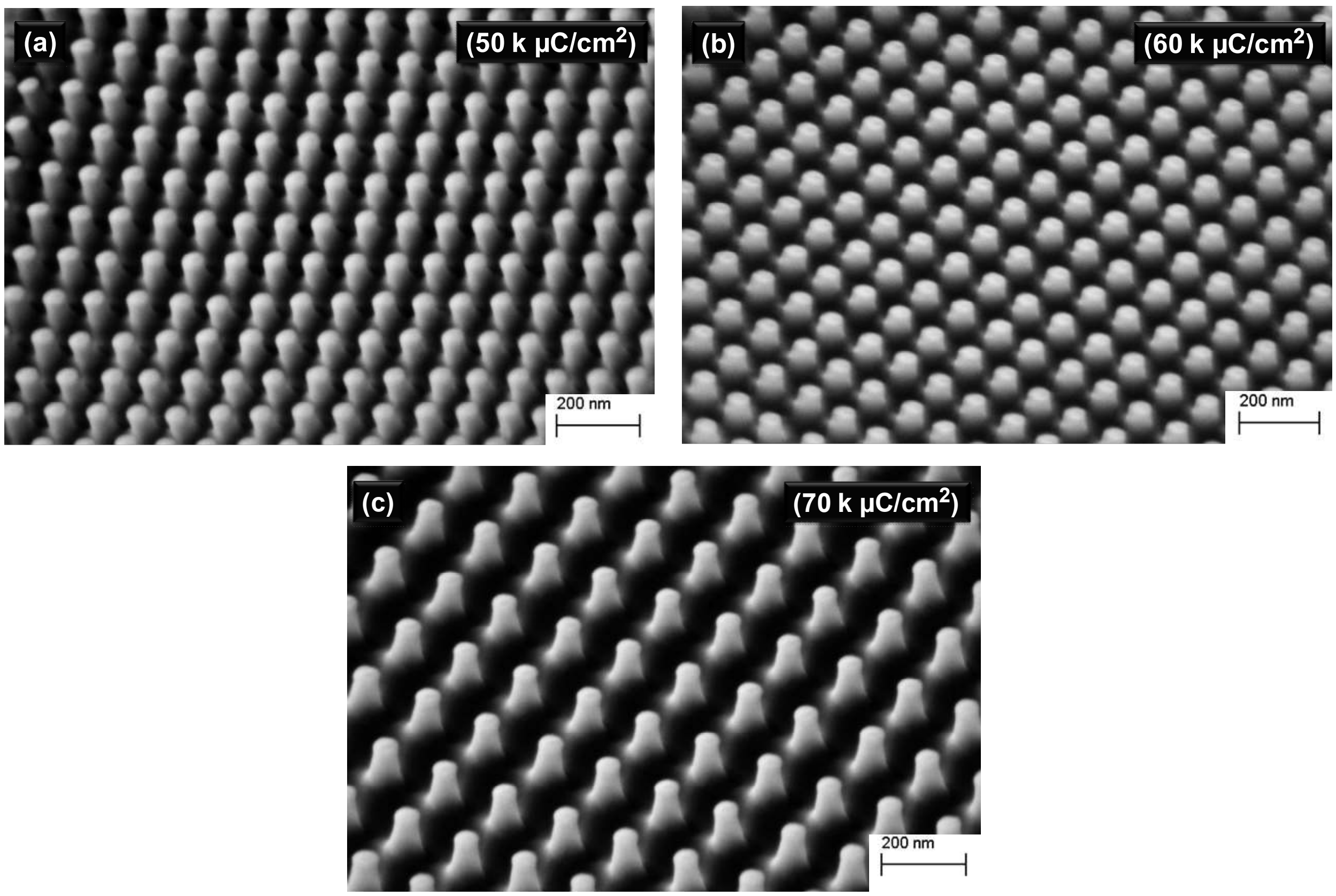

Fig. 6.18. FESEM images of the fabricated NWMOS based VTQD devices for the electron doses of: (a) 50 k μC/$cm^2$, (b) 60 k μC/$cm^2$ and (c) 70 k μC/$cm^2$.

It is apparent from the figure that such SAED spots are arranged in a circular ring which is attributed to electron diffraction from the nanowire **[473]**; and the corresponding d-value indicates the [200]-plane of Ge **[482]**. The elemental content in such single nanowire is shown in Fig. 6.19(c) from the characteristic peaks of energy-dispersive X-ray, which confirms the presence of a considerable weight percentage (~25%) of Ge along with the top electrode metal Pt (~19%); however, the X-ray peaks characteristic to Si and O are negligibly small. Further, the XRD plot of Fig. 6.19(d), exhibits a single peak at $2\theta = 62.6^0$ which confirms the formation of [400]-plane of Ge-nanowire [JCPDS 72-1089] on [400]-Si ($2\theta = 69.6^0$) substrate [JCPDS 01-0791], which also corroborates with the crystal plane obtained from the SAED pattern.

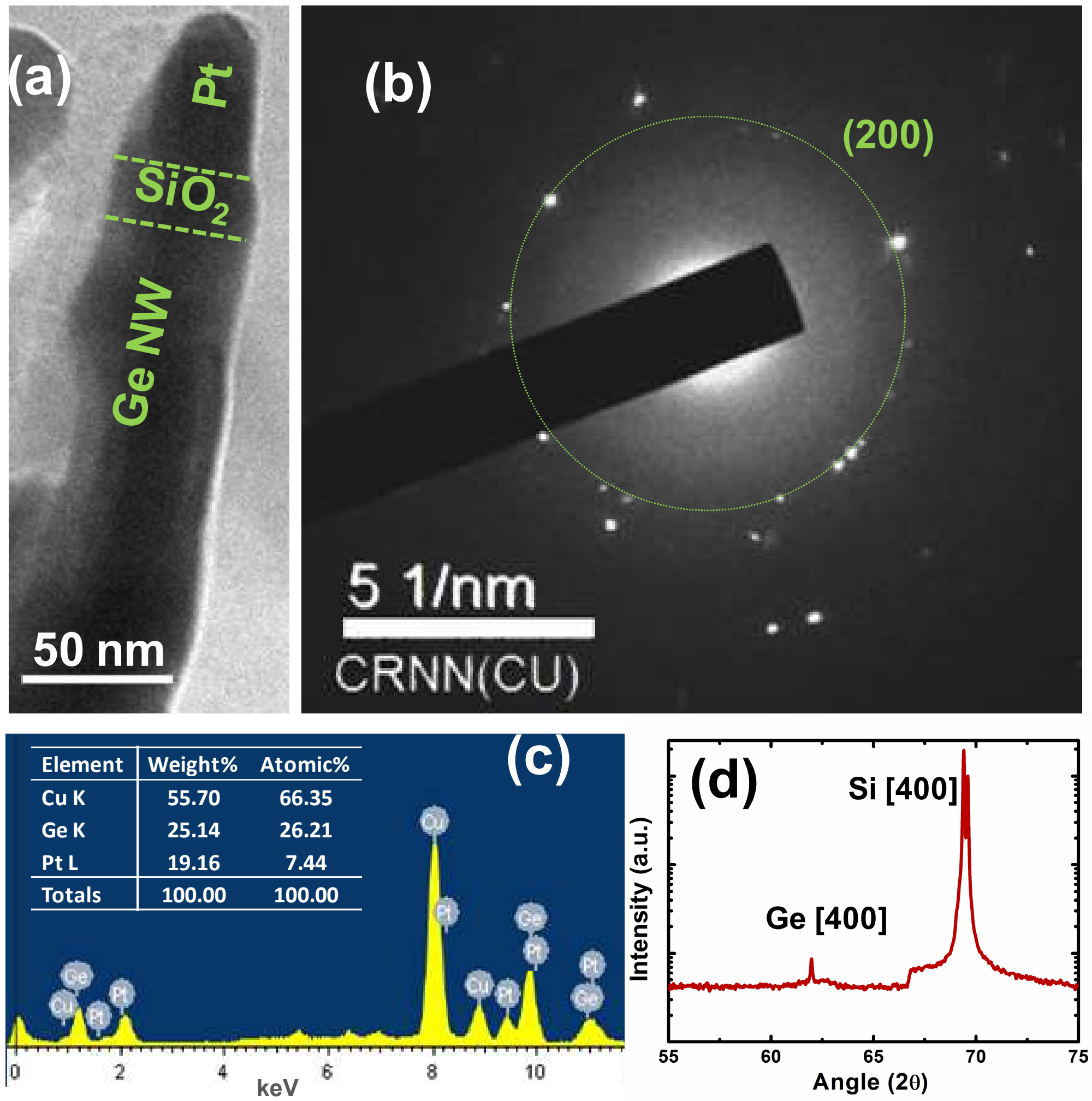

Fig. 6.19. (a) TEM image of the single nanowire MOS, showing Ge (semiconductor), $SiO_2$ (oxide) and Pt (metal) regions where nanowire radius and ~~of~~ the oxide thickness observed to be ~ 25 nm and ~20 nm, respectively; (b) SAED pattern of the Ge-nanowire showing a ring with bright spots which corresponds to the [200]-plane of Ge; (c) Result of in-situ EDS-measurement in TEM along with the atomic and weight percentages of the relevant materials; (d) Plot of XRD profile for the patterned NWMOS which confirms the Ge [400]-plane at $2\theta = 62.6^0$ on [400]-plane of Si substrate.

## 6.5. Capacitance measurement

### 6.5.1. Measurement of capacitance beyond the lower limit of SCS

In the present device scheme, the in-situ measurement of capacitance of a single nanowire MOS device by employing Keithley 4200 SCS is very critical since the capacitance of such single nanowire device will be of the order of atto-farad range whereas, such instrument can only measure efficiently up to pico/femto-farad order. Thus, in the current work, an alternate method is used to measure such atto-farad range capacitance by employing the same instrument (Keithley 4200 SCS) with utilizing frequency dispersion correction mechanism and also taking aid of the instrument operating principle.

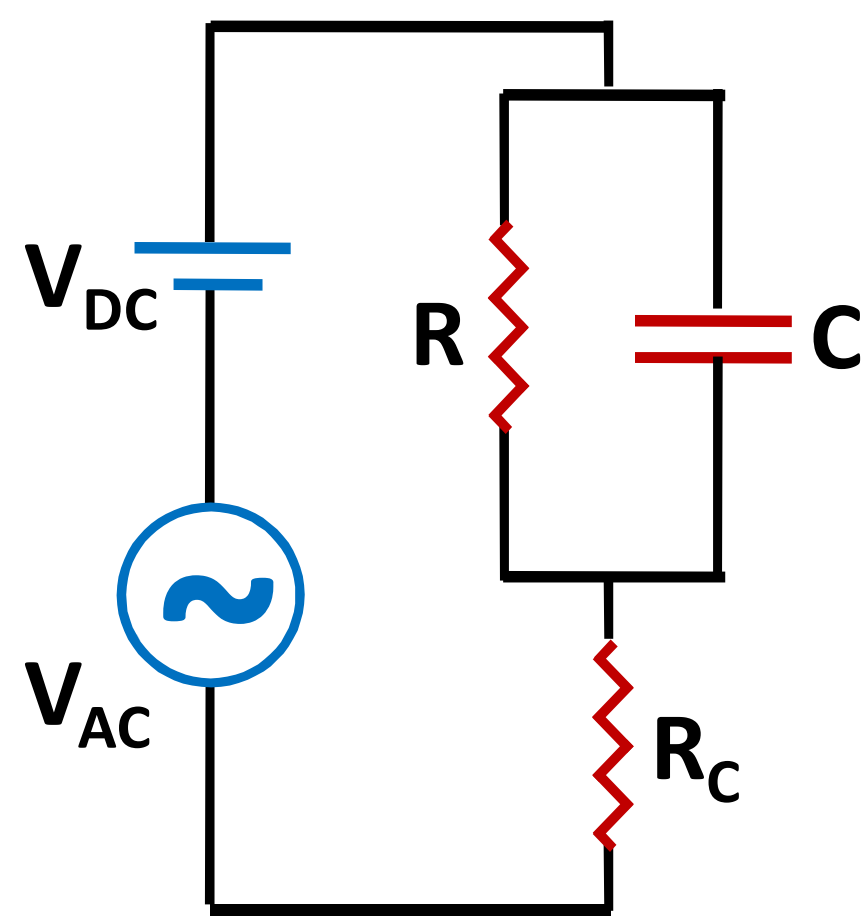

Fig. 6.20. Schematic of the equivalent circuit for measuring capacitance of the $Pt/SiO_2$/Ge-NW MOS capacitor structures.

The SCS-instrument measures C-V characteristics by applying two simultaneous biases such as, a D.C voltage ($V_{DC}$), and an A.C signal ($V_{AC} = V_0 e^{j\omega t}$) of small amplitude $V_0$ (generally in mV range) with an angular frequency of $\omega$, where, such instrument initially measures the time dependent current and further, on integration of the measured current over time gives rise to the charge and corresponding capacitance developed in the device. The equivalent circuit for MOS-capacitance measurement is shown in Fig. 6.20. The MOS-capacitance is represented by $C$ and the possible leakage is considered by the parallel path governed by resistance $R$, whereas contact resistance corresponding to the substrate back contact is represented by $R_C$.

Now, if at a time, the instantaneous current through the circuit is $i(t)$ and the instantaneous charge on capacitor is $q(t)$, then voltage across the parallel capacitor/resistor is,

$$\frac{q(t)}{C} = i(t)R \tag{6.37}$$

The equation for such instantaneous charge and current in the circuit is given by,

$$\frac{q(t)}{C} + i(t)R_C = V_{DC} + V_0 e^{j\omega t} \tag{6.38}$$

Putting the value of instantaneous current in Eqn. (6.38) from Eqn. (6.37), we get,

$$\frac{q(t)}{C}\left[1 + \frac{R_C}{R}\right] = V_{DC} + V_0 e^{j\omega t} \tag{6.39}$$

Therefore, the real capacitance ($C_R$) of the device is obtained by dividing the measured capacitance ($C_m$) by the a 'factor' $\left[1 + \frac{R_C}{R}\right]$ as,

$$C_R = C_m \Big/ \left[1 + \frac{R_C}{R}\right] \tag{6.40}$$

It is also apparent that the measured conductance follows the same 'rule' and, thus, determining such 'factor' is the key concern for developing a method to measure small capacitance values (~aF) using Keithley 4200-SCS. It is worthy to mention that the value of such 'factor' is calculated by eliminating the frequency dispersion from the measured accumulation capacitance where such frequency dispersion arises due to the presence of leakage phenomenon.

### 6.5.2. Frequency dispersion correction

In the present work, significant frequency dispersion in the accumulation region is observed from the (in-situ) measured C-V characteristics as shown in Fig. 6.21(a). Such frequency dispersion is eliminated by considering an equivalent circuit, developed in ref. **[483]**, where a resistor $R_S^{'}$ is taken to be in series with a capacitor $C_E$ to incorporate the impact of leakage phenomenon due to the presence of an unwanted lossy dielectric layer. The equivalent circuit is shown in Fig. 6.21(b) where $C_{OX}$, $C_D$ and $Y_{it}$ represent oxide capacitance, device capacitance and device admittance, respectively. The frequency dispersion can be eliminated by introducing a corrected form of capacitance termed as 'corrected capacitance ($C_c$)', developed from such equivalent circuit, as **[483]**,

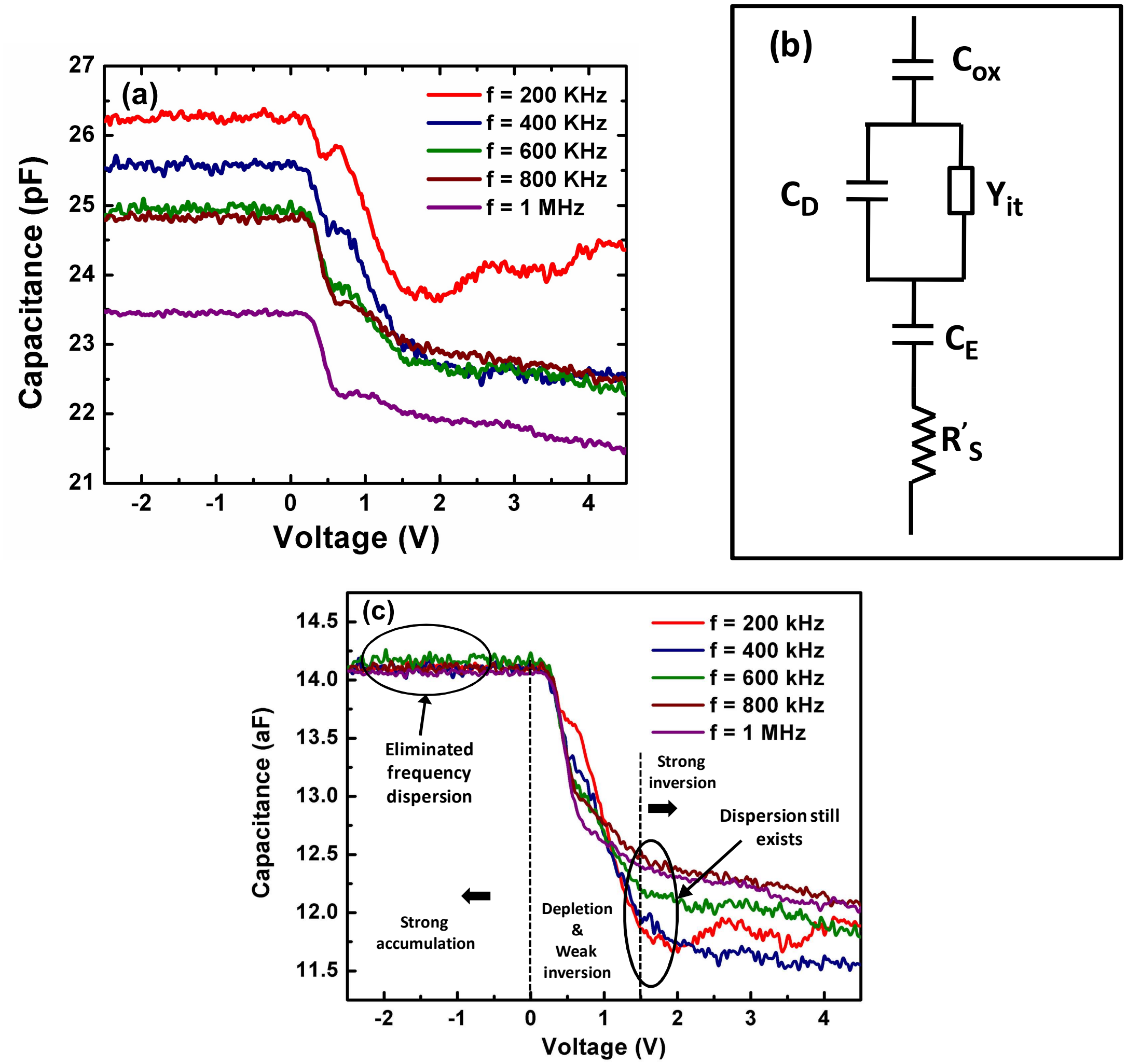

Fig. 6.21. (a) Plot of the measured C-V characteristics; (b) Equivalent circuit considered for the correction of frequency dispersion; (c) Plot of the frequency dispersion corrected C-V curves.

$$C_c = \frac{(\omega^2 C_m C_E - G_m{}^2 - \omega^2 C_m{}^2)(G_m{}^2 + \omega^2 C_m{}^2)C_E}{(\omega^2 C_E{}^2)[G_m(1 - R_s{}' G_m) - \omega^2 R_s{}' C_m{}^2]^2 + (G_m{}^2 + \omega^2 C_m{}^2 - \omega^2 C_m C_E)^2} \qquad (6.41)$$

where, $R_s{}' = \dfrac{G_{ma}}{G_{ma}{}^2 + w^2 C_{ma}{}^2}$ and $C_E = \dfrac{-C_{ox}(G_{ma}{}^2 + w^2 C_{ma}{}^2)}{w^2(C_{ma}{}^2 - C_{ma}C_{ox}) + G_{ma}{}^2}$. $C_{ma}$ and $G_{ma}$ represent the (in-situ) measured values of capacitance and conductance, respectively, at the accumulation condition. $C_m$ and $G_m$ correspond to the values of capacitance and conductance at different applied voltages, respectively.

It was previously mentioned that the measured capacitance values ($C_m$) must be divided (or can be multiplied with ($C_{ox}$) by the 'factor' ($1+R_c/R$) to remove the frequency dispersion for obtaining corrected capacitance values. In the present nanowire MOS device such value is obtained to be 2×10[6] which also suggests that the real capacitance ($C_R$) should be in the aF range although the measured capacitance ($C_m$) was in pF range. The plot of such corrected real device capacitance is shown in Fig. 6.21(c) with elimination of frequency dispersion from the accumulation region. Thus, in the current work the order of capacitance is also corrected through the frequency dispersion correction mechanism.

However, in the present work, along with such accumulation-dispersion, there exists additional frequency dispersion in the 'depletion and weak inversion → strong inversion' region, as shown in Fig. 6.21(c). Such frequency dependency arises due to the hole degeneracy in the valence band region created by presence of the hetero-junction at Ge-nanowire/p-Si substrate interface. Therefore, the present device scheme requires an additional equivalent circuit to eliminate the frequency dispersion from the 'depletion and weak inversion → strong inversion' region. In this context, another equivalent circuit is considered based on a leaky capacitance model due to the presence of such 2D-confined valence band holes.

The equivalent circuit is shown in Fig. 6.22(a) where an additional RC-component ($R_S^{''}$ in series with $C_E^{'}$) is considered to incorporate the impact of leakage phenomenon happened on the onset of strong inversion in the nanowire MOS device due to the presence of partially confined degenerated valence band holes. Here, the form of 'corrected capacitance ($C_C^f$)' is derived from such equivalent circuit to be,

$$C_C^f = \frac{(\omega^2 C_C C_E^{'} - {G_C}^2 - \omega^2 {C_C}^2)({G_C}^2 + \omega^2 {C_C}^2)C_E^{'}}{(\omega^2 {C_E^{'}}^2)[G_C(1-R_S^{''}G_C) - \omega^2 R_S^{''}{C_C}^2]^2 + ({G_C}^2 + \omega^2 {C_C}^2 - \omega^2 C_C C_E^{'})^2} \tag{6.42}$$

where, $R_S^{''} = \dfrac{G_C^{\min}}{{G_C^{\min}}^2 + w^2 {C_C^{\min}}^2}$ and $C_E^{''} = \dfrac{-C_{ox}({G_C^{\min}}^2 + w^2 {C_C^{\min}}^2)}{w^2({C_C^{\min}}^2 - C_C^{\min} C_{ox}) + {G_C^{\min}}^2}$ . The 'minimum' values of capacitance ($C_C^{\min}$) and conductance ($G_C^{\min}$) are obtained from the 'corrected capacitance ($C_c$)' values at the onset of threshold for 200 kHz operating frequency at 2 V applied bias.

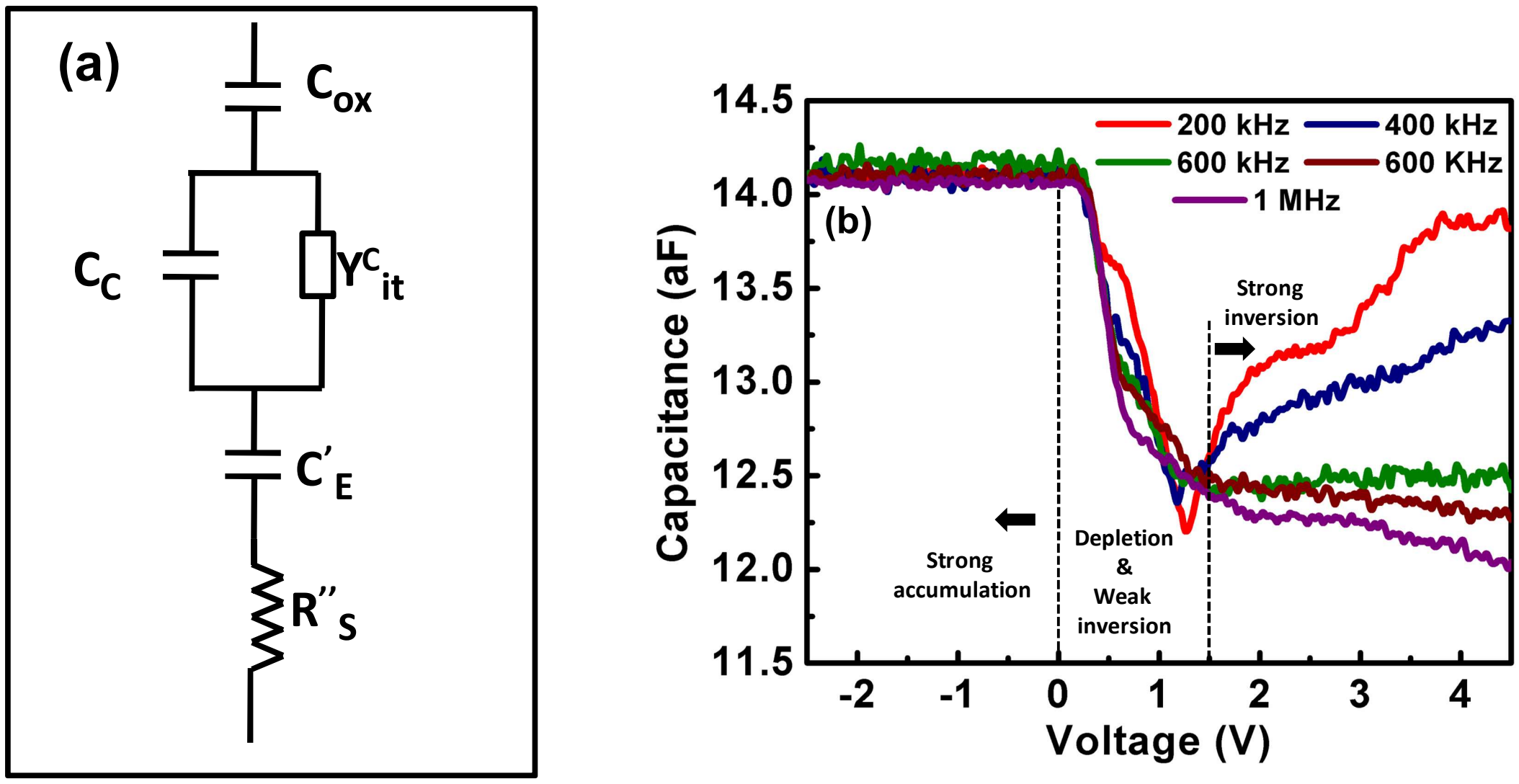

Fig. 6.22. (a) Schematic of the equivalent circuit by considering the leaky capacitance model, (b) Plot of the corrected capacitance for all the considered frequencies, with the removal of frequency dispersion in the region of from depletion to strong inversion.

The corresponding corrected capacitance values (after two successive corrections) are plotted in Fig. 6.22(b) and it is observed that the frequency dispersion in the 'depletion and weak inversion → strong inversion' region is eliminated.

### 6.5.3. C-V characteristics: confirmation of the formation of VTQD

It is apparent from Fig. 6.23(a) that the low frequency (200 kHz) room temperature C-V characteristics exhibit a 'step-like' behavior in the 'inversion' region (*e.g.*, at +2 V and +4 V). The formation of such step is attributed to the 3D-confinement of electrons in the quantum well, created at the Ge-nanowire/$SiO_2$ interface. Such capacitance steps appear due to the occupied quantum states which get gradually lowered below the Fermi level on application of voltages at the metal terminal. Careful investigation shows that each of such major steps in the C-V curve exhibits a leap of ~6 electronic charge per volt and gets partially smoothened by a number of smaller convoluted steps. The smaller steps originate due to the radial (geometrical) confinement whereas the major steps are attributed to the voltage tunable confinement along the axis (electrical). However, such 3D-quantization effect diminishes significantly at high frequency since electrons, being minority carriers, cannot respond within the time period larger than minority carrier lifetime **[484]**. On the other hand, C-V curves for both the low and high frequencies exhibit hump while transiting from ‘accumulation-to-inversion’, the latter (*i.e.,* high frequency) is observed to be reduced but not removed completely. Such hump arises from interface traps as well as the confinement of holes as shown in Fig. 6.5. It is also observed that contribution of the former is diminished at high frequency, whereas that of the latter subsists **[485]**. Such confinement effect is contributed by both of the ‘light’ and ‘heavy’ holes as shown in Fig. 6.23.

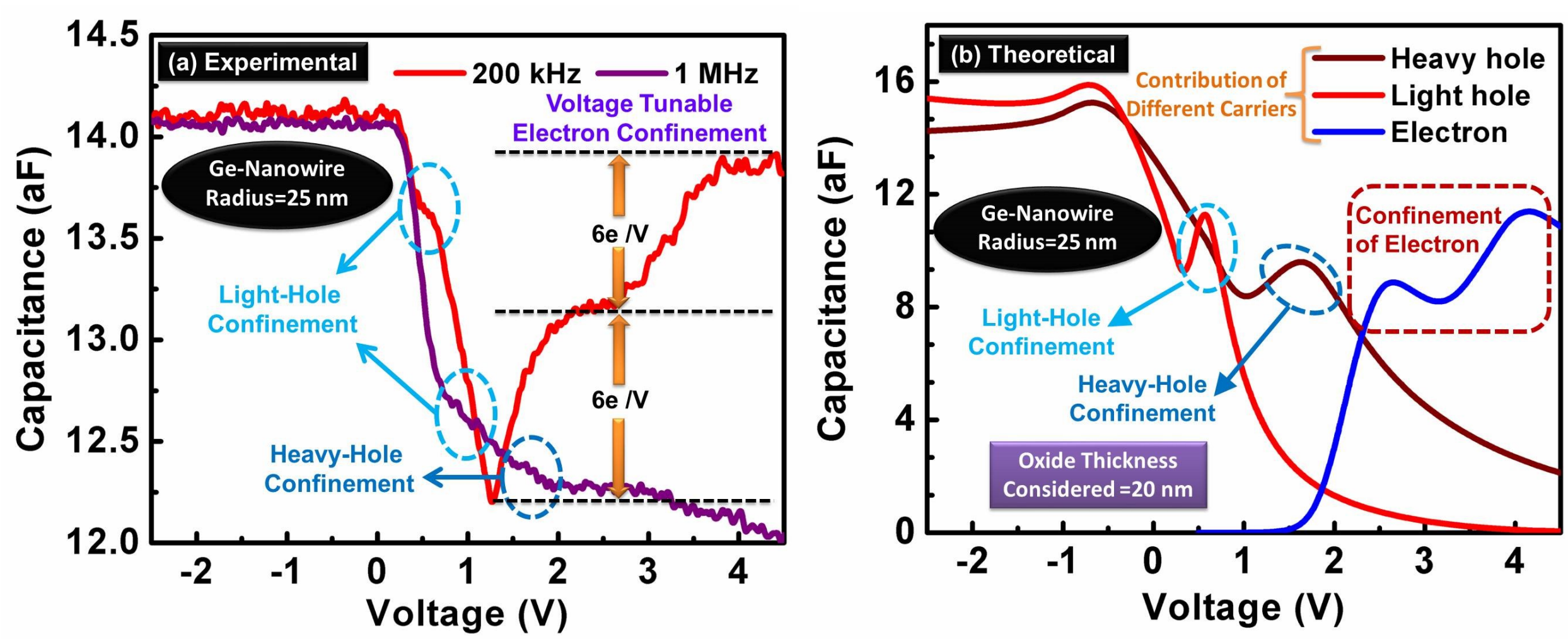

Fig. 6.23. (a) Plot of C-V characteristics measured in-situ under FESEM at room temperature for 200 kHz and 1 MHz frequencies. (b) Plot of C-V characteristics of a single nanowire MOS device obtained from the analytical model based on NEGF formalism.

Theoretical results are plotted in Fig. 6.23(b) which illustrates contributions of the electrons, heavy holes and light holes in net capacitance of the device. The occupancy of electrons significantly depends on the quantized states (below the p-Si Fermi level) which finally manifests as ‘step-like’ behavior of the device capacitance (Fig. 6.23 (a)-(b)). Comparative analysis of experimental and theoretical plots indicates that the prominent hump at ~0.5 V for low frequency is originated due to the confinement of light holes, whereas the heavy hole confinement corresponds to the convoluted hump at ~1 V exhibited for high frequency. Such frequency response is attributed to the Ge-NW↔p-Si hole transmission lifetime as shown in Eqn. (6.31) and Eqn. (6.35). The Ge-NW↔p-Si coupling is relatively stronger for ‘heavy’ holes in comparison to ‘light’ holes due to their lesser effective-mass-mismatch ( $m_h^{HH}*(Ge)=0.33$ ; $m_h^{LH}*(Ge)=0.043$ ; $m_h^{HH}*(Si)=0.49$ ; $m_h^{LH}*(Si)=0.16$ ) which results to larger broadening of hole sub-bands associated with reduced contact resistance **[481]**. The overall charge quantization (for both the 3D-confined electrons and 2D-confined 'light' and 'heavy' holes) in the energy space has already been illustrated by plotting the variation of LDOS along the nanowire axis in Fig. 6.8(a) and 6.8(b), respectively.

## 6.6. Summary

In the present chapter, fabrication of the array of Ge-nanowire based VTQD devices is discussed in detail where the nanowire radius is optimized at ~25 nm to exhibit strong quantum confinement at room temperature. The optimization process employing EBL is also extensively discussed which will further be very useful to fabricate patterned devices with various nanowire radii. In the in-situ measurement of capacitance, 'step-like' behavior is observed where the capacitance shows ~6 electrons per step to be confined per volt in such device. An alternate method is also proposed to extract aF-range device capacitance from the measured data by using leaky capacitance model to remove the frequency dispersion. The confinement properties of different types of carriers are theoretically investigated on the basis of NEGF formalism which provides physical insights about the quantum transport in such devices. The prime focus behind developing such model is to qualitatively analyze the experimental observations. The fabricated array of VTQDs can be utilized in advanced quantum technologies including q-bit generation for quantum information processing.

# Chapter: 7

# Conclusions and Future Scopes

## 7.1. Conclusions

The current dissertation comprises of the design, modeling and fabrication of vertical nanowire-MOS based voltage tunable quantum dot (VTQD) devices for advanced optoelectronic applications. The focus is to study the formation of such VTQDs at room temperature and investigate their utilization for the development of novel optoelectronic devices. Such devices include the multispectral photodetectors with high spectral sensitivity and high performance solar cells with efficiency beyond the conventional SQ-limit.

For such purpose, firstly a self-consistent quantum-electrostatic simultaneous solver is developed to study electrostatic behavior of the vertical nanowire-MOS based VTQD devices. The formation of such VTQDs is predicted by obtaining a step-like nature in the C-V characteristic curves. Further, the optoelectronic applications of such devices have been studied by developing an NEGF based analytical model which has guided to predict the performance of multispectral photodetectors and high efficiency solar cells. Finally, patterned vertical nanowire-MOS devices are fabricated by employing EBL to verify the formation of VTQDs at room temperature. In-situ measurement of such devices under FESEM shows a step-like nature in C-V characteristics which confirms the formation of VTQDs. Emphasis has also been given on selecting the appropriate materials on the basis of two specific material parameters (such as, carrier effective mass and EBR) to meet the criteria to form VTQDs at room temperature. Thus, the present dissertation shows that the vertical nanowire-MOS based VTQD devices with appropriate selection of its material and geometrical parameters will possibly emerge as a promising candidate for future digital photosensing/imaging and energy harvesting applications.

In chapter 3, a comprehensive analysis of electrostatics of the vertically aligned Si-nanowire MOS based VTQD devices is performed by formulating an analytical model with self-consistent simultaneous solution of Schrodinger and Poisson equations. The aim of this study is to obtain C-V characteristics of such devices and to investigate their electrostatic integrity by simultaneously maintaining the electrostatic control and reducing the carrier tunneling probability. The impact of high-*k* dielectrics on electrostatic control and the effect of several controlling parameters (such as applied voltage, barrier height at the insulator/semiconductor junction, carrier effective mass of the insulator and nanowire diameter) on the carrier tunneling probability are studied in detail. It is found from the study that, maintaining electrostatic integrity by using conventional high-*k* technology is not effective in the quantized nanowire MIS devices and in terms of reducing carrier tunneling probability, the insulator effective mass has appeared to be more effective. Thus, in such quantized nanowire MIS based VTQD devices, instead of high-*k* technology, the selection of high-$m^*$ materials can be useful for maintaining both the electrostatic control and reduction of carrier tunneling probability.

Chapter 4 deals with the design and modeling of Si-nanowire based vertical MOS devices for wavelength selective multi-color photosensing/imaging applications. Photogeneration phenomena in such devices is studied in detail by solving a set of coupled quantum field equations associated with second quantization field operators of electrons and photons in the non-equilibrium Green's function (NEGF) framework. The primary aim of this study is to develop a scheme for detection of the entire visible spectrum with high spectral resolution. It is found that the Si-nanowire based devices can detect the entire visible spectrum (380 nm to 700 nm) with 5 nm spectral resolution depending on different diameter-voltage combinations. However, the diameters of Si-nanowire are chosen to be ~5 nm or less to detect the desired wavelengths which may require complex fabrication process steps to realize in practice. Therefore, an InP-nanowire based device is studied for the detection of entire visible spectrum since the relatively lower electron effective mass of InP helps it to realize with relatively larger nanowire diameters (in the range of 30 nm to 50 nm). Thus, findings of the present chapter suggest that such devices with appropriate selection of material and device parameters can be the potential candidates for advanced photosensing/imaging applications.

In chapter 5, the conceptual framework of a GaAs nanowire MOS based VTQD device for solar cell applications is presented in detail where such device is capable of achieving power conversion efficiency beyond the SQ-limit. Here, the developed NEGF based model of Chapter 4 has been extended to calculate the (hole-) photocurrent by considering wavelength dependent intensities of AM 1.5G solar spectra. The study shows that higher efficiencies beyond the SQ-limit has been achieved due to the resonance between the modes of incident photons with an energy gap of (conduction band) quantum states and (valence band) sub-bands in such devices. A significantly higher value of power conversion efficiency and relatively higher values of short-circuit current, open-circuit voltage, fill-factor and photoresponsivity of ~70%, ~260 mA/cm$^2$, 850 mV, ~50% and ~12 A/W are obtained, respectively, for various nanowire diameter-oxide thickness combinations. A design window has also been proposed to obtain the desired values of solar cell performance parameters. Thus, the improved performance parameters suggest that such vertical nanowire MOS based VTQD devices may provide an efficient route to develop next generation solar cells.

In the final working chapter (Chapter 6), patterned array of vertical Ge-nanowire MOS structures are fabricated on Si-substrate to verify the formation of VTQDs at room temperature which was theoretically predicted in Chapter 3. Here, the pattern optimization process to obtain nanowire radius of the order of EBR of Ge by employing EBL has been discussed in detail. It is worthy to mention that the step-like behavior is obtained from the in-situ measurement of C-V characteristics which confirms the formation of VTQDs at room temperature. Thus, the fabricated array of such nanowire MOS based VTQD devices with relevant materials and geometric dimensions are expected to have multifaceted applications in the domain of electronics, optoelectronics and quantum technologies.

## 7.2. Key contributions of the research work

- A novel vertical nanowire-MOS based VTQD device has been proposed and the design issues regarding its electrostatic performance have been investigated by developing a self-consistent quantum-electrostatic simultaneous solver.
- The potential applications of such devices as wavelength selective multispectral photodetectors and high efficiency solar cells are extensively studied by developing an NEGF based analytical model.
- Patterned vertical nanowire-MOS based VTQD devices have been fabricated by employing EBL and in-situ measurements are performed within the FESEM chamber to study the C-V characteristics where the obtained step-like nature in capacitance values confirms the formation of VTQDs at room temperature.

## 7.3. Future scopes

In future, the impact of spherical and chromatic aberrations on developing patterned devices by employing EBL can be studied before approaching towards the large-scale production since variability plays a significant role in determining the performances of a particular batch of devices. Further, optoelectronic applications proposed in the current dissertation can be verified by performing experiments with suitable arrangement to incident light through an optical fiber into the desired location of the sample within the FESEM chamber. However, this method requires sophisticated instrumentation to be coupled with the FESEM which is indeed very expensive. In this regard, an alternative method can be designed to cover the array of patterned devices by suitable optically transparent material (probably, insulating polymers) and then RIE can be used to open up the top-contact area. The further work may be conducted to testify whether such VTQDs can be useful for generating charge qubits to develop systems enabling the room temperature quantum information processing.

# References

1. Bardeen, John, and Walter Hauser Brattain. "The transistor, a semi-conductor triode." *Physical Review* 74.2 (1948): 230.
2. Dawon, Kahng. "Electric field controlled semiconductor device." U.S. Patent No. 3,102,230. 27 Aug. 1963.
3. Sah, Chih-Tang. "A new semiconductor tetrode-the surface-potential controlled transistor." *Proceedings of the IRE* 49.11 (1961): 1623-1634.
4. Noyce, Robert N. "Semiconductor device-and-lead structure, reprint of us patent 2,981,877 (issued april 25, 1961. filed july 30, 1959)." *IEEE Solid-State Circuits Society Newsletter* 12.2 (2007): 34-40.
5. Wanlass, Frank M., and C. T. Sah. "Nanowatt logic using field-effect metal-oxide semiconductor triodes." *Semiconductor devices: pioneering papers*. 1991. 637-638.
6. Masuhara, T., et al. "A high-speed, low-power Hi-CMOS 4K static RAM." *1978 IEEE International Solid-State Circuits Conference. Digest of Technical Papers*. Vol. 21. IEEE, 1978.
7. Addanki, Satish, I. Sadegh Amiri, and P. Yupapin. "Review of optical fibers-introduction and applications in fiber lasers." *Results in Physics* 10 (2018): 743-750.
8. Kazanskiy, N. L., et al. "State-of-the-Art Optical Devices for Biomedical Sensing Applications—A Review." *Electronics* 10.8 (2021): 973.
9. Estevez, M. Carmen, Mar Alvarez, and Laura M. Lechuga. "Integrated optical devices for lab-on-a-chip biosensing applications." *Laser & Photonics Reviews* 6.4 (2012): 463-487.
10. Cheo, P. K. "Integrated optical devices and applications." *Journal of Physics E: Scientific Instruments* 12.1 (1979): 2.
11. Mardonova, Mokhinabonu, and Yosoon Choi. "Review of wearable device technology and its applications to the mining industry." *Energies* 11.3 (2018): 547.
12. Oliveira, J., et al. "Recent developments on printed photodetectors for large area and flexible applications." *Organic Electronics* 66 (2019): 216-226.
13. Castaneda, Denisse, et al. "A review on wearable photoplethysmography sensors and their potential future applications in health care." *International journal of biosensors & bioelectronics* 4.4 (2018): 195.
14. Li, Hua, et al. "Review and analysis of avionic helmet-mounted displays." *Optical Engineering* 52.11 (2013): 110901.
15. Tchon, Joseph L., et al. "Current state of OLED technology relative to military avionics requirements." *Display Technologies and Applications for Defense, Security, and Avionics VIII; and Head-and Helmet-Mounted Displays XIX*. Vol. 9086. International Society for Optics and Photonics, 2014.
16. Lu, Jiwu, et al. "Integration of solar cells on top of CMOS chips part I: A-Si solar cells." *IEEE transactions on electron devices* 58.7 (2011): 2014-2021.
17. Prabha, Rajiv Damodaran, and Gabriel A. Rincón-Mora. "CMOS photovoltaic-cell layout configurations for harvesting microsystems." *2013 IEEE 56th International Midwest Symposium on Circuits and Systems (MWSCAS)*. IEEE, 2013.

18. Goehlich, Andreas, Martin Stühlmeyer, and Holger Vogt. "D7. 4-CMOS integrated miniaturized photovoltaic cells for autonomous sensor nodes: Simulations and experimental results." *Proceedings SENSOR 2015* (2015): 635-640.
19. Cong, Jia, et al. "An efficient forward-Biased Si CMOS LED with high optical power density and nonlinear optical-power-current characteristic." *IEEE Photonics Journal* 11.2 (2019): 1-10.
20. He, Zhe, et al. "CMOS Compatible High-Performance Nanolasing Based on Perovskite–SiN Hybrid Integration." *Advanced Optical Materials* 8.15 (2020): 2000453.
21. Moore, Gordon E. "Cramming more components onto integrated circuits." (1965): 114-117.
22. https://www.tomshardware.com/reviews/intel-core-i7-8086k-cpu-8086-nniversary,5658.html
23. https://ark.intel.com/content/www/us/en/ark/products/148263/intel-core-i7-8086k-processor-12m-cache-up-to-5-00-ghz.html
24. Feynman, Richard P. "Plenty of Room at the Bottom." *APS annual meeting*. 1959.
25. Li, Yat, et al. "Nanowire electronic and optoelectronic devices." *Materials today* 9.10 (2006): 18-27.
26. Wang, Zhihuan, and Bahram Nabet. "Nanowire optoelectronics." *Nanophotonics* 4.4 (2015): 491-502.
27. Zhang, Yunyan, et al. "III–V nanowires and nanowire optoelectronic devices." *Journal of Physics D: Applied Physics* 48.46 (2015): 463001.
28. Agarwal, R., and C. M. Lieber. "Semiconductor nanowires: optics and optoelectronics." *Applied physics A* 85.3 (2006): 209-215.
29. Duan, Xiangfeng, et al. "Indium phosphide nanowires as building blocks for nanoscale electronic and optoelectronic devices." *nature* 409.6816 (2001): 66-69.
30. Deng, Hui, et al. "Growth, patterning and alignment of organolead iodide perovskite nanowires for optoelectronic devices." *Nanoscale* 7.9 (2015): 4163-4170.
31. Joyce, Hannah J., et al. "III–V semiconductor nanowires for optoelectronic device applications." *Progress in Quantum Electronics* 35.2-3 (2011): 23-75.
32. Leung, Siu-Fung, et al. "Light management with nanostructures for optoelectronic devices." *The journal of physical chemistry letters* 5.8 (2014): 1479-1495.
33. Schmid, Martina. "Review on light management by nanostructures in chalcopyrite solar cells." *Semiconductor Science and Technology* 32.4 (2017): 043003.
34. Wang, Wenhui, and Limin Qi. "Light management with patterned micro-and nanostructure arrays for photocatalysis, photovoltaics, and optoelectronic and optical devices." *Advanced Functional Materials* 29.25 (2019): 1807275.
35. Brongersma, Mark L., Yi Cui, and Shanhui Fan. "Light management for photovoltaics using high-index nanostructures." *Nature materials* 13.5 (2014): 451-460.
36. Peter Amalathas, Amalraj, and Maan M. Alkaisi. "Nanostructures for light trapping in thin film solar cells." *Micromachines* 10.9 (2019): 619.
37. Polman, Albert, and Harry A. Atwater. "Photonic design principles for ultrahigh-efficiency photovoltaics." *Nature materials* 11.3 (2012): 174-177.
38. Xu, Qiwei, et al. "On-chip colloidal quantum dot devices with a CMOS compatible architecture for near-infrared light sensing." *Optics letters* 44.2 (2019): 463-466.
39. Zheng, Li, et al. "Ambipolar graphene–quantum dot phototransistors with CMOS compatibility." *Advanced Optical Materials* 6.23 (2018): 1800985.
40. Yun, Hyeong Jin, et al. "Solution-processable integrated CMOS circuits based on colloidal CuInSe 2 quantum dots." *Nature communications* 11.1 (2020): 1-10.

41. Yuryev, Vladimir A., and Larisa V. Arapkina. "Ge quantum dot arrays grown by ultrahigh vacuum molecular-beam epitaxy on the Si (001) surface: nucleation, morphology, and CMOS compatibility." *Nanoscale research letters* 6.1 (2011): 1-15.
42. Chistyakov, A. A., et al. "Optoelectronic properties of semiconductor quantum dot solids for photovoltaic applications." *The journal of physical chemistry letters* 8.17 (2017): 4129-4139.
43. Wu, Jiang, et al. "Quantum dot optoelectronic devices: lasers, photodetectors and solar cells." *Journal of Physics D: Applied Physics* 48.36 (2015): 363001.
44. Shao, Lijia, Yanfang Gao, and Feng Yan. "Semiconductor quantum dots for biomedicial applications." *Sensors* 11.12 (2011): 11736-11751.
45. Lai, Yung-Yu, et al. "Dependence of photoresponsivity and on/off ratio on quantum dot density in quantum dot sensitized MoS2 photodetector." *Nanomaterials* 10.9 (2020): 1828.
46. Livache, Clément, et al. "A colloidal quantum dot infrared photodetector and its use for intraband detection." *Nature communications* 10.1 (2019): 1-10.
47. Wei, Yuanzhi, et al. "Hybrid organic/PbS quantum dot bilayer photodetector with low dark current and high detectivity." *Advanced Functional Materials* 28.11 (2018): 1706690.
48. Pak, Yusin, et al. "Dark-current reduction accompanied photocurrent enhancement in p-type MnO quantum-dot decorated n-type 2D-MoS2-based photodetector." *Applied Physics Letters* 116.11 (2020): 112102.
49. Chen, Baile, et al. "Low dark current high gain InAs quantum dot avalanche photodiodes monolithically grown on Si." *ACS Photonics* 7.2 (2020): 528-533.
50. Gao, Jianbo, et al. "Solution-processed, high-speed, and high-quantum-efficiency quantum dot infrared photodetectors." *ACS photonics* 3.7 (2016): 1217-1222.
51. Siontas, Stylianos, et al. "High-performance germanium quantum dot photodetectors in the visible and near infrared." *Materials Science in Semiconductor Processing* 92 (2019): 19-27.
52. García de Arquer, F. Pelayo, et al. "Semiconductor quantum dots: Technological progress and future challenges." *Science* 373.6555 (2021): eaaz8541.
53. Beard, Matthew C. "Multiple exciton generation in semiconductor quantum dots." *The Journal of Physical Chemistry Letters* 2.11 (2011): 1282-1288.
54. Nozik, Arthur J. "Multiple exciton generation in semiconductor quantum dots." *Chemical Physics Letters* 457.1-3 (2008): 3-11.
55. Liu, Zhaojun, et al. "Micro-light-emitting diodes with quantum dots in display technology." *Light: Science & Applications* 9.1 (2020): 1-23.
56. Ishida, Takeshi, et al. "How Will Quantum Dots Enable Next-Gen Display Technologies?." *Information Display* 36.6 (2020): 14-18.
57. Wu, Tingzhu, et al. "Quantum-dot-based full-color micro-LED displays." *Semiconductors and Semimetals*. Academic Press Inc., 2021.
58. Roh, Jeongkyun, et al. "Optically pumped colloidal-quantum-dot lasing in LED-like devices with an integrated optical cavity." *Nature communications* 11.1 (2020): 1-10.
59. Wang, Lei, et al. "Progress in semiconductor quantum dots-based continuous-wave laser." *SCIENCE CHINA Materials* 63.8 (2020): 1382-1397.
60. Reithmaier, Johann Peter, and Alfred Forchel. "Recent advances in semiconductor quantum-dot lasers." *Comptes Rendus Physique* 4.6 (2003): 611-619.
61. Chang, Hao, et al. "Ultrastable low-cost colloidal quantum dot microlasers of operative temperature up to 450 K." *Light: Science & Applications* 10.1 (2021): 1-10.

62. Jean, Joel, et al. "Synthesis cost dictates the commercial viability of lead sulfide and perovskite quantum dot photovoltaics." *Energy & Environmental Science* 11.9 (2018): 2295-2305.
63. Lin, Huihui, et al. "Colloidal synthesis of MoS 2 quantum dots: size-dependent tunable photoluminescence and bioimaging." *New Journal of Chemistry* 39.11 (2015): 8492-8497.
64. Griffiths, Andrew. *Exploring MBE Growth Of Quantum Dots: Low Density Growth For Quantum Information Devices*. Diss. University of Sheffield, 2014.
65. Evtikhiev, V. P., et al. "Control of density, size and size uniformity of MBE-grown InAs quantum dots by means of substrate misorientation." *Semiconductor science and technology* 17.6 (2002): 545.
66. Atkinson, P., et al. "Size evolution of site-controlled InAs quantum dots grown by molecular beam epitaxy on prepatterned GaAs substrates." *Journal of Vacuum Science & Technology B: Microelectronics and Nanometer Structures Processing, Measurement, and Phenomena* 24.3 (2006): 1523-1526.
67. Gates, Byron D., et al. "New approaches to nanofabrication: molding, printing, and other techniques." *Chemical reviews* 105.4 (2005): 1171-1196.
68. Cui, Bo, ed. "Recent advances in nanofabrication techniques and applications." (2011).
69. Liddle, J. Alexander, and Gregg M. Gallatin. "Lithography, metrology and nanomanufacturing." *Nanoscale* 3.7 (2011): 2679-2688.
70. Ariga, Katsuhiko, et al. "Challenges and breakthroughs in recent research on self-assembly." *Science and technology of advanced materials* (2008).
71. Ariga, Katsuhiko, et al. "Nanoarchitectonics: a new materials horizon for nanotechnology." *Materials Horizons* 2.4 (2015): 406-413.
72. Ariga, Katsuhiko, Jonathan P. Hill, and Qingmin Ji. "Layer-by-layer assembly as a versatile bottom-up nanofabrication technique for exploratory research and realistic application." *Physical Chemistry Chemical Physics* 9.19 (2007): 2319-2340.
73. Berkowski, Kimberly L., et al. "Introduction to photolithography: Preparation of microscale polymer silhouettes." *Journal of chemical education* 82.9 (2005): 1365.
74. Stevenson, J. T. M., and A. M. Gundlach. "The application of photolithography to the fabrication of microcircuits." *Journal of Physics E: Scientific Instruments* 19.9 (1986): 654.
75. Dill, Frederick H. "Optical lithography." *IEEE Transactions on Electron Devices* 22.7 (1975): 440-444.
76. Wang, Yuanyuan, et al. "Direct optical lithography of functional inorganic nanomaterials." *Science* 357.6349 (2017): 385-388.
77. Lamprecht, Bernhard, et al. "Organic optoelectronic device fabrication using standard UV photolithography." *physica status solidi (RRL)–Rapid Research Letters* 2.1 (2008): 16-18.
78. Chen, Yifang. "Nanofabrication by electron beam lithography and its applications: A review." *Microelectronic Engineering* 135 (2015): 57-72.
79. Tseng, Ampere A., et al. "Electron beam lithography in nanoscale fabrication: recent development." *IEEE Transactions on electronics packaging manufacturing* 26.2 (2003): 141-149.
80. Hohn, Fritz J. "Electron beam lithography-tools and applications." *Japanese journal of applied physics* 30.11S (1991): 3088.
81. Rosolen, Grahame, and Adriano Cola. "Fabrication of photonic crystal structures by electron beam lithography." *2006 Conference on Optoelectronic and Microelectronic Materials and Devices*. IEEE, 2006.

82. Chiromawa, Nura Liman, and Kamarulazizi Ibrahim. "Applications of Electron Beam Lithography (EBL) in Optoelectronics Device Fabrication." *AASCIT Journal of Physics* 4.2 (2018): 53-58.
83. Clements, S. J., M. J. Butler, and F. R. Shepherd. "Electron lithography for optoelectronics." *Proceedings of LEOS'93*. IEEE, 1993.
84. Huang, Yanyi, et al. "Fabrication and replication of polymer integrated optical devices using electron-beam lithography and soft lithography." *The Journal of Physical Chemistry B* 108.25 (2004): 8606-8613.
85. Elarde, V. C., et al. "Room-temperature operation of patterned quantum-dot lasers fabricated by electron beam lithography and selective area metal-organic chemical vapor deposition." *IEEE Photonics Technology Letters* 17.5 (2005): 935-937.
86. Tennant, D. M., et al. "Characterization of near-field holography grating masks for optoelectronics fabricated by electron beam lithography." *Journal of Vacuum Science & Technology B: Microelectronics and Nanometer Structures Processing, Measurement, and Phenomena* 10.6 (1992): 2530-2535.
87. Pham, Thach. *Fabrication of single nanowire device using electron beam lithography*. University of Arkansas, 2014.
88. Kaganskiy, Arsenty, et al. "Advanced in-situ electron-beam lithography for deterministic nanophotonic device processing." *Review of Scientific Instruments* 86.7 (2015): 073903.
89. Polyakov, A., et al. "Plasmonic light trapping in nanostructured metal surfaces." *Applied physics letters* 98.20 (2011): 203104.
90. Guo, Junpeng, Zhitong Li, and Hong Guo. "Near perfect light trapping in a 2D gold nanotrench grating at oblique angles of incidence and its application for sensing." *Optics express* 24.15 (2016): 17259-17271.
91. Bogdanov, Alexei L., Jean Lapointe, and Jens H. Schmid. "Electron-beam lithography for photonic waveguide fabrication: Measurement of the effect of field stitching errors on optical performance and evaluation of a new compensation method." *Journal of Vacuum Science & Technology B, Nanotechnology and Microelectronics: Materials, Processing, Measurement, and Phenomena* 30.3 (2012): 031606.
92. Bojko, Richard J., et al. "Electron beam lithography writing strategies for low loss, high confinement silicon optical waveguides." *Journal of Vacuum Science & Technology B, Nanotechnology and Microelectronics: Materials, Processing, Measurement, and Phenomena* 29.6 (2011): 06F309.
93. George, Steven M. "Atomic layer deposition: an overview." *Chemical reviews* 110.1 (2010): 111-131.
94. Johnson, Richard W., Adam Hultqvist, and Stacey F. Bent. "A brief review of atomic layer deposition: from fundamentals to applications." *Materials today* 17.5 (2014): 236-246.
95. Oviroh, Peter Ozaveshe, et al. "New development of atomic layer deposition: processes, methods and applications." *Science and technology of advanced materials* 20.1 (2019): 465-496.
96. Sun, Luzhao, et al. "Chemical vapour deposition." *Nature Reviews Methods Primers* 1.1 (2021): 1-20.
97. Zhao, Xinhe, et al. "Chemical vapor deposition and its application in surface modification of nanoparticles." *Chemical Papers* 74.3 (2020): 767-778.
98. Muratore, Christopher, Andrey A. Voevodin, and Nicholas R. Glavin. "Physical vapor deposition of 2D Van der Waals materials: a review." *Thin Solid Films* 688 (2019): 137500.

99. Nair, P. K., et al. "Semiconductor thin films by chemical bath deposition for solar energy related applications." *Solar Energy Materials and Solar Cells* 52.3-4 (1998): 313-344.
100. Koao, L. F., F. B. Dejene, and H. C. Swart. "Properties of flower-like ZnO nanostructures synthesized using the chemical bath deposition." *Materials science in semiconductor processing* 27 (2014): 33-40.
101. Preda, N., et al. "Synthesis of CdS nanostructures using template-assisted ammonia-free chemical bath deposition." *Journal of Physics and Chemistry of Solids* 73.9 (2012): 1082-1089.
102. Wagner, A. RS, and S. WC Ellis. "Vapor-liquid-solid mechanism of single crystal growth." *Applied physics letters* 4.5 (1964): 89-90.
103. Pinion, Christopher W., Joseph D. Christesen, and James F. Cahoon. "Understanding the vapor–liquid–solid mechanism of Si nanowire growth and doping to synthetically encode precise nanoscale morphology." *Journal of Materials Chemistry C* 4.18 (2016): 3890-3897.
104. Ambrosini, S., et al. "Vapor-liquid-solid and vapor-solid growth of self-catalyzed GaAs nanowires." *AIP Advances* 1.4 (2011): 042142.
105. Wang, Hailong, et al. "Atomistics of vapour–liquid–solid nanowire growth." *Nature communications* 4.1 (2013): 1-10.
106. Wu, Ju, and Peng Jin. "Self-assembly of InAs quantum dots on GaAs (001) by molecular beam epitaxy." *Frontiers of Physics* 10.1 (2015): 7-58.
107. Sautter, Kathryn E., Kevin D. Vallejo, and Paul J. Simmonds. "Strain-driven quantum dot self-assembly by molecular beam epitaxy." *Journal of Applied Physics* 128.3 (2020): 031101.
108. Brehm, Moritz, and Martyna Grydlik. "Site-controlled and advanced epitaxial Ge/Si quantum dots: fabrication, properties, and applications." *Nanotechnology* 28.39 (2017): 392001.
109. Biswas, Abhijit, et al. "Advances in top–down and bottom–up surface nanofabrication: Techniques, applications & future prospects." *Advances in colloid and interface science* 170.1-2 (2012): 2-27.
110. Teo, Boon K., and X. H. Sun. "From top-down to bottom-up to hybrid nanotechnologies: road to nanodevices." *Journal of cluster science* 17.4 (2006): 529-540.
111. Jüngel, Ansgar. "Macroscopic models for semiconductor devices: a review." (2000).
112. Doan, Duy-Hai, Annegret Glitzky, and Matthias Liero. "Analysis of a drift–diffusion model for organic semiconductor devices." *Zeitschrift für angewandte Mathematik und Physik* 70.2 (2019): 1-18.
113. Pisarenko, Ivan, and Eugeny Ryndin. "Drift-diffusion simulation of high-speed optoelectronic devices." *Electronics* 8.1 (2019): 106.
114. Darwish, Mohammed, and Alessio Gagliardi. "A drift-diffusion simulation model for organic field effect transistors: on the importance of the Gaussian density of states and traps." *Journal of Physics D: Applied Physics* 53.10 (2019): 105102.
115. Pinnau, René. "A review on the quantum drift diffusion model." *Transport Theory and Statistical Physics* 31.4-6 (2002): 367-395.
116. Georgiev, Vihar P., et al. "Experimental and simulation study of silicon nanowire transistors using heavily doped channels." *IEEE Transactions on Nanotechnology* 16.5 (2017): 727-735.
117. Datta, Supriyo. "Nanoscale device modeling: the Green's function method." *Superlattices and microstructures* 28.4 (2000): 253-278.

118. Datta, Supriyo. *Quantum transport: atom to transistor*. Cambridge university press, 2005.
119. Svizhenko, A., et al. "Two-dimensional quantum mechanical modeling of nanotransistors." *Journal of Applied Physics* 91.4 (2002): 2343-2354.
120. Venugopal, Rajesh, et al. "Simulating quantum transport in nanoscale transistors: Real versus mode-space approaches." *Journal of Applied physics* 92.7 (2002): 3730-3739.
121. Kolek, Andrzej. "Implementation of light–matter interaction in NEGF simulations of QCL." *Optical and Quantum Electronics* 51.6 (2019): 1-9.
122. Zheng, K., et al. "A two-dimensional van der Waals CdS/germanene heterojunction with promising electronic and optoelectronic properties: DFT+ NEGF investigations." *Physical Chemistry Chemical Physics* 19.28 (2017): 18330-18337.
123. Aeberhard, U., and R. H. Morf. "Microscopic nonequilibrium theory of quantum well solar cells." *Physical Review B* 77.12 (2008): 125343.
124. Aeberhard, U. "Spectral properties of photogenerated carriers in quantum well solar cells." *Solar energy materials and solar cells* 94.11 (2010): 1897-1902.
125. Aeberhard, U. "Quantum-kinetic theory of photocurrent generation via direct and phonon-mediated optical transitions." *Physical Review B* 84.3 (2011): 035454.
126. Miwa, Kuniyuki, and Michael Galperin. "Application of the Hubbard NEGF in Molecular Optoelectronics." *Luminescence* 1.1.7: 1-8.
127. Aeberhard, Urs. "Effective microscopic theory of quantum dot superlattice solar cells." *Optical and Quantum Electronics* 44.3 (2012): 133-140.
128. Maxwell, James Clerk. "VIII. A dynamical theory of the electromagnetic field." *Philosophical transactions of the Royal Society of London* 155 (1865): 459-512.
129. Mulligan, Joseph F. "Heinrich Hertz and the development of physics." *Physics Today* 42.3 (1989): 50-57.
130. Planck, Max. "On the law of the energy distribution in the normal spectrum." *Ann. Phys* 4.553 (1901): 1-11.
131. Einstein, A. "On a Heuristic Point of View about the Creation and Conversion of Light, 1905, translated by Wikisource."
132. Millikan, Robert Andrews. "A direct photoelectric determination of Planck's" h"." *Physical Review* 7.3 (1916): 355.
133. Bose, Satyendra Nath. "Planck's law and the light quantum hypothesis." *Journal of Astrophysics and Astronomy* 15 (1994): 3.
134. Lubsandorzhiev, Bayarto K. "On the history of photomultiplier tube invention." *Nuclear Instruments and Methods in Physics Research Section A: Accelerators, Spectrometers, Detectors and Associated Equipment* 567.1 (2006): 236-238.
135. Zworykin, V. K., G. A. Morton, and L. Malter. "The secondary emission multiplier-a new electronic device." *Proceedings of the Institute of Radio Engineers* 24.3 (1936): 351-375.
136. Flyckt, S. O., and C. Marmonier. "Photomultiplier tubes–principles & applications, Photonis, 2002." *URL http://www2. pv. infn. it/debari/doc/Flyckt Marmonier. pdf ()*.
137. Morrison, S. R. "A new type of photosensitive junction device." *Solid-State Electronics* 6.5 (1963): 485-494.
138. Horton, J. W., R. V. Mazza, and H. Dym. "The scanistor—A solid-state image scanner." *Proceedings of the IEEE* 52.12 (1964): 1513-1528.

139. Schuster, M. A., and G. Strull. "A monolithic mosaic of photon sensors for solid-state imaging applications." *IEEE Transactions on Electron Devices* 12 (1966): 907-912.
140. Weckler, Gene P. "Operation of pn junction photodetectors in a photon flux integrating mode." *IEEE Journal of Solid-State Circuits* 2.3 (1967): 65-73.
141. Dyck, Rudolph H., and Gene P. Weckler. "Integrated arrays of silicon photodetectors for image sensing." *IEEE Transactions on Electron Devices* 15.4 (1968): 196-201.
142. Weimer, Paul K., et al. "A self-scanned solid-state image sensor." *Proceedings of the IEEE* 55.9 (1967): 1591-1602.
143. Noble, Peter JW. "Self-scanned silicon image detector arrays." *IEEE Transactions on electron Devices* 15.4 (1968): 202-209.
144. Chamberlain, Savvas G. "Photosensitivity and scanning of silicon image detector arrays." *IEEE Journal of Solid-State Circuits* 4.6 (1969): 333-342.
145. Fry, Peter W., Peter JW Noble, and Robert J. Rycroft. "Fixed-pattern noise in photomatrices." *IEEE Journal of Solid-State Circuits* 5.5 (1970): 250-254.
146. Boyle, Willard S., and George E. Smith. "Charge coupled semiconductor devices." *Bell System Technical Journal* 49.4 (1970): 587-593.
147. Alfaraj, Nasir. "A review of charge-coupled device image sensors." (2017).
148. Lesser, Michael. "Charge coupled device (CCD) image sensors." *High Performance Silicon Imaging*. Woodhead Publishing, 2014. 78-97.
149. Denes, P., et al. "A fast, direct x-ray detection charge-coupled device." *Review of Scientific Instruments* 80.8 (2009): 083302.
150. Williams, E. W., et al. "Optical measurement of the charge transfer efficiency of a charge-coupled device." *Journal of Physics D: Applied Physics* 7.1 (1974): L4.
151. Boyle, W. S., and G. E. Smith. "Charge-coupled devices-a new approach to mis device structures." *IEEE spectrum* 8.7 (1971): 18-27.
152. White, M. H., et al. "Charge-coupled device (ccd) imaging at low light levels." *1972 International Electron Devices Meeting*. IEEE, 1972.
153. Younse, Jack M., John F. Breitzmann, and Jack W. Freeman. "The Modular Charge-Coupled Device (CCD) Camera." *Recent advances in TV sensors and systems*. Vol. 203. International Society for Optics and Photonics, 1979.
154. Carnes, J. E., and W. F. Kosonocky. "Charge-coupled devices and applications." *Solid State Technology* 17.4 (1974): 67-77.
155. Carnes, J. "A CMOS/buried-n-channel CCD compatible process for analog signal processing applications." *RCA Corporation* 38 (1977): 406.
156. Ong, D. "An all-implanted CCD/CMOS process." *IEEE Transactions on Electron Devices* 28.1 (1981): 6-12.
157. Ohba, Shinya, et al. "MOS area sensor: part II-low-noise MOS area sensor with antiblooming photodiodes." *IEEE Journal of Solid-State Circuits* 15.4 (1980): 747-752.
158. Senda, K. O. H. J. I., et al. "Analysis of charge-priming transfer efficiency in CPD image sensors." *IEEE transactions on electron devices* 31.9 (1984): 1324-1328.
159. Ando, Haruhisa, et al. "Design consideration and performance of a new MOS imaging device." *IEEE transactions on electron devices* 32.8 (1985): 1484-1489.
160. Mendis, Sunetra, Sabrina E. Kemeny, and Eric R. Fossum. "CMOS active pixel image sensor." *IEEE transactions on Electron Devices* 41.3 (1994): 452-453.
161. Mendis, Sunetra K., et al. "CMOS active pixel image sensors for highly integrated imaging systems." *IEEE Journal of Solid-State Circuits* 32.2 (1997): 187-197.

162. Fossum, Eric R. "CMOS image sensors: Electronic camera-on-a-chip." *IEEE transactions on electron devices* 44.10 (1997): 1689-1698.
163. Bai, Suo, et al. "High-performance integrated ZnO nanowire UV sensors on rigid and flexible substrates." *Advanced Functional Materials* 21.23 (2011): 4464-4469.
164. Lupan, Oleg, et al. "Ultraviolet photoconductive sensor based on single ZnO nanowire." *physica status solidi (a)* 207.7 (2010): 1735-1740.
165. Law, J. B. K., and J. T. L. Thong. "Simple fabrication of a ZnO nanowire photodetector with a fast photoresponse time." *Applied Physics Letters* 88.13 (2006): 133114.
166. Li, Yanbo, et al. "High-performance UV detector made of ultra-long ZnO bridging nanowires." *Nanotechnology* 20.4 (2008): 045501.
167. Kim, Daeil, et al. "Fabrication of high performance flexible micro-supercapacitor arrays with hybrid electrodes of MWNT/V2O5 nanowires integrated with a SnO2 nanowire UV sensor." *Nanoscale* 6.20 (2014): 12034-12041.
168. Han, Song, et al. "Photoconduction studies on GaN nanowire transistors under UV and polarized UV illumination." *Chemical Physics Letters* 389.1-3 (2004): 176-180.
169. Zou, Yanan, et al. "Ultraviolet detectors based on wide bandgap semiconductor nanowire: A review." *Sensors* 18.7 (2018): 2072.
170. LaPierre, R. R., et al. "A review of III–V nanowire infrared photodetectors and sensors." *Journal of Physics D: Applied Physics* 50.12 (2017): 123001.
171. Ran, Wenhao, et al. "An integrated flexible all-nanowire infrared sensing system with record photosensitivity." *Advanced Materials* 32.16 (2020): 1908419.
172. Wang, S. B., et al. "A CuO nanowire infrared photodetector." *Sensors and Actuators A: Physical* 171.2 (2011): 207-211.
173. Kuo, Cheng-Hsiang, et al. "High sensitivity of middle-wavelength infrared photodetectors based on an individual InSb nanowire." *Nanoscale research letters* 8.1 (2013): 1-8.
174. Sharma, Alka, et al. "Robust broad spectral photodetection (UV-NIR) and ultra high responsivity investigated in nanosheets and nanowires of Bi 2 Te 3 under harsh nano-milling conditions." *Scientific reports* 7.1 (2017): 1-10.
175. Cao, Fengren, et al. "Self-Powered UV–Vis–NIR Photodetector Based on Conjugated-Polymer/CsPbBr3 Nanowire Array." *Advanced Functional Materials* 29.48 (2019): 1906756.
176. Miao, Jinshui, et al. "Single InAs nanowire room-temperature near-infrared photodetectors." *ACS nano* 8.4 (2014): 3628-3635.
177. Chen, Gui, et al. "High-performance hybrid phenyl-C61-butyric acid methyl ester/Cd3P2 nanowire ultraviolet–visible–near infrared photodetectors." *ACS nano* 8.1 (2014): 787-796.
178. Li, Xiao, et al. "Optimizing GaAs nanowire-based visible-light photodetectors." *Applied Physics Letters* 119.5 (2021): 053105.
179. Gallo, Eric M., et al. "Picosecond response times in GaAs/AlGaAs core/shell nanowire-based photodetectors." *Applied Physics Letters* 98.24 (2011): 241113.
180. Chen, Gui, et al. "High performance rigid and flexible visible-light photodetectors based on aligned X (In, Ga) P nanowire arrays." *Journal of Materials Chemistry C* 2.7 (2014): 1270-1277.
181. Zhu, Yi, et al. "Self-Powered InP Nanowire Photodetector for Single-Photon Level Detection at Room Temperature." *Advanced Materials* (2021): 2105729.

182. Svensson, Johannes, et al. "Diameter-dependent photocurrent in InAsSb nanowire infrared photodetectors." *Nano letters* 13.4 (2013): 1380-1385.
183. Li, Ziyuan, et al. "Room temperature GaAsSb single nanowire infrared photodetectors." *Nanotechnology* 26.44 (2015): 445202.
184. Chen, Fei, Xiaohong Ji, and Qinyuan Zhang. "Growth and composition-dependent optoelectronic properties of AlxGa1− xN alloy nanowires arrays across the entire composition range on carbon cloth." *Materials Research Express* 3.7 (2016): 075011.
185. Zhang, Qi, et al. "Stoichiometric effect on optoelectronic properties of composition-tunable CdS1− xSex nanowires." *Advanced Optical Materials* 5.5 (2017): 1600877.
186. Ren, Pinyun, et al. "Band-selective infrared photodetectors with complete-composition-range InAsxP1-x alloy nanowires." *Advanced Materials* 26.44 (2014): 7444-7449.
187. Park, Hyunsung, and Kenneth B. Crozier. "Multispectral imaging with vertical silicon nanowires." *Scientific reports* 3.1 (2013): 1-6.
188. Pau, J. L., et al. "GaN nanostructured p-i-n photodiodes." *Applied Physics Letters* 93.22 (2008): 221104.
189. Pettersson, Håkan, et al. "Infrared photodetectors in heterostructure nanowires." *Nano Letters* 6.2 (2006): 229-232.
190. Kayes, Brendan M., Harry A. Atwater, and Nathan S. Lewis. "Comparison of the device physics principles of planar and radial p-n junction nanorod solar cells." *Journal of applied physics* 97.11 (2005): 114302.
191. Mohseni, P. K., et al. "Structural and optical analysis of GaAsP/GaP core-shell nanowires." *Journal of Applied Physics* 106.12 (2009): 124306.
192. Hayden, Oliver, Ritesh Agarwal, and Charles M. Lieber. "Nanoscale avalanche photodiodes for highly sensitive and spatially resolved photon detection." *Nature materials* 5.5 (2006): 352-356.
193. Cheng, Gang, et al. "ZnO nanowire Schottky barrier ultraviolet photodetector with high sensitivity and fast recovery speed." *Applied Physics Letters* 99.20 (2011): 203105.
194. Zhang, Arthur, et al. "Silicon nanowire detectors showing phototransistive gain." *Applied Physics Letters* 93.12 (2008): 121110.
195. Li, Yanbo, et al. "High-performance UV detector made of ultra-long ZnO bridging nanowires." *Nanotechnology* 20.4 (2008): 045501.
196. Kind, Hannes, et al. "Nanowire ultraviolet photodetectors and optical switches." *Advanced materials* 14.2 (2002): 158-160.
197. Soci, Cesare, et al. "ZnO nanowire UV photodetectors with high internal gain." *Nano letters* 7.4 (2007): 1003-1009.
198. Kar, J. P., et al. "Fabrication of UV detectors based on ZnO nanowires using silicon microchannel." *Journal of Crystal Growth* 311.12 (2009): 3305-3309.
199. Chen, Kuan-Jen, et al. "Optoelectronic characteristics of UV photodetector based on ZnO nanowire thin films." *Journal of Alloys and Compounds* 479.1-2 (2009): 674-677.
200. Kim, DaeIl, et al. "Photoconductance of aligned SnO 2 nanowire field effect transistors." *Applied Physics Letters* 95.4 (2009): 043107.
201. Wu, Jyh-Ming, and Cheng-Hsiang Kuo. "Ultraviolet photodetectors made from SnO2 nanowires." *Thin solid films* 517.14 (2009): 3870-3873.

202. Liu, Zuqin, et al. "Laser ablation synthesis and electron transport studies of tin oxide nanowires." *Advanced Materials* 15.20 (2003): 1754-1757.
203. Liu, X., et al. "Synthesis and characterization of CdO nanoneedles." (2003).
204. Feng, P., et al. "Individual β-Ga 2 O 3 nanowires as solar-blind photodetectors." *Applied physics letters* 88.15 (2006): 153107.
205. Liao, Lei, et al. "P-type electrical, photoconductive, and anomalous ferromagnetic properties of Cu 2 O nanowires." *Applied Physics Letters* 94.11 (2009): 113106.
206. Fu, X. Q., et al. "Anomalous photoconductivity of Ce O 2 nanowires in air." *Applied Physics Letters* 91.7 (2007): 073104.
207. Xue, X. Y., et al. "Individual core-shell structured ZnSnO3 nanowires as photoconductors." *Materials Letters* 62.8-9 (2008): 1356-1358.
208. Yan, Xin, et al. "A single crystalline InP nanowire photodetector." *Applied Physics Letters* 109.5 (2016): 053109.
209. Gallo, Eric M., et al. "Picosecond response times in GaAs/AlGaAs core/shell nanowire-based photodetectors." *Applied Physics Letters* 98.24 (2011): 241113.
210. Kuo, Cheng-Hsiang, et al. "High sensitivity of middle-wavelength infrared photodetectors based on an individual InSb nanowire." *Nanoscale research letters* 8.1 (2013): 1-8.
211. Tan, Huang, et al. "Single-crystalline InGaAs nanowires for room-temperature high-performance near-infrared photodetectors." *Nano-micro letters* 8.1 (2016): 29-35.
212. Ren, Aobo, et al. "Recent progress of III–V quantum dot infrared photodetectors on silicon." *Journal of Materials Chemistry C* 7.46 (2019): 14441-14453.
213. Kim, Jaehyun, et al. "A skin-like two-dimensionally pixelized full-color quantum dot photodetector." *Science advances* 5.11 (2019): eaax8801.
214. Sikorski, Ch, and U. Merkt. "Spectroscopy of electronic states in InSb quantum dots." *Physical review letters* 62.18 (1989): 2164.
215. Chen, Wei, et al. "Demonstration of InAs/InGaAs/GaAs quantum dots-in-a-well mid-wave infrared photodetectors grown on silicon substrate." *Journal of Lightwave Technology* 36.13 (2018): 2572-2581.
216. Lin, Wei-Hsun, et al. "High-temperature operation GaSb/GaAs quantum-dot infrared photodetectors." *IEEE Photonics Technology Letters* 23.2 (2010): 106-108.
217. Finkman, E., et al. "Quantum dot infrared photodetectors in new material systems." *Physica E: Low-dimensional Systems and Nanostructures* 7.1-2 (2000): 139-145.
218. Cosentino, S., et al. "High-efficiency silicon-compatible photodetectors based on Ge quantum dots." *Applied Physics Letters* 98.22 (2011): 221107.
219. Huang, Jian, et al. "Defect characterization of InAs/InGaAs quantum dot pin photodetector grown on GaAs-on-V-grooved-Si substrate." *ACS Photonics* 6.5 (2019): 1100-1105.
220. Sandall, Ian, et al. "1300 nm wavelength InAs quantum dot photodetector grown on silicon." *Optics express* 20.10 (2012): 10446-10452.
221. Inoue, D., et al. "Low-dark current 10 Gbit/s operation of InAs/InGaAs quantum dot pin photodiode grown on on-axis (001) GaP/Si." *Applied Physics Letters* 113.9 (2018): 093506.
222. Barve, A. V., and S. Krishna. "Photovoltaic quantum dot quantum cascade infrared photodetector." *Applied Physics Letters* 100.2 (2012): 021105.
223. Shieh, Jia-Min, et al. "Near-infrared silicon quantum dots metal-oxide-semiconductor field-effect transistor photodetector." *Applied Physics Letters* 94.24 (2009): 241108.

224. Shieh, Jia-Min, et al. "Enhanced photoresponse of a metal-oxide-semiconductor photodetector with silicon nanocrystals embedded in the oxide layer." *Applied physics letters* 90.5 (2007): 051105.
225. Hsu, B-C., et al. "A high efficient 820 nm MOS Ge quantum dot photodetector." *IEEE Electron Device Letters* 24.5 (2003): 318-320.
226. Tzeng, S. S., and Pei-Wen Li. "Enhanced 400–600 nm photoresponsivity of metal–oxide–semiconductor diodes with multi-stack germanium quantum dots." *Nanotechnology* 19.23 (2008): 235203.
227. Chakrabarti, S., et al. "Characteristics of a multicolor InGaAs-GaAs quantum-dot infrared photodetector." *IEEE Photonics technology letters* 17.1 (2004): 178-180.
228. Huang, G., et al. "A multicolor quantum dot intersublevel detector with photoresponse in the terahertz range." *Applied Physics Letters* 92.1 (2008): 011117.
229. Kim, Seongsin M., and James S. Harris. "Multicolor InGaAs quantum-dot infrared photodetectors." *IEEE photonics technology letters* 16.11 (2004): 2538-2540.
230. Adhikary, Sourav, et al. "A multicolor, broadband (5–20 µm), quaternary-capped InAs/GaAs quantum dot infrared photodetector." *Applied Physics Letters* 101.26 (2012): 261114.
231. Becquerel, M. E. "Mémoire sur les effets électriques produits sous l'influence des rayons solaires." *Comptes rendus hebdomadaires des séances de l'Académie des sciences* 9 (1839): 561-567.
232. Smith, Willoughby. "Effect of light on selenium during the passage of an electric current." *SPIE MILESTONE SERIES MS* 56 (1992): 3-3.
233. Fritts, Charles Edgar. "On the Fritts selenium cells and batteries." *Journal of the Franklin Institute* 119.3 (1885): 221-232.
234. Weston, Edward. "Art of utilizing solar radiant energy." *US Patent A* 389125 (1888): 1888-9.
235. Severy, M. L. "Apparatus for generating electricity by solar heat." *US Patent* 527 (1894).
236. Pearson, G., D. Chapin, and C. Fuller. "Bell labs demonstrates the first practical silicon solar cell." *American Physical Society (APS News)* 18.4 (1954).
237. Alferov, Zh I., et al. "Al x Ga 1-x As-GaAs Heterojunctions." *Physics of pn Junctions and Semiconductor Devices*. Springer, Boston, MA, 1971. 287-293.
238. Rühle, Sven. "Tabulated values of the Shockley–Queisser limit for single junction solar cells." *Solar Energy* 130 (2016): 139-147.
239. Blakers, Andrew W., et al. "22.8% efficient silicon solar cell." *Applied Physics Letters* 55.13 (1989): 1363-1365.
240. Yamaguchi, Masafumi, et al. "Multi-junction III–V solar cells: current status and future potential." *Solar Energy* 79.1 (2005): 78-85.
241. Yamaguchi, Masafumi, et al. "Multi-junction solar cells paving the way for super high-efficiency." *Journal of Applied Physics* 129.24 (2021): 240901.
242. Åberg, Ingvar, et al. "A GaAs nanowire array solar cell with 15.3% efficiency at 1 sun." *IEEE Journal of photovoltaics* 6.1 (2015): 185-190.
243. Wallentin, Jesper, et al. "InP nanowire array solar cells achieving 13.8% efficiency by exceeding the ray optics limit." *Science* 339.6123 (2013): 1057-1060.
244. Tian, Bozhi, et al. "Coaxial silicon nanowires as solar cells and nanoelectronic power sources." *nature* 449.7164 (2007): 885-889.

245. Kempa, Thomas J., et al. "Single and tandem axial pin nanowire photovoltaic devices." *Nano letters* 8.10 (2008): 3456-3460.
246. Kelzenberg, Michael D., et al. "Photovoltaic measurements in single-nanowire silicon solar cells." *Nano letters* 8.2 (2008): 710-714.
247. Tang, Jinyao, et al. "Solution-processed core–shell nanowires for efficient photovoltaic cells." *Nature nanotechnology* 6.9 (2011): 568-572.
248. Fan, Zhiyong, et al. "Three-dimensional nanopillar-array photovoltaics on low-cost and flexible substrates." *Nature materials* 8.8 (2009): 648-653.
249. Im, Jeong-Hyeok, et al. "Nanowire perovskite solar cell." *Nano letters* 15.3 (2015): 2120-2126.
250. Tang, Y. B., et al. "Vertically aligned p-type single-crystalline GaN nanorod arrays on n-type Si for heterojunction photovoltaic cells." *Nano letters* 8.12 (2008): 4191-4195.
251. Dong, Yajie, et al. "Coaxial group III− nitride nanowire photovoltaics." *Nano letters* 9.5 (2009): 2183-2187.
252. Cao, Linyou, et al. "Engineering light absorption in semiconductor nanowire devices." *Nature materials* 8.8 (2009): 643-647.
253. Wu, Phillip M., et al. "Colorful InAs nanowire arrays: from strong to weak absorption with geometrical tuning." *Nano letters* 12.4 (2012): 1990-1995.
254. Kim, Sun-Kyung, et al. "Tuning light absorption in core/shell silicon nanowire photovoltaic devices through morphological design." *Nano letters* 12.9 (2012): 4971-4976.
255. Krogstrup, Peter, et al. "Single-nanowire solar cells beyond the Shockley–Queisser limit." *Nature photonics* 7.4 (2013): 306-310.
256. Otnes, Gaute, and Magnus T. Borgström. "Towards high efficiency nanowire solar cells." *Nano Today* 12 (2017): 31-45.
257. Parkinson, Patrick, et al. "Carrier lifetime and mobility enhancement in nearly defect-free core− shell nanowires measured using time-resolved terahertz spectroscopy." *Nano letters* 9.9 (2009): 3349-3353.
258. Schroer, M. D., and J. R. Petta. "Correlating the nanostructure and electronic properties of InAs nanowires." *Nano letters* 10.5 (2010): 1618-1622.
259. Thelander, Claes, et al. "Effects of crystal phase mixing on the electrical properties of InAs nanowires." *Nano letters* 11.6 (2011): 2424-2429.
260. Joyce, Hannah J., et al. "Ultralow surface recombination velocity in InP nanowires probed by terahertz spectroscopy." *Nano letters* 12.10 (2012): 5325-5330.
261. Kitauchi, Yusuke, et al. "Structural transition in indium phosphide nanowires." *Nano letters* 10.5 (2010): 1699-1703.
262. Joyce, Hannah J., et al. "Twin-free uniform epitaxial GaAs nanowires grown by a two-temperature process." *Nano letters* 7.4 (2007): 921-926.
263. Lehmann, Sebastian, et al. "A general approach for sharp crystal phase switching in InAs, GaAs, InP, and GaP nanowires using only group V flow." *Nano letters* 13.9 (2013): 4099-4105.
264. Kayes, Brendan M., Harry A. Atwater, and Nathan S. Lewis. "Comparison of the device physics principles of planar and radial p-n junction nanorod solar cells." *Journal of applied physics* 97.11 (2005): 114302.
265. Fukui, Takashi, et al. "Position-controlled III–V compound semiconductor nanowire solar cells by selective-area metal–organic vapor phase epitaxy." *Ambio* 41.2 (2012): 119-124.

266. Mariani, Giacomo, et al. "Patterned radial GaAs nanopillar solar cells." *Nano letters* 11.6 (2011): 2490-2494.
267. Boulanger, J. P., et al. "Characterization of a Ga-assisted GaAs nanowire array solar cell on Si substrate." *IEEE Journal of Photovoltaics* 6.3 (2016): 661-667.
268. Mariani, Giacomo, et al. "Direct-bandgap epitaxial core–multishell nanopillar photovoltaics featuring subwavelength optical concentrators." *Nano letters* 13.4 (2013): 1632-1637.
269. Yao, Maoqing, et al. "GaAs nanowire array solar cells with axial p–i–n junctions." *Nano letters* 14.6 (2014): 3293-3303.
270. Cui, Yingchao, et al. "Efficiency enhancement of InP nanowire solar cells by surface cleaning." *Nano letters* 13.9 (2013): 4113-4117.
271. Mohseni, Parsian K., et al. "Monolithic III-V nanowire solar cells on graphene via direct van der waals epitaxy." *Advanced Materials* 26.22 (2014): 3755-3760.
272. Kauppinen, Christoffer, et al. "A technique for large-area position-controlled growth of GaAs nanowire arrays." *Nanotechnology* 27.13 (2016): 135601.
273. Johnston, Keith W., et al. "Schottky-quantum dot photovoltaics for efficient infrared power conversion." *Applied Physics Letters* 92.15 (2008): 151115.
274. Koleilat, Ghada I., et al. "Efficient, stable infrared photovoltaics based on solution-cast colloidal quantum dots." *ACS nano* 2.5 (2008): 833-840.
275. Pattantyus-Abraham, Andras G., et al. "Depleted-heterojunction colloidal quantum dot solar cells." *ACS nano* 4.6 (2010): 3374-3380.
276. Günes, Serap, et al. "Hybrid solar cells using PbS nanoparticles." *Solar Energy Materials and Solar Cells* 91.5 (2007): 420-423.
277. Yun, Daqin, et al. "Efficient conjugated polymer-ZnSe and-PbSe nanocrystals hybrid photovoltaic cells through full solar spectrum utilization." *Solar energy materials and solar cells* 93.8 (2009): 1208-1213.
278. Chang, Jeong Ah, et al. "High-performance nanostructured inorganic− organic heterojunction solar cells." *Nano letters* 10.7 (2010): 2609-2612.
279. Leschkies, Kurtis S., et al. "Photosensitization of ZnO nanowires with CdSe quantum dots for photovoltaic devices." *Nano letters* 7.6 (2007): 1793-1798.
280. Vogel, Ralf, Klaus Pohl, and Horst Weller. "Sensitization of highly porous, polycrystalline TiO2 electrodes by quantum sized CdS." *Chemical Physics Letters* 174.3-4 (1990): 241-246.
281. Vogel, R., P. Hoyer, and H. Weller. "Quantum-sized PbS, CdS, Ag2S, Sb2S3, and Bi2S3 particles as sensitizers for various nanoporous wide-bandgap semiconductors." *The Journal of Physical Chemistry* 98.12 (2002): 3183-3188.
282. Mulvaney, Paul, Franz Grieser, and Dan Meisel. "Electron transfer in aqueous colloidal tin dioxide solutions." *Langmuir* 6.3 (1990): 567-572.
283. Sahu, Anurag, Ashish Garg, and Ambesh Dixit. "A review on quantum dot sensitized solar cells: Past, present and future towards carrier multiplication with a possibility for higher efficiency." *Solar Energy* 203 (2020): 210-239.
284. Du, Jun, et al. "Zn–Cu–In–Se quantum dot solar cells with a certified power conversion efficiency of 11.6%." *Journal of the American Chemical Society* 138.12 (2016): 4201-4209.
285. Yu, Libo, Zhen Li, and Hai Song. "The influence of linker molecule on photovoltaic performance of CdS quantum dots sensitized translucent TiO 2 nanotube solar cells." *Journal of Materials Science: Materials in Electronics* 28.3 (2017): 2867-2876.

286. Zhang, Zhengguo, et al. "200-nm long TiO2 nanorod arrays for efficient solid-state PbS quantum dot-sensitized solar cells." *Journal of energy chemistry* 27.4 (2018): 1214-1218.
287. Zhou, Ru, et al. "Photoanodes with mesoporous TiO2 beads and nanoparticles for enhanced performance of CdS/CdSe quantum dot co-sensitized solar cells." *Electrochimica Acta* 135 (2014): 284-292.
288. Tian, Jianjun, et al. "Architectured ZnO photoelectrode for high efficiency quantum dot sensitized solar cells." *Energy & Environmental Science* 6.12 (2013): 3542-3547.
289. Tian, Jianjun, et al. "Hierarchically structured ZnO nanorods–nanosheets for improved quantum-dot-sensitized solar cells." *ACS applied materials & interfaces* 6.6 (2014): 4466-4472.
290. Hou, Juan, et al. "High performance of Mn-doped CdSe quantum dot sensitized solar cells based on the vertical ZnO nanorod arrays." *Journal of Power Sources* 325 (2016): 438-445.
291. Jean, Joel, et al. "ZnO nanowire arrays for enhanced photocurrent in PbS quantum dot solar cells." *Advanced materials* 25.20 (2013): 2790-2796.
292. Lan, Zhang, et al. "Preparation of nano-flower-like SnO 2 particles and their applications in efficient CdS quantum dots sensitized solar cells." *Journal of Materials Science: Materials in Electronics* 26.10 (2015): 7914-7920.
293. Li, Yafeng, et al. "CdS quantum-dot-sensitized Zn2SnO4 solar cell." *Electrochimica acta* 56.13 (2011): 4902-4906.
294. Yu, Jing, et al. "Application of ZnTiO3 in quantum-dot-sensitized solar cells and numerical simulations using first-principles theory." *Journal of Alloys and Compounds* 681 (2016): 88-95.
295. Chen, Cong, et al. "Strontium titanate nanoparticles as the photoanode for CdS quantum dot sensitized solar cells." *RSC Advances* 5.7 (2015): 4844-4852.
296. Ogermann, Daniel, Thorsten Wilke, and Karl Kleinermanns. "CdS x Se y/TiO 2 Solar Cell Prepared with Sintered Mixture Deposition." (2012).
297. Lee, Hyo Joong, et al. "Regenerative PbS and CdS quantum dot sensitized solar cells with a cobalt complex as hole mediator." *Langmuir* 25.13 (2009): 7602-7608.
298. Tachibana, Yasuhiro, et al. "Performance improvement of CdS quantum dots sensitized TiO2 solar cells by introducing a dense TiO2 blocking layer." *Journal of Physics D: Applied Physics* 41.10 (2008): 102002.
299. Lee, HyoJoong, et al. "Efficient CdSe quantum dot-sensitized solar cells prepared by an improved successive ionic layer adsorption and reaction process." *Nano letters* 9.12 (2009): 4221-4227.
300. Qian, Jin, et al. "P3HT as hole transport material and assistant light absorber in CdS quantum dots-sensitized solid-state solar cells." *Chemical Communications* 47.22 (2011): 6461-6463.
301. Chen, Hong-Yan, et al. "Dextran based highly conductive hydrogel polysulfide electrolyte for efficient quasi-solid-state quantum dot-sensitized solar cells." *Electrochimica Acta* 92 (2013): 117-123.
302. Pan, Zhenxiao, et al. "Near infrared absorption of CdSe x Te1–x alloyed quantum dot sensitized solar cells with more than 6% efficiency and high stability." *ACS nano* 7.6 (2013): 5215-5222.

303. Robel, Istvan, Masaru Kuno, and Prashant V. Kamat. "Size-dependent electron injection from excited CdSe quantum dots into TiO2 nanoparticles." *Journal of the American Chemical Society* 129.14 (2007): 4136-4137.
304. Mora-Seró, Iván, and Juan Bisquert. "Breakthroughs in the development of semiconductor-sensitized solar cells." *The journal of physical chemistry letters* 1.20 (2010): 3046-3052.
305. Rühle, Sven, et al. "Importance of recombination at the TCO/electrolyte interface for high efficiency quantum dot sensitized solar cells." *The Journal of Physical Chemistry C* 116.33 (2012): 17473-17478.
306. Cameron, Petra J., and Laurence M. Peter. "How does back-reaction at the conducting glass substrate influence the dynamic photovoltage response of nanocrystalline dye-sensitized solar cells?." *The Journal of Physical Chemistry B* 109.15 (2005): 7392-7398.
307. Ellingson, Randy J., et al. "Highly efficient multiple exciton generation in colloidal PbSe and PbS quantum dots." *Nano letters* 5.5 (2005): 865-871.
308. Shabaev, A. L., Al L. Efros, and A. J. Nozik. "Multiexciton generation by a single photon in nanocrystals." *Nano letters* 6.12 (2006): 2856-2863.
309. Santra, Pralay K., and Prashant V. Kamat. "Mn-doped quantum dot sensitized solar cells: a strategy to boost efficiency over 5%." *Journal of the American Chemical Society* 134.5 (2012): 2508-2511.
310. Yu, Xiao-Yun, et al. "Dynamic study of highly efficient CdS/CdSe quantum dot-sensitized solar cells fabricated by electrodeposition." *Acs Nano* 5.12 (2011): 9494-9500.
311. Wang, Jin, et al. "Core/shell colloidal quantum dot exciplex states for the development of highly efficient quantum-dot-sensitized solar cells." *Journal of the American Chemical Society* 135.42 (2013): 15913-15922.
312. Ren, Zhenwei, et al. "Amorphous TiO2 buffer layer boosts efficiency of quantum dot sensitized solar cells to over 9%." *Chemistry of Materials* 27.24 (2015): 8398-8405.
313. Jiao, Shuang, et al. "Band engineering in core/shell ZnTe/CdSe for photovoltage and efficiency enhancement in exciplex quantum dot sensitized solar cells." *ACS nano* 9.1 (2015): 908-915.
314. Holonyak Jr, Nick, and S. Fo Bevacqua. "Coherent (visible) light emission from Ga (As1− x P x) junctions." *Applied Physics Letters* 1.4 (1962): 82-83.
315. Einstein, A. "7 On the Quantum Theory of Radiation." *The Old Quantum Theory: The Commonwealth and International Library: Selected Readings in Physics* (2016): 167.
316. Bernard, Maurice GA, and Georges Duraffourg. "Laser conditions in semiconductors." *physica status solidi (b)* 1.7 (1961): 699-703.
317. Nathan, Marshall I., et al. "Stimulated emission of radiation from GaAs p-n junctions." *Applied Physics Letters* 1.3 (1962): 62-64.
318. Hall, Robert N., et al. "Coherent light emission from GaAs junctions." *Physical Review Letters* 9.9 (1962): 366.
319. Quist, Ted M., et al. "Semiconductor maser of GaAs." *Applied Physics Letters* 1.4 (1962): 91-92.
320. Hecht, Jeff. "A short history of laser development." *Applied optics* 49.25 (2010): F99-F122.
321. Nelson, H. E. R. B. E. R. T. "High-power pulsed GaAs laser diodes operating at room temperature." *Proceedings of the IEEE* 55.8 (1967): 1415-1419.

322. Alferov, Zh I., et al. "Investigation of the influence of the AlAs-GaAs heterostructure parameters on the laser threshold current and the realization of continuous emission at room temperature." *Sov. Phys. Semicond* 4.9 (1971): 1573-1575.
323. Ebeling, K. J., et al. "Single-mode operation of coupled-cavity GaInAsP/InP semiconductor lasers." *Applied Physics Letters* 42.1 (1983): 6-8.
324. Iga, Kenichi. "Surface-emitting laser-its birth and generation of new optoelectronics field." *IEEE Journal of Selected Topics in Quantum Electronics* 6.6 (2000): 1201-1215.
325. Huang, Michael H., et al. "Room-temperature ultraviolet nanowire nanolasers." *science* 292.5523 (2001): 1897-1899.
326. Law, Matt, et al. "Nanoribbon waveguides for subwavelength photonics integration." *Science* 305.5688 (2004): 1269-1273.
327. Nakayama, Yuri, et al. "Tunable nanowire nonlinear optical probe." *Nature* 447.7148 (2007): 1098-1101.
328. Duan, Xiangfeng, et al. "Single-nanowire electrically driven lasers." *Nature* 421.6920 (2003): 241-245.
329. Eaton, Samuel W., et al. "Semiconductor nanowire lasers." *Nature reviews materials* 1.6 (2016): 1-11.
330. Yan, Ruoxue, Daniel Gargas, and Peidong Yang. "Nanowire photonics." *Nature photonics* 3.10 (2009): 569-576.
331. Grzela, Grzegorz, et al. "Nanowire antenna emission." *Nano letters* 12.11 (2012): 5481-5486.
332. Qian, Fang, et al. "Gallium nitride-based nanowire radial heterostructures for nanophotonics." *Nano letters* 4.10 (2004): 1975-1979.
333. Qian, Fang, et al. "Core/multishell nanowire heterostructures as multicolor, high-efficiency light-emitting diodes." *Nano letters* 5.11 (2005): 2287-2291.
334. Tomioka, Katsuhiro, et al. "GaAs/AlGaAs core multishell nanowire-based light-emitting diodes on Si." *Nano letters* 10.5 (2010): 1639-1644.
335. Li, Shunfeng, and Andreas Waag. "GaN based nanorods for solid state lighting." *Journal of Applied Physics* 111.7 (2012): 5.
336. Waag, Andreas, et al. "The nanorod approach: GaN NanoLEDs for solid state lighting." *physica status solidi c* 8.7-8 (2011): 2296-2301.
337. Ra, Yong-Ho, et al. "Coaxial In x Ga1–x N/GaN multiple quantum well nanowire arrays on Si (111) substrate for high-performance light-emitting diodes." *Nano letters* 13.8 (2013): 3506-3516.
338. Ni, Pei-Nan, et al. "Bias-polarity dependent ultraviolet/visible switchable light-emitting devices." *ACS applied materials & interfaces* 6.11 (2014): 8257-8262.
339. Kikuchi, Akihiko, et al. "InGaN/GaN multiple quantum disk nanocolumn light-emitting diodes grown on (111) Si substrate." *Japanese Journal of Applied Physics* 43.12A (2004): L1524.
340. Nguyen, Hieu Pham Trung, et al. "p-Type modulation doped InGaN/GaN dot-in-a-wire white-light-emitting diodes monolithically grown on Si (111)." *Nano letters* 11.5 (2011): 1919-1924.
341. Hua, Bin, et al. "Characterization of Fabry-Pérot microcavity modes in GaAs nanowires fabricated by selective-area metal organic vapor phase epitaxy." *Applied Physics Letters* 91.13 (2007): 131112.

342. Yang, Lin, et al. "Fabry-Pérot microcavity modes observed in the micro-photoluminescence spectra of the single nanowire with InGaAs/GaAs heterostructure." *Optics express* 17.11 (2009): 9337-9346.
343. Zhu, Xiaolong, et al. "Bends and splitters in graphene nanoribbon waveguides." *Optics express* 21.3 (2013): 3486-3491.
344. Dasgupta, Neil P., et al. "25th anniversary article: semiconductor nanowires–synthesis, characterization, and applications." *Advanced materials* 26.14 (2014): 2137-2184.
345. Dingle, Raymond, and Charles Howard Henry. "Quantum effects in heterostructure lasers." U.S. Patent No. 3,982,207. 21 Sep. 1976.
346. Arakawa, Yasuhiko, and Hiroyuki Sakaki. "Multidimensional quantum well laser and temperature dependence of its threshold current." *Applied physics letters* 40.11 (1982): 939-941.
347. Liu, G., et al. "Extremely low room-temperature threshold current density diode lasers using InAs dots in In/sub 0.15/Ga/sub 0.85/As quantum well." *Electronics Letters* 35.14 (1999): 1163-1165.
348. Troccoli, Mariano, et al. "Mid-infrared (λ≈ 7.4 μm) quantum cascade laser amplifier for high power single-mode emission and improved beam quality." *Applied physics letters* 80.22 (2002): 4103-4105.
349. Asada, Mashiro, Yasuyuki Miyamoto, and Yasuharu Suematsu. "Gain and the threshold of three-dimensional quantum-box lasers." *IEEE Journal of quantum electronics* 22.9 (1986): 1915-1921.
350. Wasserman, D., and S. A. Lyon. "Midinfrared luminescence from InAs quantum dots in unipolar devices." *Applied physics letters* 81.15 (2002): 2848-2850.
351. Hirayama, H., et al. "Lasing action of Ga/sub 0.67/In/sub 0.33/As/GaInAsP/InP tensile-strained quantum-box laser." *Electronics Letters* 30.2 (1994): 142-143.
352. Huffaker, D. L., et al. "1.3 μm room-temperature GaAs-based quantum-dot laser." *Applied Physics Letters* 73.18 (1998): 2564-2566.
353. Liu, G., et al. "Extremely low room-temperature threshold current density diode lasers using InAs dots in In/sub 0.15/Ga/sub 0.85/As quantum well." *Electronics Letters* 35.14 (1999): 1163-1165.
354. Huang, Xiaodong, et al. "Very low threshold current density room temperature continuous-wave lasing from a single-layer InAs quantum-dot laser." *IEEE Photonics Technology Letters* 12.3 (2000): 227-229.
355. Liu, H. Y., et al. "Optimizing the growth of 1.3 μm InAs/InGaAs dots-in-a-well structure." *Journal of applied physics* 93.5 (2003): 2931-2936.
356. Walker, C. L., et al. "The role of high growth temperature GaAs spacer layers in 1.3-μm In (Ga) As quantum-dot lasers." *IEEE photonics technology letters* 17.10 (2005): 2011-2013.
357. Asryan, L. V., and R. A. Suris. "Temperature dependence of the threshold current density of a quantum dot laser." *IEEE journal of quantum electronics* 34.5 (1998): 841-850.
358. Badcock, T. J., et al. "Low threshold current density and negative characteristic temperature 1.3 μ m InAs self-assembled quantum dot lasers." *Applied physics letters* 90.11 (2007): 111102.
359. Sugawara, Mitsuru, and Michael Usami. "Handling the heat." *Nature Photonics* 3.1 (2009): 30-31.
360. Sze, Simon M., Yiming Li, and Kwok K. Ng. *Physics of semiconductor devices*. John wiley & sons, 2021.

361. Bandelow, U. "Optoelectronic Devices—Advanced Simulation and Analysis ed J Piprek." (2005): 63-85.
362. Grasser, Tibor, et al. "A review of hydrodynamic and energy-transport models for semiconductor device simulation." *Proceedings of the IEEE* 91.2 (2003): 251-274.
363. Grasser, Tibor, et al. "A Non-Parabolic Six Moments Model for the Simulation of Sub-100 nm Semiconductor Devices." *Journal of Computational Electronics* 3.3 (2004): 183-187.
364. Kumar, M. Jagadesh, Himanshu Batwani, and Mayank Gaur. "Approaches to nanoscale MOSFET compact modeling using surface potential based models." *2007 International Workshop on Physics of Semiconductor Devices*. IEEE, 2007.
365. Mugnaini, Giorgio, and Giuseppe Iannaccone. "Physics-based compact model of nanoscale MOSFETs-Part I: Transition from drift-diffusion to ballistic transport." *IEEE Transactions on Electron Devices* 52.8 (2005): 1795-1801.
366. Wang, Jing, et al. "A Fully Automated Method to Create Monte-Carlo MOSFET Model Libraries for Statistical Circuit Simulations." *Nanotech-Workshop on Compact Modeling, Sec. VIII-2*. 2012.
367. Asenov, Asen. "Advanced Monte Carlo techniques in the simulation of CMOS devices and circuits." *International Conference on Numerical Methods and Applications*. Springer, Berlin, Heidelberg, 2010.
368. Soroosh, M., and Y. Amiri. "An Ensemble Monte Carlo Model to Calculate Photocurrent of MSM Photodetector." *International Journal of Computer Science Issues (IJCSI)* 10.2 Part 2 (2013): 319.
369. Martin, Paul C., and Julian Schwinger. "Theory of many-particle systems. I." *Physical Review* 115.6 (1959): 1342.
370. Bonch-Bruevich, Viktor Leopolʹdovich, and Sergeĭ Vladimirovich Tyablikov. *The Green function method in statistical mechanics*. Courier Dover Publications, 2015.
371. Keldysh, Leonid V. "Diagram technique for nonequilibrium processes." *Sov. Phys. JETP* 20.4 (1965): 1018-1026.
372. Lake, Roger, et al. "Single and multiband modeling of quantum electron transport through layered semiconductor devices." *Journal of Applied Physics* 81.12 (1997): 7845-7869.
373. Venugopal, Rajesh, et al. "A simple quantum mechanical treatment of scattering in nanoscale transistors." *Journal of Applied Physics* 93.9 (2003): 5613-5625.
374. Ren, Zhibin, et al. "nanoMOS 2.5: A two-dimensional simulator for quantum transport in double-gate MOSFETs." *IEEE Transactions on Electron Devices* 50.9 (2003): 1914-1925.
375. Wang, Jing, Eric Polizzi, and Mark Lundstrom. "A three-dimensional quantum simulation of silicon nanowire transistors with the effective-mass approximation." *Journal of Applied Physics* 96.4 (2004): 2192-2203.
376. Kryjevski, Andrei, Deyan Mihaylov, and Dmitri Kilin. "Dynamics of charge transfer and multiple exciton generation in the doped silicon quantum dot–carbon nanotube system: Density functional theory-based computation." *The journal of physical chemistry letters* 9.19 (2018): 5759-5764.
377. Aeberhard, Urs. "Quantum-kinetic perspective on photovoltaic device operation in nanostructure-based solar cells." *Journal of Materials Research* 33.4 (2018): 373-386.
378. Soo, Mun Teng, Kuan Yew Cheong, and Ahmad Fauzi Mohd Noor. "Advances of SiC-based MOS capacitor hydrogen sensors for harsh environment applications." *Sensors and Actuators B: Chemical* 151.1 (2010): 39-55.

379. Zhou, Guangdong, et al. "Two-bit memory and quantized storage phenomenon in conventional MOS structures with double-stacked Pt-NCs in an HfAlO matrix." *Physical Chemistry Chemical Physics* 18.9 (2016): 6509-6514.
380. Yasue, Toshio, et al. "A 1.7-in, 33-Mpixel, 120-frames/s CMOS image sensor with depletion-mode MOS capacitor-based 14-b two-stage cyclic A/D converters." *IEEE Transactions on Electron Devices* 63.1 (2015): 153-161.
381. Liu, Ansheng, et al. "A high-speed silicon optical modulator based on a metal–oxide–semiconductor capacitor." *Nature* 427.6975 (2004): 615-618.
382. Chatbouri, S., et al. "The important contribution of photo-generated charges to the silicon nanocrystals photo-charging/discharging-response time at room temperature in MOS-photodetectors." *Superlattices and Microstructures* 94 (2016): 93-100.
383. Tsai, P. C., W. R. Chen, and Y. K. Su. "Enhanced ESD properties of GaN-based light-emitting diodes with various MOS capacitor designs." *Superlattices and Microstructures* 48.1 (2010): 23-30.
384. Ho, Wen-Jeng, et al. "High efficiency textured silicon solar cells based on an ITO/TiO2/Si MOS structure and biasing effects." *Computational Materials Science* 117 (2016): 596-601.
385. Bhatia, Deepak, et al. "Observation of temperature effect on electrical properties of novel Au/Bi0. 7Dy0. 3FeO3/ZnO/p-Si thin film MIS capacitor for MEMS applications." *Microelectronic Engineering* 168 (2017): 55-61.
386. Chand, Rohit, et al. "Detection of protein kinase using an aptamer on a microchip integrated electrolyte-insulator-semiconductor sensor." *Sensors and Actuators B: Chemical* 248 (2017): 973-979.
387. Kao, Chyuan-Haur, et al. "Influence of NH 3 plasma and Ti doping on pH-sensitive CeO 2 electrolyte-insulator-semiconductor biosensors." *Scientific reports* 7.1 (2017): 1-9.
388. Prakash, Amit, et al. "Resistive switching memory characteristics of Ge/GeO x nanowires and evidence of oxygen ion migration." *Nanoscale research letters* 8.1 (2013): 1-10.
389. Cho, Soo-Yeon, et al. "High-resolution p-type metal oxide semiconductor nanowire array as an ultrasensitive sensor for volatile organic compounds." *Nano letters* 16.7 (2016): 4508-4515.
390. Bae, Joonho, et al. "Si nanowire metal–insulator–semiconductor photodetectors as efficient light harvesters." *Nanotechnology* 21.9 (2010): 095502.
391. Oener, Sebastian Z., et al. "Metal–insulator–semiconductor nanowire network solar cells." *Nano letters* 16.6 (2016): 3689-3695.
392. Ng, Hou T., et al. "Single crystal nanowire vertical surround-gate field-effect transistor." *Nano Letters* 4.7 (2004): 1247-1252.
393. Larrieu, G., and X-L. Han. "Vertical nanowire array-based field effect transistors for ultimate scaling." *Nanoscale* 5.6 (2013): 2437-2441.
394. Hochbaum, Allon I., et al. "Controlled growth of Si nanowire arrays for device integration." *Nano letters* 5.3 (2005): 457-460.
395. Mohan, Premila, Junichi Motohisa, and Takashi Fukui. "Controlled growth of highly uniform, axial/radial direction-defined, individually addressable InP nanowire arrays." *Nanotechnology* 16.12 (2005): 2903.
396. Ngo-Duc, Tam-Triet, et al. "Controlled growth of vertical ZnO nanowires on copper substrate." *Applied Physics Letters* 102.8 (2013): 083105.

397. Tomioka, Katsuhiro, et al. "Control of InAs nanowire growth directions on Si." *Nano letters* 8.10 (2008): 3475-3480.
398. Zhao, Xin, et al. "Vertical nanowire InGaAs MOSFETs fabricated by a top-down approach." *2013 IEEE International Electron Devices Meeting*. IEEE, 2013.
399. Tomioka, Katsuhiro, Masatoshi Yoshimura, and Takashi Fukui. "A III–V nanowire channel on silicon for high-performance vertical transistors." *Nature* 488.7410 (2012): 189-192.
400. Hourdakis, E., et al. "Three-dimensional vertical Si nanowire MOS capacitor model structure for the study of electrical versus geometrical Si nanowire characteristics." *Solid-State Electronics* 143 (2018): 77-82.
401. Yeo, Yee-Chia, Tsu-Jae King, and Chenming Hu. "MOSFET gate leakage modeling and selection guide for alternative gate dielectrics based on leakage considerations." *IEEE Transactions on Electron Devices* 50.4 (2003): 1027-1035.
402. Locquet, Jean-Pierre, et al. "High-K dielectrics for the gate stack." *Journal of Applied Physics* 100.5 (2006): 051610.
403. Rahman, Anisur, et al. "Theory of ballistic nanotransistors." *IEEE Transactions on Electron devices* 50.9 (2003): 1853-1864.
404. Canham, Leigh T. "Silicon quantum wire array fabrication by electrochemical and chemical dissolution of wafers." *Applied physics letters* 57.10 (1990): 1046-1048.
405. Neophytou, Neophytos, et al. "Bandstructure effects in silicon nanowire electron transport." *IEEE Transactions on Electron Devices* 55.6 (2008): 1286-1297.
406. Lo, S-H., et al. "Quantum-mechanical modeling of electron tunneling current from the inversion layer of ultra-thin-oxide nMOSFET's." *IEEE Electron Device Letters* 18.5 (1997): 209-211.
407. Ochiai, Masaru, et al. "AlGaN/GaN heterostructure metal-insulator-semiconductor high-electron-mobility transistors with Si3N4 gate insulator." *Japanese Journal of Applied Physics* 42.4S (2003): 2278.
408. Ye, P. D., et al. "GaAs metal–oxide–semiconductor field-effect transistor with nanometer-thin dielectric grown by atomic layer deposition." *Applied Physics Letters* 83.1 (2003): 180-182.
409. Rastogi, A. C., and S. B. Desu. "Current conduction and dielectric behavior of high kY 2 O 3 films integrated with Si using chemical vapor deposition as a gate dielectric for metal-oxide-semiconductor devices." *Journal of electroceramics* 13.1 (2004): 121-127.
410. Kang, Laegu, et al. "Electrical characteristics of highly reliable ultrathin hafnium oxide gate dielectric." *IEEE Electron Device Letters* 21.4 (2000): 181-183.
411. Lee, Byoung Hun, et al. "Thermal stability and electrical characteristics of ultrathin hafnium oxide gate dielectric reoxidized with rapid thermal annealing." *Applied Physics Letters* 76.14 (2000): 1926-1928.
412. Wu, Y. H., et al. "Electrical characteristics of high quality La 2 O 3 gate dielectric with equivalent oxide thickness of 5/spl Aring." *IEEE Electron Device Letters* 21.7 (2000): 341-343.
413. Kakushima, Kuniyuki, et al. "Advantage of further scaling in gate dielectrics below 0.5 nm of equivalent oxide thickness with La2O3 gate dielectrics." *Microelectronics Reliability* 50.6 (2010): 790-793.
414. Yeo, Yee Chia, et al. "Direct tunneling gate leakage current in transistors with ultrathin silicon nitride gate dielectric." *IEEE Electron Device Letters* 21.11 (2000): 540-542.

415. Hinkle, C. L., et al. "A novel approach for determining the effective tunneling mass of electrons in HfO2 and other high-K alternative gate dielectrics for advanced CMOS devices." *Microelectronic engineering* 72.1-4 (2004): 257-262.
416. Monaghan, S., et al. "Determination of electron effective mass and electron affinity in HfO2 using MOS and MOSFET structures." *Solid-State Electronics* 53.4 (2009): 438-444.
417. Roy, Kaushik, Saibal Mukhopadhyay, and Hamid Mahmoodi-Meimand. "Leakage current mechanisms and leakage reduction techniques in deep-submicrometer CMOS circuits." *Proceedings of the IEEE* 91.2 (2003): 305-327.
418. Yeo, Yee-Chia, Tsu-Jae King, and Chenming Hu. "Direct tunneling leakage current and scalability of alternative gate dielectrics." *Applied Physics Letters* 81.11 (2002): 2091-2093.
419. Holms, Ann, and Alvin Quach. "Complementary Metal-Oxide Semiconductor Sensors." (2010).
420. Park, Hyunsung, et al. "Filter-free image sensor pixels comprising silicon nanowires with selective color absorption." *Nano letters* 14.4 (2014): 1804-1809.
421. Nakamura, Junichi, ed. *Image sensors and signal processing for digital still cameras*. CRC press, 2017.
422. Foster, David H. "Color constancy." *Vision research* 51.7 (2011): 674-700.
423. Al-Asadi, Ahmed S., et al. "Fabrication and characterization of ultraviolet photosensors from ZnO nanowires prepared using chemical bath deposition method." *Journal of Applied Physics* 119.8 (2016): 084306.
424. Fernandes, F. M., E. C. F. da Silva, and A. A. Quivy. "Mid-infrared photodetection in an AlGaAs/GaAs quantum-well infrared photodetector using photoinduced noise." *Journal of Applied Physics* 118.20 (2015): 204507.
425. Dai, Xing, et al. "GaAs/AlGaAs nanowire photodetector." *Nano letters* 14.5 (2014): 2688-2693.
426. Wang, Hao. "High gain single GaAs nanowire photodetector." *Applied Physics Letters* 103.9 (2013): 093101.
427. Mulazimoglu, Emre, et al. "Silicon nanowire network metal-semiconductor-metal photodetectors." *Applied Physics Letters* 103.8 (2013): 083114.
428. Wang, S. B., et al. "A CuO nanowire infrared photodetector." *Sensors and Actuators A: Physical* 171.2 (2011): 207-211.
429. Bandyopadhyay, Saumil, Pratik Agnihotri, and Supriyo Bandyopadhyay. "A self-assembled room temperature nanowire infrared photodetector based on quantum mechanical wavefunction engineering." *Physica E: Low-dimensional Systems and Nanostructures* 44.7-8 (2012): 1478-1485.
430. Kim, Cheol-Joo, et al. "Diameter-dependent internal gain in ohmic Ge nanowire photodetectors." *Nano letters* 10.6 (2010): 2043-2048.
431. Seo, Young-Kyo, Sanjeev Kumar, and Gil-Ho Kim. "Photoconductivity characteristics of ZnO nanoparticles assembled in nanogap electrodes for portable photodetector applications." *Physica E: Low-dimensional Systems and Nanostructures* 42.4 (2010): 1163-1166.
432. Hu, Linfeng, et al. "Ultrahigh external quantum efficiency from thin SnO2 nanowire ultraviolet photodetectors." *small* 7.8 (2011): 1012-1017.
433. Lim, Seon-Jeong, et al. "Organic-on-silicon complementary metal–oxide–semiconductor colour image sensors." *Scientific reports* 5.1 (2015): 1-7.

434. Park, Hyunsung, and Kenneth B. Crozier. "Multispectral imaging with vertical silicon nanowires." *Scientific reports* 3.1 (2013): 1-6.
435. Park, Hyunsung, Kwanyong Seo, and Kenneth B. Crozier. "Adding colors to polydimethylsiloxane by embedding vertical silicon nanowires." *Applied Physics Letters* 101.19 (2012): 193107.
436. Vanligten, Raoul F., and Harold Osterberg. "Holographic microscopy." *Nature* 211.5046 (1966): 282-283.
437. Tokunaga, A. T., D. A. Simons, and W. D. Vacca. "The Mauna Kea Observatories Near-Infrared Filter Set. II. Specifications for a New JHKL′ M′ Filter Set for Infrared Astronomy." *Publications of the Astronomical Society of the Pacific* 114.792 (2002): 180.
438. Liu, Dan, Da-Wen Sun, and Xin-An Zeng. "Recent advances in wavelength selection techniques for hyperspectral image processing in the food industry." *Food and Bioprocess Technology* 7.2 (2014): 307-323.
439. Gao, Bo-Cai. "NDWI—A normalized difference water index for remote sensing of vegetation liquid water from space." *Remote sensing of environment* 58.3 (1996): 257-266.
440. Tessler, Nir, et al. "Efficient near-infrared polymer nanocrystal light-emitting diodes." *Science* 295.5559 (2002): 1506-1508.
441. Weissleder, Ralph, et al. "In vivo imaging of tumors with protease-activated near-infrared fluorescent probes." *Nature biotechnology* 17.4 (1999): 375-378.
442. Aeberhard, Urs. "Theory and simulation of quantum photovoltaic devices based on the non-equilibrium Green’s function formalism." *Journal of computational electronics* 10.4 (2011): 394-413.
443. Aeberhard, Urs. "Theory and simulation of photogeneration and transport in Si-SiO x superlattice absorbers." *Nanoscale research letters* 6.1 (2011): 1-10.
444. Kurniawan, Oka, Ping Bai, and Er Ping Li. "Non-Equilibrium Green's Function Calculation of Optical Absorption in Nano Optoelectronic Devices." *2009 13th International Workshop on Computational Electronics*. IEEE, 2009.
445. Pourfath, Mahdi, Oskar Baumgartner, and Hans Kosina. "On the non-locality of the electron-photon self-energy: Application to carbon nanotube photo-detectors." *2008 International Conference on Numerical Simulation of Optoelectronic Devices (NUSOD)*. IEEE, 2008.
446. Guo, Jing, Muhammad A. Alam, and Youngki Yoon. "Theoretical investigation on photoconductivity of single intrinsic carbon nanotubes." *Applied physics letters* 88.13 (2006): 133111.
447. Armin, Ardalan, et al. "Narrowband light detection via internal quantum efficiency manipulation of organic photodiodes." *Nature communications* 6.1 (2015): 1-8.
448. Campos, Leonardo C., et al. "ZnO UV photodetector with controllable quality factor and photosensitivity." *AIP Advances* 3.2 (2013): 022104.
449. Hruska, Ryan, et al. "Radiometric and geometric analysis of hyperspectral imagery acquired from an unmanned aerial vehicle." *Remote Sensing* 4.9 (2012): 2736-2752.
450. Sanchez, Stephanie Michelle, and Miguel Velez-Reyes. "Skin detection in hyperspectral images." *Algorithms and Technologies for Multispectral, Hyperspectral, and Ultraspectral Imagery XXI*. Vol. 9472. International Society for Optics and Photonics, 2015.
451. Lambrechts, Andy, et al. "A CMOS-compatible, integrated approach to hyper-and multispectral imaging." *2014 IEEE International Electron Devices Meeting*. IEEE, 2014.

452. Geelen, Bert, Nicolaas Tack, and Andy Lambrechts. "A compact snapshot multispectral imager with a monolithically integrated per-pixel filter mosaic." *Advanced fabrication technologies for micro/nano optics and photonics VII*. Vol. 8974. International Society for Optics and Photonics, 2014.
453. Kwan, Chiman. "Methods and challenges using multispectral and Hyperspectral images for practical change detection applications." *Information* 10.11 (2019): 353.
454. Mokkapati, Sudha, et al. "Optical design of nanowire absorbers for wavelength selective photodetectors." *Scientific reports* 5.1 (2015): 1-7.
455. Lin, Qianqian, et al. "Filterless narrowband visible photodetectors." *Nature Photonics* 9.10 (2015): 687-694.
456. Seo, Kwanyong, et al. "Filter-Free Image Sensor Pixels Comprising Silicon Nanowires with Selective Color Absorption." (2014).
457. Qin, Liqiao, et al. "Wavelength selective p-GaN/ZnO colloidal nanoparticle heterojunction photodiode." *Applied Physics Letters* 102.7 (2013): 071106.
458. Hou, Y. N., et al. "Monolithic color-selective ultraviolet (266–315 nm) photodetector based on a wurtzite MgxZn1− xO film." *Applied Physics Letters* 105.13 (2014): 133510.
459. Nayak, Pabitra K., et al. "Photovoltaic solar cell technologies: analysing the state of the art." *Nature Reviews Materials* 4.4 (2019): 269-285.
460. Bergmann, R. B. "Crystalline Si thin-film solar cells: a review." *Applied physics A* 69.2 (1999): 187-194.
461. Beaucarne, G., et al. "Epitaxial thin-film Si solar cells." *Thin Solid Films* 511 (2006): 533-542.
462. Liu, Sha, et al. "High-efficiency organic solar cells with low non-radiative recombination loss and low energetic disorder." *Nature Photonics* 14.5 (2020): 300-305.
463. Fan, Baobing, et al. "Surpassing the 10% efficiency milestone for 1-cm 2 all-polymer solar cells." *Nature communications* 10.1 (2019): 1-8.
464. Cui, Yingchao. "High-efficiency nanowire solar cells." *Eindhoven University of Technology* (2015).
465. Garnett, Erik C., et al. "Nanowire solar cells." *Annual review of materials research* 41 (2011): 269-295.
466. Zhang, Yunyan, and Huiyun Liu. "Nanowires for high-efficiency, low-cost solar photovoltaics." *Crystals* 9.2 (2019): 87.
467. Tong, H., and M. W. Wu. "Theory of excitons in cubic III-V semiconductor GaAs, InAs and GaN quantum dots: Fine structure and spin relaxation." *Physical Review B* 83.23 (2011): 235323.
468. Lindell, Linda, et al. "Transparent, plastic, low-work-function poly (3, 4-ethylenedioxythiophene) electrodes." *Chemistry of materials* 18.18 (2006): 4246-4252.
469. Chowdhury, Basudev Nag, and Sanatan Chattopadhyay. "Unusual impact of electron-phonon scattering in Si nanowire field-effect-transistors: A possible route for energy harvesting." *Superlattices and Microstructures* 97 (2016): 548-555.
470. Zhou, Weidong, and James J. Coleman. "Semiconductor quantum dots." *Current Opinion in Solid State and Materials Science* 20.6 (2016): 352-360.
471. Gibson, Sandra J., et al. "Tapered InP nanowire arrays for efficient broadband high-speed single-photon detection." *nature nanotechnology* 14.5 (2019): 473-479.
472. Oehler, Fabrice, et al. "Measuring and modeling the growth dynamics of self-catalyzed GaP nanowire arrays." *Nano letters* 18.2 (2018): 701-708.

473. Kosloff, Alon, et al. "Nanodicing Single Crystalline Silicon Nanowire Arrays." *Nano letters* 16.11 (2016): 6960-6966.
474. Anttu, Nicklas, et al. "Absorption of light in InP nanowire arrays." *Nano Research* 7.6 (2014): 816-823.
475. Seo, Kwanyong, et al. "Multicolored vertical silicon nanowires." *Nano letters* 11.4 (2011): 1851-1856.
476. Alsmeier, J., E. Batke, and J. P. Kotthaus. "Voltage-tunable quantum dots on silicon." *Physical Review B* 41.3 (1990): 1699.
477. Bennett, Anthony J., et al. "Voltage tunability of single-spin states in a quantum dot." *Nature communications* 4.1 (2013): 1-5.
478. Wang, Rui, et al. "Gate tunable hole charge qubit formed in a Ge/Si nanowire double quantum dot coupled to microwave photons." *Nano letters* 19.2 (2019): 1052-1060.
479. Veldhorst, M., et al. "An addressable quantum dot qubit with fault-tolerant control-fidelity." *Nature nanotechnology* 9.12 (2014): 981-985.
480. Penthorn, Nicholas E., et al. "Two-axis quantum control of a fast valley qubit in silicon." *npj Quantum Information* 5.1 (2019): 1-6.
481. Nag Chowdhury, Basudev, and Sanatan Chattopadhyay. "Investigating the impact of source/drain doping dependent effective masses on the transport characteristics of ballistic Si-nanowire field-effect-transistors." *Journal of Applied Physics* 115.12 (2014): 124502.
482. Kasper, J. S., and S. M. Richards. "The crystal structures of new forms of silicon and germanium." *Acta Crystallographica* 17.6 (1964): 752-755.
483. Kwa, K. S. K., et al. "A model for capacitance reconstruction from measured lossy MOS capacitance–voltage characteristics." *Semiconductor science and technology* 18.2 (2002): 82.
484. Chattopadhyay, S., et al. "C–V characterization of strained Si/SiGe multiple heterojunction capacitors as a tool for heterojunction MOSFET channel design." *Semiconductor science and technology* 18.8 (2003): 738.
485. Dalapati, Goutam Kumar, et al. "Impact of strained-Si thickness and Ge out-diffusion on gate oxide quality for strained-Si surface channel n-MOSFETs." *IEEE transactions on electron devices* 53.5 (2006): 1142-1152.